\documentclass{aa}  
\usepackage{graphicx}
\usepackage[varg]{txfonts}
\usepackage{natbib}
\bibpunct{(}{)}{;}{a}{}{,} 
\usepackage{multirow}
\usepackage{url}
\usepackage[implicit=false,hidelinks]{hyperref}
\usepackage[colorinlistoftodos]{todonotes}
\usepackage{subcaption,rotating}
\usepackage{lscape}
\usepackage{verbatim}
\usepackage{adjustbox}
\usepackage{siunitx}
\defcitealias{2017MNRAS.472..483F}{F17}
\defcitealias{2018MNRAS.476.3991M}{M18}
\defcitealias{2014MNRAS.437.1268H}{H14}
\defcitealias{2016ApJ...833...72B}{B16}
\defcitealias{1999ApJ...521...64M}{M99}

\begin{document}

   \title{The ALMA Frontier Fields Survey}

   \subtitle{V: ALMA Stacking of Lyman-Break Galaxies in Abell 2744, Abell 370, Abell S1063, MACSJ0416.1-2403 and MACSJ1149.5+2223}

   \author{R. Carvajal\inst{1,2}
          \and
          F. E. Bauer\inst{1,2,3,4}
          \and
          R. J. Bouwens\inst{5}
          \and
          P. A. Oesch\inst{6}
          \and
          J. Gonz\'{a}lez-L\'{o}pez\inst{7,1}
          \and
          T. Anguita\inst{8,3}
          \and
          M. Aravena\inst{7}
          \and
          R. Demarco\inst{9}
          \and
          L. Guaita\inst{1,7}
          \and
          L. Infante\inst{1,2,10}
          \and
          S. Kim\inst{1}
          \and
          R. Kneissl\inst{12,13}
          \and
          A. M. Koekemoer\inst{11}
          \and
          H. Messias\inst{12,13}
          \and
          E. Treister\inst{1}
          \and
          E. Villard\inst{12,13}
          \and
          A. Zitrin\inst{14}
          \and
          P. Troncoso\inst{15}
          }

   \institute{Instituto de Astrof\'{i}sica, Pontificia Universidad Cat\'{o}lica de Chile,
              Casilla 306, Santiago 22, Chile\\
              \email{rcarvaja@astro.puc.cl}
        \and
            Centro de Astroingenier{\'{\i}}a, Pontificia Universidad Cat\'{o}lica de Chile, Casilla 306, Santiago 22, Chile
        \and
            Millennium Institute of Astrophysics (MAS), Av. Vicu\~{n}a Mackenna 4860, Macul, Santiago, Chile
        \and
            Space Science Institute, 4750 Walnut Street, Suite 205, Boulder, Colorado 80301
        \and
            Leiden Observatory, Leiden University, NL-2300 RA Leiden, Netherlands
		\and
            Geneva Observatory, University of Geneva, Ch. des Maillettes 51, 1290 Versoix, Switzerland
        \and
            N\'{u}cleo de Astronom\'{i}a de la Facultad de Ingenier\'{i}a y Ciencias, Universidad Diego Portales, Av. Ej\'{e}rcito Libertador 441, Santiago, Chile
        \and
            Departamento de Ciencias F\'{i}sicas, Universidad Andr\'{e}s Bello, Fern\'{a}ndez Concha 700, Las Condes, Santiago, Chile
        \and
            Departamento de Astronom\'ia, Facultad de Ciencias F\'isicas y Matem\'aticas, Universidad de Concepci\'{o}n, Concepci\'on, Chile
        \and
            Carnegie Institution for Science, Las Campanas Observatory, Casilla 601, Colina El Pino S/N, La Serena, Chile
        \and
            Space Telescope Science Institute, Baltimore, MD 21218, USA
        \and
            Joint ALMA Observatory, Alonso de C\'{o}rdova 3107, Vitacura 763-0355, Santiago, Chile
        \and
            European Southern Observatory, Alonso de C\'{o}rdova 3107, Vitacura, Casilla 19001, Santiago, Chile
        \and
            Physics Department, Ben-Gurion University of the Negev, P.O. Box 653, Be’er-Sheva 8410501, Israel
        \and
            Universidad Autónoma de Chile, Chile. Av. Pedro de Valdivia 425, Santiago, Chile
        }

   \date{Draft version. December 2019}

 
  \abstract
   {The Hubble Frontier Fields offer an exceptionally deep window into the high-redshift universe, covering a substantially larger area than the Hubble Ultra-Deep field at low magnification and probing 1--2 mags deeper in exceptional high-magnification regions. This unique parameter space, coupled with the exceptional multi-wavelength ancillary data, can facilitate for useful insights into distant galaxy populations.}
   {We aim to leverage Atacama Large Millimetre Array (ALMA) band 6 ($\approx$263\,GHz) mosaics in the central portions of five Frontier Fields to characterize the infrared (IR) properties of $1582$ ultraviolet (UV)-selected Lyman-Break Galaxies (LBGs) at redshifts of $z {\sim}$2--8. We investigated individual and stacked fluxes and IR excess (IRX) values of the LBG sample as functions of stellar mass ($\mathrm{M}_{\bigstar}$), redshift, UV luminosity and slope $\beta$, and lensing magnification.}
   {LBG samples were derived from color-selection and photometric redshift estimation with {\it Hubble} Space Telescope photometry. Spectral energy distributions (SED)-templates were fit to obtain luminosities, stellar masses, and star formation rates for the LBG candidates. We obtained individual IR flux and IRX estimates, as well as stacked averages, using both ALMA images and $u$--$v$ visibilities.}
   {Two (2) LBG candidates were individually detected above a significance of $4.1{-}\sigma$, while stacked samples of the remaining LBG candidates yielded no significant detections. We investigated our detections and upper limits in the context of the IRX-$\mathrm{M}_{\bigstar}$ and IRX-$\beta$ relations, probing at least one dex lower in stellar mass than past studies have done. Our upper limits exclude substantial portions of parameter space and they are sufficiently deep in a handful of cases to create mild tension with the typically assumed attenuation and consensus relations. We  observe a clear and smooth trend between $\mathrm{M}_{\bigstar}$ and $\beta$, which extends to low masses and blue (low) $\beta$ values, consistent with expectations from previous works.}
   {}

   \keywords{galaxies: high-redshift -- galaxies  -- 
             galaxies: clusters: general  -- submillimetre: galaxies -- gravitational lensing: strong
             }

   \maketitle
%

\section{Introduction}

	The detailed determination of the conditions that led to the formation of the first galaxies in the early Universe and their subsequent evolution remains a key issue in modern astronomy \citep[e.g.,][]{2016ARA&A..54..761S}. A truly broadband multi-wavelength perspective is likely required to robustly account for a galaxy's growth and energy production. However, obtaining such multi-wavelength properties can be challenging due to their faint fluxes and the large distances involved. 

    A good example of this is the assessment of star formation rates (SFRs) in galaxies, where we must account for extinction by gas and dust in order to extract the intrinsic amount of the ultraviolet (UV) light emitted by the underlying stellar population. Deep near-infrared (NIR), optical, and UV surveys now routinely allow us to estimate unobscured SFRs down to a few $M_{\odot}$\,yr$^{-1}$ in galaxies out to $z{\sim}6$--10 \citep[e.g.,][]{2015ApJ...803...34B, 2016MNRAS.459.3812M, 2017ApJ...847...76S, 2018ApJ...855..105O}. A straightforward way to measure the extinction from these sources is to estimate the steepness of their UV spectra \citep[e.g.,][]{2012ApJ...754...83B, 2014ApJ...793..115B}, generally characterized by fitting a power law ($f_{\lambda} {\sim} \lambda^{\beta}$) to two or more rest-frame UV bands. A synthetic stellar population with solar metallicity and an age of ${\gtrsim} 100$ Myr is expected to have intrinsic $\beta$ values in the range of ${\sim} {-} 2.0$ to ${-} 2.2$. Redward (higher $\beta$) deviations from this are thought to relate to the amount of dust extinction (reddening) and scattering that light from massive stars suffers after its emission. Blueward (lower $\beta$) deviations likely imply a very young or metal-deficient stellar population \citep[e.g.,][]{2012IAUS..284...49H, 2016ARA&A..54..761S}.
    
    Detailed spectroscopic observations are generally required to break degeneracies between extinction, stellar age, and metallicity \citep[e.g.,][]{2013ApJ...763..129S}, all of which ultimately contribute to the observed UV stellar slope $\beta$. However, for fainter or more distant galaxies, this remains quite challenging \citep[e.g.,][]{2017ApJ...837L..21L, 2017MNRAS.469..448B, 2018ApJ...854...39H, 2018Natur.557..392H}. Such degeneracies become particularly problematic at high redshifts, where the likelihood of young, metal-poor stellar populations and, hence, the uncertainties, are largest \citep[e.g.,][]{2003A&A...401.1063A, 2009A&A...502..423S, 2017PASA...34...58E}. 
    
	A second approach for assessing extinction/absorption, as well as to examine the potential for highly or entirely obscured regions of star formation, is to measure the IR luminosity. Until recently, such observations were strongly limited in sensitivity and resolution (spatial and spectral), effectively only probing down to SFRs of $\sim$10--100 $M_{\odot}$\,yr$^{-1}$ at $z {\sim}1$--2 \citep[e.g.,][]{2013A&A...553A.132M}. The advent of the Atacama Large Millimetre Array  (ALMA), with its large collecting area and high spatial resolution capabilities, now provides the opportunity to narrow considerably the SFR gap between the UV and optical, and FIR and mm bands for galaxies across a large redshift range and, hence, make a fairer comparison between the obscured and visible light being generated.
 
    Numerous observational studies of $z {\gtrsim}1 $ star-forming galaxies have been made over the years, comparing the two approaches above to well-known correlations for local galaxies (e.g., 
    \citealp{1999ApJ...521...64M}, hereafter M99; 
    \citealp{2006ApJ...644..792R}; 
    \citealp{2009ApJ...705..936B, 2016ApJ...833...72B}, hereafter B16; 
    \citealp{2012A&A...539A.145B}; 
    \citealp{2015Natur.522..455C}; 
    \citealp{2016A&A...587A.122A}; 
    \citealp{2018MNRAS.476.3991M}; 
    \citealp{2018MNRAS.479.4355K}). 
    Many observers have focused on the relationships between the so-called ``infrared excess'' (IRX${\equiv} L_{\rm IR}$/$L_{\rm UV}$) and UV-continuum slope ($\beta$) or stellar mass ($M_{\bigstar}$); such relations are often invoked to make dust attenuation corrections out to high redshifts. Most critically, while such correlations appear to be confirmed out to $z{\sim}1$-2, based on a variety of multi-wavelength data \citep[e.g.,][]{2006ApJ...644..792R, 2008ApJS..175...48R, 2010ApJ...712.1070R, 2007ApJ...670..156D, 2007ApJ...670..173D, 2009ApJ...698L.116P}, it remains unclear how applicable they are at earlier times \citepalias[e.g.,][]{2016ApJ...833...72B}.
    
	The goal of our work here is to characterize the IR emission (individually and, given the low number of expected detections, as stacked-averages) for robust samples of Lyman-Break Galaxy (LBG) candidates at $z{=}2$--8 found in the Frontier Fields (FFs) survey.\footnote{\url{http://www.stsci.edu/hst/campaigns/frontier-fields/}} The FFs were initiated as {\it Hubble} ({\it HST}) and {\it Spitzer} Space Telescope Director's discretionary campaigns to peer as deeply as possible into the distant universe, leveraging the power of gravitational lensing from six massive high-magnification clusters of galaxies to probe to extremely faint emission levels in the most highly magnified regions \citep{2015ApJ...800...84C, 2017ApJ...837...97L}. 
     
    These fields have since been observed across the electromagnetic spectrum with, for example, {\it Chandra}, VLT/MUSE, JVLA and, of course, ALMA. We aim here to assess the IR and UV emission, stellar masses, and star formation properties of these LBG candidates, and to investigate how they compare to $z{\sim}0$ objects and correlations.

    This paper is organized as follows. In $\S$\ref{sec:Data}, we describe the ALMA FFs observations, the LBG candidates, and their derived properties. In $\S$\ref{sec:Methods}, we explain the selection criteria we applied to our candidates and the stacking procedures we utilized (ALMA image stacking and IRX stacking). In $\S$\ref{sec:Results}, we present the individual properties that we obtain for our sample, as well as the stacked values for luminosities and IRXs. $\S$\ref{sec:Discussion} provides a comparison of our results with previously published works, as well as results not covered fully in preceding sections. Finally, we summarize our work and present our conclusions in $\S$\ref{sec:Summary}. Throughout this work, we assume a cosmology with $H_{0} {=} 70$ km s$^{-1}$ Mpc$^{-1}$, $\Omega_{m} {=} 0.3$, and $\Omega_{\Lambda} {=} 0.7$.


\section{Data and derived quantities}\label{sec:Data}

    \subsection{ALMA data}\label{subsec:ALMA_data}

    The inner ${\sim}2\arcmin {\times} 2\arcmin$ regions of the FFs, centered on the massive clusters to benefit most strongly from the boost from gravitational lensing, were observed in band 6 by ALMA through two projects, 2013.1.00999.S (PI Bauer; cycle 2) and 2015.1.01425.S (PI Bauer; cycle 3). Only five FFs clusters were completely observed by ALMA and, thus, used here. These include, from cycle 2, \object{Abell 2744}, \object{MACSJ0416.1$-$2403}, and \object{MACSJ1149.5$+$2223} observed in 2014 and 2015 (hereafter A2744, MACSJ0416 and MACSJ1149, respectively) and, from cycle 3, \object{Abell 370} and \object{Abell S1063} ---also designed as RXJ2248$-$4431--- observed in 2016 (hereafter A370 and AS1063, respectively). As stated in \citet{2017A&A...608A.138G}, {MACSJ0717.5$+$3745} was only partially observed (just 1 out of 9 planned executions) and, given its substantially worse sensitivity and calibration, is not useful for this work.
    
    The mosaic data were reduced and calibrated using the Common Astronomy Software Applications (\texttt{CASA} v4.2.2; \citealt{2007ASPC..376..127M});\footnote{\url{https://casa.nrao.edu}} details can be found in \citet[][]{2017A&A...597A..41G}. Automatic reduction with the \texttt{CASA}-generated pipelines for A2744 and MACSJ1149 presented problems and, hence, manual and ad-hoc pipelines were used to reduce the data. For MACSJ0416, A370, and AS1063, the \texttt{CASA}-generated pipelines worked smoothly and were used. Observations from ALMA are characterized as visibilities ($u$--$v$ plane), which must be Fourier-transformed to obtain image files (image-plane). Each visibility corresponds to an antenna pair or baseline. The visibilities (or baselines) can be weighted to produce different synthetic beamsizes and shapes. To assess the results, we applied two nominal weighting schemes, natural and taper, to the imaged (or \texttt{CLEAN}ed) datasets using \texttt{CASA}.\footnote{
    Natural weighting assigns equal weights to every visibility in the deconvolution process. It corresponds to $1/\sigma^{2}$, where $\sigma$ is the noise variance of the data (visibility) and maximizes sensitivity for point sources. Alternatively, $u$--$v$ tapering creates an adjustable gaussian-like window function ($W(u, v) {=} \exp(-(u^{2} + v^{2})/t^{2})$) with $t$ being the taper parameter. As it gives more weight to shorter baselines, it can offer additional sensitivity to extended sources (the flux of which is missed in long baselines).} For this work, we adopted a taper parameter of $t{=}1$\farcs5. Employing both weighting schemes offers more flexibility (and sensitivity) when searching for point-like and extended detections.

    Our reductions achieved natural-weight $rms$\footnote{$rms$ defined as $\sqrt{\sum_{i}\left(x_{i}^{2}\right)}$ with $x_{i}$ being the elements of the set or, in this case, observed fluxes over the maps.} errors of $55$, $61$, $67$, $59$ and $71\, \mu \mathrm{Jy\,beam^{-1}}$ for FFs A2744, A370, AS1063, MACSJ0416 and MACSJ1149, respectively. The resulting maps have relatively uniform $rms$ properties over the central regions due to Nyquist sampling, but exhibit strong attenuation at the edges from the primary beam (PB) pattern. For the purposes of this work, we limited our analysis to regions of each mosaic with a PB-correction factor $pbcor {>} 0.5$, designated hereafter as the field of view (FoV) of each observation; regions with $pbcor {<} 0.5$ have substantially elevated $rms$ values that are not very constraining. Notably, portions of the MACSJ0416 and MACSJ1149 mosaics exhibit $rms$ variations by as much as $\sim$15--20\% \citep[for details, see $\S$2.4 and Fig. 4 of][]{2017A&A...597A..41G}. These variations were captured in the $pbcor$ values used to weight individual sources in our stacking procedure (see $\S$\ref{subsec:Stacking}). 

    Some basic properties of each dataset, including central position, are listed on Table~\ref{tab:ALMA_maps_props}. For reference, the ALMA maps of the FFs are all sufficiently deep to detect exceptional LBGs like Abell 1689-zD1, which has a band 6 flux of $0.56{\pm}0.1$\,mJy \citep{2017MNRAS.466..138K}, with S/N$\sim$8--10.
    
\begin{table*}[t]
\begin{center}
\caption{ALMA Properties of observed clusters}
\label{tab:ALMA_maps_props}
\resizebox{.90\linewidth}{!}{\begin{tabular}{@{}l@{\hspace{1.25\tabcolsep}}c@{\hspace{1.25\tabcolsep}}c@{\hspace{1.25\tabcolsep}}c@{\hspace{1.25\tabcolsep}}c@{\hspace{1.9\tabcolsep}}c@{\hspace{1.25\tabcolsep}}c@{\hspace{1.25\tabcolsep}}c@{}}
\hline
Cluster Name & R.A. [J2000]\tablefootmark{a} & Dec. [J2000] 		 & $z$ & Observation Date Range& $rms$ & $b_{\mathrm{max}} \times b_{\mathrm{min}}$\tablefootmark{b}  &   \# Pointings\tablefootmark{c}\\
			 & [hh:mm:ss.s] & [$\pm$ dd:mm:ss.s] & 	 &		& [$\mu$Jy]		& [$\arcsec \times \arcsec$]	&   \\
\hline
Abell 2744 & 00:14:21.2 & -30:23:50.1 & 0.308 & 29-Jun-2014/31-Dec-2014 & 55 & 0.63 $\times$ 0.49   &   126\\
Abell 370  & 02:39:52.9 & -01:34:36.5 & 0.375 & 05-Jan-2016/17-Jan-2016 & 61 & 1.25 $\times$ 0.99   &   126\\
Abell S1063 & 22:48:44.4 & -44:31:48.8 & 0.348 & 16-Jan-2016/02-Apr-2016 & 67 & 0.96 $\times$ 0.79  &   126\\
MACSJ0416.1-2403 & 04:16:08.9 & -24:04:28.7 & 0.396 & 04-Jan-2015/02-May-2015 & 59 & 1.52 $\times$ 0.85 &   126\\
MACSJ1149.5+2223 & 11:49:36.3 & +22:23:58.1 & 0.543 & 14-Jan-2015/22-Apr-2015 & 71 & 1.22 $\times$ 1.08 &   126\\
\hline
\end{tabular}}
\tablefoot{
\tablefoottext{a}{Position of mosaic center.}
\tablefoottext{b}{Major and minor axes of synthesized beam, in arcseconds.}
\tablefoottext{c}{Number of pointings that compose the final ALMA maps.}
}
\end{center}
\end{table*}

    \subsection{LBG candidates}\label{subsec:LBG_candidates}
        
    Deep \textit{HST} images are available in seven broadband filters as part of the FFs campaign \citep{2017ApJ...837...97L}: Advanced Camera for Surveys (ACS) filters $F435W$, $F606W$, $F814W$ (with 0\farcs4 aperture 5-$\sigma$ depths of 28.8, 28.8 and 29.1 ABmag, respectively); Wide Field Camera 3 (WFC3) IR filters $F105W$, $F125W$, $F140W$, $F160W$ (with 0\farcs4 aperture 5-$\sigma$ depths of 28.9, 28.6, 28.6 and 28.7 ABmag, respectively). Two additional deep images were obtained with WFC3 UVIS filters F275W and F336W (with 0\farcs4 aperture 5-$\sigma$ depths of $\approx$27.5--28.0 ABmag, depending on the cluster) as part of a supporting UV campaign \citep[PI: Siana;][]{2016ApJ...832...56A}. 
    
    Bouwens et al. 2019 (in prep; hereafter B19) use these images to identify large samples of $z {\sim} 2, 3, 4, 5, 6, 7, 8$, and $9$ star-forming galaxies through the LBG selection technique in the FFs. Light from the foreground cluster galaxies and the intracluster medium was removed using \textsc{galfit} \citep{2002AJ....124..266P} and fitting the background light via median filtering routines, respectively, as described in B19. Source catalogs were then produced using SExtractor \citep{1996A&AS..117..393B} by detecting sources in the coadded images of the four WFC3/IR filters.  Colors were measured in small scalable apertures using a \citet{1980ApJS...43..305K} factor of 1.2. The small scalable aperture magnitudes were then corrected to total ones based on (1) the relative extra flux seen in larger versus small scalable apertures (Kron factor of 2.5 vs. Kron factor of 1.2) and (2) the point-source encircled energy estimated to lie outside the larger scalable apertures. The correction to the total magnitude was performed based on the detection image constructed by coadding all four WFC3/IR bands. See \citet{2015ApJ...803...34B} for more details on the applied photometric procedure. Finally, B19 applied several color and signal-to-noise ratio (S/N) criteria to select LBG candidates in crude redshift bins as well as remove obvious point-like ("stellar") contamination.
    
    For our purposes, we did not use the B19 $z {\sim} 4$ LBG sample, due to the lack of photometric coverage around $\sim$5500\AA\ (e.g., F555W) coupled with the potential for strong contamination by foreground galaxies in four of the five clusters considered. 
    
    B19 produced a final list of $3050$ LBG candidates based on the {\it HST} cluster and parallel observations of the six FFs across all their drop-out bands, with $3029$ candidates selected in the bands we use for our study ($z {\sim} 2$, 3, 5, 6, 7, and 8). From this parent sample, we investigate the properties of the $1582$ candidates located within the FoVs of five ALMA-observed FFs. Thus, all of our final results are drawn from this subset. We expect the spatial distribution of our LBG candidates to be roughly uniform over the source plane of the selected FFs. This will translate to fewer sources in highly magnified regions (near critical lines on the magnification maps) in the image plane, as we are sampling smaller intrinsic space densities. However, in a critical sense, the magnification means we probe further down the luminosity function in these regions. Thus, we expect the targets to span an interesting range in properties (e.g., magnification, SFR, $M_{\star}$, redshift, etc.). This helps to build a statistically diverse set of LBG properties to study. Distributions of their attributes can be seen in $\S$\ref{subsec:zphot} and later sections.
    
    \subsection{ALMA stacking considerations}\label{subsec:PrepTargets}
    
    We used \texttt{STACKER} \citep{2015MNRAS.446.3502L} to perform the stacking of our candidates in the ALMA images (see $\S$\ref{subsec:Stacking}). This program takes, as input, the lists of target positions (R.A., Dec.) and weights (for the actual stacking process). Weights are drawn from the \texttt{CASA} \texttt{clean} PB-correction map, which corresponds to the sky sensitivity over the field. This initial definition of the weight can be modified by further criteria (see $\S$\ref{subsec:Stacking}). For this work, two schemes were used to weight the stacked signal according to the observed properties of the LBG candidates.
    
    One important issue to consider is that we used information from \textit{HST} and ALMA. It is possible that potential mm and submm emission in the ALMA maps may arise from a somewhat different position than the optical one, given the large span in observed wavelengths and distinct emission and extinction mechanisms at work \citep[e.g.,][]{2002ApJ...568..651G}. In particular, the more dust-rich regions that could give rise to submm continuum emission would tend to attenuate embedded stars, while nearby stars in less dust-rich regions might contribute more to the observed near-IR light.
    
    We argue, however, that such offsets were unlikely to affect our final results (i.e., the stacked flux). For one, the angular sizes of the LBG candidates are generally similar or smaller than the beam sizes of our ALMA observations ($\sim$1$\arcsec$). Secondly, spatial offsets between securely detected bright mm/submm sources (e.g., submillimeter galaxies, or SMGs) and optical/NIR counterparts are generally small \citep[e.g., 0\farcs17$\pm$0\farcs02][]{2017A&A...597A..41G}. The offsets in SMGs, where extinction is so high that the optical emission is not detected, probably represent one extreme, while offsets in less extreme UV-selected LBGs should relatively minimal. 
    
    For the reasons above and to simplify calculations, no correction was performed regarding relative positional offsets. That being said, to obtain actually stacked fluxes from the ALMA maps in $\S$\ref{subsec:ALMA_up_lims}, we will consider values at larger distances due to the influence of the synthesized beam.
    
\subsection{Photometric redshifts}\label{subsec:zphot}

	As a cross-check on our LBG candidate selection, we used the photometry for each LBG candidate to obtain a photometric redshift estimate. For this purpose, we used the \texttt{C++} version\footnote{\texttt{FAST++}. \url{https://github.com/cschreib/fastpp}} of the code \texttt{FAST} \citep[Fitting and Assessment of Synthetic Templates;][]{2009ApJ...700..221K} with a bin size of $\Delta z_{ph} {=} 0.001$ and $500$ Monte Carlo simulations per source to derive confidence levels. 
	
\begin{figure}
    \centering
    \includegraphics[width=1.\columnwidth]{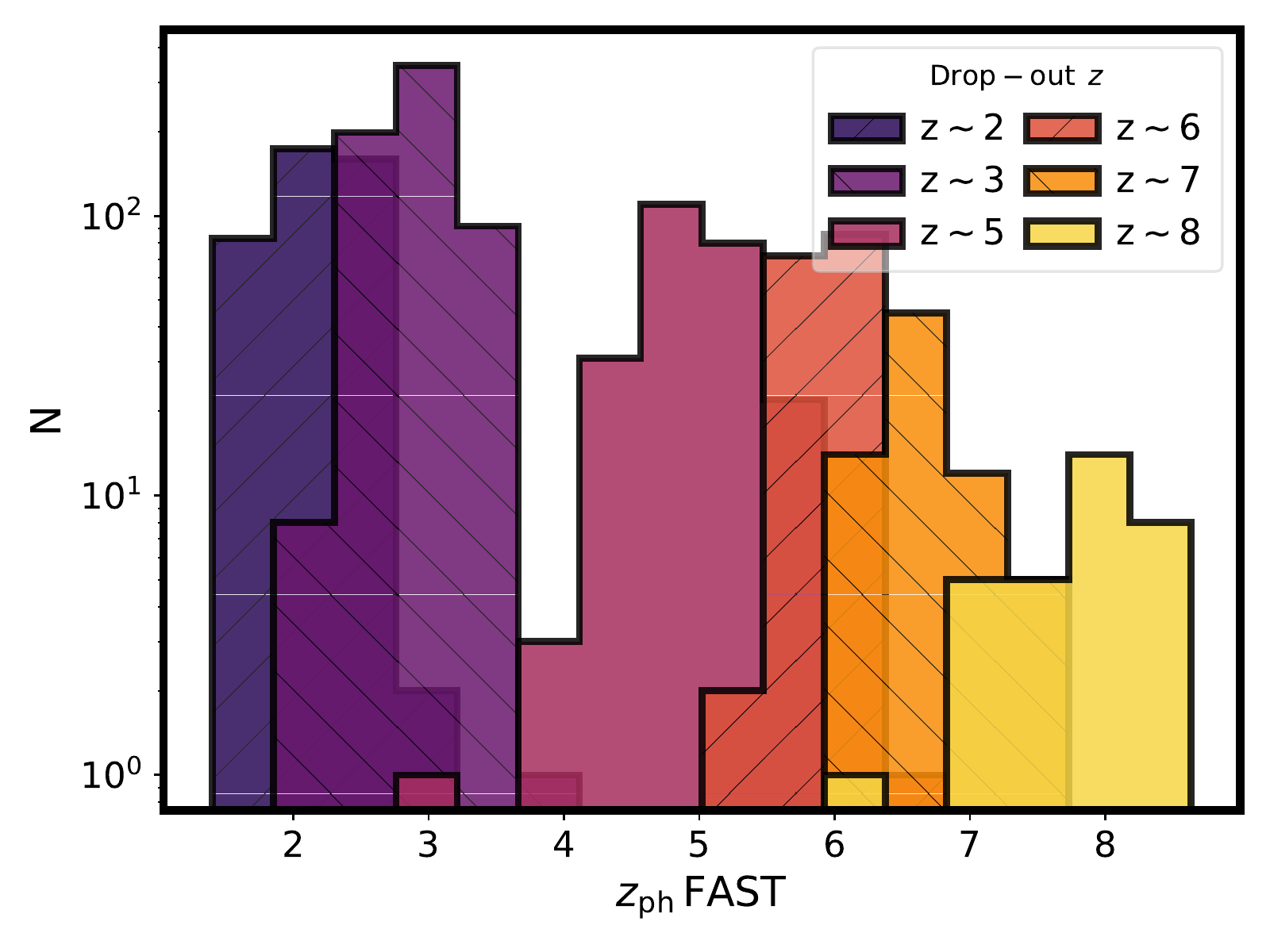}
    \caption{Photometric redshift ($z_{\mathrm{ph}}$) values in our sample (see $\S$\ref{subsec:Filtering}) calculated from \texttt{FAST++}. Colors represent each drop-out band.}
    \label{fig:hist_z_ph_fast}
\end{figure}

	The distribution of the photometric redshifts from our candidates calculated with \texttt{FAST++} are shown in Fig.~\ref{fig:hist_z_ph_fast}, color-coded by the drop-out band used to detect them. We find that the sub-samples do not overlap strongly and show roughly flat distributions. Only three sources, all $z {\sim} 5$ dropouts, exhibit strong deviations between their drop-out selection band and \texttt{FAST++} estimate, with $z_{\rm ph} {\sim} 1.5$; these three sources were excluded. We did not consider here the error distribution provided by \texttt{FAST++}, which would extend the drop-out distributions shown in Fig.~\ref{fig:hist_z_ph_fast} by ${\sim} 25{\%}$. 
	
	We also assessed the dropout candidates by comparing them to published spectroscopic redshifts. Unfortunately, only a few fields have extensive published redshift catalogs and most of our candidates are either too faint or were not targetted at appropriate wavelengths to confirm their redshifts in such surveys. Nonetheless, we compared our dropout catalogs with the VLT/MUSE redshift catalogs of \citet{2018MNRAS.473..663M}, \citet{2019MNRAS.485.3738L}, \citet{2017A&A...599A..28K}, \citet{2017A&A...600A..90C}, \citet{2016ApJ...822...78G}, and \citet{2016ApJ...817...60T} resulting in matches for ${\sim} 10\%$ of our candidates (per cluster) within a $0\farcs5$ circle. Among the 238 matches, only 14 candidates have strong differences between their photometric and spectroscopic redshifts; all 14 were removed.
	
	Given the gap in the photometric redshift distribution of the LBG candidates around $z {\sim} 4$ shown in Fig.~\ref{fig:hist_z_ph_fast}, for convenience we separated our candidates into two main sub-samples: high ($z_{ph} {\geq} 4.0$) and low ($z_{ph} {<} 4.0$) redshift. Additionally, we further subdivided the high-redshift sample into two parts: $4.0 {\geq} z_{ph} {\leq} 7.0$ and $z_{ph} {>} 7.0$. These divisions are used for the rest of the work.

\subsection{Magnification factors}\label{subsec:mufactor}
    
    Magnification factors were obtained following the procedure from  \citet{2015ApJ...800...84C}, coded as a public \texttt{Python} script.\footnote{\url{https://archive.stsci.edu/prepds/frontier/lensmodels/\#magcalc}} This code obtains the values from the lensing shear ($\gamma$) and mass surface density ($\kappa$) maps that are part of the lens models products, to calculate the magnification map for each redshift. Based on FFs mass model comparisons \citep[e.g.,][]{2017MNRAS.472.3177M,2018ApJ...863...60R}, we adopted the CATS (Clusters As TelescopeS) team models for our work \citep[\texttt{v4};][]{2014MNRAS.443.1549J, 2014MNRAS.444..268R}, as their methodology is well-documented and they appear to be among the most reliable mass models and magnifications maps of the publicly available models. With the CATS models and the photometric redshifts of the candidates as input, magnification factors ($\mu$) were obtained from the expression:
    \begin{eqnarray}
		\frac{1}{\mu} = |(1 - \kappa)^{2} - \gamma^{2}|.
	\end{eqnarray}
    To assess uncertainties associated with the magnification factors, we calculated both statistical errors using the limits of the 1-$\sigma$ confidence levels of the photometric redshifts and systematic uncertainties based on the standard deviation of the magnifications of each source using four different version \texttt{v4} FFs models: CATS, GLAFIC \citep{2010PASJ...62.1017O, 2016ApJ...819..114K}, Diego \citep{2005MNRAS.360..477D, 2007MNRAS.375..958D}, and Williams \citep{2006MNRAS.367.1209L, 2014MNRAS.443.1549J}. These uncertainties are presented in Table~\ref{tab:model_props}. We note that the dispersion can be large and asymmetric since some models are not as robust as others; for this reason, we chose to incorporate a systematic error coupled with the CATS team model, rather than find a representative $\mu$ value from all the models. 

\begin{figure}
    \centering
    \includegraphics[width=1.\columnwidth]{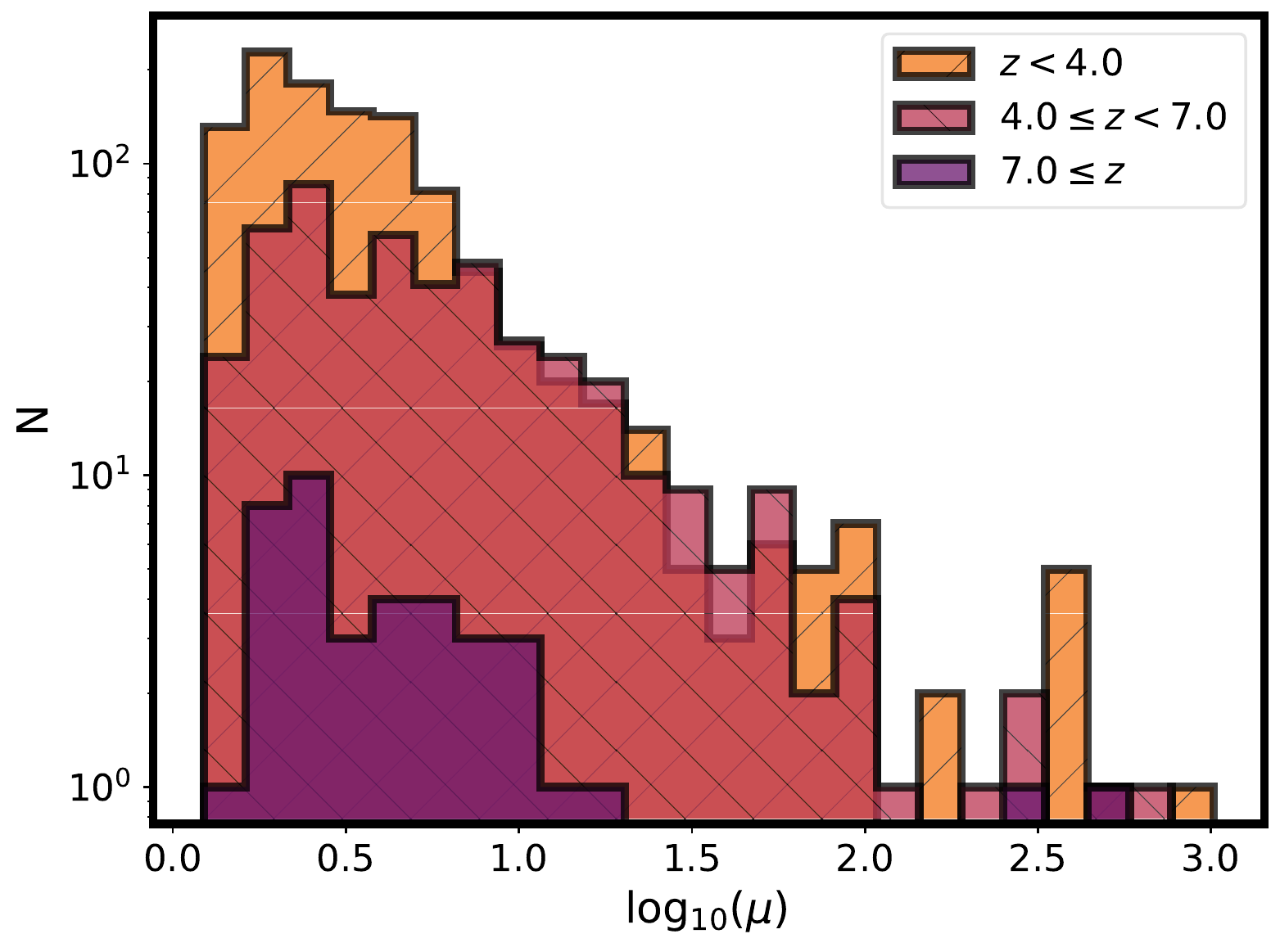}
    \caption{Distribution of magnification factors ($\mu$) for the three photometric redshift bin samples.}
    \label{fig:hist_magnific_original}
\end{figure}
    
\begin{figure}
    \centering
    \includegraphics[width=1.\columnwidth]{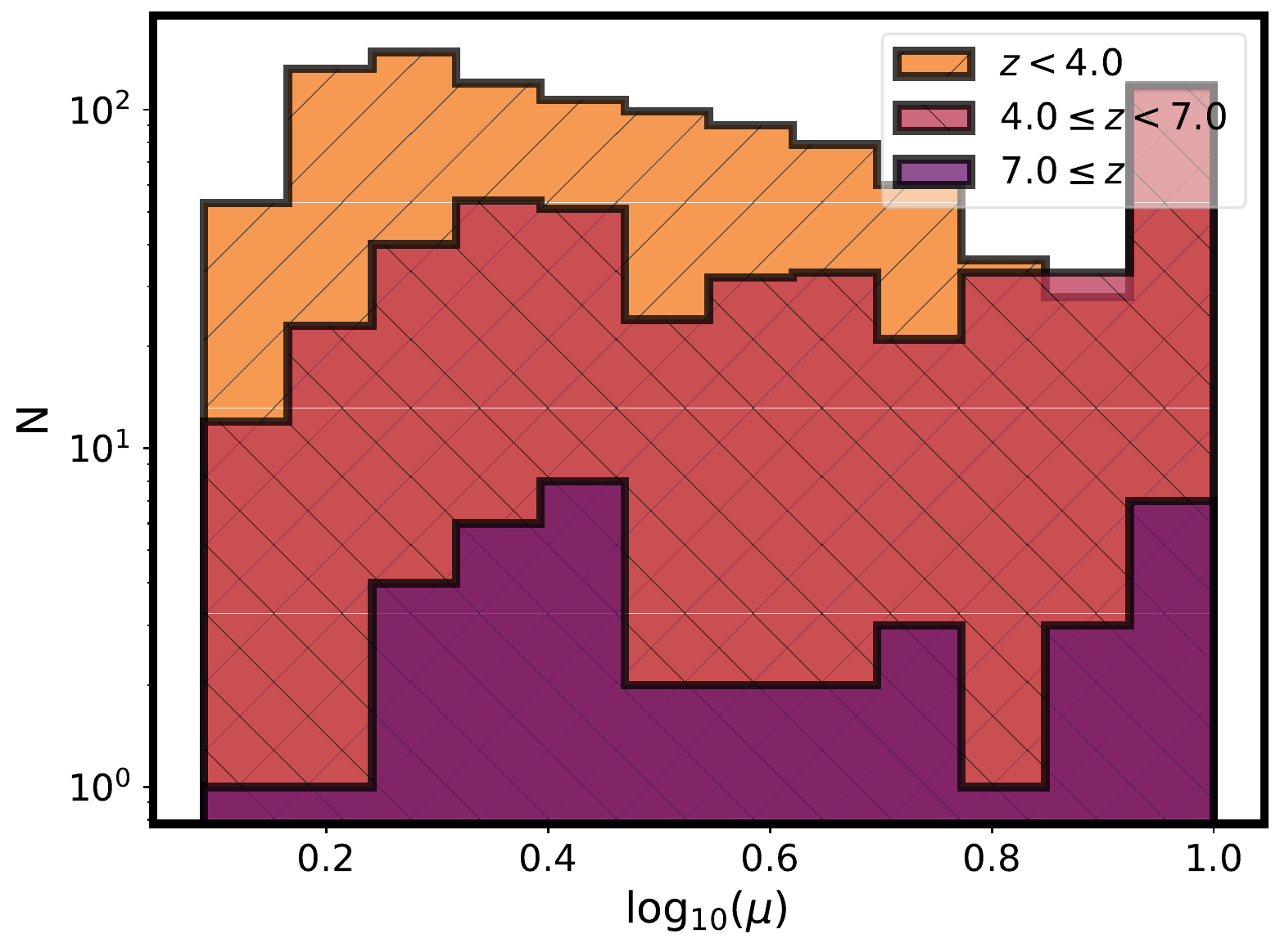}
    \caption{Capped distribution of magnification factors ($\mu$) for the three photometric redshift bin samples. Compare to the full distributions shown in Fig.~\ref{fig:hist_magnific_original}.}
    \label{fig:hist_magnific}
\end{figure}

    Some targets can lie in positions very close to the critical curves for a given lens model and redshift, leading to extreme magnifications ($\mu {\ga} 1000$); see Fig.~\ref{fig:hist_magnific_original}. Given the photometric redshift and lens model uncertainties, as well as the observed compact sizes of most candidates, extreme magnifications should be far less probable than moderate ones. Thus, to avoid possible spurious results when using these targets in calculations (e.g., when stacking with magnification factors as weights; see $\S$\ref{subsec:Stacking}), we capped the magnification factors at $\mu {=} 10$, even when models predicted larger values. This choice was driven by the fact that, after accounting for both the statistical and systematic uncertainties, ${>}60{\%}$ of our $\mu {>} 10$ candidates are compatible with magnifications of $\mu {\leq} 10$ at 1-$\sigma$ confidence and ${>} 88 {\%}$ at 2-$\sigma$ confidence level. Coupled with the small probability that candidates can have $\mu {>} 10$, we consider lower magnifications to be far more likely. As a result of this imposed ceiling, the magnification values in our full sample range from $\mu {=} 1.23$ to $\mu {=} 10$, with a manifest over-population at $\mu {=} 10$ (due to our cap) for all three redshift bins, as seen in Fig.~\ref{fig:hist_magnific}. For a comparison, we present a histogram of the unmodified $\mu$ values in Fig.~\ref{fig:hist_magnific_original}. Finally, we note that this magnification cap should have no strong effect on our results, as higher magnifications result in no change in IRX or $\beta$ values, and will only lower stellar masses, pushing candidates into a regime where we expect few detections (see $\S$\ref{subsec:IRXrelations}); thus, some care should be taken in evaluating detections at lower stellar masses due to highly magnified sources, 
    
\subsection{UV-continuum slope}\label{subsec:UV-slope}
    
    The observed UV-continuum slope, $\beta$, is often used to assess the amount of extinction/absorption that a particular stellar population suffers, under the assumption that a nominal intrinsic UV slope is typically $\beta_{0} {\approx}{-2.0}$ to $-2.2$, for constant star formation, with high values indicating higher attenuation. For each candidate in our sample, nine-band {\it HST} photometry was used to obtain the observed values of $\beta$. Several methods have been developed to calculate $\beta$ from different photometric bands (for a review, see $\S$2 of \citealt{2013MNRAS.429.2456R} and $\S$2.7 of \citealt{2018MNRAS.476.3991M}). Here, we adopted a simplistic approach using the bands (and the flux in them) that fall in the expected UV-continuum spectral region (e.g., $\lesssim$3000\AA) assuming the previously derived redshift. A power law ($F_{\lambda} \propto \lambda^{\beta}$) was fit to the rest-frame UV photometry using the \texttt{Python} implementation of the Affine Invariant Markov chain Monte Carlo Ensemble sampler  \citep[\texttt{emcee;}][]{2013PASP..125..306F}. In particular, we adopted the functional form chosen by \citet{2012A&A...540A..39C}, which is:
    \begin{eqnarray}\label{eq:uv_slope}
    	m_{i} = -2.5 \times (\beta + 2.0) \times \log{(\lambda_{i})} + c,
    \end{eqnarray}
    \noindent where $m_{i}$ is the AB magnitude in the i-th band \citep{1983ApJ...266..713O} at an effective wavelength $\lambda_{i}$ and $c$ is the intercept. As priors for the model fitting, we used the outputs of a simple maximum likelihood estimator with Eq. \ref{eq:uv_slope}. For each LBG candidate, $2500$ iterations were performed per each one of the $100$ "random-walkers" which were set for this procedure. From them, we obtained the most probable $\beta$ values and the limits of their $1{-}\sigma$ credible intervals.
    
\begin{figure}
    \centering
    \includegraphics[width=1.\columnwidth]{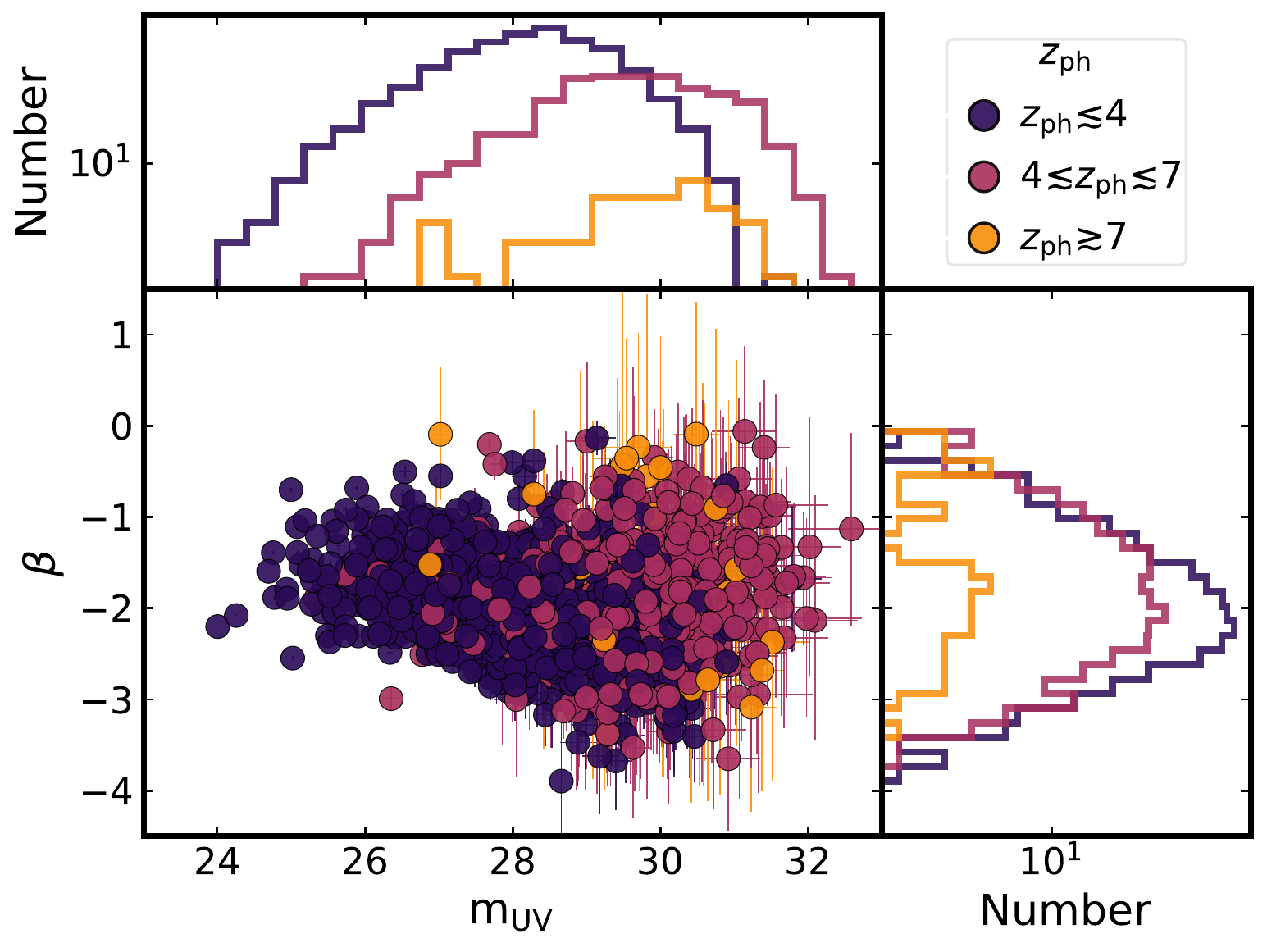}
    \caption{Comparison of $\beta$ and magnification-corrected magnitudes for selected sources. Top and right panels present histograms of UV slope and magnitude distributions. Colors represent photometric redshift subsamples, as described in the legend and in $\S$\ref{sec:Methods}.}
    \label{fig:uv_beta_mag}
	\end{figure}

\begin{figure}
    \centering
    \includegraphics[width=1.\columnwidth]{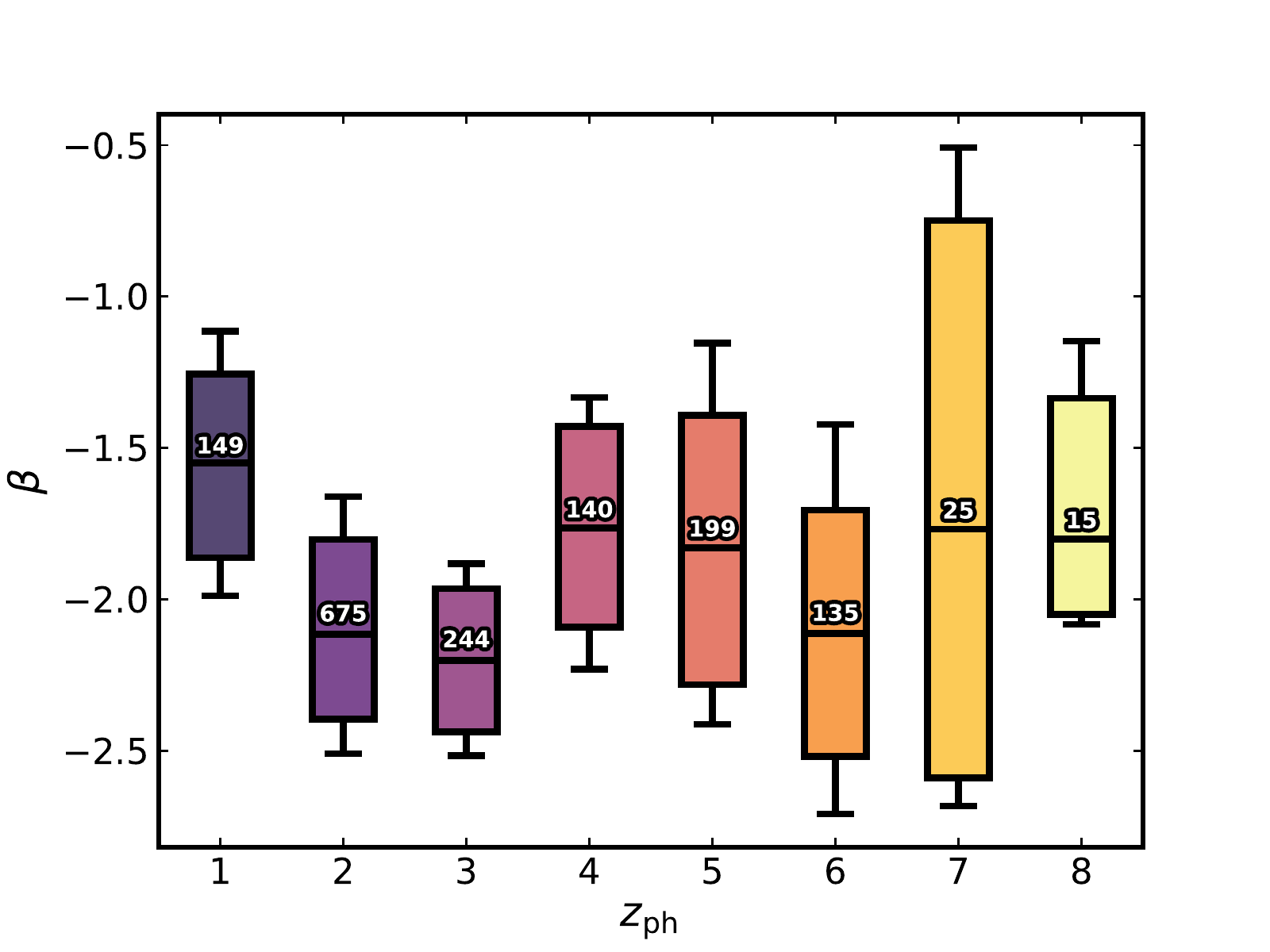}
    \caption{Distribution of $\beta$ values according to photometric redshift. Heights of boxes represent the $25\%$ and $75\%$ quartiles of the data. Horizontal lines inside the box indicate the median value for each redshift bin. Vertical error bars span  the central 2${-}\sigma$ of the data. Numbers above the median in each box state the number of LBG candidates assigned to each bin. Even though there is not a $z_{\mathrm{ph}} {\sim} 4$ band from drop-out selection, there are candidates in that bin.}
    \label{fig:beta_z_ph_boxplot}
	\end{figure}

    A comparison of the UV slopes ($\beta$) and magnification-corrected magnitudes, as well as their overall distributions, is presented in Fig.~\ref{fig:uv_beta_mag}. The three broad divisions in photometric redshift do not show any particular trend between $\beta$ and redshift. A finer binning of the targets according to photometric redshift is shown in Fig.~\ref{fig:beta_z_ph_boxplot}, where it can be seen that the UV-slopes of our LBG candidates are generally consistent with being more or less constant between $z {\sim}$1--8, within the large dispersion. Previous works such as \citet{2012ApJ...754...83B, 2012ApJ...756..164F, 2014ApJ...793..115B} have reported mild evolution in $\beta$ for $z_{\mathrm{ph}} {\gtrsim} 4$ LBG candidates due to a possible increase in dust extinction with time. This weak evolution lies within the dispersion of our sample and, thus, we can neither confirm nor reject it. 
    
    For the purposes of this work, we defined the UV flux or luminosity ($F_{\mathrm{UV}}$, $L_{\mathrm{UV}}$) to be that measured at $1600\,\mbox{\AA}$ (following, among others,  \citealt{2014ARA&A..52..415M}, who suggested that UV wavelengths between $1400\,\mbox{\AA}$ and $1700\,\mbox{\AA}$ provided a reasonable estimate). In our case, we used the photometric band which lies closest to that rest-frame wavelength.
    
	The UV slope can also be related to the dust attenuation factor, $A_{\lambda}$, as in \citetalias{1999ApJ...521...64M} and \citet{2000ApJ...533..682C}. For this work, we favored the relation found by \citet{2000ApJ...533..682C}:
	\begin{eqnarray}\label{eq:calzetti_law}
    	A_{1600} = 2.31 \times \beta + 4.85,
    \end{eqnarray}
	\noindent which was similarly assessed at $1600\,\mbox{\AA}$ for low-redshift galaxies.
    
\subsection{Stellar masses}\label{subsec:Mstar_calc}
    
    Stellar masses were estimated using \texttt{FAST++}, which fits stellar population synthesis templates to photometric data. The input values were the magnitudes from our LBG catalogs as well as the photometric redshifts (also determined with \texttt{FAST++}). For this work, we assumed \citet{2003MNRAS.344.1000B} stellar spectral energy distributions (SEDs) with a Chabrier Stellar IMF \citep{2003PASP..115..763C}. We assumed an approximately constant SFR in modeling the star formation history, effectively realized by setting $\log_{10}{(\tau / \mathrm{yr})}{=}11$ with an exponentially declining star formation history (SFR ${\propto} \exp(-t/\tau)$) and a metallicity of $0.2 Z/Z_{\odot}$. Finally, a \citet{2000ApJ...533..682C} dust attenuation law with a range of $0.0 {\leq} A_{V} {\leq} 1.0$ was adopted. The code outputs, apart from other relevant properties, a stellar mass estimate for each target. The above parameter choices have a sizeable impact on inferred quantities such as the stellar population age (${>}$0.3-–0.5 dex) but do not strongly impact the inferred stellar masses (${\ga}$0.2 dex).
	
    To obtain the magnification-corrected stellar masses, the values given by \texttt{FAST++} were divided by the magnification factors.
	
\begin{figure}
    \centering
    \includegraphics[width=1.\columnwidth]{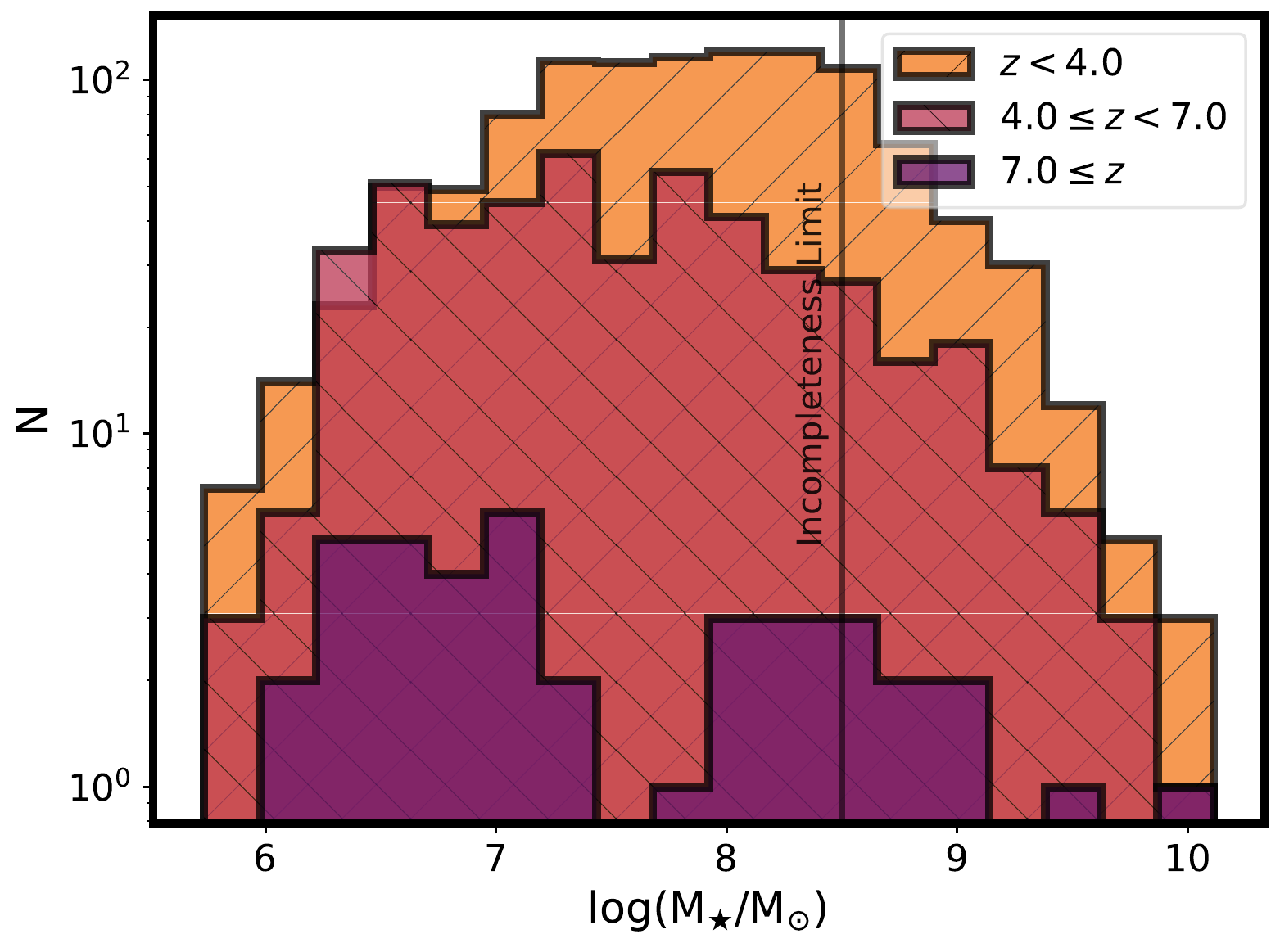}
    \caption{Stellar masses in our sample. Sample has been divided according to the photometric redshift bins defined in $\S$\ref{sec:Methods}. Vertical dark line represents the approximate completeness 
    limit from \citetalias{2018MNRAS.476.3991M} (See 
    $\S$\ref{subsubsec:Mstar_beta_corr}).}
    \label{fig:hist_stell_mass}
\end{figure}

    For the rest of this work, we refer to the magnification-corrected stellar mass simply as stellar mass. The distribution of best-fit values for our three photometric redshift bins can be seen in Fig.~\ref{fig:hist_stell_mass}. Factoring in the $1{-}\sigma$ confidence intervals on the stellar mass (see Fig.~\ref{fig:irx_vs_smass}), the full range spans ${\sim} 10^{5.6} \mathrm{M}_{\odot}$ to ${\sim} 10^{10.2} \mathrm{M}_{\odot}$.

\subsection{(Specific) Star formation rates}\label{subsec:SFR_calc}

	One of the by-products of \texttt{FAST++} is a SFR estimation. As with stellar mass, SFR values were corrected by the magnification factor ($\mu$).

    For the rest of this work, we refer to the magnification-corrected SFR as SFR. From the SFRs and stellar masses, specific SFRs, or sSFRs, are obtained as:
    \begin{eqnarray}
		\log_{10}{\left(\mathrm{sSFR} / \mathrm{yr}^{-1}\right)} = \log_{10}\left(\frac{\mathrm{SFR}/ \mathrm{M}_{\odot} \mathrm{yr}^{-1}}{\mathrm{M}_{\bigstar}/ \mathrm{M}_{\odot}}\right).
	\end{eqnarray}
    
\subsection{ALMA primary-beam corrections}\label{subsec:ALMA_pbcor}

	To obtain images from mosaics of interferometric data, each element of the observation has to be corrected by a combination of the sensitivity of every pointing in the observation and the change of sensitivity across the mosaic. These two elements constitute the PB corrections of the observed maps. 
	
	With interferometric data, the deconvolution from the $u$--$v$ visibility plane to the image plane includes a division (deconvolution) by these primary-beam correction factors. One of the ALMA pipeline data products is a normalized map of sensitivities, which incorporates the primary-beam correction factors, ranging from no (0) to full (1) sensitivity \citep[see, for instance,][]{2017isra.book.....T, 2013tra..book.....W}. 

\subsection{ALMA peak fluxes}\label{subsec:ALMA_up_lims}

	For simplicity, we adopted peak flux measurements, $F\mathrm{^{indiv,obs}_{ALMA,peak}}$, since integrated fluxes require an assumption about the flux distribution shape. To assess these peak fluxes within the ALMA maps, we searched for the pixel with the maximum value within a 0\farcs5$\times$0\farcs5 box (i.e., comparable to one synthesized beam) centered at the position of each LBG candidate. This procedure attempts to account for the influence of the synthesized beam, as well as possible extended emission, in the ALMA maps. We corrected this flux for the PB attenuation  (i.e., accounting only for the properties of the observations) as follows:
    \begin{eqnarray}\label{eq:peak_flux_detect}
    F\mathrm{^{indiv,obs}_{ALMA,peak,pbcor}} = \frac{F\mathrm{^{indiv,obs}_{ALMA,peak}}}{pbcor\mathrm{^{indiv}_{ALMA}}}.
    \end{eqnarray}
    \noindent Likewise, we related the $rms$ error at the position of an individual source to the field $rms$ ($rms\mathrm{^{cluster}_{ALMA}}$) listed in $\S$\ref{sec:Data} for each studied cluster, as
    \begin{eqnarray}\label{eq:rms_detect}
    rms\mathrm{^{indiv}_{ALMA,pbcor}} =   \frac{rms\mathrm{^{cluster}_{ALMA}}}{pbcor\mathrm{^{indiv}_{ALMA}}}.
    \end{eqnarray}
	The bulk of our candidates have ALMA fluxes comparable to the $rms$ values of their respective maps, but a few are associated with brighter peak fluxes. For this reason, we want to define clearly which targets are detected and for which we only have upper limits. As a first conservative approach, we searched for LBG candidates with S/N above $5.0$ in each image, which roughly corresponds to the blind detection limit for the ALMA-FF maps \citep{2017A&A...597A..41G}. This high S/N limit arises in the context of having large maps with ${\approx} 1.7{\times}10^{7}$ pixels yet only a handful of highly secure detections per field. The map noise is approximately Gaussian \citep{2017A&A...597A..41G}, meaning that there should be roughly 45896, 1077, and 9 pixels above 3, 4, and 5 times the $rms$, respectively, in each map. Excess numbers of pixels above these expectations imply real sources. We defined here the S/N as:
    \begin{eqnarray}\label{eq:signal_to_noise}
    \mathrm{S/N} 
    = \frac{F\mathrm{^{indiv,obs}_{ALMA,peak}}}{rms\mathrm{^{cluster}_{ALMA}}}
    = \frac{F\mathrm{^{indiv,obs}_{ALMA,peak,pbcor}}}{rms\mathrm{^{indiv}_{ALMA,pbcor}}}.
    \end{eqnarray}
    \noindent None of our targets fulfills this first condition, with a maximum value of S/N$=4.21$ for a candidate in AS1063.
    
    The blind detection limit, however, is with respect to a search of all positions on the map. Nevertheless, since we know the positions of the $1582$ LBG candidates and they comprise only a small fraction of the overall map area (${\approx}1.1{\times}10^{5}$ pixels),\footnote{Naively, we expect roughly 297, 7, and 0.06 pixels above 3, 4, and 5 times the $rms$, respectively, in each map.} a more realistic estimate of the detection significance is to evaluate the False Detection Rate \citep[FDR or {$p_{\mathrm{FDR}}$},][]{benjamini1995, benjamini2001} for each ALMA map. As described in \citet{2001AJ....122.3492M} and \citet{2002AJ....123.1086H}, the FDR is different from other thresholding methods in that it constrains the fraction of false detections compared with the total number of detections rather than the fraction of pixels falsely detected over the total number of pixels. Given its definition, the FDR does not depend on the distribution of sources and, thus, we are not forced to assume a specific behavior for them.
    
\begin{figure}
    \centering
    \includegraphics[width=1.\columnwidth]{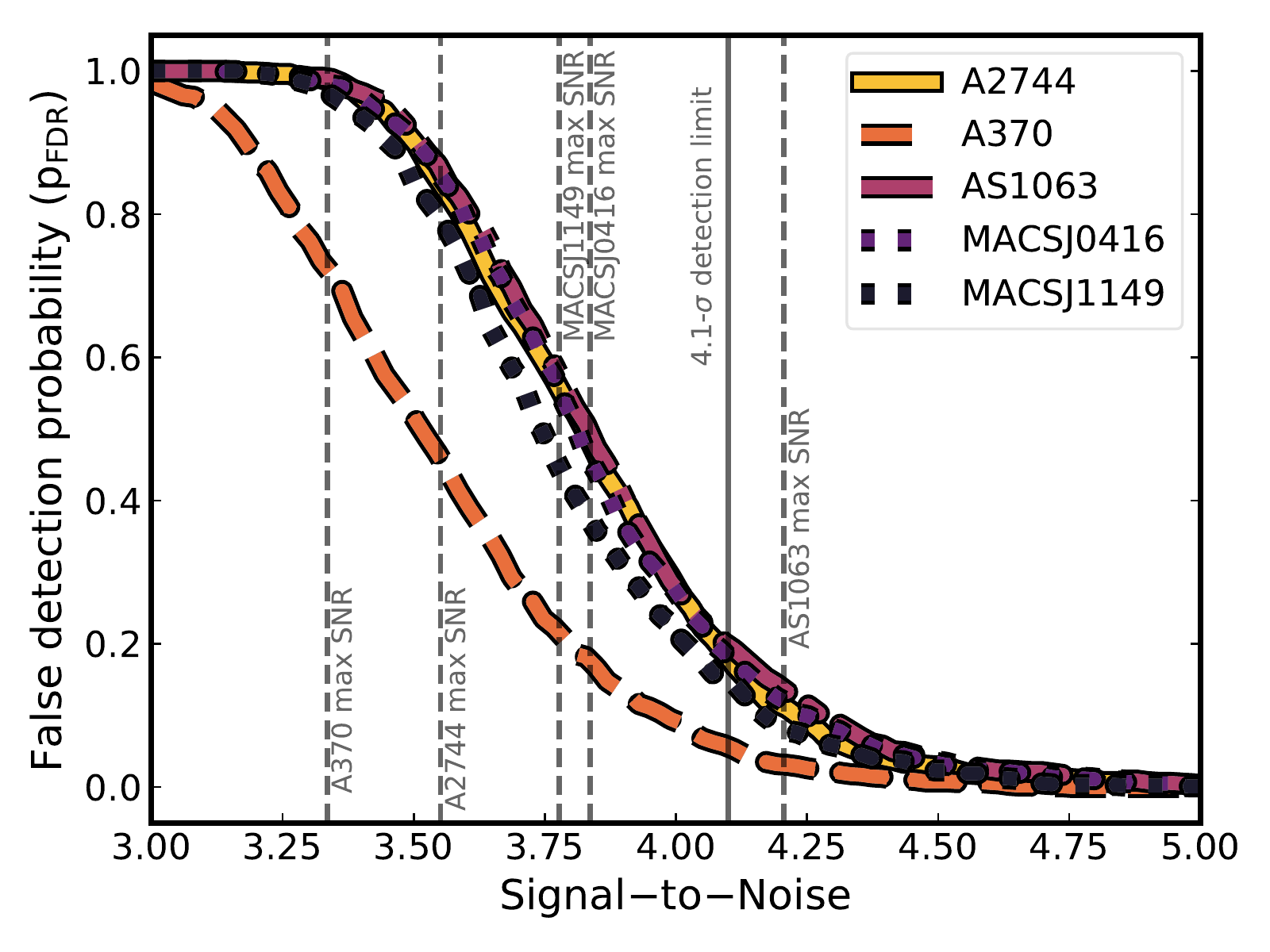}
    \caption{False detection rate, $p_{\mathrm{FDR}}$, for the five ALMA maps. Vertical dashed lines denote the highest detected S/N among the LBG candidates in each ALMA map. The vertical solid line denotes our adopted S/N cutoff of $4.1$, which equates to a FDR around $15 \%$ among the cluster fields.}
    \label{fig:fdr_alma_maps}
	\end{figure}

    To this end, following the procedures outlined in \citet{2018A&A...620A.125M}, we generated 1000 simulated maps for each ALMA field with a normal distribution in units of signal-to-noise. From these we extracted the same number of simulated peak fluxes per cluster as we did for the LBG candidates, again choosing the highest peak flux within a square of 0\farcs5 on a side. We defined $p_{\mathrm{FDR}}(\mathrm{S/N})$ to be the fraction of simulated maps of a specific cluster where at least one sampled pixel was found above a given S/N. Fig.~\ref{fig:fdr_alma_maps} shows the FDRs for our five ALMA maps. 
    
	Based on the FDRs, we find that sources with S/N${\gtrsim} 4.1$ have a relatively low ($\lesssim$15\%) chance of being false. For simplicity and uniformity, we considered all LBG candidates above this limit to be detected, while the LBG candidates below this were treated as upper limits. We calculated individual detected peak fluxes following Eq.~\ref{eq:peak_flux_detect}, while n-$\sigma$ upper limits were calculated as
    \begin{eqnarray}\label{eq:ALMA_flux_nsigma}
    F\mathrm{^{indiv,obs\,n-\sigma\,lim}_{ALMA,peak,pbcor}} = \frac{(F\mathrm{^{indiv,obs}_{ALMA,peak}}>0) + \mathrm{n} \times rms\mathrm{^{cluster}_{ALMA}}}{pbcor\mathrm{^{indiv}_{ALMA}}},
    \end{eqnarray}
	\noindent where the $>$0 expression indicates the fact that the observed peak flux from the ALMA map is only used if it is greater than zero. This implies that no single candidate will have a 1-$\sigma$ upper limit lower than the noise level of the map to which it belongs. The incorporation of local map noise, in addition to the average $rms$, yields a more conservative upper limit.
	
\subsection{UV luminosities}\label{subsec:UVlums}
	
	As in $\S$\ref{subsec:UV-slope}, we defined the UV flux and luminosity as that at $1600\,\mbox{\AA}$, ensuring that appropriate rest-frame and magnification corrections are applied for the best-fit photometric redshift.
	
\subsection{IR luminosities}\label{subsec:IRlums}
    
	With only one point of constraint from ALMA, the procedure to estimate the IR luminosity is more model-dependent than for the UV bands. For this, we fit a graybody spectrum (e.g., \citealt{2012MNRAS.425.3094C,  2013A&A...549A...4S}) to the ALMA photometric data. We adopted evolving values for dust temperature following the redshift-dependent formulation from \citet{2018A&A...609A..30S}. We caution, however, that this relation is only fit up to $z {=} 4$ and, thus, extrapolations may be problematic. The distribution of dust temperatures of our candidates ranges from ${\sim} 30$\,K to ${\sim} 65$\,K, which is in line with the $z {\geq} 5$ simulations from \citet{2019MNRAS.487.1844M} and the $2 \leq z \leq 4$ simulations from \citet{2019MNRAS.489.1397L}. We also considered typical fixed values of 2.0 for the mid-IR power-law slope, and 1.6 for the emissivity \citep[e.g., best-fit values for the GOALS survey;][]{2012MNRAS.425.3094C}. For simplicity, we adopted the same shape for every LBG candidate. The best-fitted rest-frame SED is integrated between 8\,$\mu$m and 1000\,$\mu$m to yield the rest-frame IR luminosity. In practical terms, we defined a scale factor $f^{\mathrm{ALMA,peak,pbcor,\mu\,cor}}_{\mathrm{IR}}$ to convert observed ALMA peak flux to the magnification-corrected, rest-frame IR luminosity as    
	\begin{eqnarray}
		f^{\mathrm{ALMA,peak,pbcor,\mu\,cor}}_{\mathrm{IR}} = \left(\frac{F_{\mathrm{ALMA,peak,pbcor}}/\mu}{F_{\mathrm{SED}[1.14 \mathrm{mm}/(1+z)]}}\right).
	\end{eqnarray}

	We chose this method over \texttt{FAST++} or \texttt{magphys} \citep[Multi-wavelength Analysis of Galaxy Physical Properties;][]{2008MNRAS.388.1595D} SED fitting to obtain IR luminosity estimates due to the fewer number of free parameters (e.g., dust temperature, SED templates), which made for a more straightforward implementation and interpretation. In general, the luminosities derived from the best-fit modified blackbody to the ALMA data are factors of 10--100 higher than rest-frame UV/optical based estimates from \texttt{FAST++} or \texttt{magphys}. Our estimates are presumably more robust for the few detections, while the upper limits should be considered as very conservative.
	
	To test this method, we calculated IR luminosities for the sources reported by \citet{2016ApJ...833...68A} using their ALMA (Band 6) flux measurements. Our results lie within ${\sim} 0.5$ dex of theirs, which were obtained with \texttt{magphys}. These results demonstrate that we can obtain relatively reliable IR luminosities from the graybody spectrum.
    
	In addition to the aforementioned corrections for redshift and magnification, the IR luminosities (or fluxes) have an additional dependence on the redshift of the candidate due to the impact of the CMB temperature on the dust properties. Following the procedure of \citet{2013ApJ...766...13D}, the derived IR luminosities were divided by the factor
	\begin{eqnarray}
		g_{\nu}^{\mathrm{CMB}} = \left[ 1 - \frac{B_{\nu}(T_{\mathrm{CMB}}(z))}{B_{\nu}(T_{\mathrm{dust}}, z)} \right],
	\end{eqnarray}
	
    \noindent where $B_{\nu}(T_{\mathrm{dust}})$ and $B_{\nu}(T_{\mathrm{CMB}},z)$ correspond to the source and CMB blackbody contributions at the observed frequency and redshift of the source, respectively.
    
	Errors were propagated according to Eq. \ref{eq:rms_detect} (in units of luminosity), applying the same corrections (e.g., redshift, magnification, and CMB temperature). We calculated IR luminosity upper limits as the peak value at source location plus $n$-sigma:
	\begin{eqnarray}\label{eq:L_ir_upper}
		\mathrm{L}\mathrm{^{indiv,obs\,n-\sigma\,lim}_{IR,peak,pbcor}} = \mathrm{L}^{\mathrm{peak,pbcor}}_{\mathrm{IR,\mu\,cor}} + n \times \mathrm{rms}^{\mathrm{pbcor,\mu\,cor}}_{\mathrm{IR,up\,lim}}[\mathrm{L}_{\odot}],
	\end{eqnarray}
	
	\noindent and generally adopted $1 {-} \sigma$ as the credible interval used. This upper limit formalism is also adopted for other quantities throughout this work (e.g., IRX).

\subsection{IRX relations}\label{subsec:IRXrelations}

	Sensitive millimeter facilities such as {\it Herschel} and ALMA have only become available in the last decade. Prior to these, it was generally difficult to measure IR luminosities for distant galaxies, and indirect methods were employed to understand and predict the IR emission. Principal among these is the so-called IR excess ratio (IRX), which is loosely defined as the ratio between the IR and UV luminosities (or fluxes) of a source (in this case, a galaxy). One of the most utilized definitions was developed by \citetalias{1999ApJ...521...64M}, which relates the UV and IR fluxes as:
	\begin{eqnarray}
		\mathrm{IRX} = \frac{F_{\rm IR}}{F_{\mathrm{UV}}}
	\end{eqnarray}
	\noindent where $F_{\mathrm{ IR}}$ is the rest-frame 8--1000\,$\mu$m IR flux and $F_{\mathrm{UV}}$ is the rest-frame 1600\,\mbox{\AA} UV flux, both of them corrected for magnification factors. This can be trivially extended for rest-frame luminosities instead of fluxes. These relations were developed using local galaxy data, but have been tested on a variety of distant (mostly massive) galaxy samples.
    
    Similar to the IRX-$\beta$ relations, there have been a large number of studies arguing that the total stellar mass of a galaxy is strongly related to the degree of dust extinction and, hence, IRX. We highlight four recent published correlations between IRX and stellar mass by \citet[][hereafter H14]{2014MNRAS.437.1268H}, \citet[][hereafter F17]{2017MNRAS.472..483F}, \citet[][hereafter M18]{2018MNRAS.476.3991M},  and \citetalias[][]{2016ApJ...833...72B}.
    
    Finally, \citetalias{2016ApJ...833...72B} also derived a "consensus" IRX-$\mathrm{M}_{\bigstar}$ relation from a variety of previous studies in the redshift range $z {\sim} 0$ to $z {\sim} 3$ \citep[e.g,][]{2009ApJ...698L.116P, 2010ApJ...712.1070R}.
	
    The various IRX-$\mathrm{M}_{\bigstar}$ relations have relatively similar slopes and exhibit a typical dispersion of up to $\sim$1 dex, excluding the strong deviation of  \citetalias{2014MNRAS.437.1268H} above $10^{10} \mathrm{M}_{\odot}$. As such, they provide a potentially useful means of predicting dust attenuation as a function of stellar mass.


\section{Methods}\label{sec:Methods}

\subsection{Target final sample}\label{subsec:Filtering}

	With all of the derived quantities in hand ($\S$\ref{sec:Data}), we now address the selection of the LBG candidate sample, in order to improve the reliability and trustworthiness of the estimated physical properties and stacking results.
    
    We began by discarding a handful ($7$) of LBG candidates with UV-slopes $\beta {<} {-} 4.0$ or $\beta {\geq} 1.5$ (see Fig.~\ref{fig:uv_beta_mag}). These extremely low or high values arise at faint magnitudes, have large error bars, and are physically implausible. This is qualitatively comparable to a (UV) color selection. 

	Before stacking, we also excluded $408$ LBG candidates in close proximity but unrelated to any $\ge$4-$\sigma$ detected sources in the ALMA maps in order to avoid contamination in the stacked signal. We conservatively adopted a circular exclusion region equal to five times the major axis of the natural-weighted synthesized beam for each map (i.e., 3\farcs2--7\farcs6). We additionally removed all LBG candidates with primary-beam correction factors lower than 0.5 (see $\S$\ref{subsec:ALMA_pbcor}), as the edges of the ALMA maps have considerably higher noise and other observational artifacts that can adversely affect the sensitivity of the stacking. 

	Based on the FDR assessment in $\S$\ref{subsec:ALMA_up_lims}, we also identified two LBG candidates associated with ALMA detections at S/N$\gtrsim$4.1, adopting a matching radius of 0.5 times the major axis of the natural-weighted synthesized beam for each map (i.e., 0\farcs3--0\farcs8). These sources, along with their key attributes are listed in Table~\ref{tab:detect_props} and they were not included in the main stacking and have been treated separately. For comparison, the typical positional uncertainties between ALMA and {\it HST} sources are $\lesssim$0\farcs1 \citep[e.g., $\lesssim$10\% of the beam size in][]{2017A&A...597A..41G}.
  
    Finally, we considered whether LBG candidates were multiply imaged. We did not want to double-count the same source, as this could have potentially distorted our stacking results. Thus, we removed all multiple images. To determine whether a candidate was multiply imaged, we matched the positions of our LBG candidates against the multiple-image catalogs from the CATS team (\texttt{v4}; see $\S$\ref{subsec:mufactor}), which comprise a compilation of secure multiple images found via {\it HST} or ground-based spectroscopic confirmation \citep[e.g.,][]{2009ApJ...707L.163S, 2011MNRAS.417..333M, 2011MNRAS.410.1939Z, 2013ApJ...762L..30Z, 2014MNRAS.443.1549J, 2014MNRAS.444..268R, 2016ApJ...819..114K, 2017A&A...600A..90C, 2017MNRAS.469.3946L, 2018ApJ...855....4K,2018MNRAS.473..663M}. In total, we removed $53$ LBG candidates with positions conservatively lying within 0\farcs5 radius of a known multiple images ($23$ lie within 0\farcs25).
    
    We summarize our selection criteria in Table \ref{tab:filter}, which resulted in a sample of $1580$ undetected LBG candidates to stack: $383$ from A2744; $369$ from MACSJ0416; $315$ from MACSJ1149; $121$ from A370; and $391$ from AS1063. For some specific results below, to avoid problems related to combining values spanning several orders of magnitude (e.g., the weights from $\S$~\ref{subsec:Stacking}), we restricted the sample even further; for instance, when considering stacking in bins of $M_{\bigstar}$, we discarded a handful of very low-mass LBGs and only considered $1569$ candidates.
    
    \begin{table}
        \caption{LBG Candidate Selection Criteria}
            \label{tab:filter}
        \centering
        \resizebox{.99\linewidth}{!}{
            \begin{tabular}{@{}l@{\hspace{1.25\tabcolsep}}r@{\hspace{1.25\tabcolsep}}c@{}}
            \hline
            \noalign{\smallskip}
            \multicolumn{1}{c}{Property}                &  \multicolumn{1}{c}{Criterion}                                    & \multicolumn{1}{c}{Discarded \#\tablefootmark{a}} \\
            \noalign{\smallskip}
            \hline
            \noalign{\smallskip}
            Well-observed clusters                      & Cluster ${\ne}$ MACSJ0717                                            & 379\\
            Magnification                               & $\mu > 1.0$                                                       & 2\\
            UV slope                                    & $1.5 > \beta \geq {-}4$                                           & 7\\
            ALMA PB-correction                          & $pbcor_{\rm ALMA} > 0.5$                                          & 970\\
            Bright source contamination                 & $dist_{\rm\, S/N>4} > 5 {\times} b_{\rm maj}$                     & 408\\
            FDR detections                              & $dist_{\rm\, S/N>3.5} >= 0.5 {\times} b_{\rm maj}$                 & 2\\
            Multiple images                             & $dist_{\rm\, mult} <$ 0\farcs5                                    & 53\\
            Low stellar mass                            & $\log_{10}{(\mathrm{M}_{\bigstar} / \mathrm{M}_{\odot})} > 6.0$   & 16\\
            Match drop-out and $z_{\mathrm{ph}}$            & $z_{\mathrm{drop-out}} - z_{\mathrm{ph}} < 2.0$                   & 9\\
            Match $z_{\mathrm{spec}}$ and $z_{\mathrm{ph}}$ & $z_{\mathrm{spec}} - z_{\mathrm{ph}} < 2.75$                      & 14\\
            \noalign{\smallskip}
            \hline
        \end{tabular}
        }
        \tablefoot{
        \tablefoottext{a}{We begin with an initial sample of $3050$ LBG candidates from all six FFs (see $\S$\ref{subsec:LBG_candidates}), but refine the sample for the various reasons listed above (see $\S$\ref{sec:Data} for details). The final number of candidates studied results from a mixture of all these criteria, ranging from $1569$ to $1582$, depending on our goals.}
        }
    \end{table}

\subsection{Stacking}\label{subsec:Stacking}

	To perform the stacking process for our ALMA data, we used the \texttt{STACKER} code developed by \citet{2015MNRAS.446.3502L}. It can stack interferometric data in both the $u$--$v$ (visibilities) and image domains. For the image domain, the code uses median or mean stacking with weights. These weights can be fixed a priori or obtained from the PB-correction data present in ALMA datasets. The product of this stacking process is an ALMA image file. In the $u$--$v$ domain, the stack aligns the phases and then adds up the weighted visibilities.
    
    We adopted four different weighting schemes for the stacking code and further analysis: no or equal weights for all sources; PB correction $pbccor$-weighting; (magnification-corrected) UV flux $F_{\rm UV}$ and $pbcor$ weighting; and magnification $\mu$ and $pbcor$-weighting. For the equal weight scenario, the weight factor ($W_{k}^{no}$) is simply a constant of unity for all $k$ sources. 
    
    For the $pbcor$-weighting scenario, the sensitivity maps were used, with the weight factor given by:
    \begin{eqnarray}\label{eq:pb_weight}
		W_{k}^{pbcor} = \left(pbcor\mathrm{^{indiv}_{ALMA}}\right)^{2}.
	\end{eqnarray}
	\noindent This scheme simply counteracts the effects of the primary-beam correction on the determination of ALMA peak fluxes and, hence, enhances the contributions from the sources with the lowest $rms$ values. 
	
    For the UV-flux $F_{\rm UV}$ weighting scenario, the factor has the form:
    \begin{eqnarray}
		W_{k}^{UV} = \left(pbcor\mathrm{^{indiv}_{ALMA}}\right)^{2} \times F_{UV}^{2}.
	\end{eqnarray}
    \noindent This scheme should enhance the contribution from sources that show a higher ultraviolet flux and, by extension, higher star formation activity (and possibly stellar masses due to the star-formation main sequence), in addition to the $pbcor$ correction. We caution that this scheme could bias the stacking results toward sources that are less obscured and are more likely to lie closer to the \citetalias{1999ApJ...521...64M} IRX-$\beta$ relation. 
    
    Likewise, for the magnification $\mu$-weighting scenario, the weight factor is:
    \begin{eqnarray}
		W_{k}^{\mu} = \left(pbcor\mathrm{^{indiv}_{ALMA}}\right)^{2} \times \mu^{2}.
	\end{eqnarray}
    \noindent This weight configuration takes advantage of the magnification power of the galaxy clusters, which can amplify the influence of faint or less obscured sources in the final results, in addition to the $pbcor$ correction. 
    
    We expected some S/N variations among the different weighted stacks since they include different contributions of ALMA flux into the final results. The adopted weighting schemes might have inadvertently downweighted contributions from LBG candidates with higher individual S/N values. For instance, by favoring properties that are not directly expressed in the ALMA data, we may have been selecting against the most dust enshrouded candidates. 
    
    This stacking produces, ultimately, an image file. In this image, the stacked flux from the candidates is present in the central pixel if the objects are point-like. If highly extended or offset sources are part of stacked targets, other considerations must be taken into account; for instance, if extended, we would want to adopt an appropriate beam shape, or if offset, we would want to calculate the center of each target from the ALMA observation itself, rather than adopting the {\it HST} catalog position. As stated in $\S$\ref{subsec:PrepTargets}, we did not expect UV and IR offsets to be a preponderant issue here and, thus, calculated the stacking results adopting the individual UV ({\it HST}) positions of the LBG candidates.

    After \texttt{STACKER} was run for each data configuration, every stacked image was inspected to determine if a detection has been achieved. We calculated the detection levels for each stacked image using the procedure described by \citet{2017A&A...597A..41G}, in which peaks (sources) with S/N $>$ $5{-}\sigma$ are iteratively discarded until we arrive at a stable rms noise value.
    
    On the other hand, to obtain stacked values of IRX, a different method must be employed in which the stacking of ALMA observations is not directly utilized. 
     
    Following previous discussions from \citet{2017MNRAS.467.1360B} and \citet{2018MNRAS.479.4355K}, and taking into account the weights we are using, the appropriate method to determine stacked IRX values is
    \begin{eqnarray}\label{eq:stack_irx_avg}
		\overline{\mathrm{IRX}} = \overline{\left(\frac{ L_{\mathrm{IR}}}{L_{\mathrm{UV}}}\right)}
	\end{eqnarray}
	\noindent for each subsample in bins of redshift, stellar mass and UV-slope.
    We adopted this indicator since it is non-trivial to know, a priori, how the UV and IR luminosities are related. Thus, we stacked the individual IRX values and not the separate luminosities. The calculated IR luminosities were provided using the procedure described in $\S$\ref{subsec:IRlums}.
    
    In the case of upper limits, we stacked, separately, the peak IRX values and their $3 {-} \sigma$ error values. Then, we combined them to obtain the final stacked upper limits. That is:
    \begin{eqnarray}\label{eq:stack_irx_avg_ul}
		\overline{\mathrm{IRX}^{n {-} \sigma \mathrm{lim}}} = \overline{\left(\frac{ L^{\mathrm{peak,pbcor}}_{\mathrm{IR,\mu\,cor}}}{L_{\mathrm{UV}}}\right)} + n \times \Delta \overline{\left(\frac{ L^{\mathrm{peak,pbcor}}_{\mathrm{IR,\mu\,cor}}}{L_{\mathrm{UV}}}\right)}
	\end{eqnarray}

    Finally, to investigate the relation between IRX and other parameters, the target stacking was binned as a function of three different quantities; UV-slope, stellar mass, and redshift.

    With UV-slope, targets were stacked in five bins and, for stellar mass, in nine bins. Candidates with stellar masses less than $10^{6.0}$ M$_{\odot}$ were excluded from stacking calculations because of their very low expected luminosities and low numbers. For redshift, three sub-samples were utilized. These divisions were adopted considering the apparent distribution of redshift values shown in Fig.~\ref{fig:hist_z_ph_fast}. 
    The choice of bin widths was made as a compromise between having sufficient numbers of sources to reap the benefits of stacking and using equal-width bins in parameter space to facilitate interpretation. For the latter reason, we did not attempt to have a similar number of elements per bin. The bins are presented in Table~\ref{tab:LBG_bins}, while the number of sources per bin are presented in column 3 of Tables \ref{tab:stack_results_full_beta_z_bin} \ref{tab:stack_results_mid_mass_z_bin}, \ref{tab:stack_results_full_mass_z_bin}, and \ref{tab:ALMA_props_low_mass}. We can see that the uncertainties for the $\beta$ and $\mathrm{M}_{\bigstar}$ (Tables~\ref{tab:model_props} and \ref{tab:fast_props}) are small enough to not pose major problems to the binning of the sources.
    
    \begin{table}
        \caption{LBG Candidates binning}
            \label{tab:LBG_bins}
        \centering
        \resizebox{.95\linewidth}{!}{
            \begin{tabular}{@{}p{0.3\textwidth}r@{}}
            \hline
            \noalign{\smallskip}
            \multicolumn{1}{c}{$\mathrm{M}_{\bigstar}$ bins} &  \multicolumn{1}{c}{$\beta$ bins} \\
            \noalign{\smallskip}
            \hline
            \noalign{\smallskip}
            $6.0 \leq \log{(M_{\bigstar} / M_{\odot})} < 6.5$   &   $-4.0 \leq \beta < -3.0$\\
            $6.5 \leq \log{(M_{\bigstar} / M_{\odot})} < 7.0$   &   $-3.0 \leq \beta < -2.0$\\
            $7.0 \leq \log{(M_{\bigstar} / M_{\odot})} < 7.5$   &   $-2.0 \leq \beta < -1.0$\\
            $7.5 \leq \log{(M_{\bigstar} / M_{\odot})} < 8.0$   &   $-1.0 \leq \beta < 0.0$ \\
            $8.0 \leq \log{(M_{\bigstar} / M_{\odot})} < 8.5$   &   $0.0 \leq \beta < 1.5$  \\
            $8.5 \leq \log{(M_{\bigstar} / M_{\odot})} < 9.0$   &                           \\
            $9.0 \leq \log{(M_{\bigstar} / M_{\odot})} < 9.5$   &                           \\
            $9.5 \leq \log{(M_{\bigstar} / M_{\odot})} < 10.0$  &                           \\
            $\log{(M_{\bigstar} / M_{\odot})} \geq 10.0$        &                           \\
            \noalign{\smallskip}
            \hline
            \noalign{\smallskip}
            \multicolumn{2}{c}{$z_{\mathrm{ph}}$ bins}\\
            \noalign{\smallskip}
            \hline
            \noalign{\smallskip}
            \multicolumn{2}{c}{$z_{\rm ph} < 4.0$}\\
            \multicolumn{2}{c}{$4.0 \leq z_{\rm ph} < 7.0$}\\
            \multicolumn{2}{c}{$z_{\rm ph} \geq 7.0$}\\
            \noalign{\smallskip}
            \hline
        \end{tabular}
      }
    \end{table}

\subsection{Considerations on stacking weighting}\label{subsec:stacking_weights}

    Stacking of the ALMA data and IRX values can potentially constrain the average properties of a sample well below the formal detection limits for individual sources. The obtained values, however, should be regarded with some reservations. For one, the average properties can be skewed by a few outliers, since we are {\it not} individually detecting objects. Secondly, we employed $\mu$ and $F_{\rm UV}$ weighting schemes (see $\S$\ref{subsec:Stacking}) with the aim to improve our sensitivity. The downside of weighting, however, is that our stacked result can be biased toward the candidates with the highest weights.
    
    As an example, consider the case of IRX stacking with $F_{\rm UV}$ weighting. We can expect that stacking results will be skewed toward candidates with higher UV luminosities and, hence, lower stacked IRX values, which is not, necessarily, an expression of the behavior of most LBG candidates. Thus, any stacked IRX value has to be considered as a manifestation of the influence of the candidates with the highest weights and not as a true expression of the overall trend from the full studied sample.
    
	
\section{Results}\label{sec:Results}
    
    We describe below the main results obtained both for the individually detected sources reported in $\S$\ref{sec:Data} and from the stacking of the ALMA and IRX values of our sample.

\subsection{Individual results}

	Based on the individual luminosities obtained using the graybody SED and our \textit{HST} photometry, we derive IRX values (or upper limits) and compare them with previously calculated properties for each LBG candidate. We focus our comparisons on the UV slopes and stellar masses of the candidates. Some key properties for our ALMA detections are listed in Table~\ref{tab:detect_props}. A broader set of properties for all our LBG candidates are listed in the tables of Appendix~\ref{tab:indiv_props}.

\begin{table*}
\caption{Observed and derived properties for detected LBG candidates. Further properties and errors can be found in the appendices.}\label{tab:detect_props}
\centering          
\resizebox{1.0\textwidth}{!}{
\begin{tabular}{@{}c@{\hspace{1.2\tabcolsep}}c@{\hspace{1.2\tabcolsep}}c@{\hspace{1.2\tabcolsep}}l@{\hspace{1\tabcolsep}}c@{\hspace{1.2\tabcolsep}}c@{\hspace{1.2\tabcolsep}}c@{\hspace{1.2\tabcolsep}}c@{\hspace{1.2\tabcolsep}}c@{\hspace{1.2\tabcolsep}}c@{\hspace{1.2\tabcolsep}}c@{\hspace{1.2\tabcolsep}}S[separate-uncertainty=true,table-figures-decimal=0]@{\hspace{1.2\tabcolsep}}S@{}}     
\hline
ID & R.A. [J2000] & Dec. [J2000] & \multicolumn{1}{l}{cluster} & $z_{\rm ph}$ & $\mu$ & $\beta$ & \textrm{$\log{(L_{\rm UV}} / L_{\odot})$} & \textrm{$\log{(L_{\rm IR} / L_{\odot})}$} & \textrm{$\log{(L_{\rm IR} / L_{\rm UV})}$} & \textrm{$\log{(M_{\bigstar} / M_{\odot})}$} & $F\mathrm{^{indiv,obs}_{ALMA,peak,pbcor}}$ & S/N$\mathrm{^{indiv}_{peak}}$\\
 &    [hh:mm:ss.ss] & [$\pm$dd:mm:ss.s] &  &  &  &  &  &  & & & [$\mu$Jy] & \\
 \noalign{\smallskip}
\hline
\noalign{\smallskip}
2155 & 22:48:47.67 & -44:32:09.80 & AS1063 & $5.49^{+0.50}_{-4.09}$ & $2.88^{+0.02}_{-0.67}$ & $-1.22^{+0.80}_{-0.82}$ & $8.80^{+0.42}_{-0.34}$ & $11.88^{+0.11}_{-0.16}$ & $3.03^{+0.11}_{-0.16}$ & $6.89^{+1.83}_{-0.36}$ & 285 \pm 68 & 4.21 \\
\noalign{\smallskip}
2212 & 22:48:46.22 & -44:31:12.90 & AS1063 & $5.35^{+0.31}_{-0.57}$ & $39.00^{+14.92}_{-4.38}$ & $-1.62^{+0.59}_{-0.59}$ & $8.68^{+0.76}_{-0.39}$ & $11.33^{+0.66}_{-0.22}$ & $3.17^{+0.66}_{-0.22}$ & $6.62^{+1.86}_{-0.68}$ & 287 \pm 70 & 4.11 \\
\noalign{\smallskip}
\hline
\end{tabular}}
\end{table*}

\subsubsection{ALMA peak fluxes}\label{ALMA_fluxes}

\begin{figure*}
    	\centering
    	\includegraphics[width=.495\textwidth]{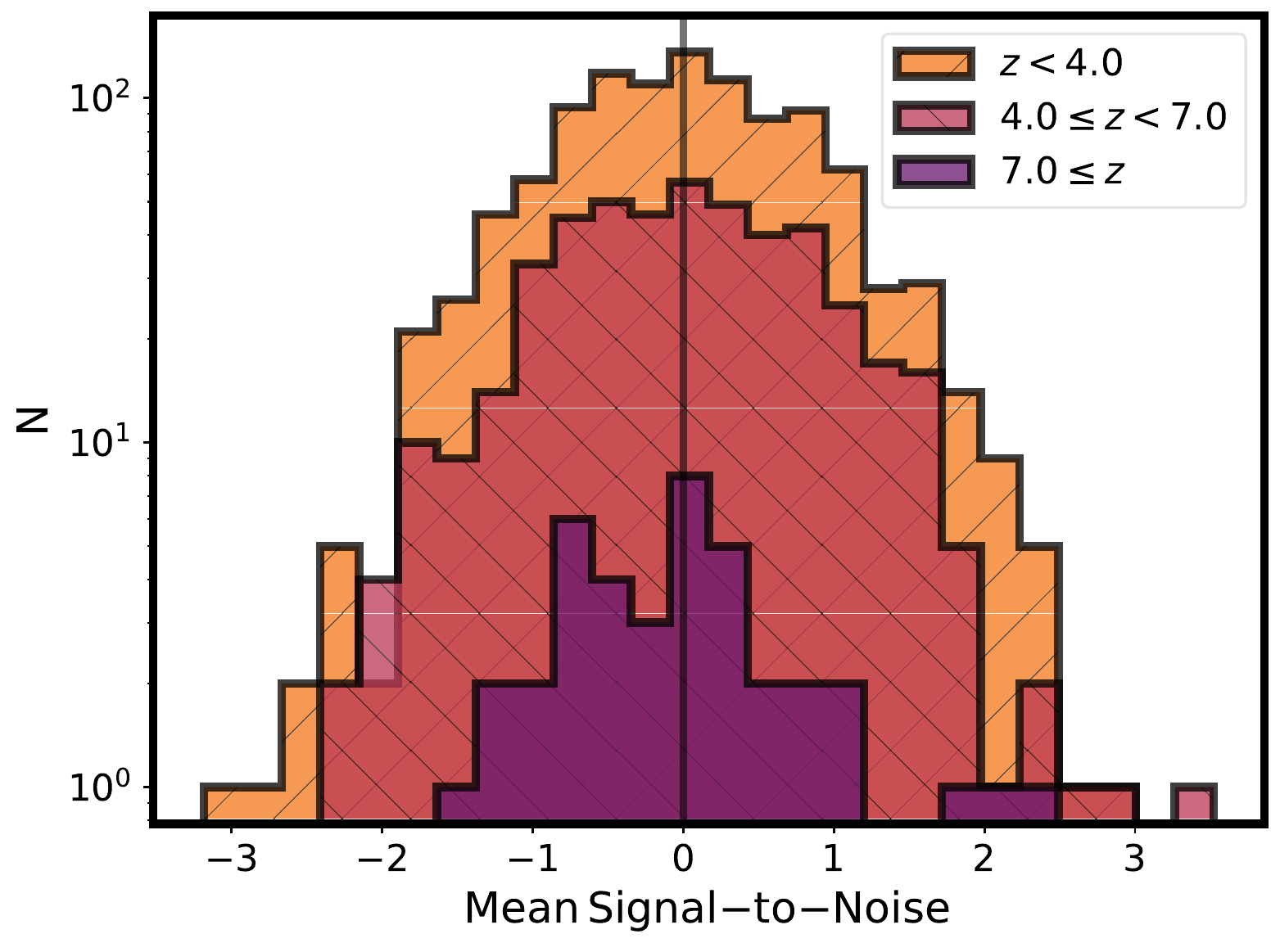}
    	\includegraphics[width=.495\textwidth]{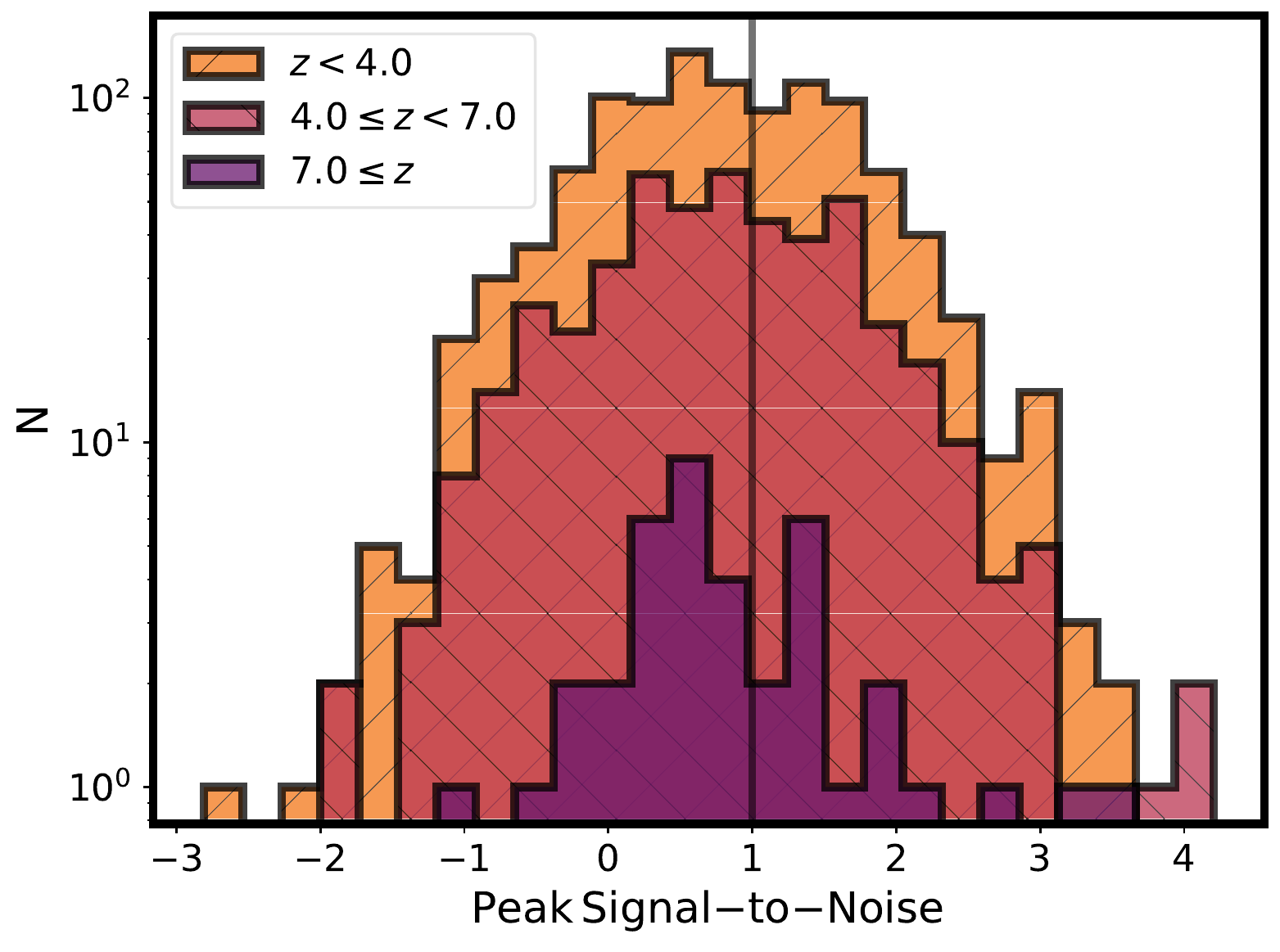}
    	\caption{Mean ({\it left}) and peak ({\it right}) signal-to-noise ratios (S/N) for our candidates in the ALMA maps. The LBG candidates are separated into three photometric redshift sub-samples, represented by distinct colors. The mean value is centered around S/N$\sim$0 (vertical dark line) and is roughly Gaussian. The peak values are centered around S/N$\sim$1, rather than S/N$\sim$0 (vertical dark line, due to the selection of the peak pixel which arises within half a beamwidth; this conservatively biases the maximum flux associated with a candidate to higher values.}
    	\label{fig:hist_alma_snr}
	\end{figure*}

	The mean and peak S/N distributions for the $1582$ LBG candidates are shown in  Fig.~\ref{fig:hist_alma_snr}. As already mentioned in $\S$\ref{subsec:ALMA_up_lims}, all our targets exhibit S/N values lower than $|{\pm}5.0|$. The mean S/N distributions for each redshift bin are centered around $\sim$0 as expected, while the peak S/N distributions are centered around $\sim$1 as a result of selecting the peak pixel which arises within half a beamwidth; this conservatively biases the maximum flux associated with a candidate to higher values. Both distributions appear roughly Gaussian.
    
    From our sample, we find two ($2$) candidates with $\mathrm{S/N}_{\mathrm{peak}}^{\mathrm{indiv}} {>} 4.1$ (see Table \ref{tab:detect_props}). Based on the results from $\S$\ref{subsec:ALMA_up_lims}, we expect $\approx$0.3 candidates to be false positives at this $\mathrm{S/N}$ ($\rho_{\mathrm{FDR}} {=} 0.15$) and, thus, consider the two detections to be real.
    
\subsubsection{UV and IR luminosities}

\begin{figure}
    	\centering
    	\includegraphics[width=1.\columnwidth]{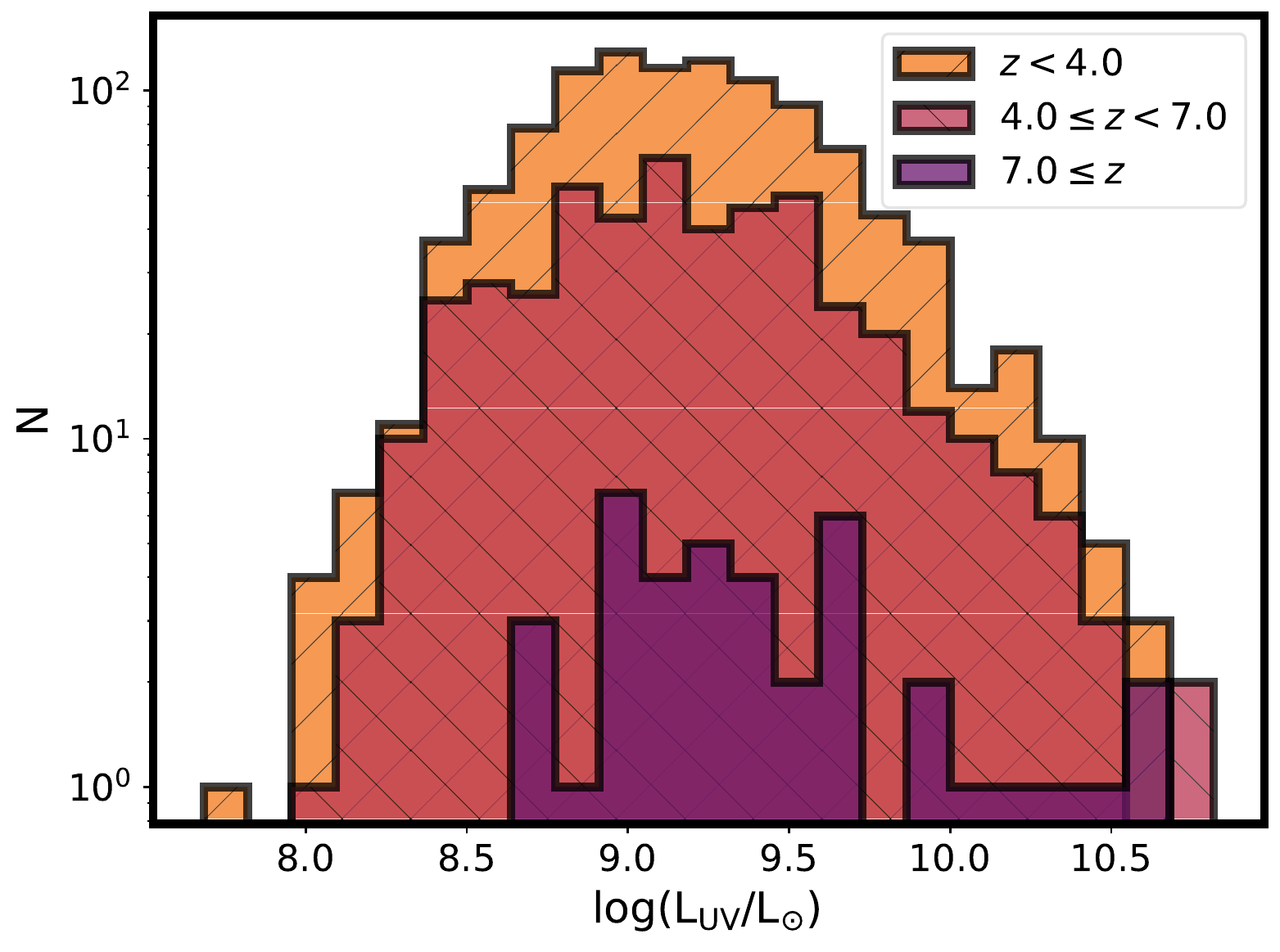}
    	\caption{UV luminosities in our sample. The LBG candidates are separated into three photometric redshift sub-samples, represented by distinct colors.}
    	\label{fig:hist_UV_lums}
	\end{figure}

\begin{figure}
    	\centering
    	\includegraphics[width=1.\columnwidth]{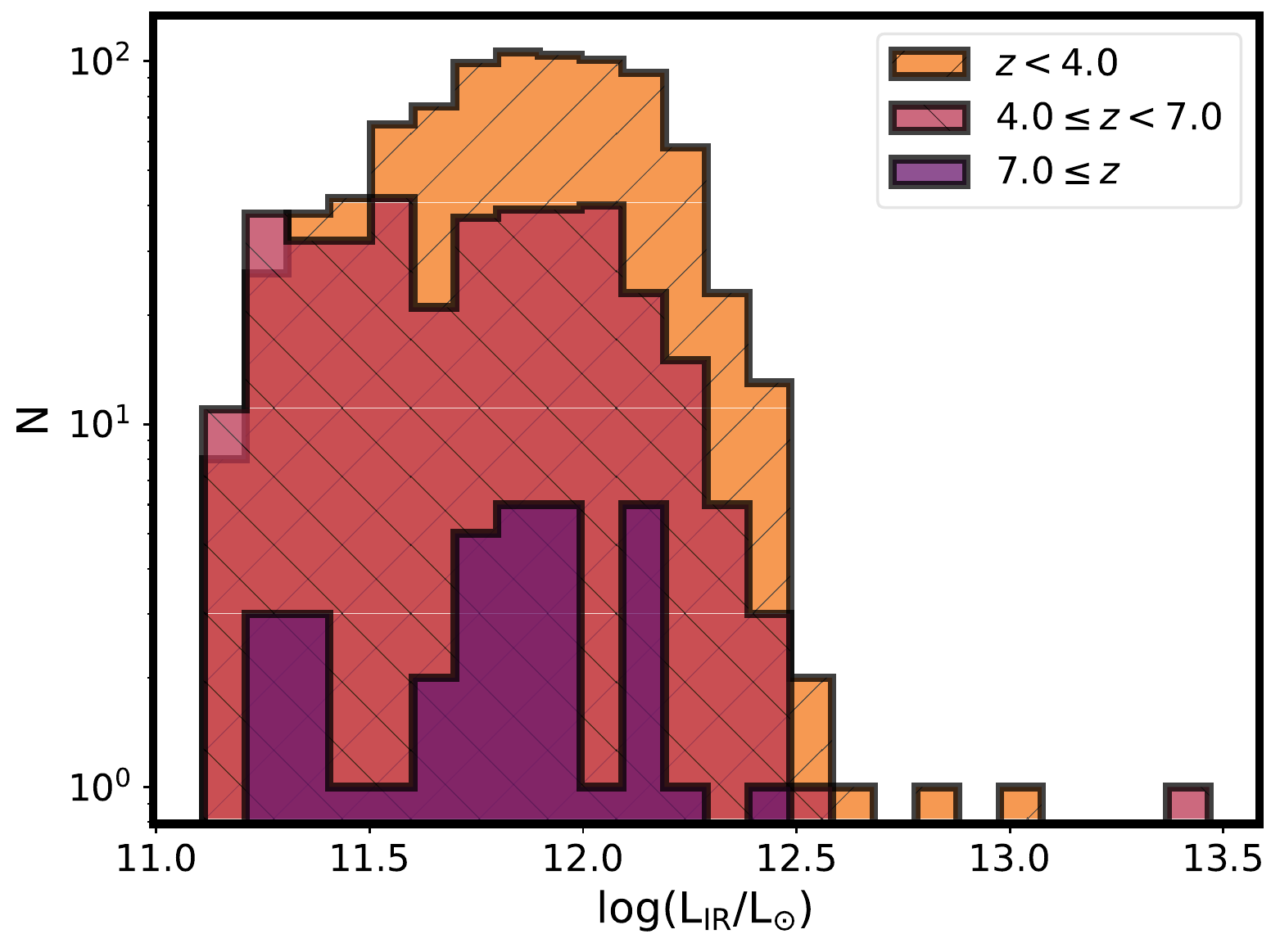}
    	\caption{IR luminosities ($3{-}\sigma$ upper limits from our ALMA maps) in our sample. The LBG candidates are separated into three photometric redshift sub-samples, represented by distinct colors.}
    	\label{fig:hist_IR_lums}
	\end{figure}

	Following the steps described in $\S$\ref{subsec:UVlums} and $\S$\ref{subsec:IRlums}, we utilized {\it HST} photometry to calculate UV luminosities for each LBG candidate and a graybody SED to calculate the IR luminosities, re-scaled by the individual ALMA peak fluxes. The vast majority of the latter are upper limits. The distributions of the individual UV and IR luminosities ($3{-}\sigma$ upper limits) are shown in Figs.~\ref{fig:hist_UV_lums} and \ref{fig:hist_IR_lums}, respectively.
    
    The magnification-corrected observed UV luminosity $3 {-} \sigma$ upper limits of the LBG candidates span a range from ${\sim}10^{7.8}$--$10^{10.8}\,\mathrm{L_{\odot}}$, effectively probing apparent SFRs between ${\sim}$0.02--20\,$M_{\odot}$\,yr$^{-1}$ \citep[e.g.,][]{2013seg..book..419C}. We see a peak at around ${\sim}10^{9}\,\mathrm{L_{\odot}}$ for the two lower redshift bins ($z {<} 4$ and $4 {\leq} z {<} 7$), while we see a relatively flat distribution between ${\sim}10^{8.5}$--$10^{10.5}\, \mathrm{L_{\odot}}$ for the higher redshift bin. In general, the UV luminosities probed here are lower than the values presented in other works \citep[e.g.,][]{2017arXiv170505858N, 2017arXiv170509302R}.
	
	The magnification-corrected IR luminosity limits of the LBG candidates exhibit a somewhat different behavior from the UV luminosities. Due to the nature of the K-correction on the long-wavelength side of the graybody SED, high redshifts probe somewhat lower IR luminosities. Specifically, we find that the $z {<} 4$, $4 {\leq} z {<} 7$, and $z {\ge} 7$ bins are centered around values of ${\sim}10^{11.9}$, ${\sim}10^{11.7}$, and ${\sim}10^{11.7}$\, $\mathrm{L_{\odot}}$, respectively. Given our imposed maximum magnification of 10, coupled with the relatively uniform $rms$ limits, we see that each photometric redshift subsample spans roughly 1.5 dex in luminosity (without accounting for outliers). Thus, all our redshift bins probe IR luminosity limits of ${\sim}10^{11.1}$--$10^{12.5}$\,$\mathrm{L_{\odot}}$, or equivalently 20--400 $M_{\odot}$\,yr$^{-1}$ \citep[e.g.,][]{2011ApJ...741..124H, 2013seg..book..419C}. 
    
    Comparing the UV and IR luminosity limits, it is clear that the UV data generally probes to much lower effective SFRs. Thus, our current individual ALMA constraints are only able to rule out the possibility of rather extreme obscured star formation events associated with any of the LBG candidates. 
    
    The two detected LBGs have UV and IR luminosities in the range $L_{\mathrm{UV}}{\sim}10^{8.7}$--$10^{8.8}$\, $\mathrm{L_{\odot}}$ and $L_{\mathrm{IR}}{\sim}10^{11.3}$--$10^{11.9}$\,$\mathrm{L_{\odot}}$, respectively. Relating these in terms of SFRs, the detected LBGs have $\sim$2--3 dex more obscured than unobscured star formation present.

\subsubsection{IRX-$\beta$ relation}\label{subsec:IRXbeta}

\begin{figure}
    \centering
    \includegraphics[scale=0.67]{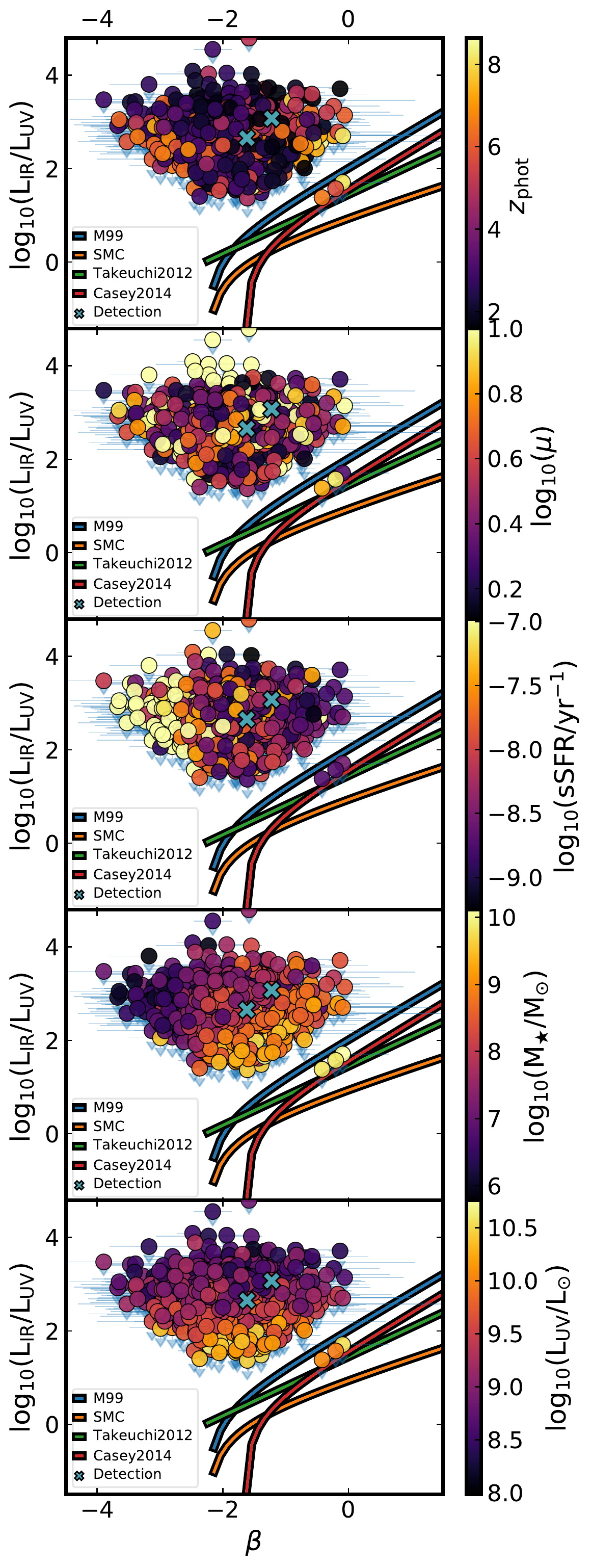}
    \caption{Comparison of nfrared xcess (IRX) $3{-}\sigma$ upper limits and UV-slopes ($\beta$) for our LBG candidates. Downward arrows have 1-$\sigma$ length. From top to bottom panels, colors represent: photometric redshift ($z_{ph}$), magnification factor ($\mu$), star formation rate (SFR), stellar mass (M$\mathrm{_{\bigstar}}$) and UV Luminosity (L$\mathrm{_{\rm UV}}$). Local IRX-$\beta$ relations presented in $\S$\ref{subsec:IRXrelations} are shown for reference. Blue crosses represent the two detections.} \label{fig:irx_vs_beta}
	\end{figure}

	With the UV and IR luminosities in hand, we can compare IRX limits to the UV-slope $\beta$, as shown in Fig.~\ref{fig:irx_vs_beta}. We color-code the LBG candidates as functions of redshift, magnification, sSFR, $M_{\bigstar}$, and $L_{\mathrm{UV}}$, as well as show the local IRX-$\beta$ relations presented in $\S$\ref{subsec:IRXrelations}.  
	
	The main trends we see in the IRX-$\beta$ diagram are with the \texttt{FAST}-derived quantities SFR and $M_{\bigstar}$ (third and fourth panels), where stronger upper limits tend to lie to the lower right, closer to the local relations (and weaker limits tend to lie further away from local relations). This is due in part to observation bias, coupled with the $M_{\bigstar}$-SFR main sequence relation. We detect LBG candidates spanning $\sim$3\,dex in m$_{\rm UV}$ or $L\mathrm{_{UV}}$ (bottom panel), while our IR limits only span 1\,dex. Thus, the highest $M_{\bigstar}$-SFR sources have the lowest IRX limits, and the lowest ones have the highest IRX limits. This trend extends into the $z_{ph}$ and $\mu$ panels with lower redshift and higher $\mu$ sources (i.e., lower $L\mathrm{_{UV}}$ candidates) having higher IRX limits. There appears to be a mild intrinsic trend between higher (redder) $\beta$ values and higher $M_{\bigstar}$ (see $\S$\ref{subsubsec:Mstar_beta_corr} for further details).
	
    While the vast majority of limits lie above the local relations, we find $3$ LBG candidates located completely below at least one relation. Given the dispersion in these local relations, however, all we can say is that our individual limits remain consistent with the relations.  

\subsubsection{IRX-$M_{\bigstar}$ relation}\label{subsec:IRXMstar}
    
\begin{figure}
    \centering
    \includegraphics[scale=0.67]{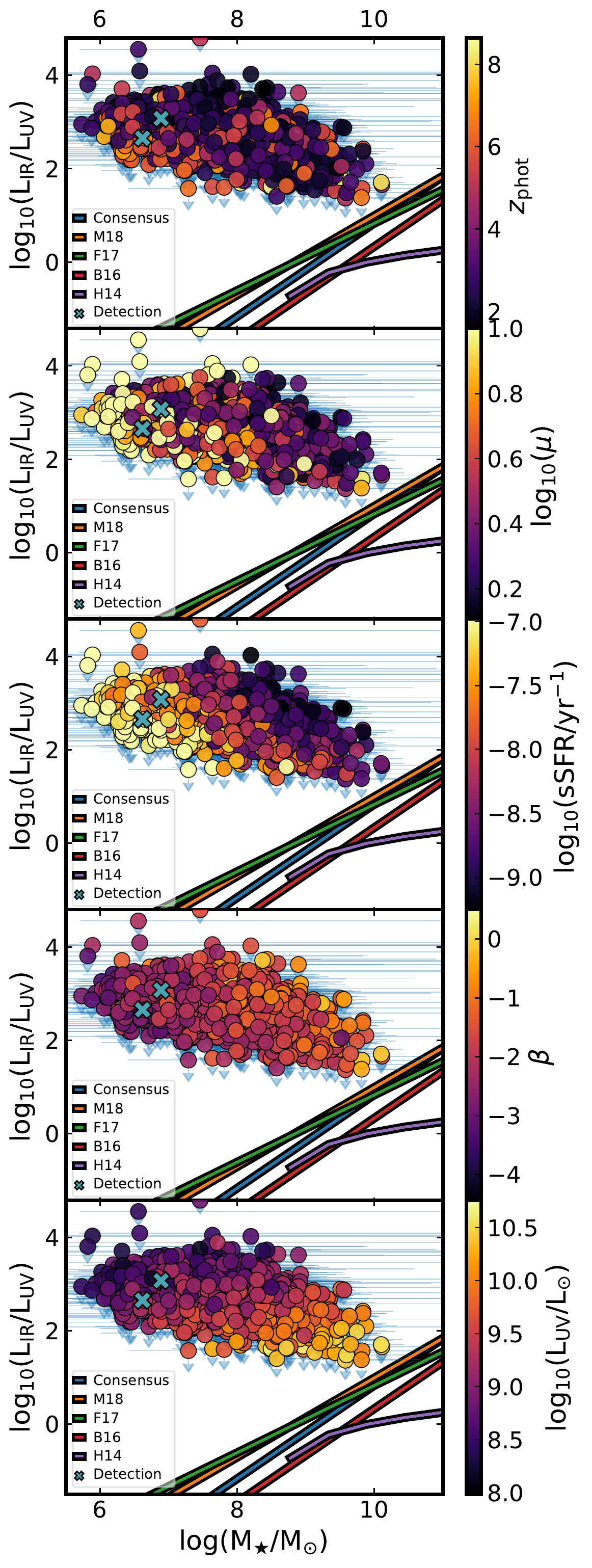}
    \caption{Comparison of infrared excess (IRX) $3{-}\sigma$ upper limits and stellar masses $M_{\bigstar}$ for our LBG candidates. Downward arrows have 1-$\sigma$ length. From top to bottom panels, colors represent: photometric redshift ($z_{ph}$), magnification factor ($\mu$), star formation rate (SFR), UV slope ($\beta$) and UV Luminosity (L$\mathrm{_{UV}}$). Local IRX-$M_{\bigstar}$ relations presented in $\S$\ref{subsec:IRXrelations} are shown for reference. Blue crosses represent the two detections.}\label{fig:irx_vs_smass}
	\end{figure}

	We can also compare the IRX limits and stellar masses $M_{\bigstar}$ of our LBG candidates, depicted in Fig.~\ref{fig:irx_vs_smass}. Again, we color-code the LBG candidates as functions of redshift, magnification, sSFR, $\beta$, and $L\mathrm{_{UV}}$, and show several IRX-$M_{\bigstar}$ relations from $\S$\ref{subsec:IRXrelations}.
    
	We see a number of trends in the IRX-$M_{\bigstar}$ diagram as functions of $\mu$ (second panel), sSFR (third panel), $\beta$ (fourth panel), and $L\mathrm{_{UV}}$ (fifth panel). Unsurprisingly, higher magnifications allow us to probe lower stellar masses. $M_{\bigstar}$ is related to sSFR and $L\mathrm{_{UV}}$ following the star-formation main sequence. Here we now see more clearly a $M_{\bigstar}$ and $\beta$ trend, such that more massive systems (which have built up more metals and dust) tend to show higher extinction.
	
	In this case, unlike the IRX-$\beta$ trends, all of our $3 {-} \sigma$ upper limits lie completely above the relations. Factoring in the dispersion in these relations, our individual limits remain consistent with the relations.  The massive and luminous LBG candidates that lie closest to the relations all have high ($z \ga 5$) photometric redshifts and low magnifications and, hence, comprise the rare, bright end of the high-z population.

\subsection{Stacking results}\label{subsec:Stack_results}

	To gain further insights into the LBG population, we used \texttt{STACKER} to perform $u$--$v$ stacking on all five ALMA cluster datasets. Some tests were applied to the stacking method and their details are presented in Appendix~\ref{app0}.
	
	Importantly, our tests demonstrate the capabilities of \texttt{STACKER} in substantially reducing the noise levels compared to the nominal natural-weight \textit{CLEAN}ing $rms$ (e.g., from $55 \mu$Jy - $90 \mu$Jy to stacked $rms$ errors as low as 2\,$\mu$Jy, which is close to the theoretical limit). Comparable results are achieved with image stacking, and give us confidence in the LBG stacking results presented below. 
	 
\begin{figure*}[htb]
		\centering
		\begin{subfigure}[b]{.245\linewidth}
			\centering
			\includegraphics[width=1.\textwidth]{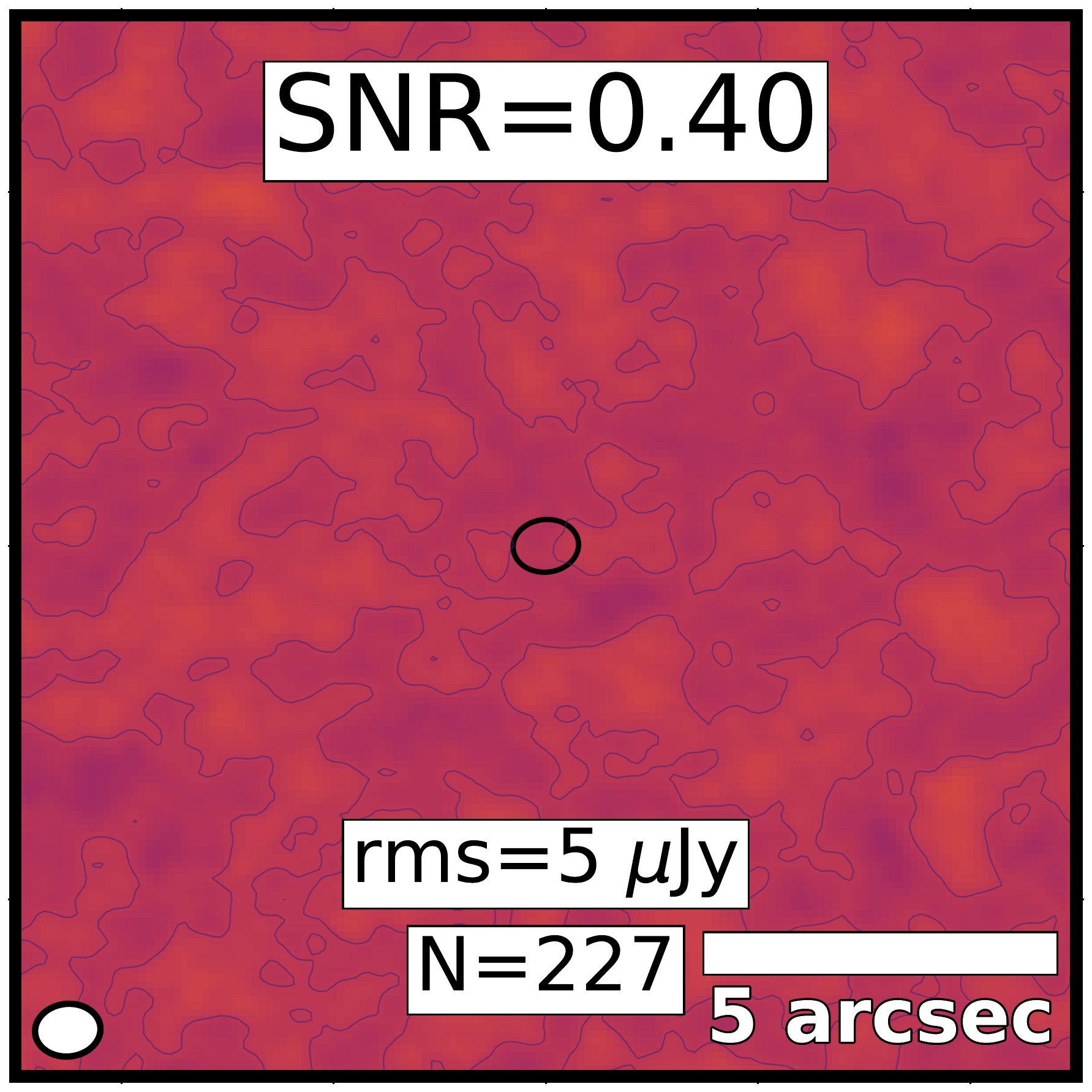}
  		\end{subfigure}%
		\begin{subfigure}[b]{.245\linewidth}
			\centering
			\includegraphics[width=1.\textwidth]{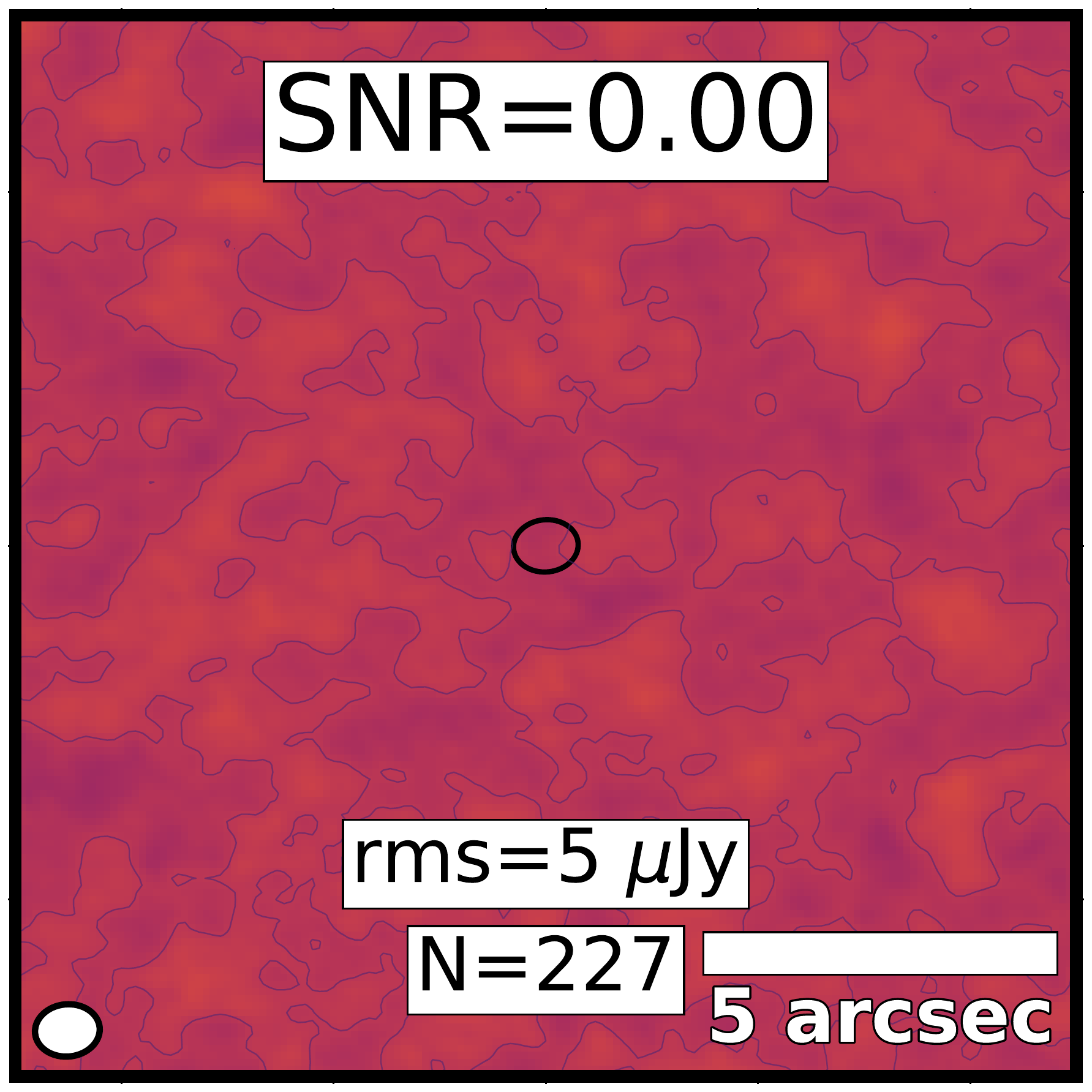}
		\end{subfigure}%
		\begin{subfigure}[b]{.245\linewidth}
			\centering
			\includegraphics[width=1.\textwidth]{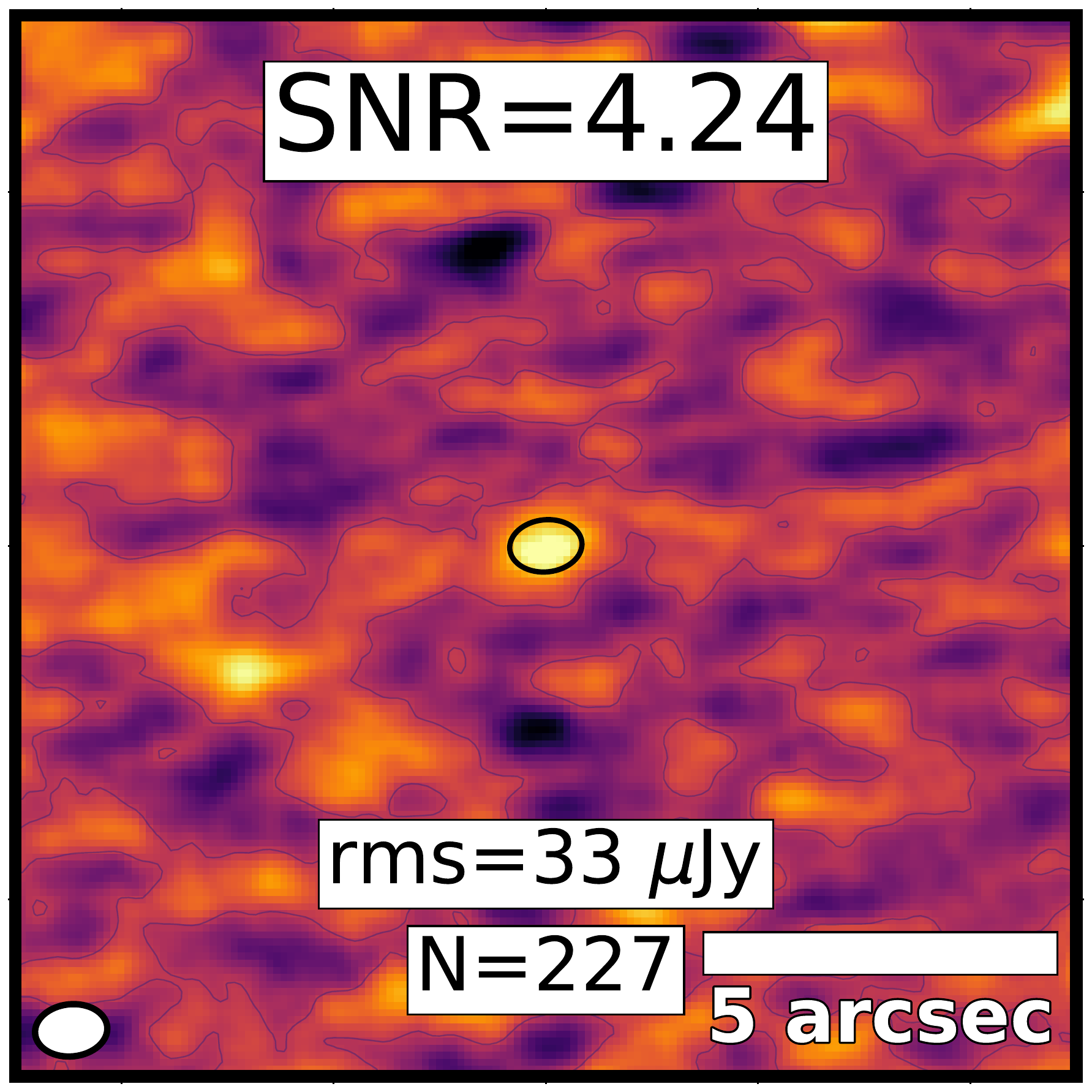}
  		\end{subfigure}%
		\begin{subfigure}[b]{.245\linewidth}
			\centering
			\includegraphics[width=1.\textwidth]{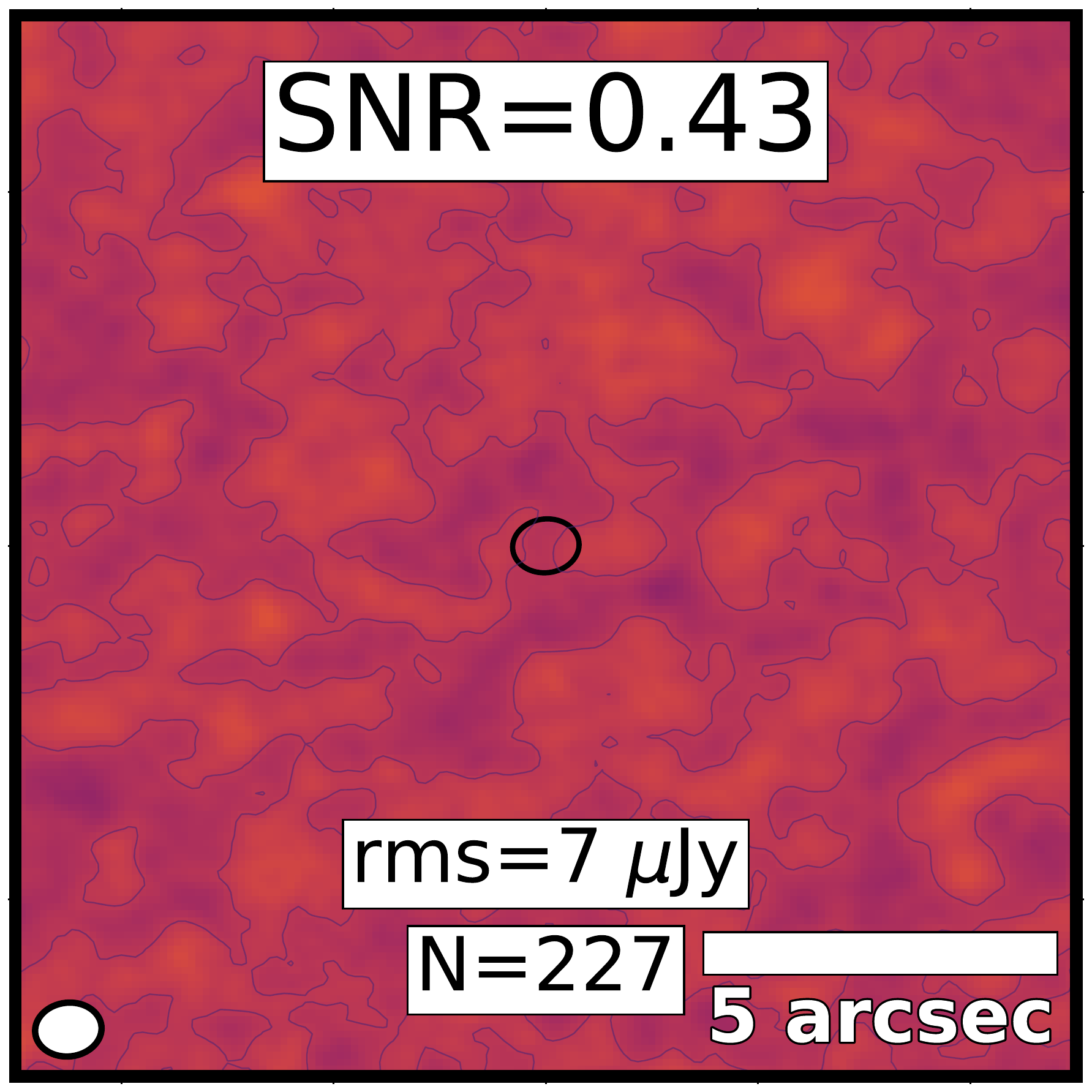}
		\end{subfigure}\\%
		\begin{subfigure}[b]{.245\linewidth}
			\centering
			\includegraphics[width=1.\textwidth]{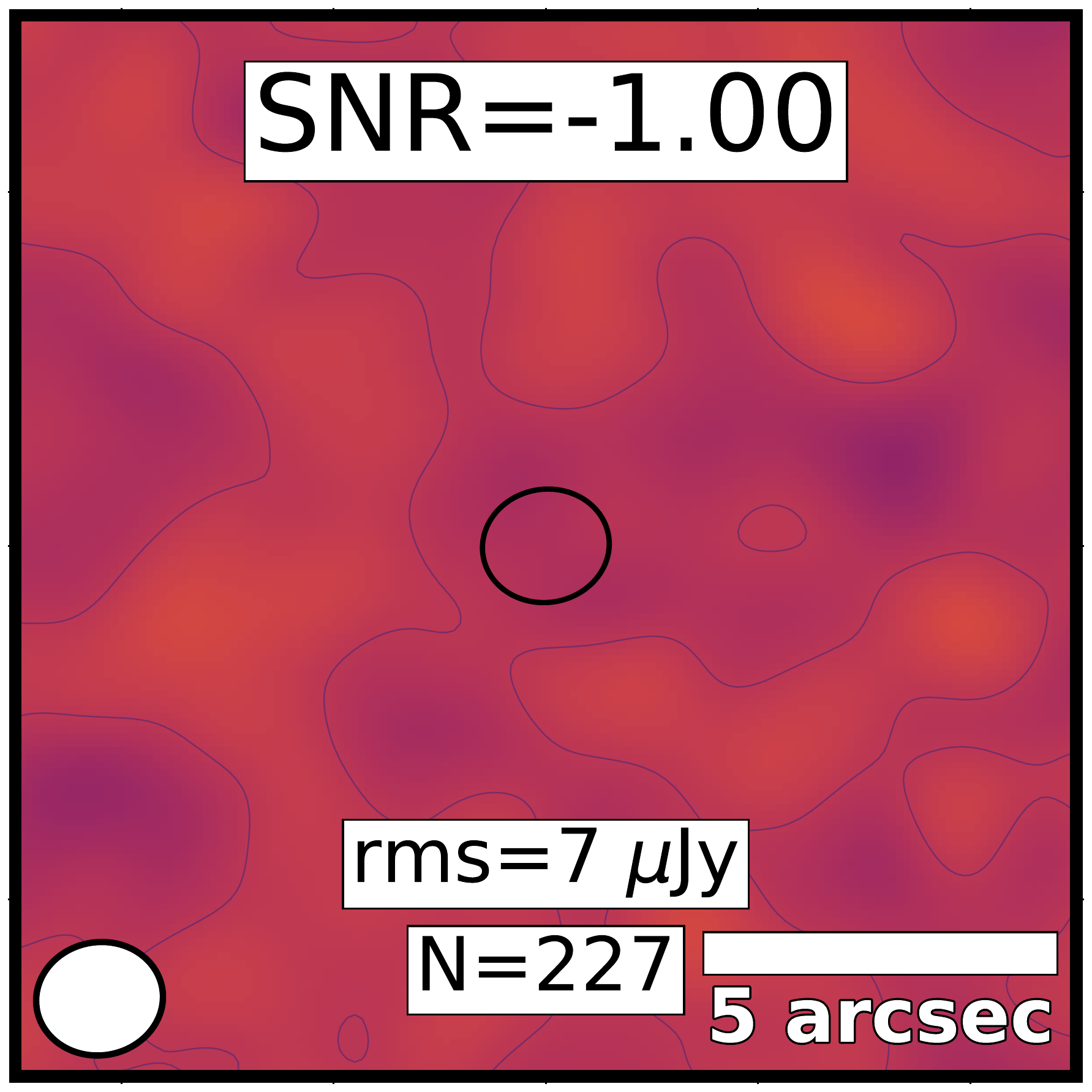}
			\caption{Weight: equal}\label{fig:beta_nw}
		\end{subfigure}%
		\begin{subfigure}[b]{.245\linewidth}
			\centering
			\includegraphics[width=1.\textwidth]{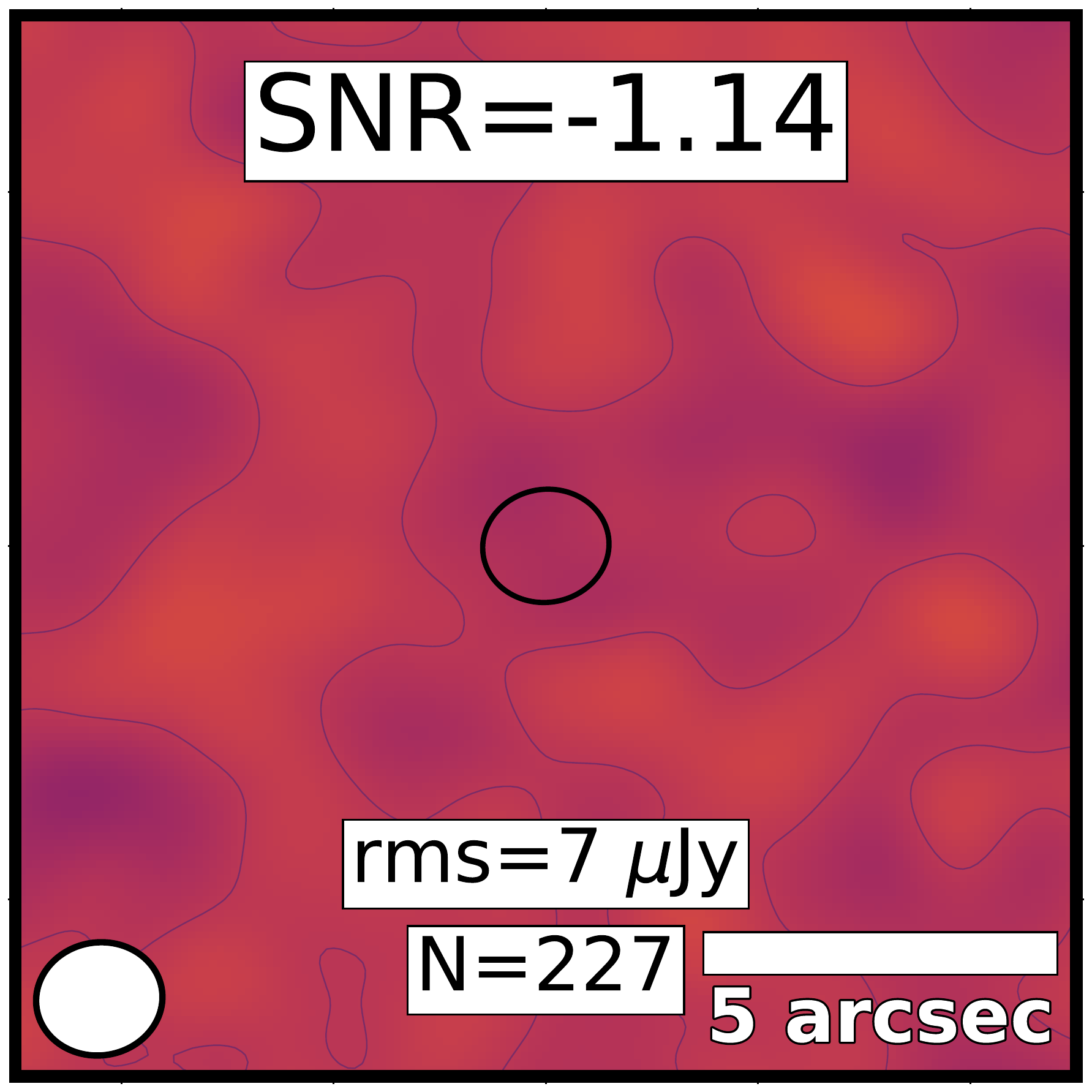}
			\caption{Weight: $pbcor$} \label{fig:beta_pb}
		\end{subfigure}%
		\begin{subfigure}[b]{.245\linewidth}
			\centering
			\includegraphics[width=1.\textwidth]{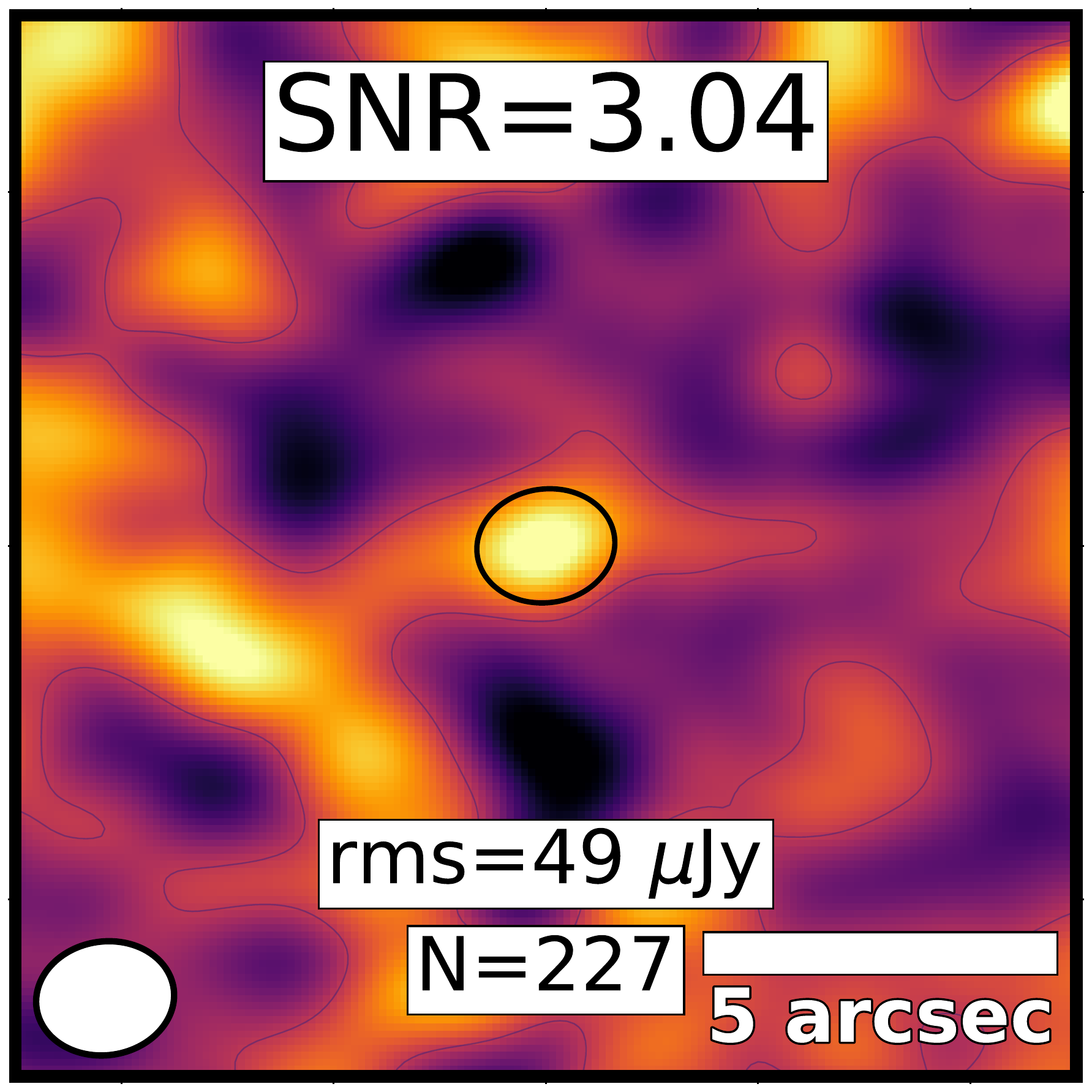}
			\caption{Weight: $F_{\rm UV}$}\label{fig:beta_fuv}
		\end{subfigure}%
		\begin{subfigure}[b]{.245\linewidth}
			\centering
			\includegraphics[width=1.\textwidth]{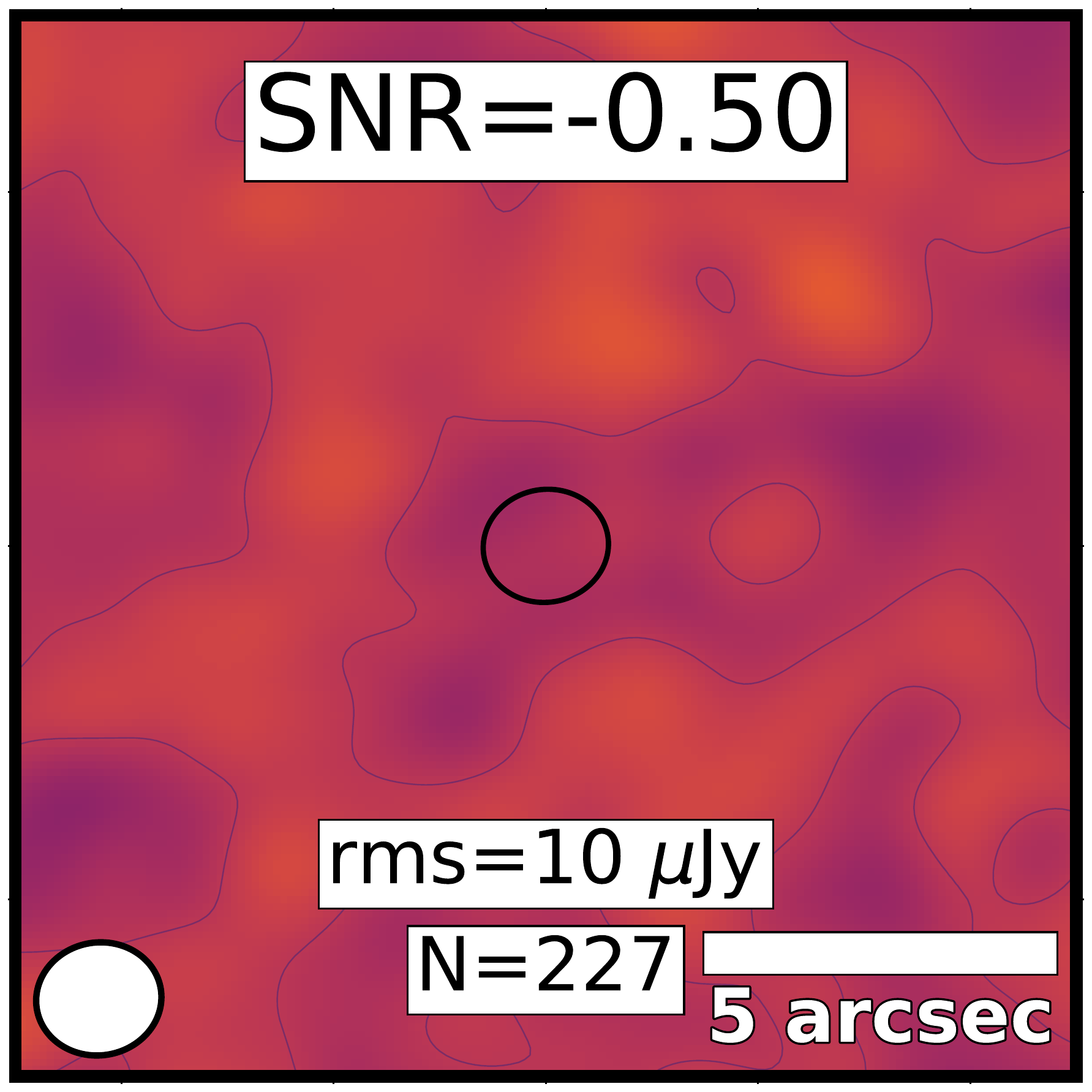}
			\caption{Weight: $\mu$} \label{fig:beta_mag}
		\end{subfigure}%
		\caption{Example $u$--$v$ stacked image stamps for $227$ undetected LBG candidates in the range of ${-}2.0 {\leq} \beta {<} {-}1.0$ and $4.0 {\leq} z {<} 7.0$. Details same as Fig.~\ref{fig:stack_FFI_uv}. Color scale spans $-$125\,$\mu$Jy to $+$125\,$\mu$Jy range.}\label{fig:beta_stack_example}
	\end{figure*}

\begin{figure*}[htb]
		\centering
		\begin{subfigure}[b]{.245\linewidth}
			\centering
		    \includegraphics[width=1.\textwidth]{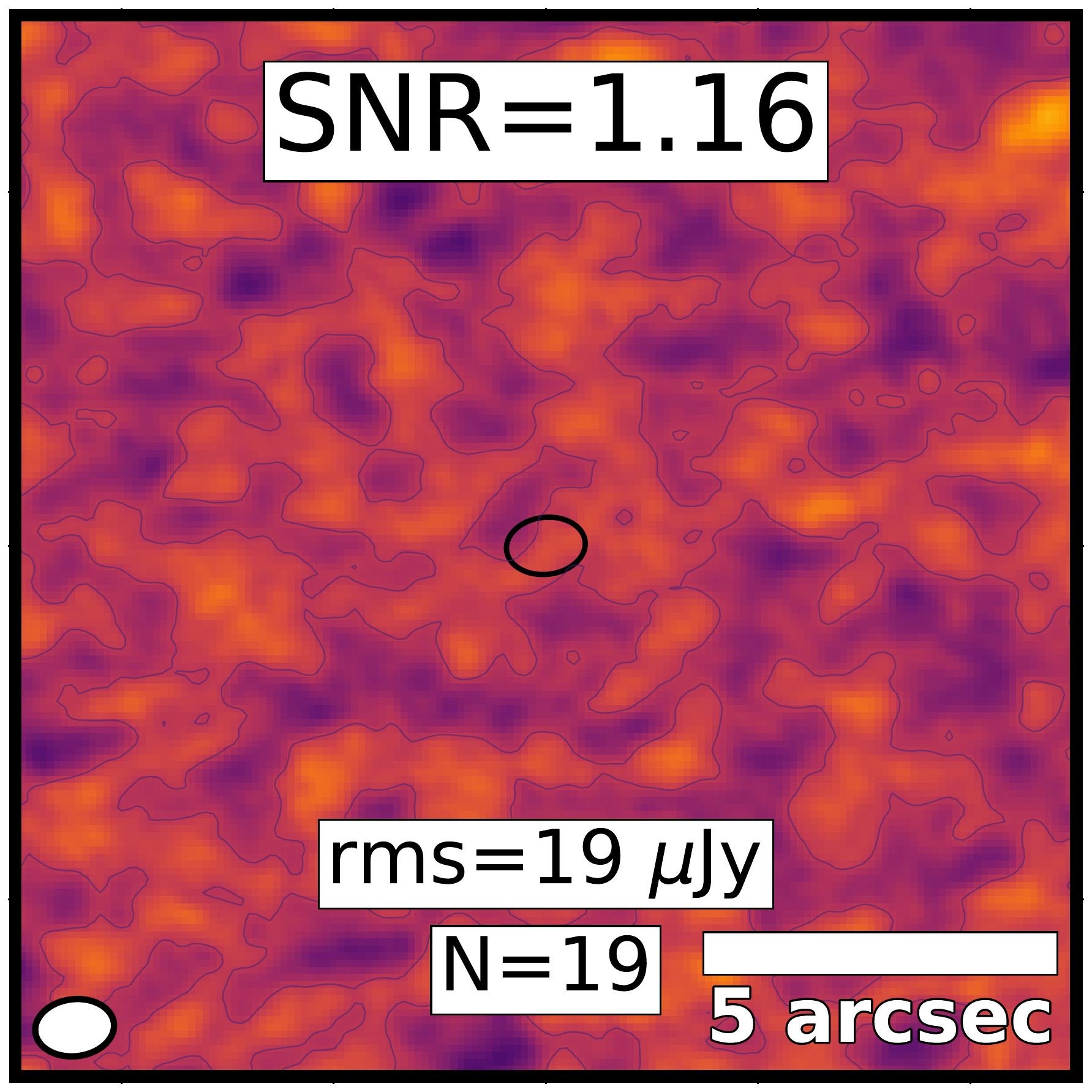}
  		\end{subfigure}%
		\begin{subfigure}[b]{.245\linewidth}
			\centering
			\includegraphics[width=1.\textwidth]{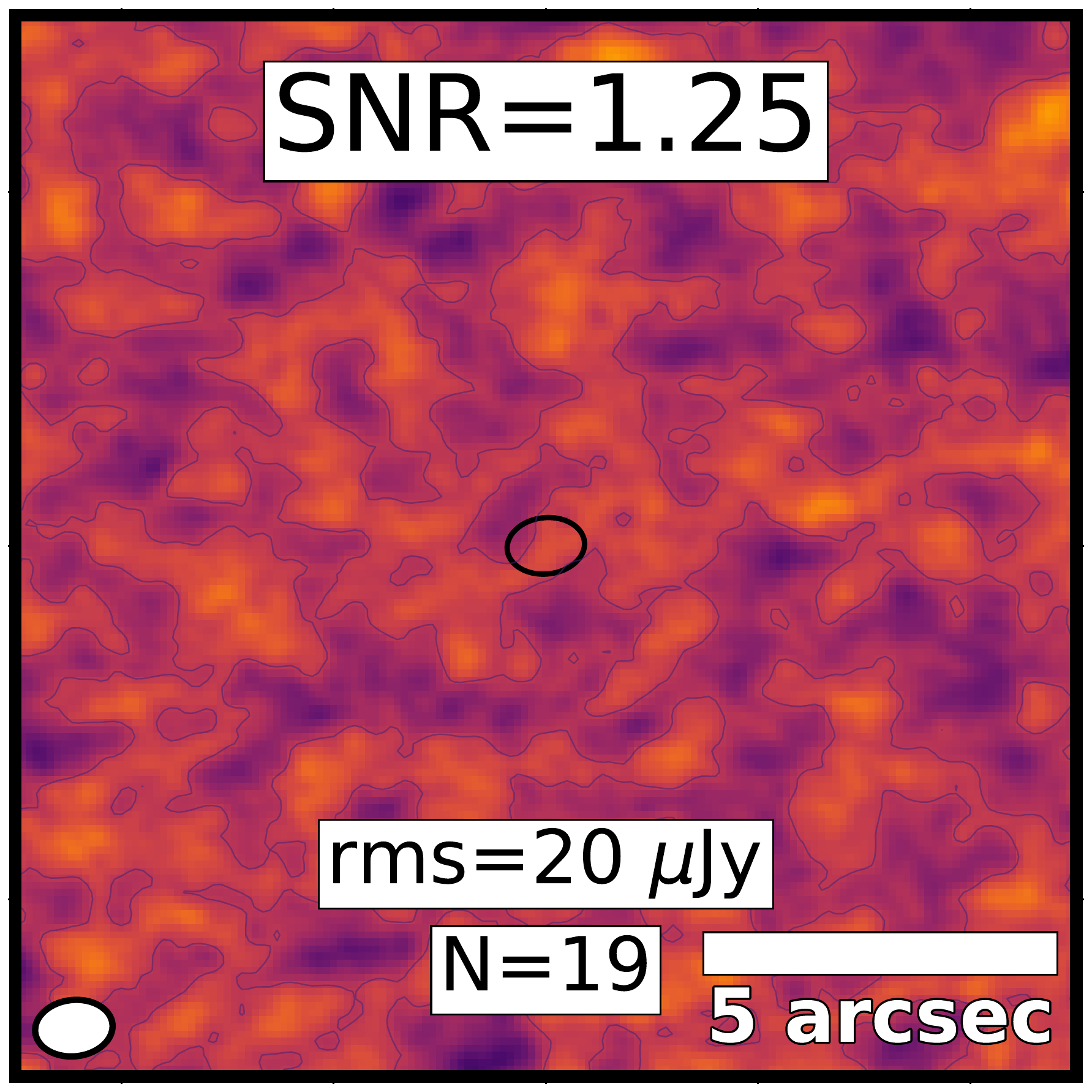}
		\end{subfigure}%
		\begin{subfigure}[b]{.245\linewidth}
			\centering
			\includegraphics[width=1.\textwidth]{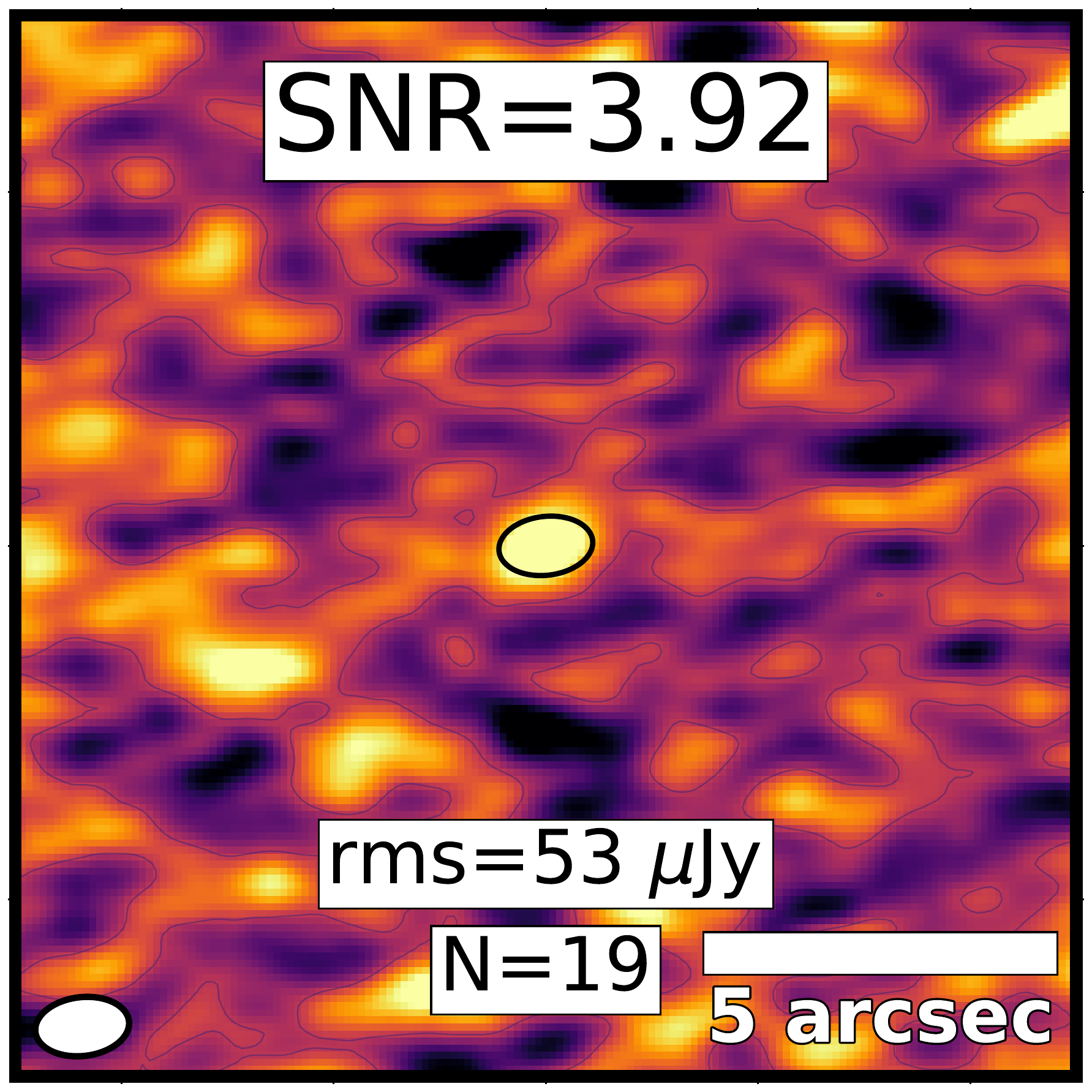}
  		\end{subfigure}%
		\begin{subfigure}[b]{.245\linewidth}
			\centering
			\includegraphics[width=1.\textwidth]{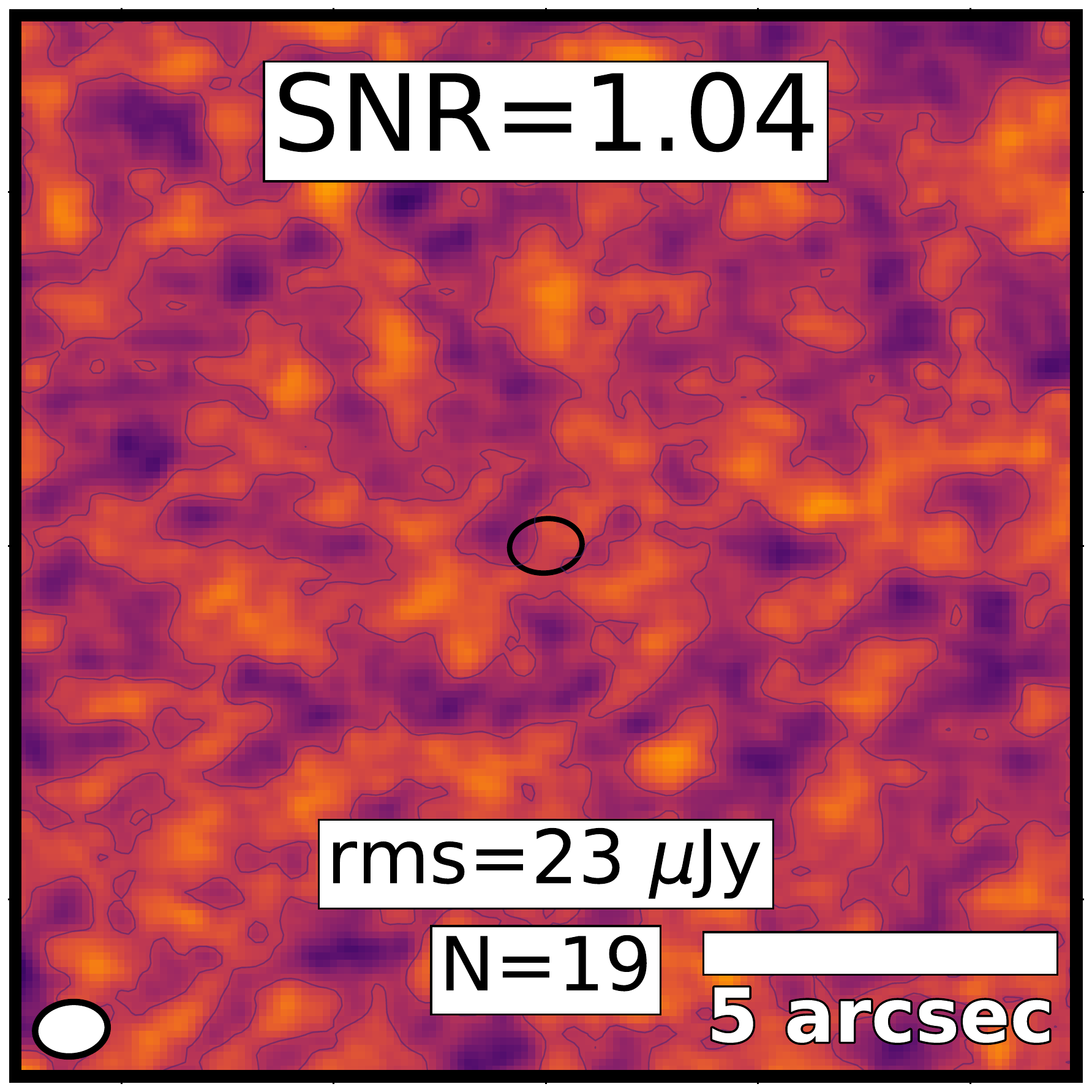}
		\end{subfigure}\\%
		\begin{subfigure}[b]{.245\linewidth}
			\centering
			\includegraphics[width=1.\textwidth]{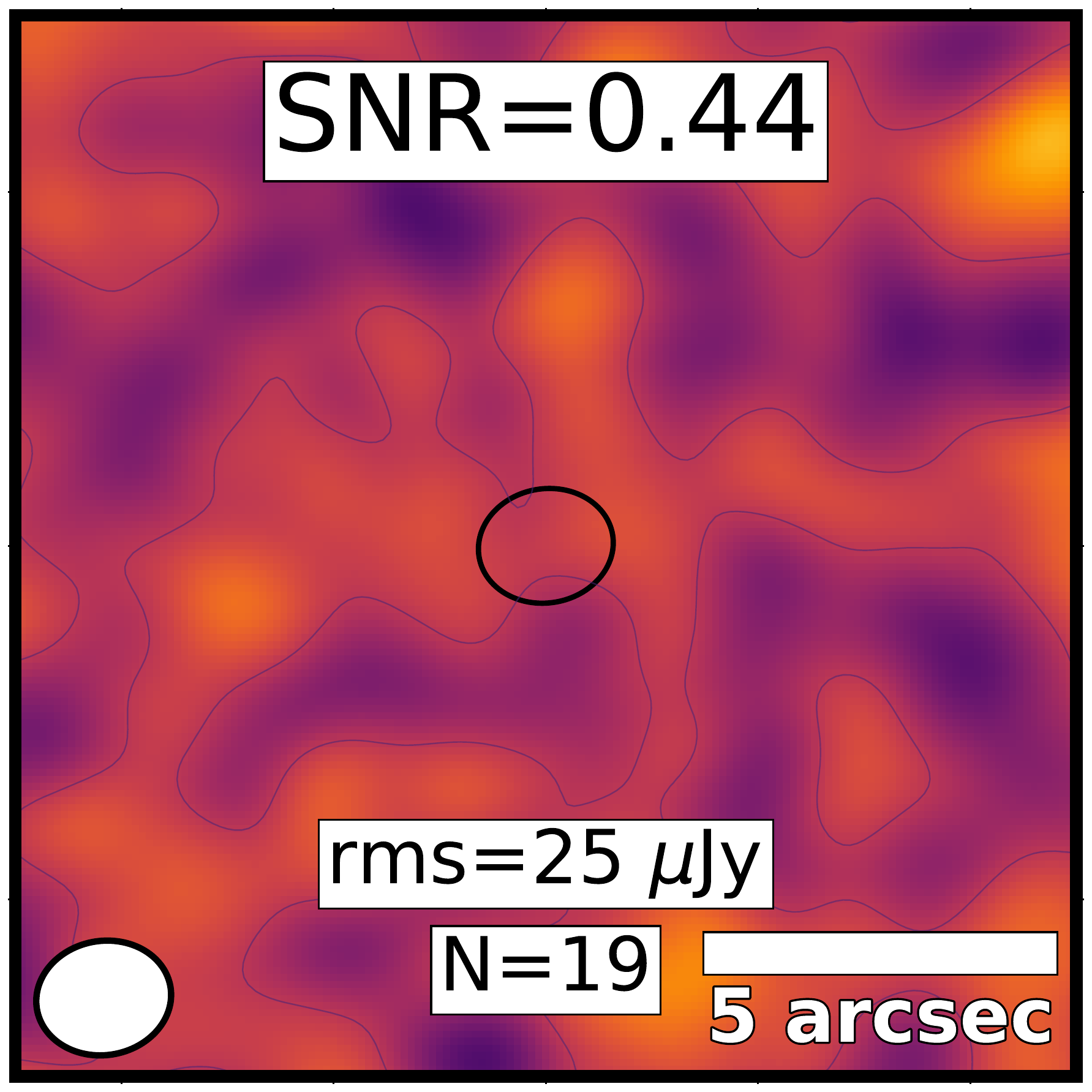}
			\caption{Weight: equal}\label{fig:mass_nw}
		\end{subfigure}%
		\begin{subfigure}[b]{.245\linewidth}
			\centering
			\includegraphics[width=1.\textwidth]{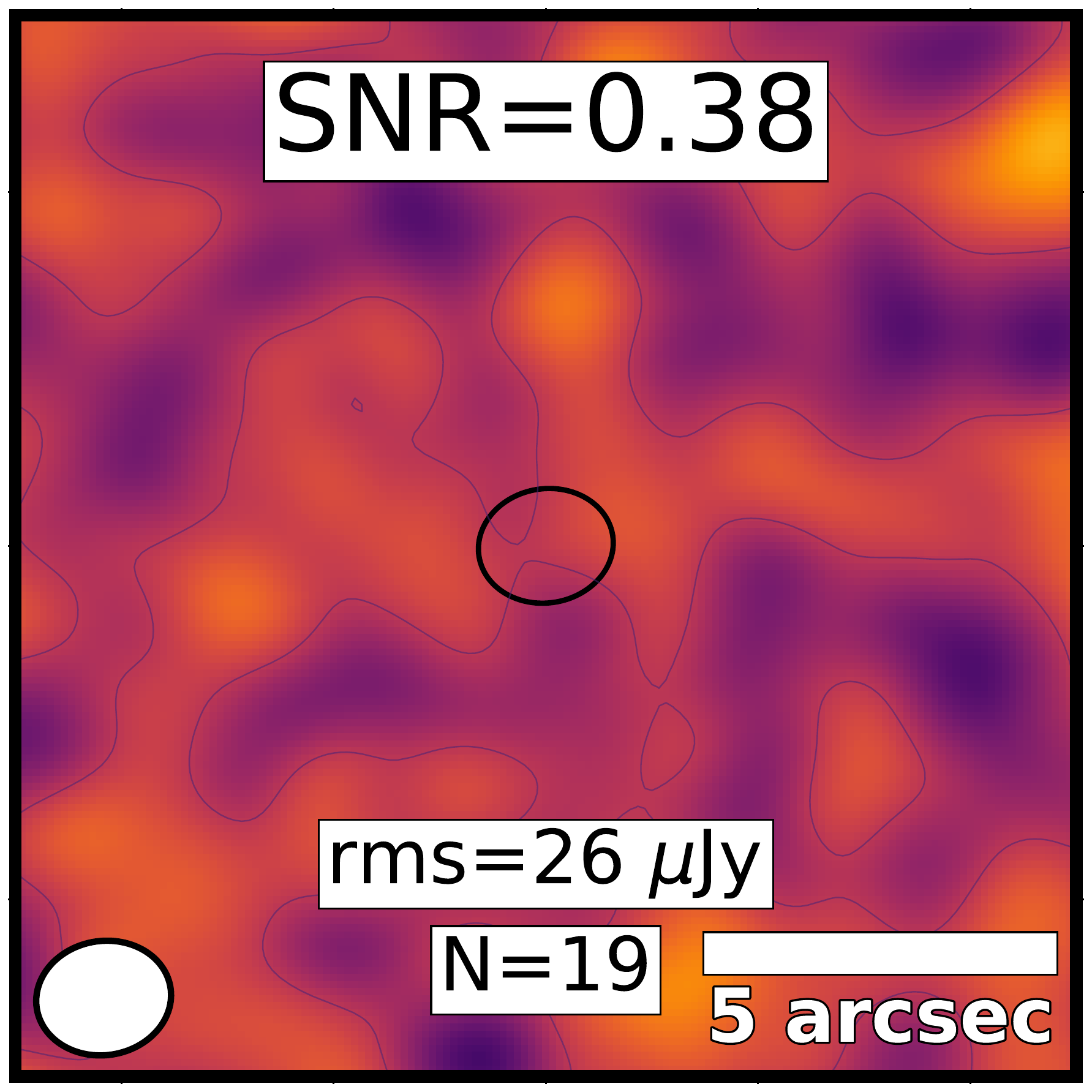}
			\caption{Weight: $pbcor$ }\label{fig:mass_pb}
		\end{subfigure}%
		\begin{subfigure}[b]{.245\linewidth}
			\centering
			\includegraphics[width=1.\textwidth]{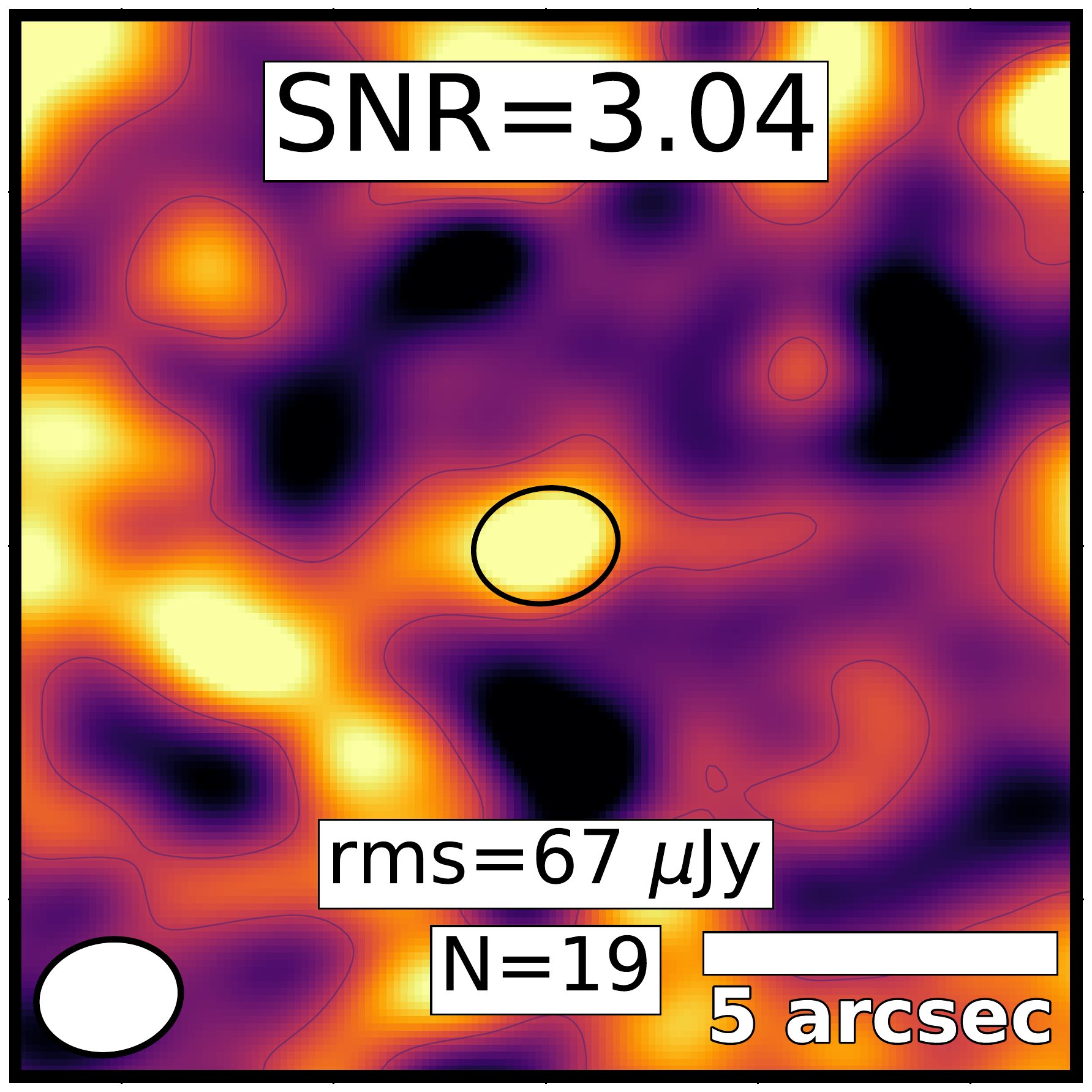}
			\caption{Weight: $F_{\rm UV}$}\label{fig:mass_fuv}
		\end{subfigure}%
		\begin{subfigure}[b]{.245\linewidth}
			\centering
			\includegraphics[width=1.\textwidth]{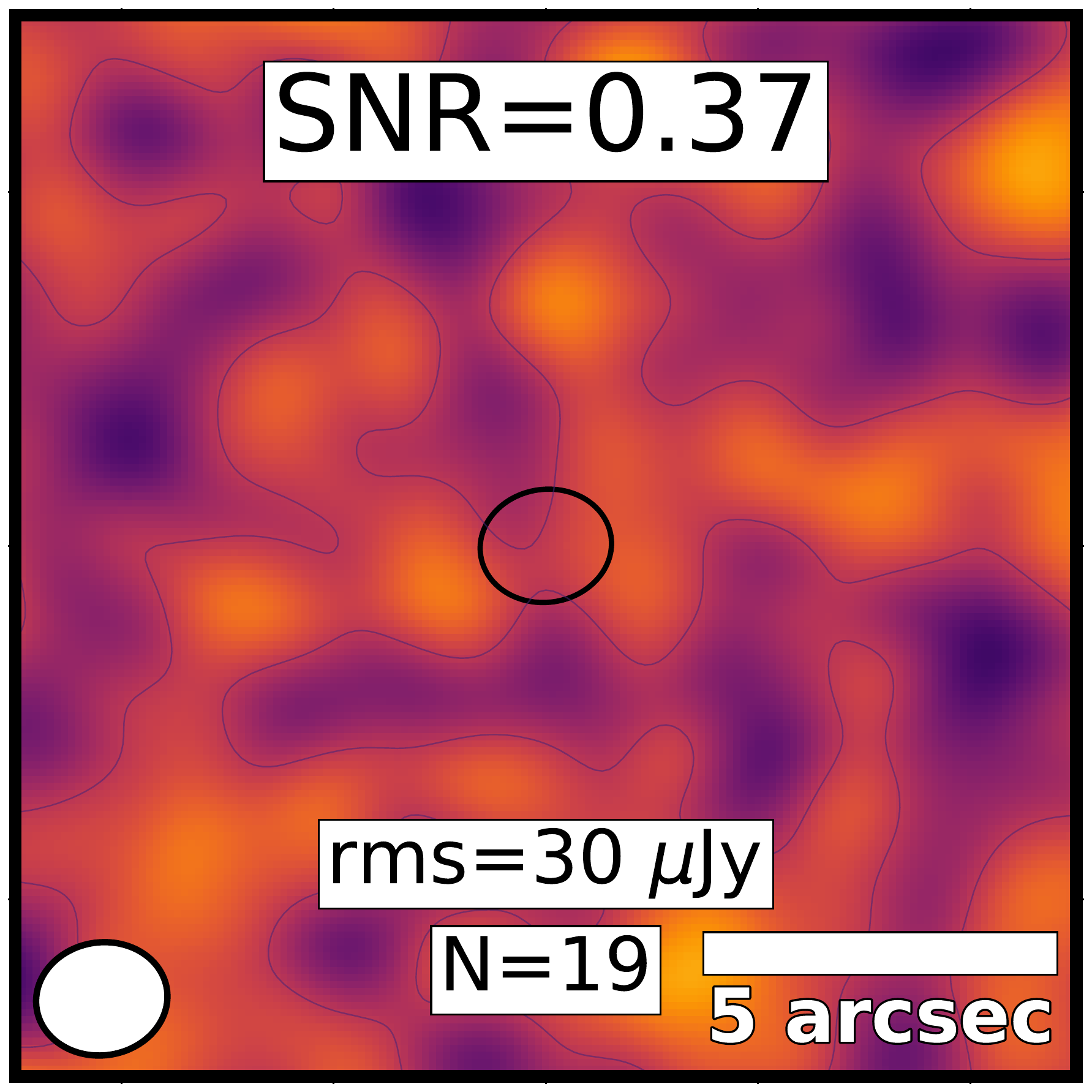}
			\caption{Weight: $\mu$ }\label{fig:mass_mag}
		\end{subfigure}%
		\caption{Example $u$--$v$ stacked image stamps for $19$ undetected LBG candidates in the range of $9.0 {\leq} \log{(M_{\star} / M_{\odot})} {<} 9.5$ and $4.0 {\leq} z {<} 7.0$. Details same as Fig.~\ref{fig:stack_FFI_uv}. Color scale spans $-$125\,$\mu$Jy to $+$125\,$\mu$Jy range.}\label{fig:mass_stack_example}
	\end{figure*}

    From here, we turned to stacking the undetected LBG candidates in the three broad photometric redshift bins as functions of UV-slope binning and stellar mass binning. The $u$--$v$ stacking results are presented in Tables~\ref{tab:stack_results_full_beta_z_bin}, \ref{tab:stack_results_mid_mass_z_bin}, and \ref{tab:stack_results_full_mass_z_bin} of appendices \ref{tab:beta_stack} and \ref{tab:mass_stack}, respectively. Stacked image stamps for two example bins are presented in Fig.~\ref{fig:beta_stack_example} ($4.0 {\leq} z {<} 7.0$ and $-2.0 {\leq} \beta {<} -1.0$) and Fig.~\ref{fig:mass_stack_example} ($4.0 {\leq} z {<} 7.0$ and $9.0 {\leq} \log{(M_{\star} / M_{\odot})} {<} 9.5$). With the large number of undetected LBG candidates in some bins, we achieve stacked $rms$ values as low as $\approx$5\,$\mu$Jy. This highlights the power of stacking to reduce the errors and increase the signal strength (S/N) accordingly by ${\sim} \sqrt{N}$. 
    
    Overall, only one bin among all of the stacks achieves a S/N high enough to be considered a detection ($227$ $F_{\mathrm{UV}}$-weighted sources in the range $4.0 {\leq} z {<} 7.0$ and $-2.0 {\leq} \beta {<} {-}1.0$ with $\mathrm{S/N}_{\mathrm{peak}}^{\mathrm{stack}} {=} 4.24$ for the natural-weight \textit{CLEAN}ing (Fig.~\ref{fig:beta_stack_example}). We treat this result with caution, however, since is not replicated in any other weighting schemes and \textit{CLEAN}ing configurations for the same targets, which yield  $\mathrm{S/N}_{\mathrm{peak}}^{\mathrm{stack}} {=} {-}1.14\, \text{to}\, 3.04$. As seen in Tables~\ref{tab:stack_results_full_beta_z_bin}, \ref{tab:stack_results_mid_mass_z_bin}, and \ref{tab:stack_results_full_mass_z_bin}, there are only a few bins with even marginally significant signal (i.e., S/N${\geq} 3.00$), the highest being $\mathrm{S/N}_{\mathrm{peak}}^{\mathrm{stack}} {=} 3.92$ for $19$ $F_{\mathrm{UV}}$-weighted candidates with stellar masses and reshifts in the ranges $9.0 \leq \log{(M_{\star} / M_{\odot})} < 9.5$ and $4.0 {\leq} z {<} 7.0$ (Fig.~\ref{fig:mass_stack_example}). In general, the equal and $pbcor$ weighting schemes achieve lower $rms$ values in each bin, but the S/N values are modestly higher in some bins with $F_{\mathrm{UV}}$ and $\mu$-weighting, mirroring the results from stacking all sources combined. For instance, when using $F_{\mathrm{UV}}$ ($\mu$) weighting, we find that ${\sim}$4\% (21\%) of the binning configurations with more than one candidate deliver better S/N values than the $pbcor$ or equal weighting cases.
    
    Given past efforts \citepalias[e.g.,][]{2016ApJ...833...72B}, it is somewhat surprising that we do not find significant stacked signal from LBG candidates with stellar masses in excess of or close to $10^{10}M_{\odot}$ (Table \ref{tab:stack_results_full_mass_z_bin}). In part, this is a consequence of the small number of sources in the highest mass bin (only three candidates, one per each redshift bin). Moreover, for most of the configurations in this range, the stacks show relatively high noise levels ($rms {\gtrsim} 115\, \mu$Jy), which arise because the targets are rare and generally lie close to the border of the ALMA maps and, hence, have higher noise due to beam attenuation. For this reason, the stacking results for these bins provide only relatively weak constraints.

\subsubsection{Stacked IRX-$\beta$ relation}

\begin{figure*}
    \centering
    \includegraphics[width=1.\textwidth]{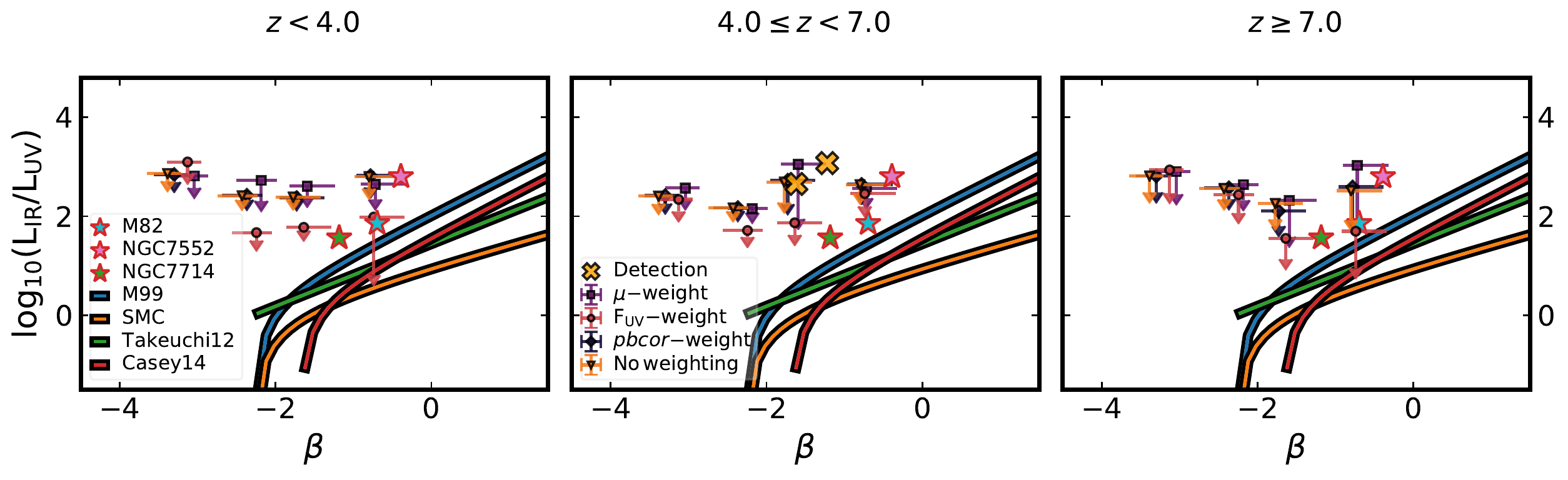}
    \caption{Stacked observed infrared-excess (IRX) $3{-}\sigma$ limits as a function of UV-slope ($\beta$). For each $\beta$ bin, the weighted median IRX upper limit is shown (orange triangles for equal weighting, black diamonds for $pbcor$-weighting, purple squares for magnification-weighting, and light rose circles for UV flux weighting; see $\S$\ref{subsec:Stacking}). Results are separated into three photometric redshift bins all of which are upper limits. For comparison, we also show the local IRX-$\beta$ relations (\citetalias{1999ApJ...521...64M}, SMC, \citealt{2016ApJ...833..254S}, \citealt{2012ApJ...755..144T} and \citealt{2014ApJ...796...95C}) and the locations of three well-known local star-forming galaxies (\object{M82}, \object{NGC7552} and \object{NGC7714}). The downward arrows have 1-$\sigma$ length. Horizontal errorbars indicate the 16th and 84th percentiles of the distribution of LBG candidates for each UV-slope bin. Yellow crosses show the two individual ALMA detections.} \label{fig:irx_expect_stack_beta}
	\end{figure*}

	We next consider the stacking constraints on the IRX-$\beta$ relation, which are presented in Fig.~\ref{fig:irx_expect_stack_beta}. We apply all four weighting schemes and list the full results in Tables~\ref{tab:stack_results_full_beta_z_bin} of appendix \ref{tab:beta_stack}. Here we split the sources into several $\beta$ bins for three distinct photometric redshift ranges. For completeness, we plot the ALMA detected LBG candidates alongside the stacking results. We omit $\beta$ bins which contain no LBG candidates or resulted in a negative IRX stacked value.
	
    In all three redshift bins, we see that the $F_{\rm UV}$-weighting generally produces much lower average IRX constraints than the other weighting schemes. This is perhaps no surprise, given the previously mentioned correlation between stellar mass and $L_{\rm UV}$ (or equivalently $F_{\rm UV}$ over limited redshift ranges) in $\S$\ref{subsec:IRXMstar}. Indeed, the most massive and UV-luminous LBG candidates in Fig.~\ref{fig:irx_vs_beta} were the ones with limits closest to the local relations. In contrast, the $pbcor$ and equal weighting schemes generate substantially weaker IRX constraints because they average in more of the high individual IRX constraints, which result from low $L_{\rm UV}$ detections and high $L_{\rm IR}$ limits. Finally, as can be seen from the $\mu$ color-coded panel of Fig.~\ref{fig:irx_vs_beta}, the individual high-magnification LBG candidates generally have some of the highest IRX limits, which translates into high IRX limits for the $\mu$-weighted stacked bins too. We additionally see that the lowest $\beta$ bins ($\beta$${<}$$-2$) have systematically higher IRX limits, mirroring the trend seen in the individual limits of Fig.~\ref{fig:irx_vs_beta}.
    
    For comparison, we show the two ALMA-detected LBG candidates and three local star-forming galaxies: \object{M82} \citep{2010PASP..122.1397S, 2003ApJ...599..193F, 2007ApJ...655..863D, 2012ApJ...757...24G}, \object{NGC7552} \citep{2010PASP..122.1397S, 2007ApJ...655..863D, 2015MNRAS.452.2712W} and \object{NGC7714} \citep{2010PASP..122.1397S, 1999ApJ...513..707G, 2004ApJS..154..188B, 2014ApJS..212...18B}. These local galaxies have range of M$_{\bigstar} {\sim} 10^{10.3}$--10$^{10.7}$ M$_{\odot}$ and SFR$\sim$1--10\,M$_{\odot}$\,yr$^{-1}$, with M82 being perhaps the most reasonable counterpart to the more massive LBG candidates. The detected LBG candidates generally have lower UV-slopes (less extinction), much lower stellar masses, and higher or comparable IRX values to the local objects. The limits for the $F_{\rm UV}$-weighted limits are systematically lower than the two ALMA detections and, in the higher redshift bins, show similar or lower IRX values than the local objects despite having similar stellar masses.
    
    The limits at $z {<} 4$ lie well above the local IRX-$\beta$ relations, demonstrating that at least 1-dex deeper IR observations are needed to start placing meaningful constraints on even the most luminous $z {\sim} 2$--4 LBGs, and 2--3 dex more for the bulk of the population. At higher redshifts, the results appear more auspicious, as the limits on the most UV-luminous objects are approaching those of the local relations. Unfortunately, the low numbers of sources in these high-redshift bins mean the results are subject to small number statistical uncertainties and, thus, we can only say that they remain marginally consistent with the local IRX-$\beta$ relations at the depths we probe.
    
\subsubsection{Stacked IRX-$\mathrm{M}_{\bigstar}$ relation}

\begin{figure*}
    \centering
    \includegraphics[width=1.\textwidth]{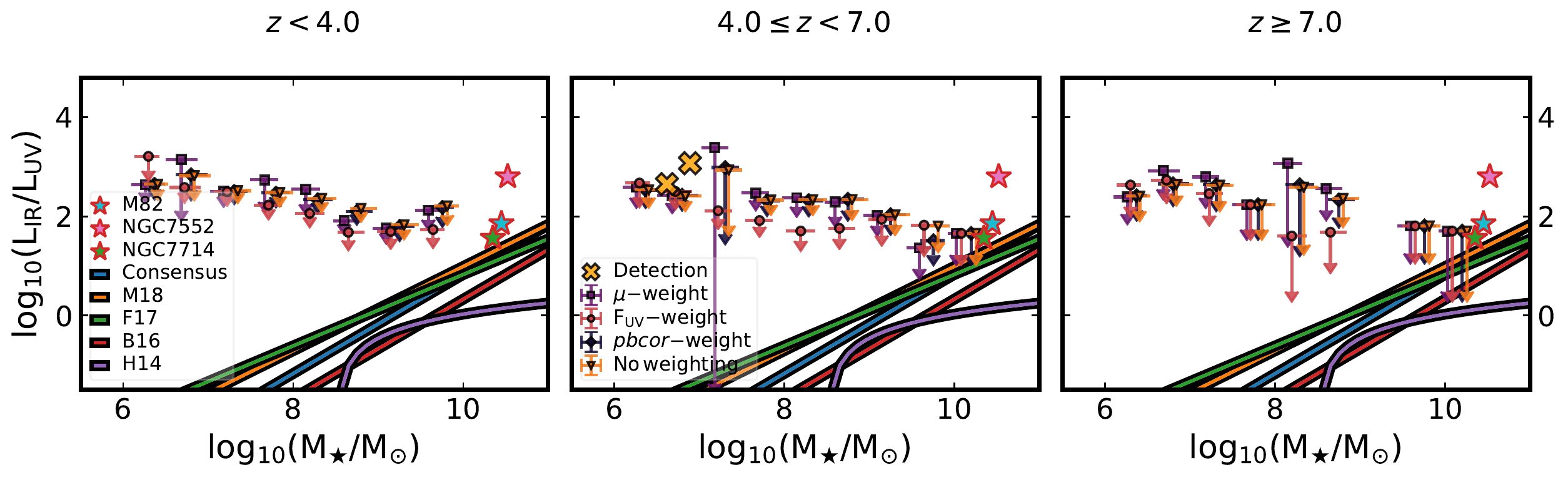}
    \caption{Stacked observed infrared-excess (IRX) $3{-}\sigma$ limits as a function of stellar mass ($M_{\bigstar}$). For each $M_{\bigstar}$ bin, the weighted median IRX upper limit is shown (same colors and markers as in Fig.~\ref{fig:irx_expect_stack_beta}). Results are separated into three photometric redshift bins, all of which are upper limits. We also show a number of previously reported IRX-$M_{\bigstar}$ relations (\textit{Consensus}, \citetalias{2017MNRAS.472..483F}, \citetalias{2018MNRAS.476.3991M}, \citetalias{2016ApJ...833...72B} and \citetalias{2014MNRAS.437.1268H}) and the locations of three well-known local star-forming galaxies (\object{M82}, \object{NGC7552} and \object{NGC7714}). Downward arrows have 1-$\sigma$ length. Horizontal errorbars indicate the 16th and 84th percentiles of the distribution of LBG candidates for each $M_{\bigstar}$ bin. Yellow crosses show the two individual ALMA detections.}
    \label{fig:irx_expect_stack_mass}
	\end{figure*}

	Finally, we consider the IRX-$M_{\bigstar}$ relation, the results of which are presented in Fig.~\ref{fig:irx_expect_stack_mass}. Again, we apply all four weighting schemes and list the full results in Tables~ \ref{tab:stack_results_mid_mass_z_bin} and \ref{tab:stack_results_full_mass_z_bin} of appendix \ref{tab:mass_stack}. Here we split the sources into several $M_{\bigstar}$ bins spanning three photometric redshift ranges. We omit stellar mass bins that contain no LBG candidates or resulted in a negative IRX stacked value. For comparison, we plot the ALMA-detected LBG candidates and local galaxies alongside the stacking results. 
    
    As with the IRX-$\beta$ results, we find that the $F_{\rm UV}$-weighting produces lower median IRX constraints compared to the other weighting schemes, although the distinction between these is, in general, less pronounced than with $\beta$. In most bins, the limits lie 1--2 dex above the consensus relations, although we do see a trend wherein the highest mass bins (e.g., ${\ga}10^{9}$\,$M_{\odot}$) have lower IRX limits than the lower $M_{\bigstar}$ bins and fall near or below the \citetalias{2017MNRAS.472..483F}, \citetalias{2018MNRAS.476.3991M} and \textit{consensus} relations. It is important to note here, however that these limits comprise only 1--2 of the most extreme individual LBG candidates and, hence, cannot be considered representative of the full sample.
    
    The ALMA-detected LBG candidates have comparable IRX and stellar mass values to the stacked limits (regardless of weighting scheme) and are presumably the extreme tail of the distribution.

\section{Discussion}\label{sec:Discussion}

\subsection{Individual constraints}\label{subsec:disc_indiv}

    We compare the properties from our sample (UV-slope, stellar mass, UV magnitude) with the distributions from \citetalias{2016ApJ...833...72B}, where $330$ LBGs were studied and six 2-$\sigma$ tentative ALMA detections were obtained. In particular, \citetalias{2016ApJ...833...72B}  present histograms for these properties as a function of drop-out bins in their Fig.~2, while we present our sample distributions in Figs.~\ref{fig:uv_beta_mag} and \ref{fig:hist_stell_mass}.
    
    The $\beta$ values in both samples show a peak near \hbox{$\beta {\sim} {-}2.2$} and similar distribution shapes, with the bulk of sources located in the range ${-}3.5 {\lesssim} \beta {\lesssim} 0.5$. The stellar mass distributions both peak at around ${\log{(\mathrm{M}_{\bigstar} / \mathrm{M}_{\odot})} = 8}$. However, our sample effectively probes one dex lower in mass due to the magnifying power of the galaxy clusters. Finally, the (magnification-corrected) apparent UV magnitude distributions share similar maximum (${m}_{\mathrm{UV}} {\sim} 24$) and peak values ($28 \lesssim {m}_{\mathrm{UV}} \lesssim 29.5$) but our candidates probe two magnitudes deeper (${m}_{\mathrm{UV}} {\sim} 32$) than in  \citetalias{2016ApJ...833...72B}, again due to the galaxy cluster lensing. Notably, without the capped magnification factors, our distributions would extend to even smaller values.

\subsubsection{ALMA expectations}\label{subsubsec:alma_expect}

    Our ALMA observations (both detections and upper limits) can be contrasted with previous studies of mm and submm emission from LBG candidates over comparable redshift ranges. Several works have stacked multi-band IR photometry to generate empirical SEDs or fit against templates to derive average physical properties \citep[e.g.,][]{2011A&A...533A.119E, 2011ApJ...740L..15M, 2011A&A...534A..15M, 2013A&A...554L...3O, 2013MNRAS.435..158O, 2015MNRAS.446.1293C, 2017A&A...599A.134S, 2017ApJ...847...21F, 2018MNRAS.481.1631B}.
    These works studied star-forming galaxies or LBGs with redshifts ranging from $z {\sim} 1$ to $z {\sim} 7$ and derived IR luminosities ranging from $\log{(L_\mathrm{IR}/L_{\odot})} {=} 9.9$ \citep[in][]{2011ApJ...740L..15M} to $\log{(L_\mathrm{IR}/L_{\odot})} {=} 12.5$ \citep[in][]{2013MNRAS.435..158O}. The majority of these works focus on extreme or rare LBG candidates with masses above $(\log{M_{\bigstar}/{M}_{\odot}}{\ga}$10.0 and \hbox{SFRs ${\ga}$100}. In these cases, there is essentially no overlap in $M_{\bigstar}$ or SFR distributions compared to our sample, making comparisons and interpretations of detections or limits impossible. Two exceptions that share some overlap are \citet{2018MNRAS.481.1631B} and \citet{2015MNRAS.446.1293C}, which we discuss below.
    
    \citet{2018MNRAS.481.1631B} reported on ALMA band 6 observations of $6$ $z {\sim} 7$ LBGs with  $\log{M_{\bigstar}/{M}_{\odot}}{=}$9.0--9.6 and $\log{L_{\rm UV}/L_{\odot}}{=}$11.3--11.6, selected from 1.0 deg$^{2}$ of UltraVISTA imaging in the COSMOS field; one confirmed LBG is marginally detected in ALMA Band $6$ with 168$\pm$56 $\mu$Jy while the stacked limit for the remaining five candidates is 100$\pm$50 $\mu$Jy. We only have 1 LBG candidate which overlaps with this stellar mass at $z>7$, but has a much fainter UV luminosity; our limits are consistent with the ones reported by \citet{2018MNRAS.481.1631B}.
    
    \citet{2015MNRAS.446.1293C} stacked $850 \mu$m SCUBA data for 5138 $z {\sim} 3 {-} 5$ LBG candidates with $\log{M_{\bigstar}/{M}_{\odot}}{=}$9.0--11.0) and $\log{L_{\rm UV}/L_{\odot}}{=}$10.0--11.6, selected from 0.62 deg$^{2}$ in the UKIDSS-UDS field; they reported 850\,$\mu$m detections of 
    250$\pm$29, 411$\pm$64,  875$\pm$229 $\mu$Jy for their $z{\sim}$3, 4, and 5 samples, respectively (adopting an emissivity of 1.6, these equate to 1.1\,mm detections of ${\approx} 165$, 270, and 580\,$\mu$Jy). Our stacked limits for 72 ($z{<}4$) and 25 ($4{<}z{<}7$) similarly massive LBG candidates $(\log{M_{\bigstar}/{M}_{\odot}}{=}$9.0--10.0; Table \ref{tab:stack_results_full_mass_z_bin}) lie between 24$\pm$9 and 62$\pm$23 for $z{<}4$ and 22$\pm$19 to 27$\pm$30 $\mu$Jy for $4{<}z{<}7$. Even our two tentative detections at ${\approx}$285 $\mu$Jy (Table \ref{tab:detect_props}) lie a factor of 2 below the stacked average at $z{\sim}5$, and have masses that are more than two orders of magnitude lower than those from \citet{2015MNRAS.446.1293C}. 
    This strong discrepancy, by factors $>$ 3 to 20, implies that a few strong sources (e.g., obscured AGN or dusty star-forming galaxies), are very likely biasing their results. \citet{2015MNRAS.446.1293C} do note that the fitted SED models show lower radio fluxes than the measured values, possibly implying an AGN contribution in part of their sample.
    
    Another method to test our IR expectations involves deriving IR luminosities with the use of the UV luminosities (see $\S$\ref{subsec:UVlums}) and their SFR values ($\S$\ref{subsec:SFR_calc}) separately. Several calibrations have been developed to extract a star formation rate estimate, with the one from \citet{1998ApJ...498..541K} and the reformulation from \citet{2001ApJ...548..681B} being among the most utilized. The difference between the corrected and uncorrected UV SFRs corresponds to the IR SFR values \citep{2007ApJ...670..156D}. From the derivations of \citet{2013MNRAS.433.2706O} and \citet{2018A&A...616A.110E}, we derived the corrected and uncorrected UV SFRs using the \citet{2000ApJ...533..682C} law (see $\S$\ref{eq:calzetti_law}).
    
    Thus, we obtained an estimate for the $3 {-} \sigma$ IR luminosity upper limits and compared them with our values derived from ALMA observations. We find that UV-based $L_{\mathrm{IR}}$ estimates spread along a wider range (${\sim}10^{8.5}$--$10^{13.2}$\,$\mathrm{L_{\odot}}$) than the ALMA ones (${\sim}10^{11.1}$--$10^{13.4}$\,$\mathrm{L_{\odot}}$). For the vast majority of LBG candidates, the UV-based estimates lie well below the ALMA-based ones, essentially in lines with their locations on the IRX diagrams. For a handful of a few high-luminosity and high-redshift LBG candidates at high redshift, the constraints are similar; these are the same sources that lie near the local IRX relations. In summary, we find good consistency between this method and our IRX analysis.

\subsubsection{Detected LBGs}\label{subsubsec:disc_detections}

    Overall, we detect only two sources with S/N$\gtrsim$4.1, both from the AS1063 field. The latter fact may imply that these detections are not representative of the LBG population as a whole. One source is highly magnified ($\mu{=}39.0$), while the other is not ($\mu{=}2.9)$. The best-fit UV properties of the detected sources appear normal for LBGs, but quite atypical of known ALMA sources, with high redshifts ($z{\sim}$5.4--5.5), low stellar masses (${\approx}6.6$--6.9\,$\log{M_{\bigstar}/M_{\odot}}$), low UV fluxes (${\approx}$8.7--8.8\,$\log{L_{\rm UV}/L_{\odot}}$), and modest UV slopes ($\beta{\approx}{-}1.2\,{\rm to}\,{-}1.6$) yet relatively high IR luminosities (${\approx}11.3$--$11.9\,\log{(L_{\rm IR}/L_{\odot})}$). These properties imply that both sources may be spurious, or that the ALMA detections are coming from highly obscured regions within these galaxies that are not accounted for by the {\it HST} data. Despite the low number of individual detections, stacking provides better insights into our LBG candidates and their properties. 
    
    With so few detected sources, stacking does not offer any further insight. The non-detections in all but one of the ALMA-FFs, despite the potential for strong lensing, indicate that the intrinsic IR emission from all LBGs is faint. Our results are consistent with many past works (\citetalias{2016ApJ...833...72B, 2017MNRAS.472..483F, 2018MNRAS.476.3991M, 2014MNRAS.437.1268H}; \citealt{2014ApJ...796...95C, 2012ApJ...755..144T, 2016A&A...587A.122A, 2017ApJ...845...41B, 2018MNRAS.481.1631B, 2017MNRAS.467.1360B}).
    
    Regardless of their scarcity, we can compare the detections to the stacked IRX upper limits depicted in Figs. \ref{fig:irx_expect_stack_beta} and \ref{fig:irx_expect_stack_mass}. The IR luminosities of the detections are higher than an important fraction of our sample but seem to lie within the 
    sample distribution, consistent with expectations.
    
\subsubsection{IRX-\texorpdfstring{$\beta$}{beta} upper limits and previous works}\label{subsubsec:disc_up_lims_irx_beta}

	Figure \ref{fig:irx_vs_beta} shows the distribution of our sample over the IRX-$\beta$ plane color-coded by five different quantities. Along with this distribution, we include four different relations from past works: M99 IRX-$\beta$ relation by \citet[][also introduced in $\S$\ref{subsec:IRXrelations}]{1999ApJ...521...64M}, Small Magellanic Cloud (SMC) IRX-$\beta$ relation \citep{2016ApJ...833..254S}, \citet{2012ApJ...755..144T} and \citet{2014ApJ...796...95C} relations.
    
    We can see that the vast majority of our candidates have limits well above the relations already mentioned. Only three ($3$) points are located below the M99 IRX relation. Thus, most of our upper limits remain compatible with all of the studied IRX-$\beta$ relations. It is pertinent to mention that the candidates situated between the M99 and SMC relations have high photometric redshifts; they lie in the interval $z_{\mathrm{ph}} {=} 5.69$--8.12 with a mean value of $z_{\mathrm{ph}} {=} 6.77$. The UV luminosities from these candidates all skew toward the high end of the distribution, while their IR luminosities skew toward the low end. This combination leads to lower IRX values, pushing them below several published IRX-$\beta$ relations.
    
    Given the known dispersion in the local relations, we would have naively expected to detect at least a few among the handful of sources that lie between the M99 and SMC relations. The fact that we have no detections at least hints at the possibility that high-redshift relations may have systematically lower IRX values.
    
    Targets that lie below $\beta {=} {-}2.23$, which represents the intrinsic, non-dust-obscured, UV-slope value from \citet{1999ApJ...521...64M}, cannot be compared directly with the mentioned relations as they do not cover the same region of the parameter space. Extremely low $\beta$ sources must be compared with previous relations, as mentioned in $\S$\ref{subsec:IRXrelations}, developed specifically with such galaxies in mind. Leaving these issues aside, it is clear that with current instrumentation, we are unable to probe to sufficiently low IR luminosities yet to study the behavior of typical LBGs with extremely low $\beta$ values and, thus, extend known relations. Furthermore, having obtained only upper limits makes it impossible to search for a meaningful correlation between IRX and either $\beta$ or $M_{\bigstar}$.
    
    Most of the recent works mentioned here agree on the fact that star-forming galaxies up to $z {\sim} 3{\mbox{--}}4$ follow the \citetalias{1999ApJ...521...64M} IRX-$\mathrm{M}_{\bigstar}$ relation more closely than an SMC-like curve. Our IRX-$\mathrm{M}_{\bigstar}$ upper limits (both individual and $F_{\mathrm{UV}}$-weighted stacks) suggest that the most luminous and highest redshift sources ($\beta {\gtrsim} {-}1.0$ and $z {\gtrsim} 4$) are pushing below the
	\citetalias{1999ApJ...521...64M} relation and may be more compatible with an SMC-like curve 
	\citep[as suggested by, e.g.,][F17]{2018MNRAS.479.4355K}.
    
    Regarding a possible evolution of the IRX relations with redshift as mentioned by, for example, F17, we are unable to establish this given that we only find IRX upper limits in the stacked emission. We can only observe (uppermost panels in Figs. \ref{fig:irx_vs_beta} and \ref{fig:irx_vs_smass}) that, roughly, upper limits with higher redshifts tend to exhibit lower IRX ratios, and the lack of any individual detections hints at some evolution, but the statistics are currently too limited to say more.

    Finally, we compare our upper limits in the IRX-$\beta$ space with the recent values presented by \citet{2019ApJ...872...23S}, who examined more than $20,000$ low-redshift galaxies ($z {<} 0.3$) from GALEX-SDSS-WISE Legacy Catalog 2 (GSWLC-D2). Notably, they find that a majority of their sources in the range ${-}2.0 {\lesssim} \beta {\lesssim} {-}0.5$ lie below the \citetalias{1999ApJ...521...64M} relation, while nearly all $\beta {\gtrsim} {-} 0.5$ sources lie well below it, indicating a less abrupt slope comparable to the other relations presented in $\S$\ref{subsec:IRXrelations}. We see that our best $10$ upper limits are currently consistent with the IRXs of these possible low redshift analogs, although given that our constraints are only limits, there remains the potential for some mild tension with even the SMC-like relation, which appears to act as a lower bound on the local \citet{2019ApJ...872...23S} sample.
	
\subsubsection{IRX-\texorpdfstring{$M_{\bigstar}$}{Mstar} upper limits and previous works}\label{subsubsec:disc_up_lims_irx_mass}
	
    Individual IRX and stellar mass values are shown in Fig.~\ref{fig:irx_vs_smass}, along with five relations from previous works \citepalias[see $\S$\ref{subsec:IRXrelations} for details;][]{2017MNRAS.472..483F, 2018MNRAS.476.3991M, 2016ApJ...833...72B, 2014MNRAS.437.1268H}. These relations are extrapolated down to stellar masses of $\log{(M_{\star} / M_{\odot})} {=} 6.0$ where applicable, to match our lowest stacking bins. For the same reasons exposed in $\S$\ref{subsubsec:disc_up_lims_irx_beta}, we do not fit any relations to our upper limits.
    
    All our upper limits lie above the curves. The LBGs that fall very close to the higher IRX-$M_{\bigstar}$ relations correspond to the candidates with the highest stellar masses. Similar to the observed behavior with the UV-slope, lower stellar mass candidates do not probe to as low IRX values as their higher mass counterparts.
    
	While our candidates share similar distributions in several important properties with \citetalias{2016ApJ...833...72B}, the resulting IRX limits for the majority of candidates remain ${\approx}0.3$ \texttt{dex} higher due to our shallower ALMA maps (factor of $\approx$4--5 higher rms) and the low average magnifications (${<}\mu{>} {\approx}4.3$) of our candidates. For the rare high-magnification targets, predominantly low-mass LBG candidates (as seen in panel 2 of Fig.~\ref{fig:irx_vs_smass}), our maps provide modestly deeper IRX constraints.
	
	With respect to the lower part of the IRX distribution, our $z {<} 4.0$ sample lies ${\sim} 1.0$~\texttt{dex} above the respective bins from \citetalias{2016ApJ...833...72B}.
	
	Comparing our results with those of \citetalias{2017MNRAS.472..483F}, we find that our IRX upper limits, for	comparable $\beta$ and $\mathrm{M}_{\bigstar}$ ranges, are ${\sim} 0.5$ \texttt{dex} higher than their sources. Since both ALMA observations reach similar noise levels, the discrepancy must lie in the fact that their sample is composed of sources up to much higher stellar masses 
	($\mathrm{M}_{\bigstar} {\sim} 10^{10.7} \mathrm{M}_{\odot}$) 
	and much higher $\mathrm{L}_{UV}$ values \citep[i.e., their most 
	stringent constraints arise from the sample of][who targeted the 
	brightest LBGs over the much larger 2 deg$^{2}$ COSMOS field]{2015Natur.522..455C}.
    
\subsubsection{\texorpdfstring{$M_{\bigstar}$}{Mstar}-\texorpdfstring{$\beta$}{beta} correlation}\label{subsubsec:Mstar_beta_corr}

\begin{figure}
    \centering
    \includegraphics[width=1.\columnwidth]{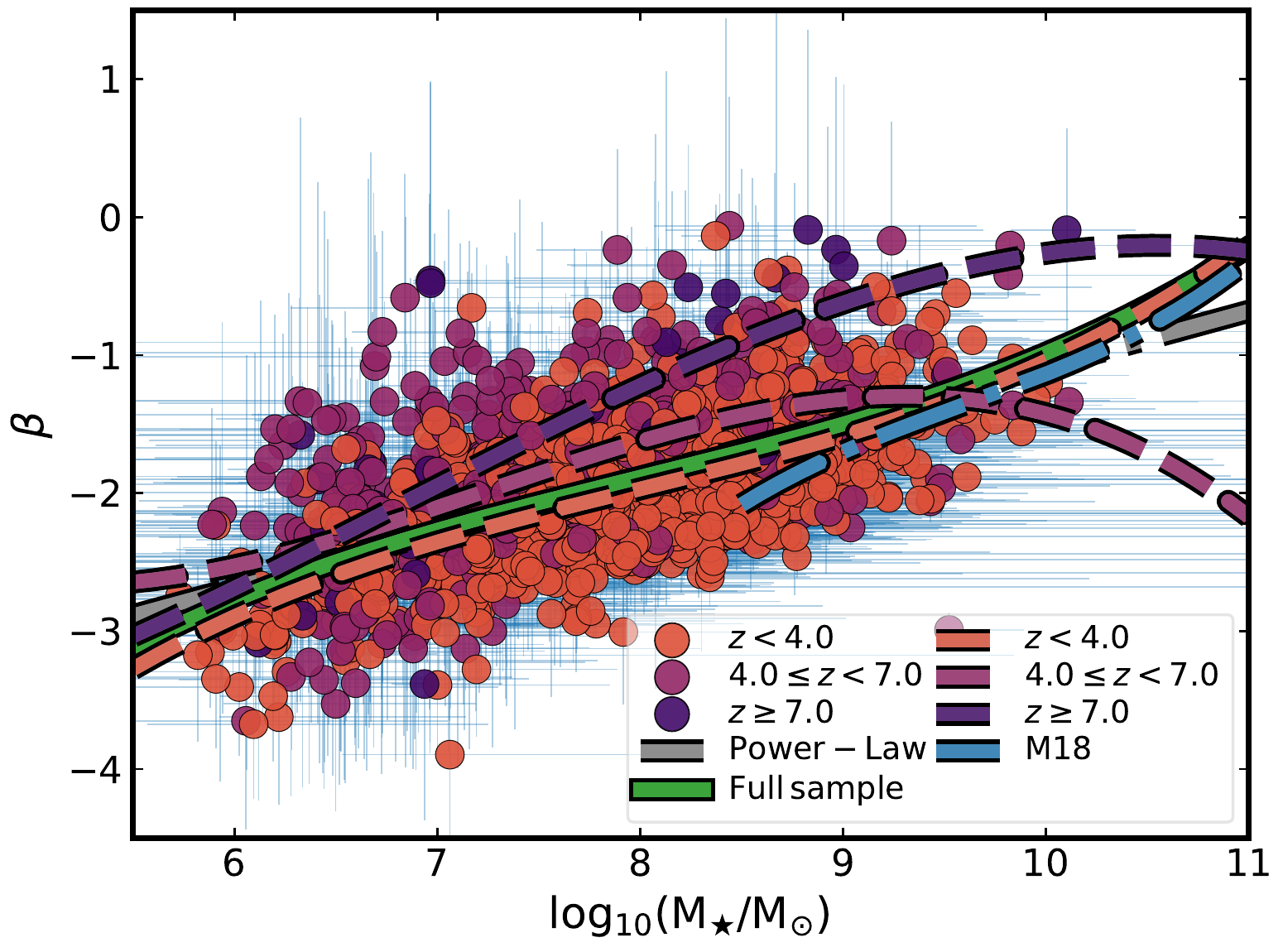}
    \caption[UV-slope vs. stellar mass for the studied sample]{UV-slope ($\beta$) vs. stellar mass ($\mathrm{M}_{\bigstar}$) for our selected LBG candidates. Colors represent our three photometric redshift bins (both point and polynomial fitting lines). Blue dashed line represents fit from \citetalias{2018MNRAS.476.3991M} for their sample with stellar masses $\log{(\mathrm{M}_{\bigstar} / \mathrm{M}_{\odot})} {\geq} 8.5$ and green solid line shows our third-order polynomial fit (Eq. \ref{eq:beta_vs_smass}). We also include the first-order polynomial from Eq.~\ref{eq:beta_vs_smass_power} in gray.}\label{fig:beta_vs_smass}
	\end{figure}

    As mentioned in $\S$\ref{subsec:IRXMstar}, we see a fairly clear trend between $M_{\bigstar}$ and $\beta$. To place this in better context, we plot in Fig.~\ref{fig:beta_vs_smass} the relation between UV-slope and stellar mass directly for our sample. Under the assumption that all star-forming galaxies have similar intrinsic UV slopes, \citetalias{2018MNRAS.476.3991M} used the values of $\beta$ as a proxy for the UV attenuation ($\mathrm{A}_{1600}$). They fit a polynomial to a mass-complete sample of star-forming galaxies selected from the Hubble Ultra Deep Field \citep[HUDF;][]{2006AJ....132.1729B, 2017MNRAS.466..861D} and obtained the third-order relation plotted in the dashed blue line in Fig.~\ref{fig:beta_vs_smass} (for stellar masses in the range $8.5 {<} \log{(M_{\star} / M_{\odot})} {<} 11.5$, with a 1.1\,mm $rms$ of $35 \mu$Jy $\mathrm{beam}^{-1}$).
	
    We performed the same experiment using our full LBG sample down to a mass of $\log{(\mathrm{M}_{\bigstar} / \mathrm{M}_{\odot})} {=} 6.0$. We caution that this limit is likely substantially below the nominal mass completeness threshold in the Hubble Frontier Fields (HFFs), which should be similar to that of the HUDF at $z {\sim} 3$ (e.g.,${\gtrsim}10^{8.5}$\,$\mathrm{M}_{\odot}$; see Fig.~\ref{fig:hist_stell_mass}). Due to the lensing amplification, we do expect to find at least some representative sources among the lower mass LBGs in our sample, but we could have strong selection effects that bias the resulting fitted relations at low stellar masses.
    
    Nonetheless, applying a third-order polynomial fit to our $\mathrm{M}_{\bigstar}$ and $\beta$	values, we find
	\begin{eqnarray}\label{eq:beta_vs_smass}
		\beta = -0.99 + 0.62X + 0.13X^{2} + 0.02X^{3},
	\end{eqnarray}
	\noindent in which $X {=} \log{(\mathrm{M}_{\bigstar} / 10^{10}\mathrm{M}_{\odot})}$. Similar exercises were performed binning the sample in our adopted redshift bins. The ${z {<} 4.0}$ trend is nearly identical to the full sample, due in large part to the fact that such low-redshift sources account for the majority of our sample ($1068$ candidates). However, the trends found for the higher redshift bins remain consistent within the dispersion. Within the mass-complete range of $8.5 {\lesssim} \log{(M_{\star} / M_{\odot})} {\lesssim} 10$, our fits appear to be consistent with that of \citetalias{2018MNRAS.476.3991M}, particularly in the low-redshift bin (within $0.25$ \texttt{dex}), which is most comparable to the range they studied.
	
	We note that fitting a simpler first-degree polynomial (power law) to our full sample ($\mathrm{M}_{\bigstar} {-} \beta$ space) yields the following line:
	\begin{eqnarray}\label{eq:beta_vs_smass_power}
		\beta = -1.08 + 0.40X,
	\end{eqnarray}
    \noindent with $X {=} \log{(\mathrm{M}_{\bigstar} / 10^{10}\mathrm{M}_{\odot})}$. This line, shown in grey in  Fig.~\ref{fig:beta_vs_smass}, is nearly identical to the third-order polynominal fit above, demonstrating that there is no unexpected oscillatory behavior in the former curve and, thus, corroborating the trend seen in \citetalias{2018MNRAS.476.3991M}.

	Pushing below stellar masses of ${\sim} 10^{8.5}$\,$\mathrm{M}_{\odot}$, we observe a smooth trend toward lower (bluer) $\beta$ values, consistent with expectations from increasingly metal-poor stellar populations. Some caution must be exercised nonetheless since both properties, stellar mass and UV-slope, have been derived from the same data (\textit{HST} photometry) and they are not, consequently, completely independent.

\subsection{The role of dust temperature}\label{subsubsec:dust_temp}

    To understand the effects of our evolving dust temperature prescription on our results, we compared against a model assuming that all LBG candidates have a constant temperature of $\mathrm{T}_{\mathrm{dust}} {=} 35 \mathrm{K}$ \citep[following, for example,][]{2006ApJ...650..592K, 2008MNRAS.384.1597C}. Such low dust temperatures predominantly affect the properties of high-redshift ($z > 4$) candidates, resulting in IR luminosity and IRX values drops of ${\sim} 0.5$ dex, such that more candidate upper limits (both individually and stacked) are pushed below the \citetalias{1999ApJ...521...64M} IRX-$\beta$ relation (with some even approaching the SMC relation) and a few candidates fall below the IRX-$\mathrm{M}_{\bigstar}$ consensus relation.
	
\subsection{ALMA-FF LBG sample overview}\label{subsec:LBGdensity}
    
    To place our LBG sample in context with other samples, we estimated in very crude terms the source density of LBGs per angular area in the source plane. We shifted to the source plane because the high magnification by massive clusters strongly affects the number density of observed background sources. This is non-trivial, however, since the exact magnification depends on the redshifts of the sources. 

    We estimated the source-plane area as follows. Following the methodology presented in $\S$\ref{subsec:mufactor}, we calculated the source-plane beam area for each candidate LBG as the area of the synthesized beam for the observed ALMA map centered on the LBG position, divided by the adopted magnification factor. We then summed the individual LBG source and image-plane areas and divided the totals to obtain a "demagnification" factor, which in our case is $0.35$. Finally, we multiplied the total (image-plane) area of the ALMA FFs observations by this ratio to estimate crudely the total source-plane area covered by our ALMA observations. Each FF cluster was observed over a ${\approx} 2\farcm1 {\times} 2\farcm2$ ALMA mosaic \citep{2017A&A...597A..41G}, summing up to an image-plane area of ${\sim} 23$ arcmin${}^{2}$ used in this work. Applying the factor of $0.35$, we should have, on average, an effective source-plane region of ${\approx} 1\farcm24 {\times} 1\farcm3$ per cluster, and a total source-plane area of ${\sim} 8$ arcmin${}^{2}$ over the five FFs clusters. 
    
    With this, we can obtain an estimate for the intrinsic density of LBGs (regardless of their redshift) per unit area. A simple ratio of our $1582$ studied sources over the effective area covered by ALMA gives a value of ${\approx} 200$ LBGs per arcmin${}^{2}$. To establish a reference with other ALMA observations, \citetalias{2016ApJ...833...72B} studied $330$ LBGs over a $1$ arcmin${}^{2}$ region of HUDF. And \citetalias{2017MNRAS.472..483F} examined $67$ star-forming galaxies in an area of $39\arcsec {\times} 39\arcsec$, which corresponds to a density of ${\approx} 160$ sources per arcmin${}^{2}$. Our survey appears to be intermediate between these two.


\section{Summary and conclusions}\label{sec:Summary}

	In this paper, we utilize ALMA 1.1\,mm mosaic images for five of the six FFs clusters, with {\it rms} values between $\approx$55--71$\mu$Jy, to place constraints on the IR excesses of $1582$ UV-selected LBGs as functions of their UV-slopes ($\beta$), stellar masses ($M_{\bigstar}$), sSFRs and photometric redshifts. After correcting for magnification, the source plane area of the five clusters is ${\sim} 8$ arcmin${}^{2}$, probing LBG candidates with rest-frame UV magnitudes ranging from ${\sim}$23--32 ABmag. We summarize our results as follows:
    
    \begin{enumerate}
    
    \item The rms levels in the ALMA maps, coupled with the likely faint intrinsic fluxes of the LBG candidates, result in very few outright detections. With a detection threshold of $4.1{-}\sigma$ (equivalent to a $15\%$ false detection rate), only two LBG candidates are considered detected (both located in AS1063). The rest are treated as upper limits.
    Comparing our $1580$ IRX $3{-}\sigma$ upper limits with previous IRX relations (IRX-$\beta$ and IRX-$M_{\bigstar}$), the vast majority lie above the local and \textit{Consensus} relations; only $\mathbf{3}$ LBGs are constrained to lie below the M99 IRX-$\beta$ relation and none of them below any of the discussed IRX-$M_{\bigstar}$ relations. Our lowest IRX limits appear consistent with the known dispersion around these relations.
    
    \item We divided the $1580$ undetected LBG candidates into bins of stellar mass ($6.0$$\le$$\log{(M_{\bigstar}/M_{\odot})}$$\le$$11$), UV-slope ($-4.0$$\le$$\beta$$\le$$1.5$) and photometric redshift ($<$4.0, 4.0--7.0, $>$7.0), and stacked their 
    ALMA data using the \texttt{STACKER} software.
    We implemented four weighting schemes for the \texttt{uv}-stacking: equal weighting; $pbcor$-weighting only; $pbcor$ and UV-flux $F_{\rm UV}$-weighting; and $pbcor$ and magnification $\mu$-weighting. 
    With these configurations, we stacked the ALMA observations and computed the stacked IRX values, obtaining upper limits in all but one bin (i.e., S/Ns${\lesssim} 3.5$). This single detected bin yields a S/N${\approx} 4.24$, although this was only obtained for one weighting scheme and \textit{CLEAN}ing configuration; as such we do not consider it to be a robust result.
    The bulk of our stacked IRX values remain above those cited in most of the literature for IRX-$\beta$ and IRX-M$_{\bigstar}$ relations, although a few bins ($\beta {>} {-}1.0$ with $z_{ph} \geq 7.0$ and M$_{\bigstar} {>} 10^{9.5} M_{\odot}$ with $z_{ph} \geq 7.0$) push below the \citetalias{1999ApJ...521...64M} IRX-$\beta$ and \textit{Consensus} relations. Since these limits only represent a small portion of the overall LBG population and there is substantial known dispersion in the relations themselves, we can only say at present that these high-redshift LBGs appear consistent with the relations. That being said, the lack of any detections in the vicinity of these relations hints a possible evolution of these relations, which could be linked to an evolution in metallicity.
    
    \item We also investigated the correlation between $\beta$ and stellar mass for our candidates. Despite the significant dispersion, as well as growing incompleteness below stellar masses of ${\sim}10^{8.5}$\,$M_{\odot}$, we observe a clear and smooth trend that extends to lower masses and bluer (lower) $\beta$ values, consistent with expectations from previous works regarding metallicity-driven evolution.
    
    \end{enumerate}

    To improve upon our results would require either the stacking of substantially larger LBG samples or ALMA observations with at least a factor of several and ideally ${>} 1$ dex lower average rms. Since covering larger areas or reducing noise levels to this extent would be costly in terms of ALMA observation time (e.g., \citealp{2018A&A...620A.152F} achieve ${\approx}180$\,$\mu$Jy $rms$ in the $\sim$69 arcmin$^{2}$ GOODS-S mosaic at 1.1\,mm in 18.5 hrs; Gonzalez-Lopez et al. 2019b, in preparation, achieve ${\approx}14$\,$\mu$Jy $rms$ in a $\sim$5 arcmin$^{2}$ HUDF mosaic at 1.1\,mm in 85 hrs), a better strategy may be to target the most highly magnified LBG candidates with deep, single-pointing observations, as opposed to the mosaic observations used for this work. In this way, significantly fewer pointings would be required but for longer durations. 

\begin{acknowledgements}
    We acknowledge support from CONICYT grants Basal-CATA AFB-170002 (RC, FEB, ET, RD, LG),
    FONDECYT Regular 1141218 (RC, FEB) 1160999 (ET) and 1190818 (ET),
    Programa de Astronomia FONDO ALMA 2016 31160033 (LG),
    and the Ministry of Economy, Development, and Tourism's Millennium Science Initiative through grant IC120009, awarded to The Millennium Institute of Astrophysics, MAS (FEB).

    This paper makes use of the following ALMA data: 
    ADS/JAO.ALMA\#2013.1.00999.S, ADS/JAO.ALMA\#2015.1.01425.S. 
    ALMA is a partnership of ESO (representing its member states), NSF (USA) and NINS (Japan), together with NRC (Canada), NSC and ASIAA (Taiwan), and KASI (Republic of Korea), in cooperation with the Republic of Chile. The Joint ALMA Observatory is operated by ESO, AUI/NRAO and NAOJ.
      
    Part of our work is based on observations obtained with the NASA/ESA Hubble Space Telescope, retrieved from the Mikulski Archive for Space Telescopes (MAST) at the Space Telescope Science Institute (STScI). STScI is operated by the Association of Universities for Research in Astronomy, Inc. under NASA contract NAS 5-26555.
      
    This work utilizes gravitational lensing models produced by PIs Brada\v{c}, Natarajan \& Kneib (CATS), Merten \& Zitrin, Sharon, Williams, Keeton, Bernstein and Diego, and the GLAFIC group. 
    This lens modeling was partially funded by the HST Frontier Fields program conducted by STScI. STScI is operated by the Association of Universities for Research in Astronomy, Inc. under NASA contract NAS 5-26555. The lens models were obtained from the Mikulski Archive for Space Telescopes (MAST).
      
    This research made use of \texttt{Astropy}, a community-developed core \texttt{Python} package for Astronomy \citep{2013A&A...558A..33A, 2018AJ....156..123A}, \texttt{matplotlib}, a \texttt{Python} library for publication quality graphics \citep{Hunter:2007}, \texttt{APLpy}, an open-source plotting package for \texttt{Python} \citep{2012ascl.soft08017R}, and NASA's Astrophysics Data System.
\end{acknowledgements}

\bibliographystyle{aa}
\bibliography{StackBib.bib}

\begin{appendix}

\section{$u$--$v$ stacking tests}\label{app0}

\begin{figure*}[htb]
		\centering 
		\begin{subfigure}[t]{.245\linewidth}
			\centering
			\includegraphics[width=1.\textwidth]{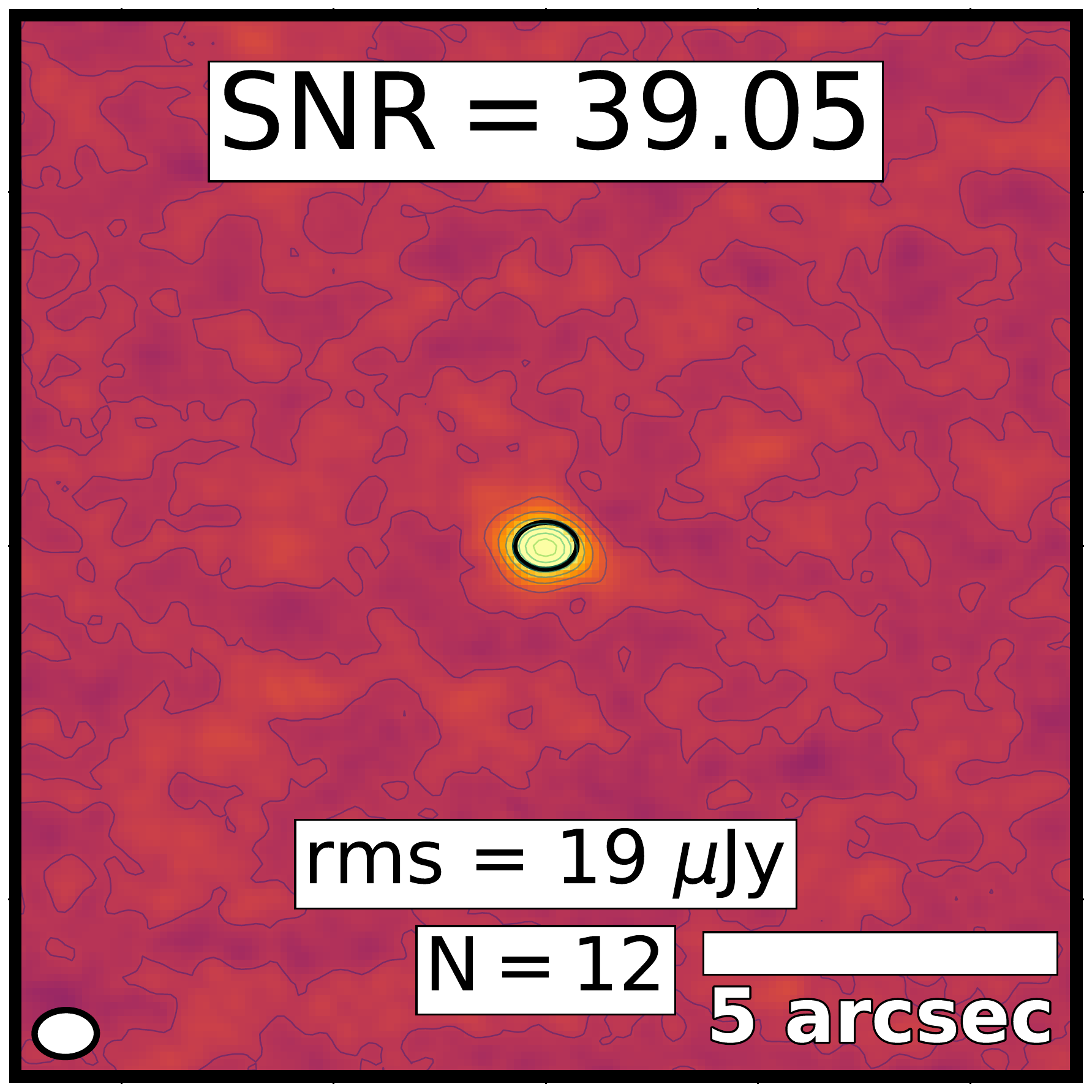}
		\end{subfigure}%
		\begin{subfigure}[t]{.245\linewidth}
			\centering
			\includegraphics[width=1.\textwidth]{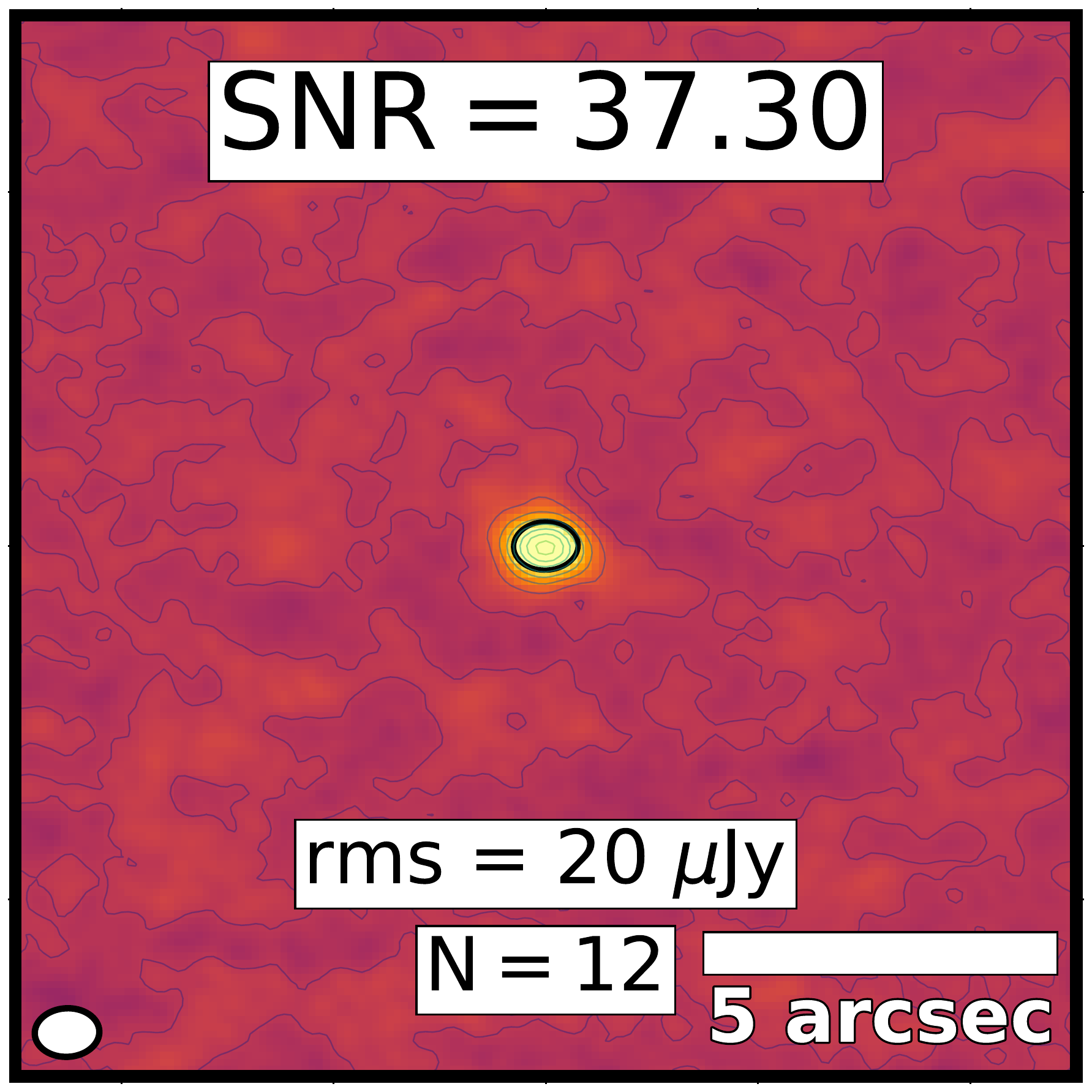}
		\end{subfigure}%
		\begin{subfigure}[t]{.245\linewidth}
			\centering
			\includegraphics[width=1.\textwidth]{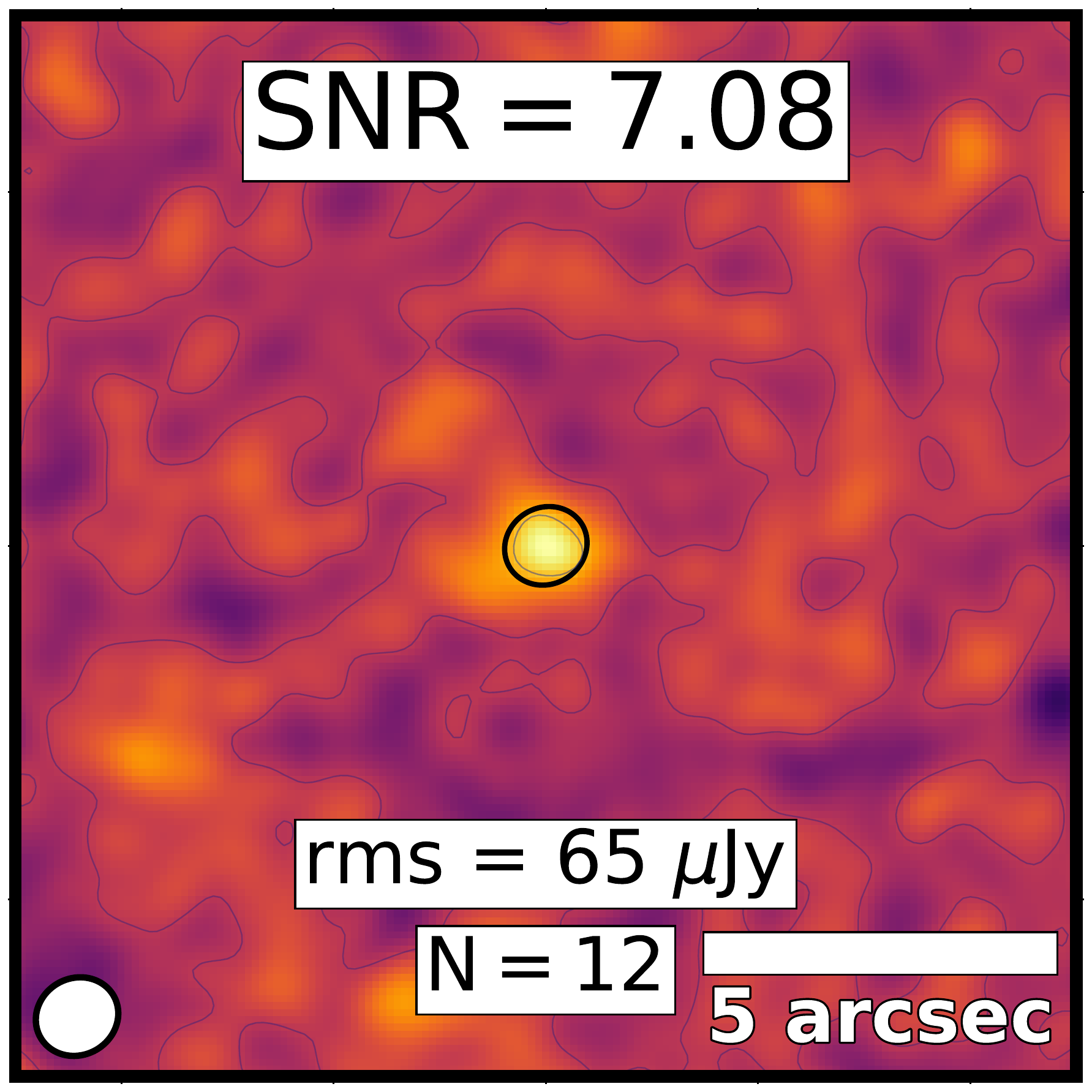}
		\end{subfigure}%
		\begin{subfigure}[t]{.245\linewidth}
			\centering
            \includegraphics[width=1.\textwidth]{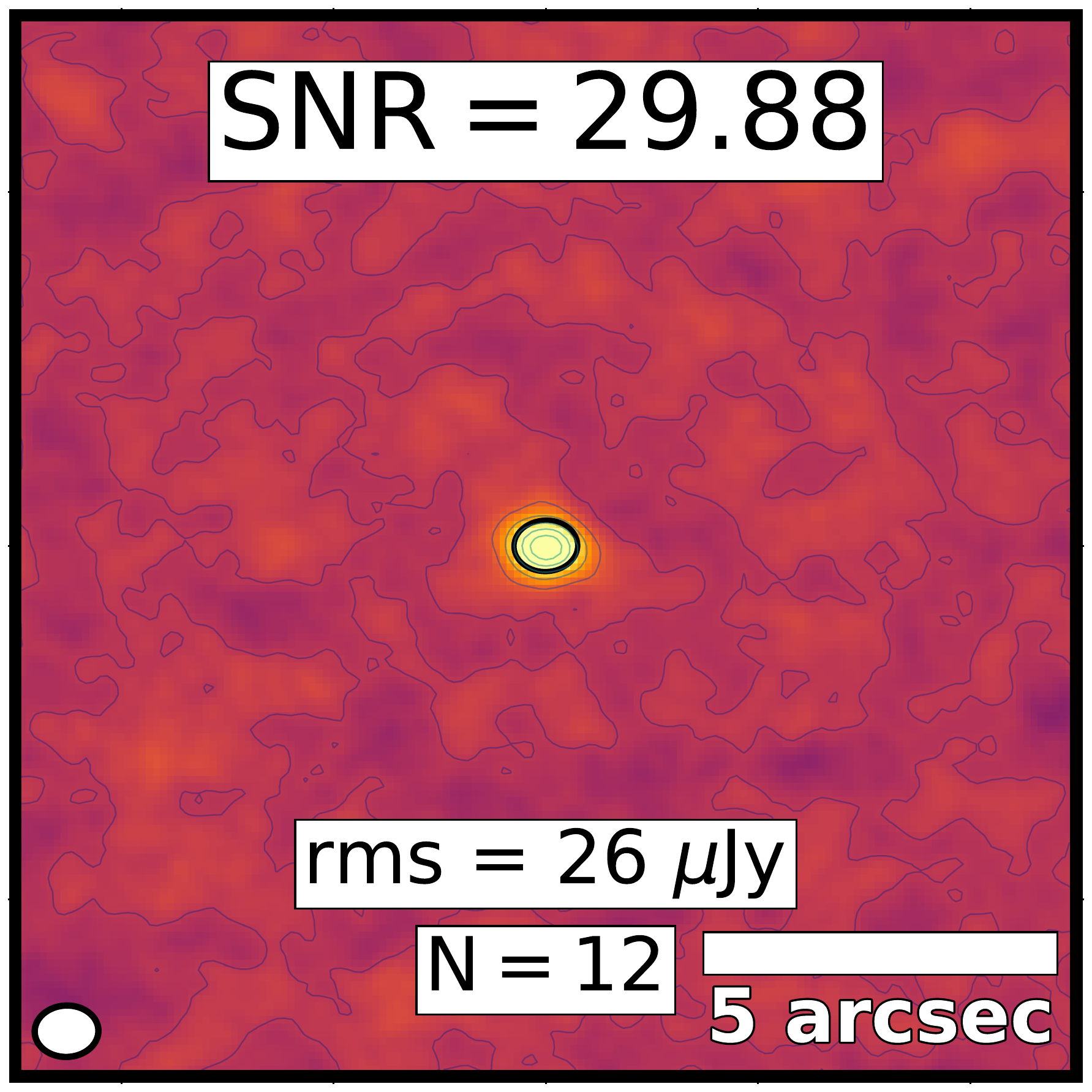}
		\end{subfigure}\\%
		\begin{subfigure}[t]{.245\linewidth}
			\centering
			\includegraphics[width=1.\textwidth]{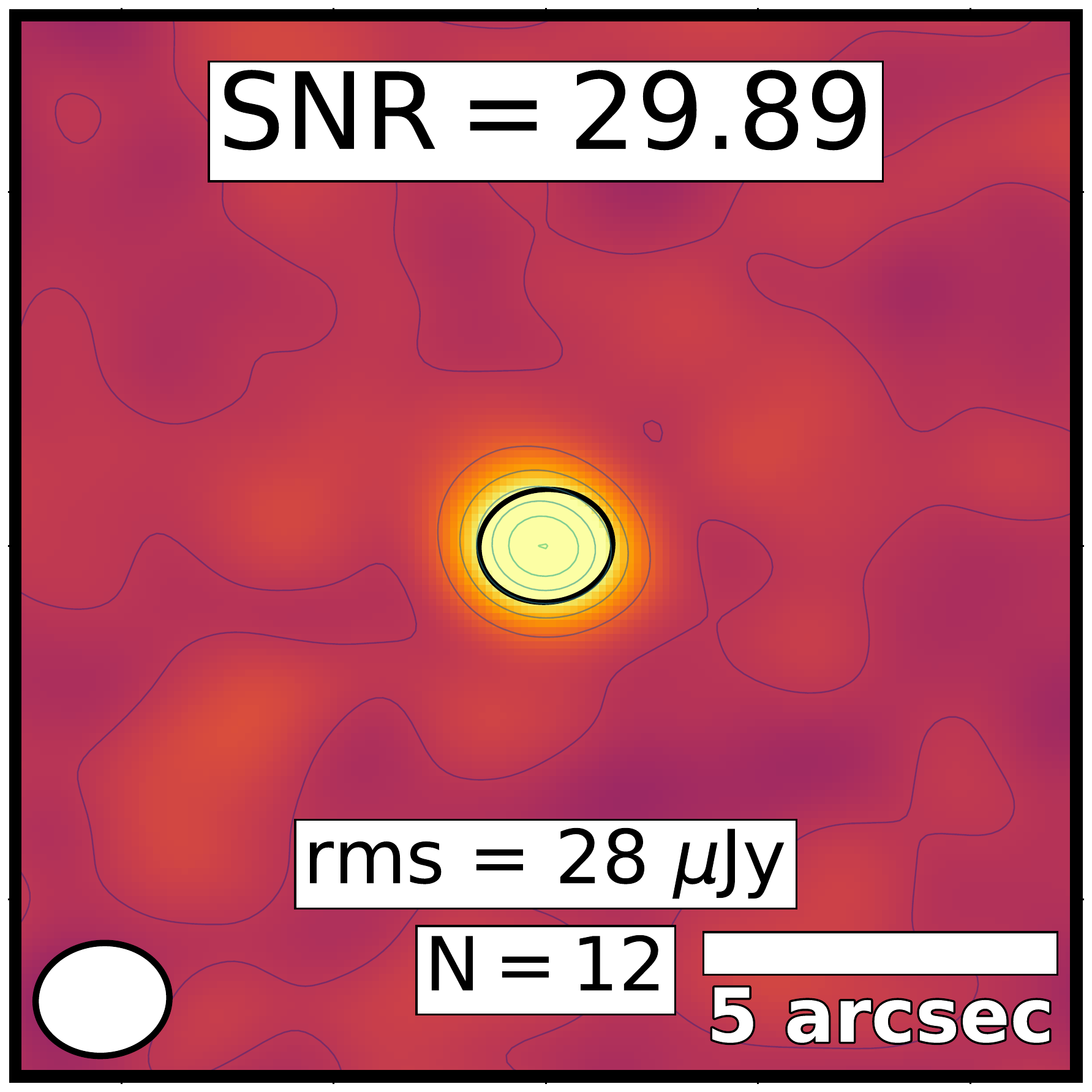}
			\caption{Weight: equal}\label{fig:FFI_uv_nw}
		\end{subfigure}%
		\begin{subfigure}[t]{.245\linewidth}
			\centering
			\includegraphics[width=1.\textwidth]{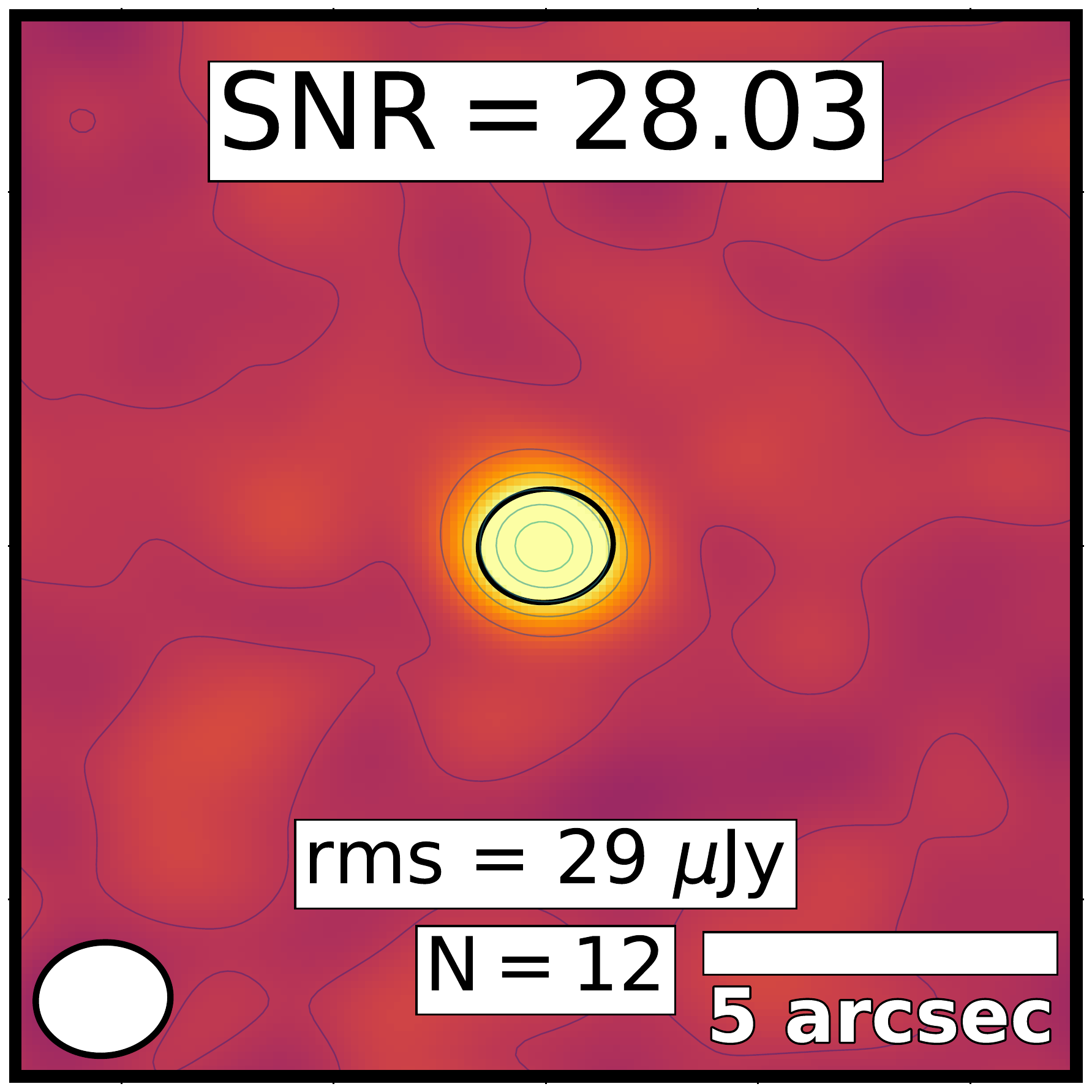}
			\caption{Weight: $pbcor$}\label{fig:FFI_uv_pb}
		\end{subfigure}%
		\begin{subfigure}[t]{.245\linewidth}
			\centering
			\includegraphics[width=1.\textwidth]{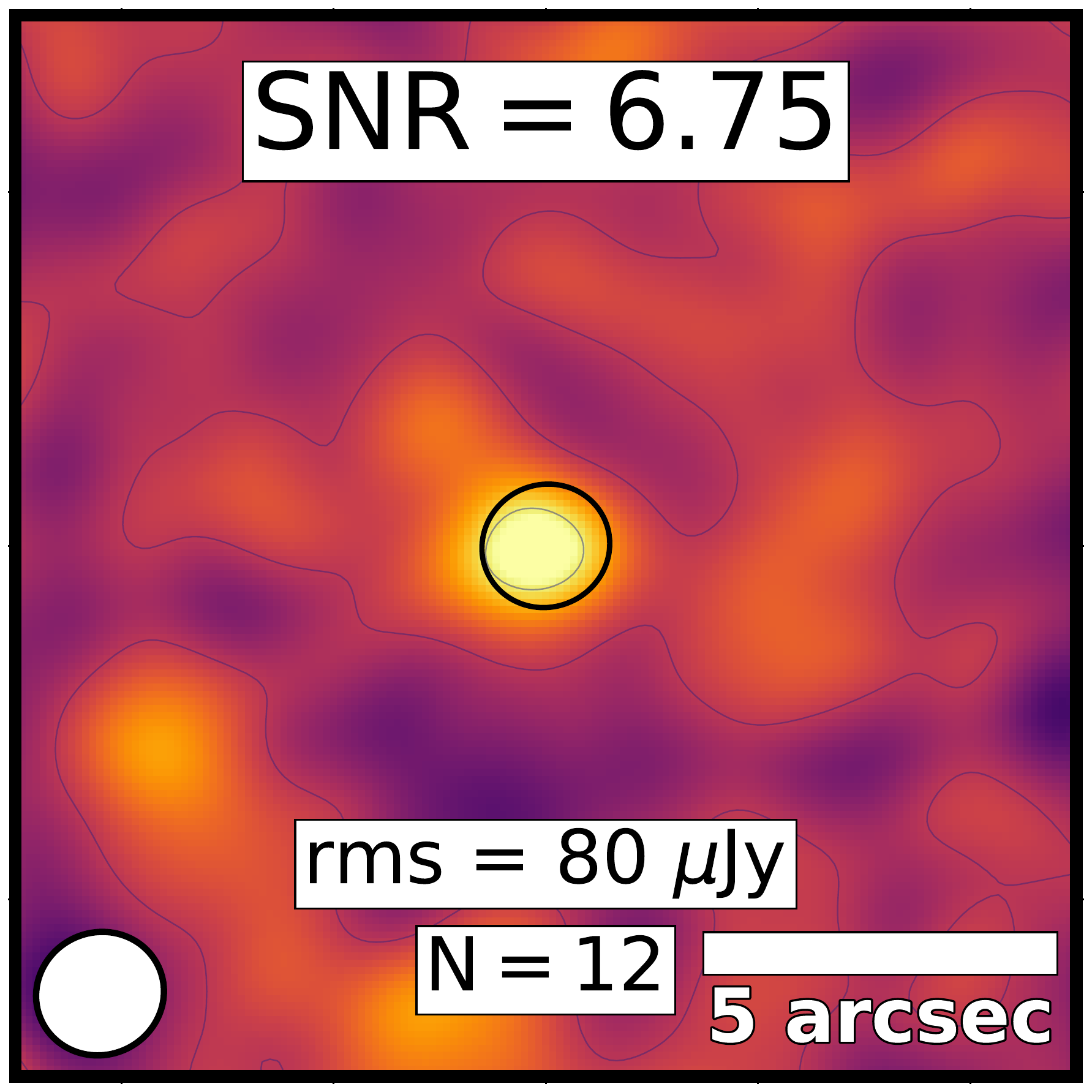}
			\caption{Weight: $F_{\rm UV}$}\label{fig:FFI_uv_fuv}
		\end{subfigure}%
		\begin{subfigure}[t]{.245\linewidth}
			\centering
			\includegraphics[width=1.\textwidth]{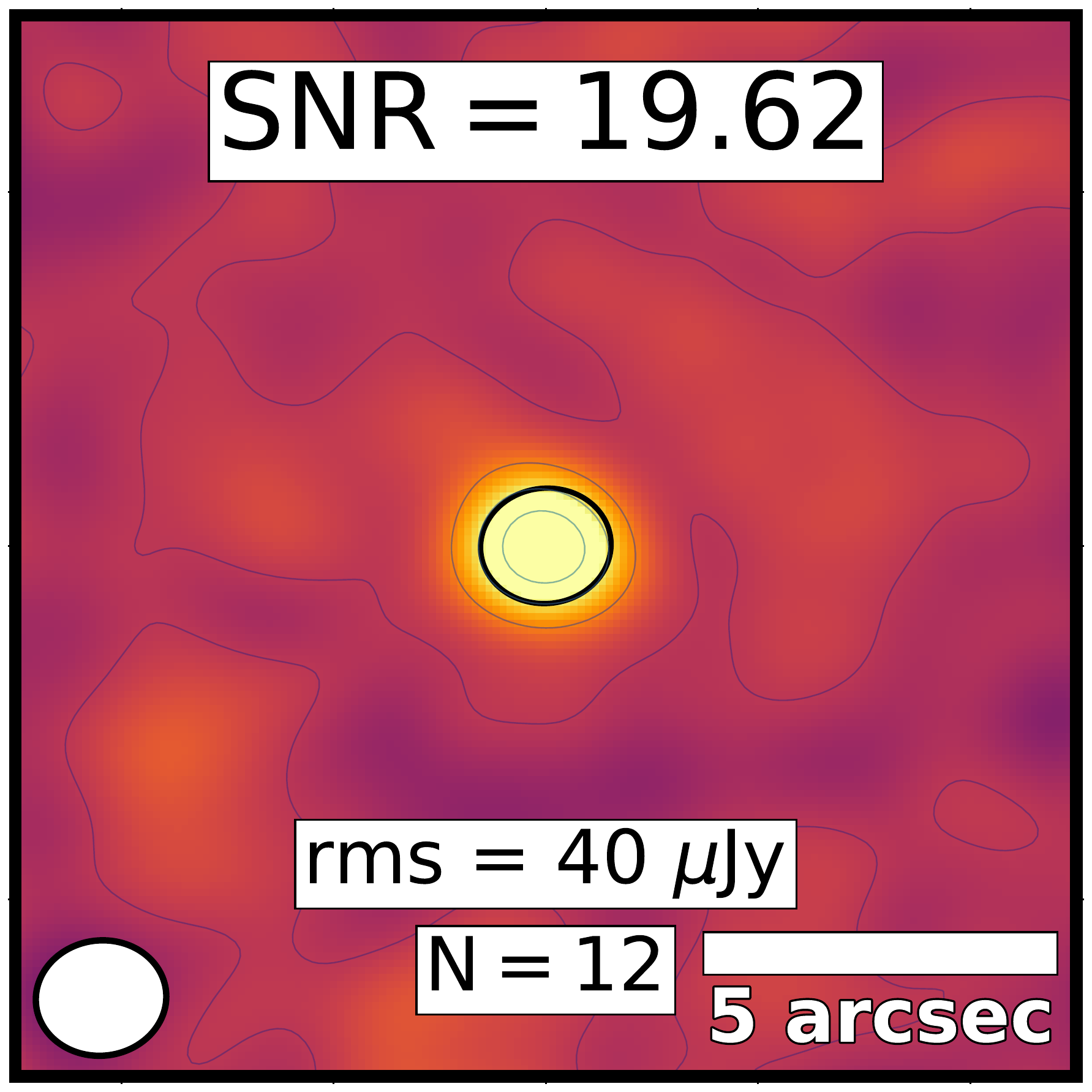}
			\caption{Weight: $\mu$}\label{fig:FFI_uv_mu}
		\end{subfigure}%
		\caption{$u$--$v$ stacked image stamps for the 12 detected sources from \citet{2017A&A...597A..41G}. Panels denote specific weighting configurations (from left to right: equal, $pbcor$, $F_{\rm UV}$, and $\mu$) and \texttt{CASA} \textit{CLEAN}ing procedures ({\it upper}: Natural; {\it lower}: Taper). Color scale spans $-$455\,$\mu$Jy to $+$455\,$\mu$Jy range and contours are drawn for every $5 {\times} rms$ level. White and black ellipses represent the synthesized beam size, while white bars in the right corner denote $5 \arcsec$ scale. The number of sources used for the stacked bin is denoted at bottom, as well as the resultant $rms$.}\label{fig:stack_FFI_uv}
	\end{figure*}

\begin{figure*}[htb]
		\centering 
		\begin{subfigure}[t]{.245\linewidth}
			\centering
			\includegraphics[width=1.\textwidth]{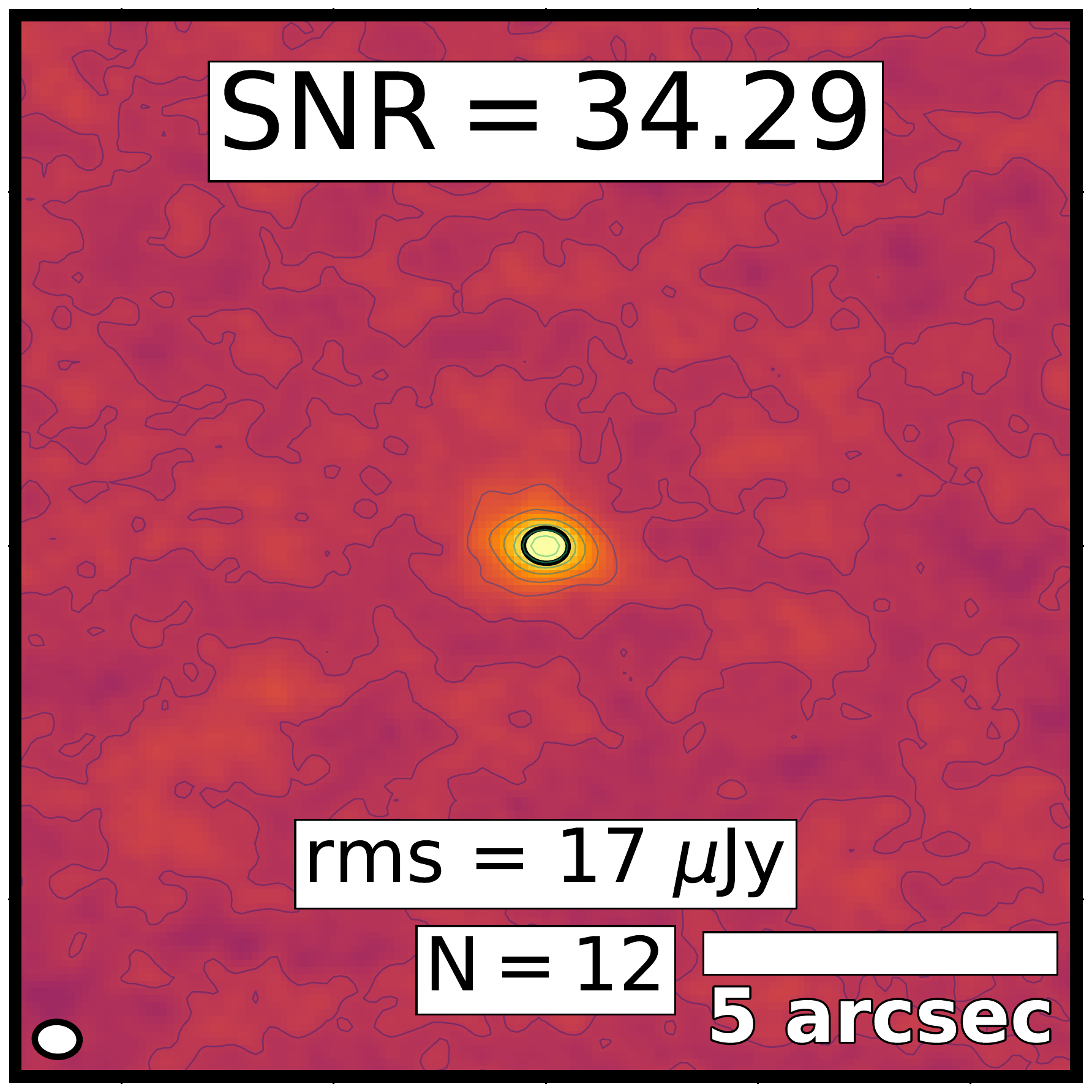}
		\end{subfigure}%
		\begin{subfigure}[t]{.245\linewidth}
			\centering
			\includegraphics[width=1.\textwidth]{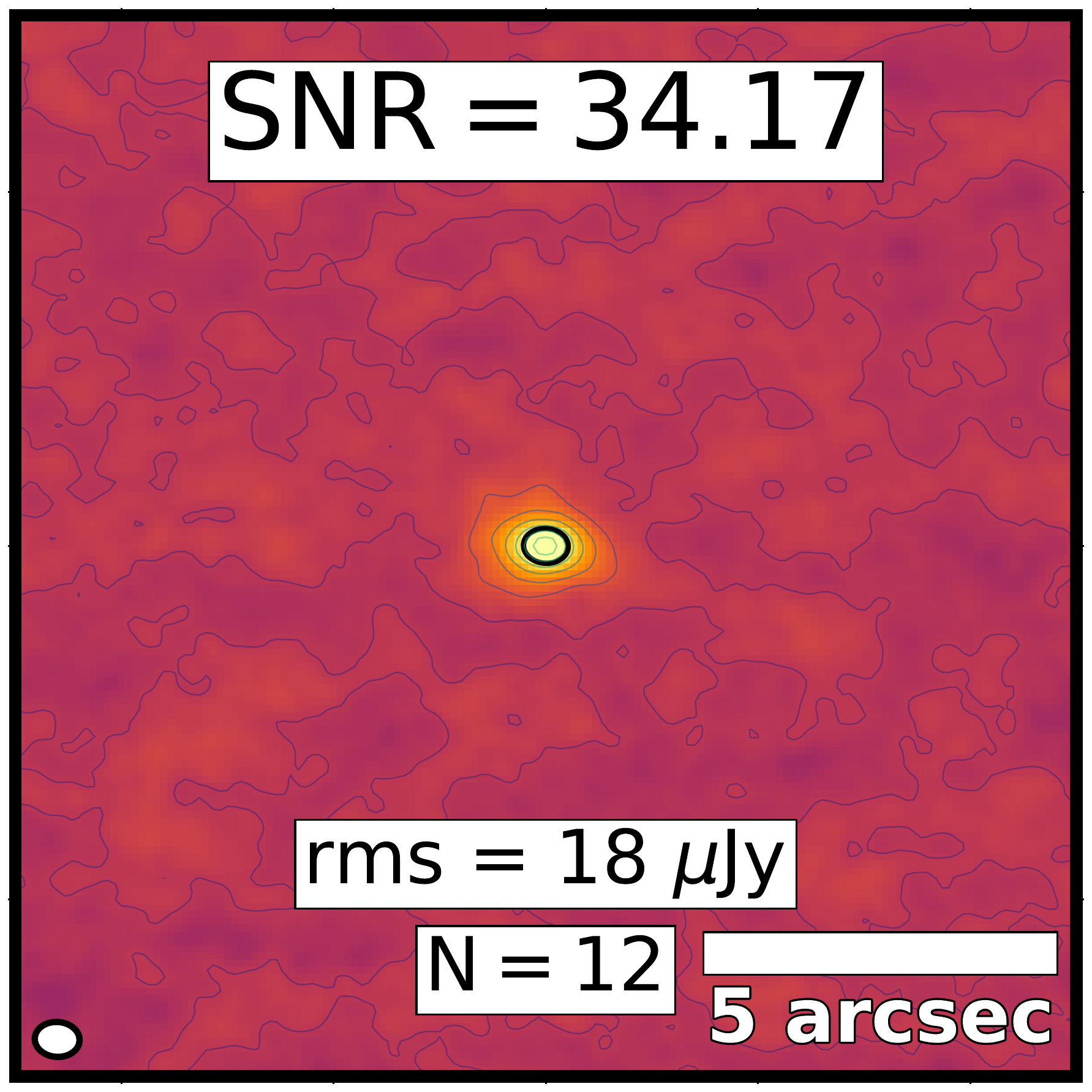}
		\end{subfigure}%
		\begin{subfigure}[t]{.245\linewidth}
			\centering
			\includegraphics[width=1.\textwidth]{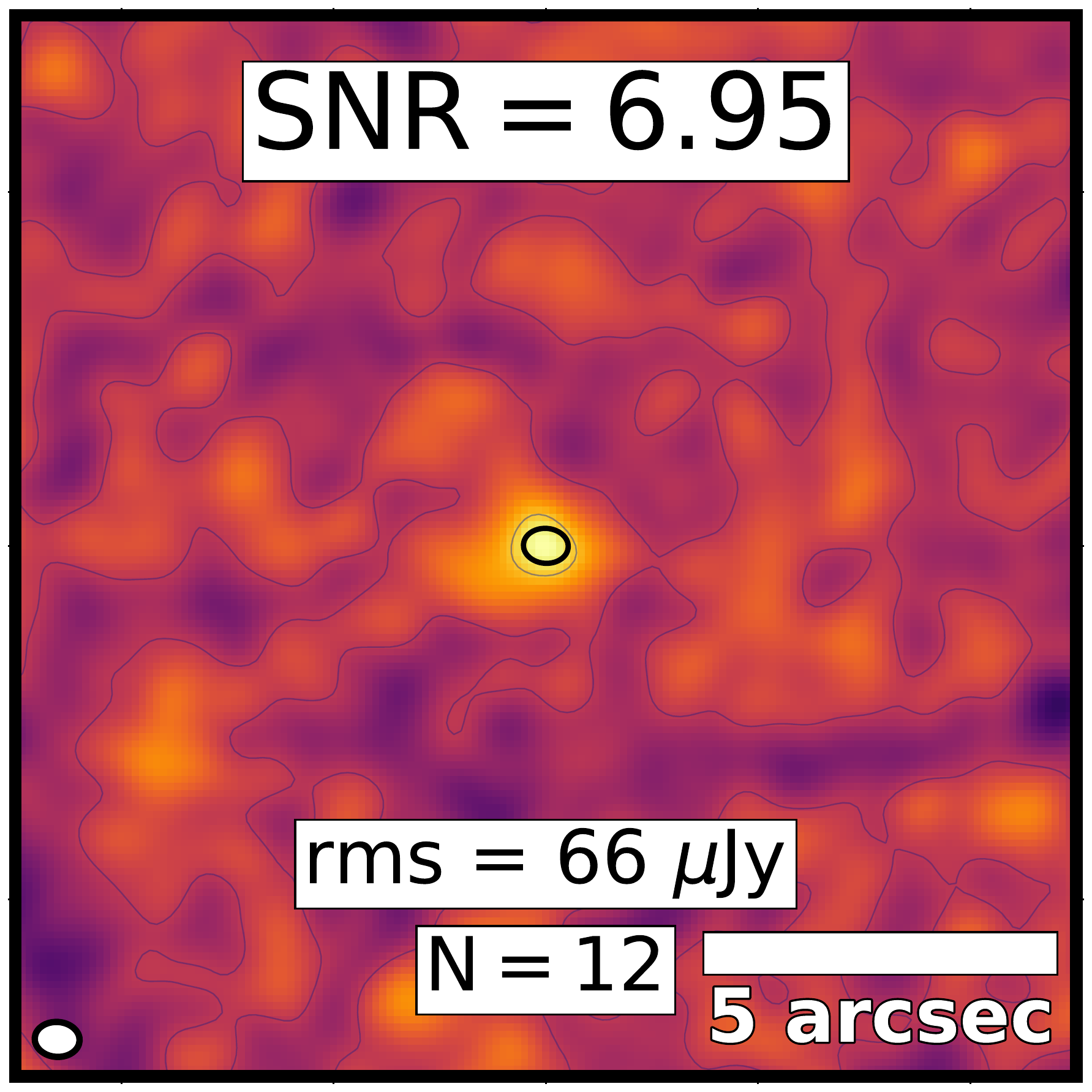}
		\end{subfigure}%
		\begin{subfigure}[t]{.245\linewidth}
			\centering
            \includegraphics[width=1.\textwidth]{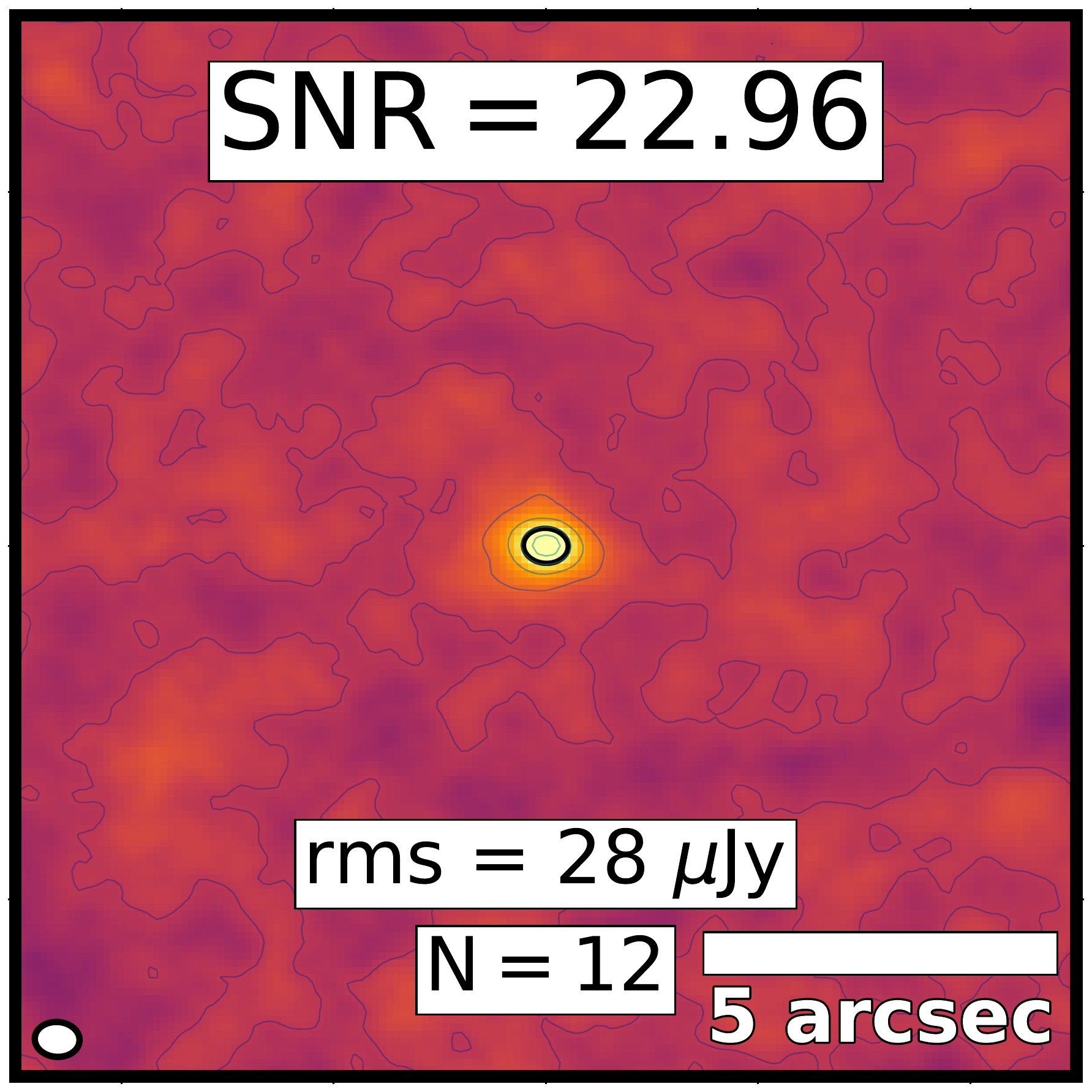}
		\end{subfigure}\\%
		\begin{subfigure}[t]{.245\linewidth}
			\centering
			\includegraphics[width=1.\textwidth]{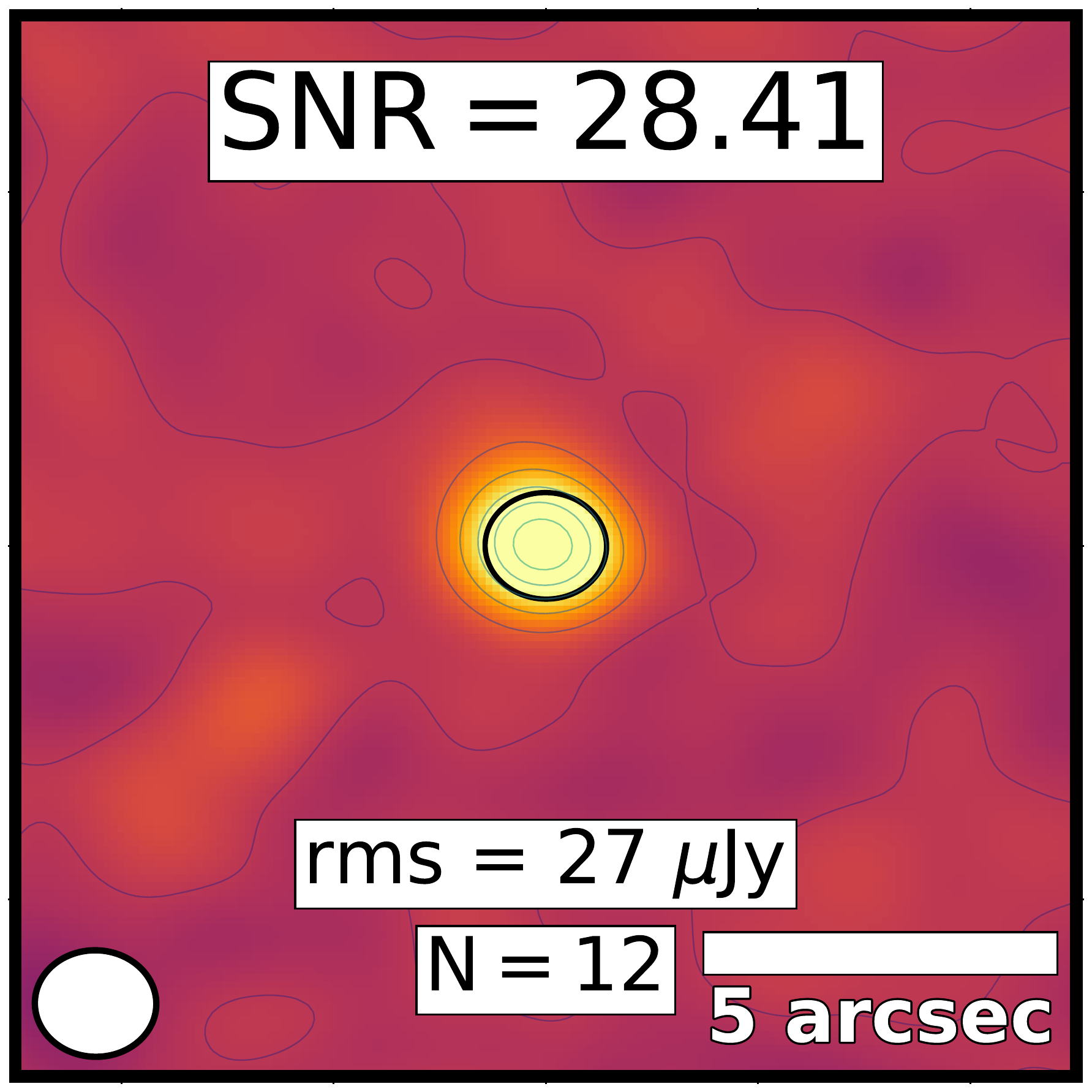}
			\caption{Weight: equal}\label{fig:FFI_im_nw}
		\end{subfigure}%
		\begin{subfigure}[t]{.245\linewidth}
			\centering
			\includegraphics[width=1.\textwidth]{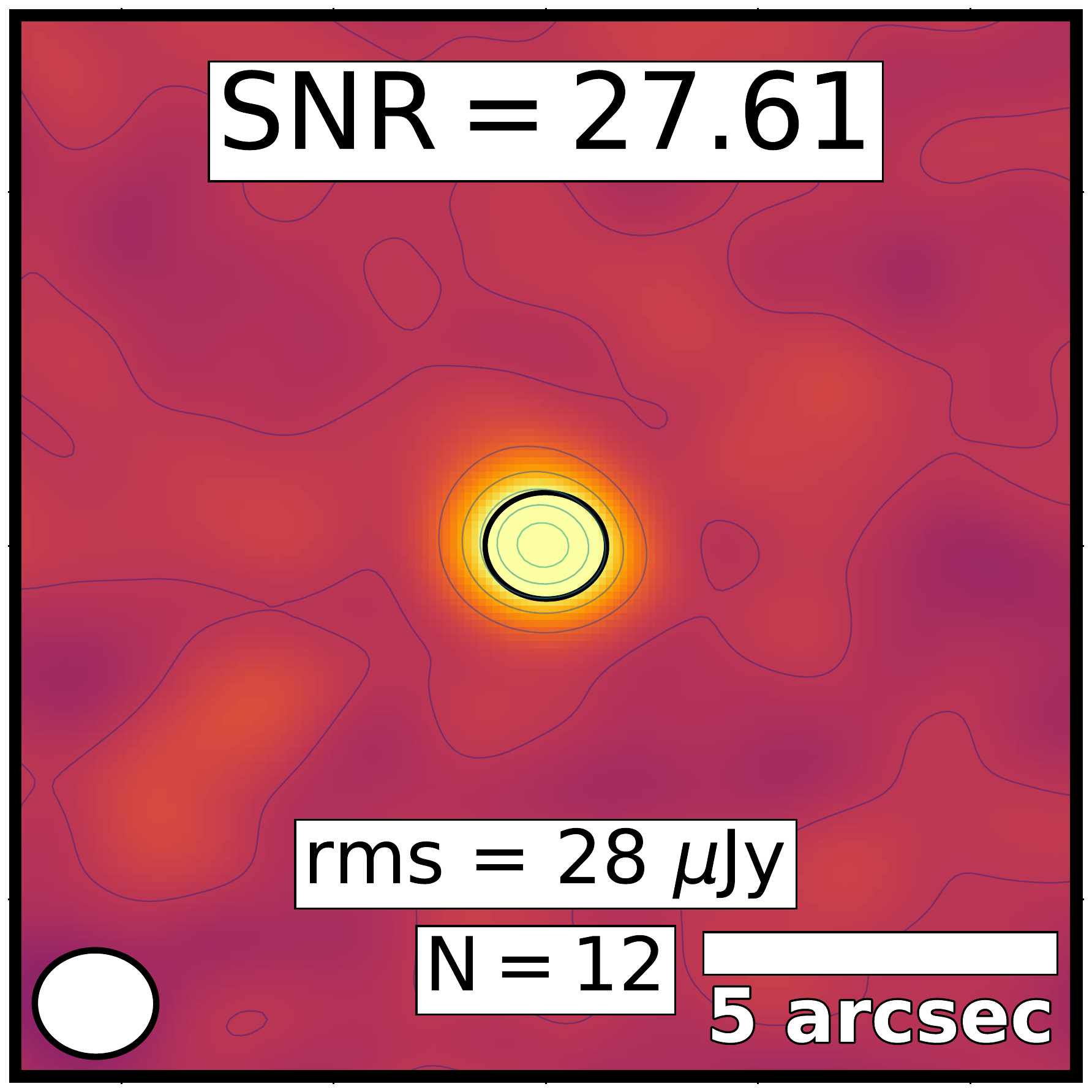}
			\caption{Weight: $pbcor$}\label{fig:FFI_im_pb}
		\end{subfigure}%
		\begin{subfigure}[t]{.245\linewidth}
			\centering
			\includegraphics[width=1.\textwidth]{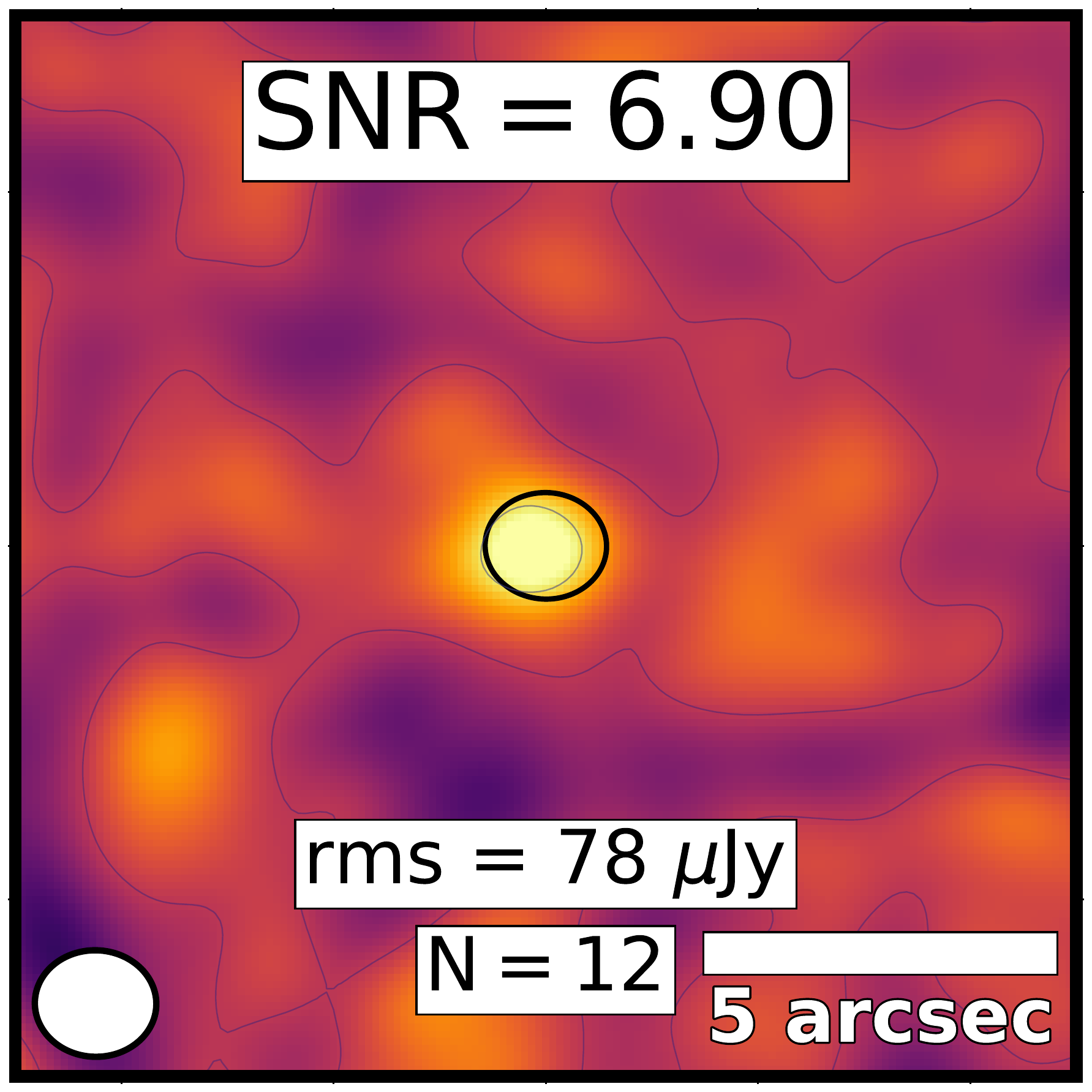}
			\caption{Weight: $F_{\rm UV}$}\label{fig:FFI_im_fuv}
		\end{subfigure}%
		\begin{subfigure}[t]{.245\linewidth}
			\centering
			\includegraphics[width=1.\textwidth]{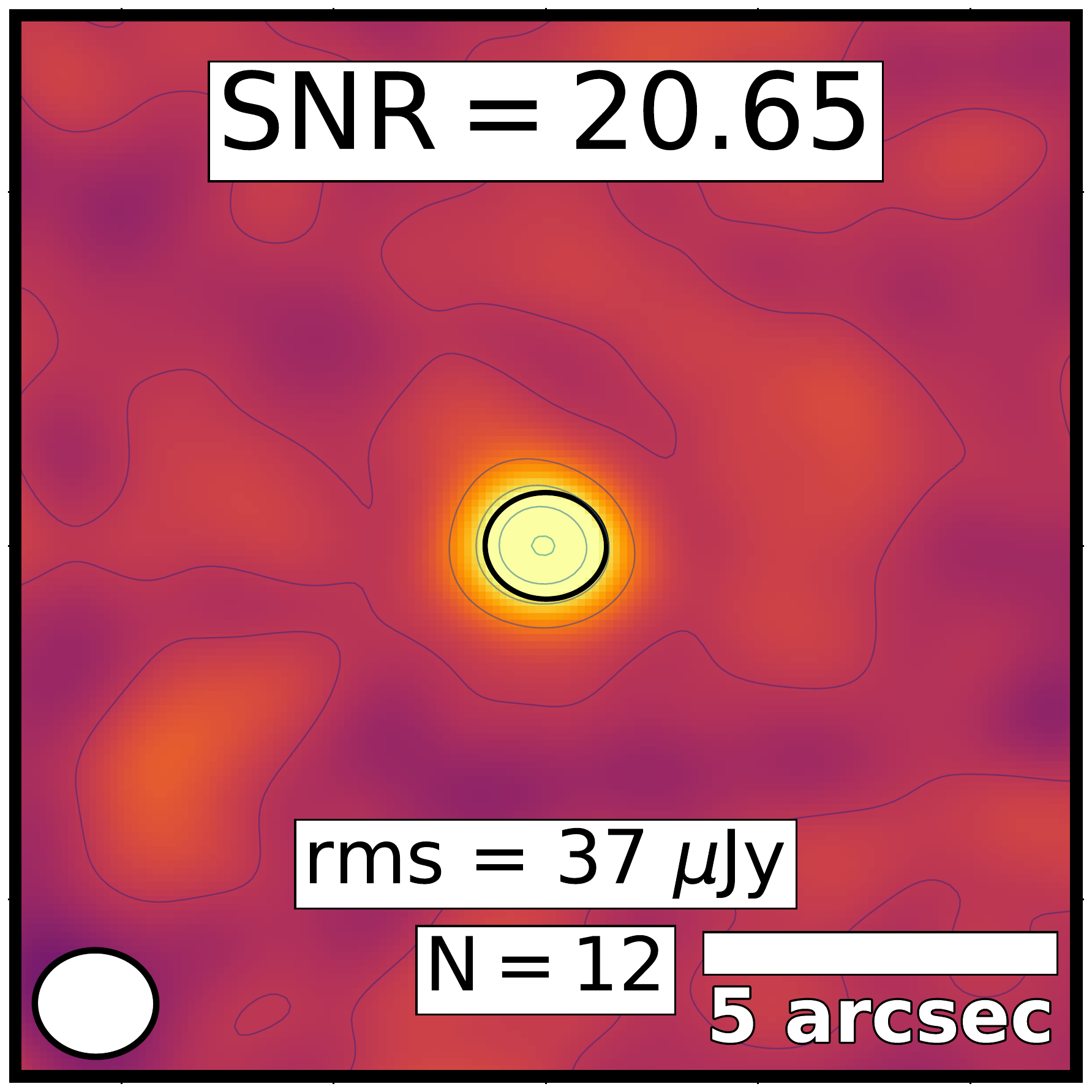}
			\caption{Weight: $\mu$}\label{fig:FFI_im_mu}
		\end{subfigure}%
		\caption{Image stacked stamps for the 12 detected sources from \citet{2017A&A...597A..41G}. Details same as Fig.~\ref{fig:stack_FFI_uv}.}\label{fig:stack_FFI_im}
	\end{figure*}

As a cross-check to confirm that \texttt{STACKER} behaves as expected, we first performed image and $u$--$v$ stacking for the twelve (12) robustly detected dusty star{-}forming galaxies from A2744, MACSJ0416, and MACSJ1149, which are described in \citet{2017A&A...597A..41G} and in \citet{2017A&A...604A.132L}; the individual detections range from 5.1--$\sigma$ to 25.9--$\sigma$. These stacking results are presented in Figs. \ref{fig:stack_FFI_uv} ($u$--$v$ stacking) and \ref{fig:stack_FFI_im} (image stacking).
	
	The $u$--$v$ stacking (Fig. \ref{fig:stack_FFI_uv}), adopting natural-weight \textit{CLEAN}ing, yields stacked detections of 
	39.1--$\sigma$, 37.3--$\sigma$, 29.9--$\sigma$, and 7.1--$\sigma$, for equal, $pbcor$, $F_{\mathrm{UV}}$, and $\mu$-weighting, respectively. Unsurprisingly the unweighted and $pbcor$-weighted stacks achieve much better stacked S/N and {\it rms} in this scenario because all of the sources have high S/N and lie within the high $pbcor$ regions. The strong variation between the $\mu$ and $F_{\rm UV}$ weighted stacks arises because of the dominating presence of the UV-bright, low-magnification source M1149-ID01. 

\begin{figure*}[htb]
		\centering 
		\begin{subfigure}[t]{.245\linewidth}
			\centering
			\includegraphics[width=1.\textwidth]{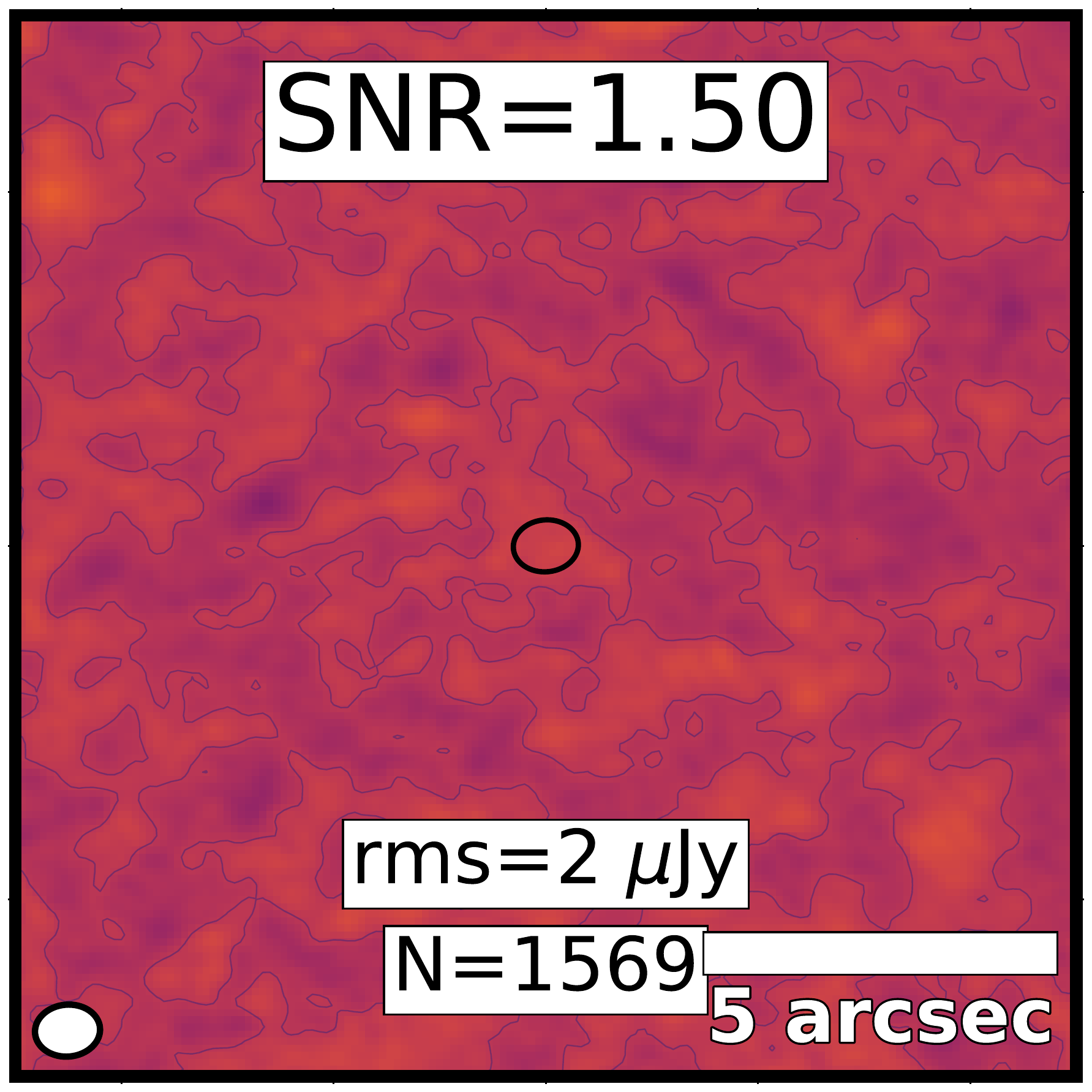}
		\end{subfigure}%
		\begin{subfigure}[t]{.245\linewidth}
			\centering
			\includegraphics[width=1.\textwidth]{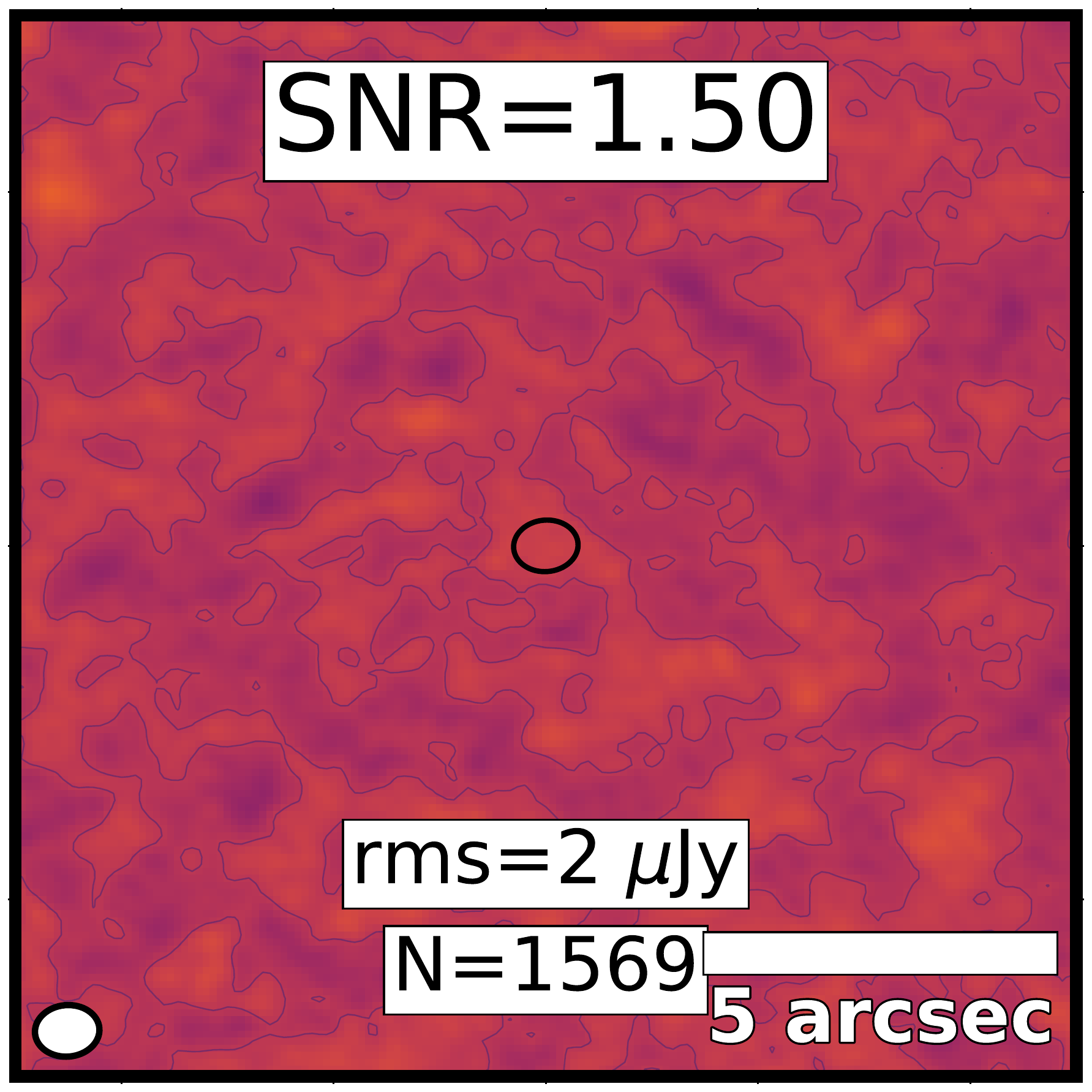}
		\end{subfigure}%
		\begin{subfigure}[t]{.245\linewidth}
			\centering
			\includegraphics[width=1.\textwidth]{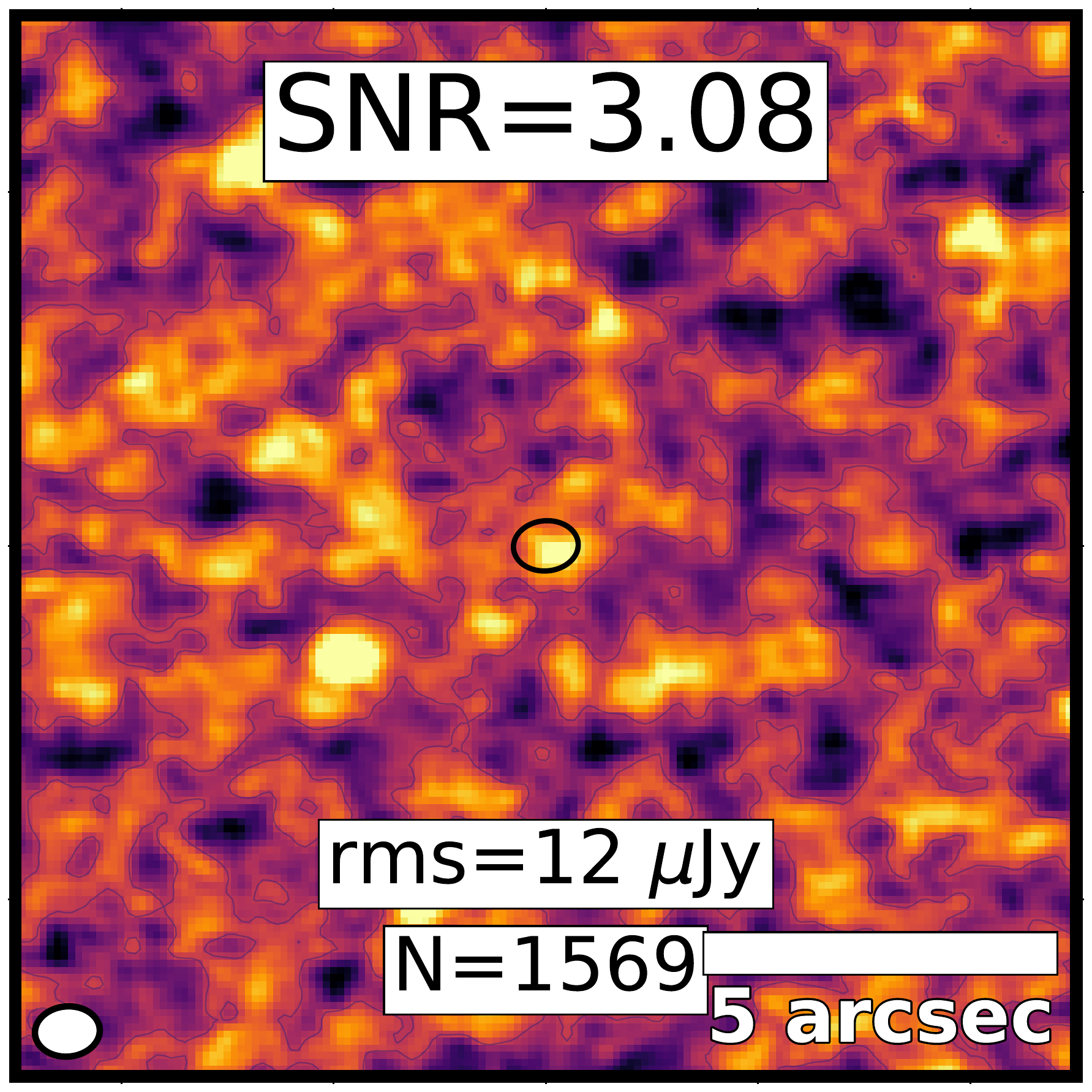}
		\end{subfigure}%
		\begin{subfigure}[t]{.245\linewidth}
			\centering
            \includegraphics[width=1.\textwidth]{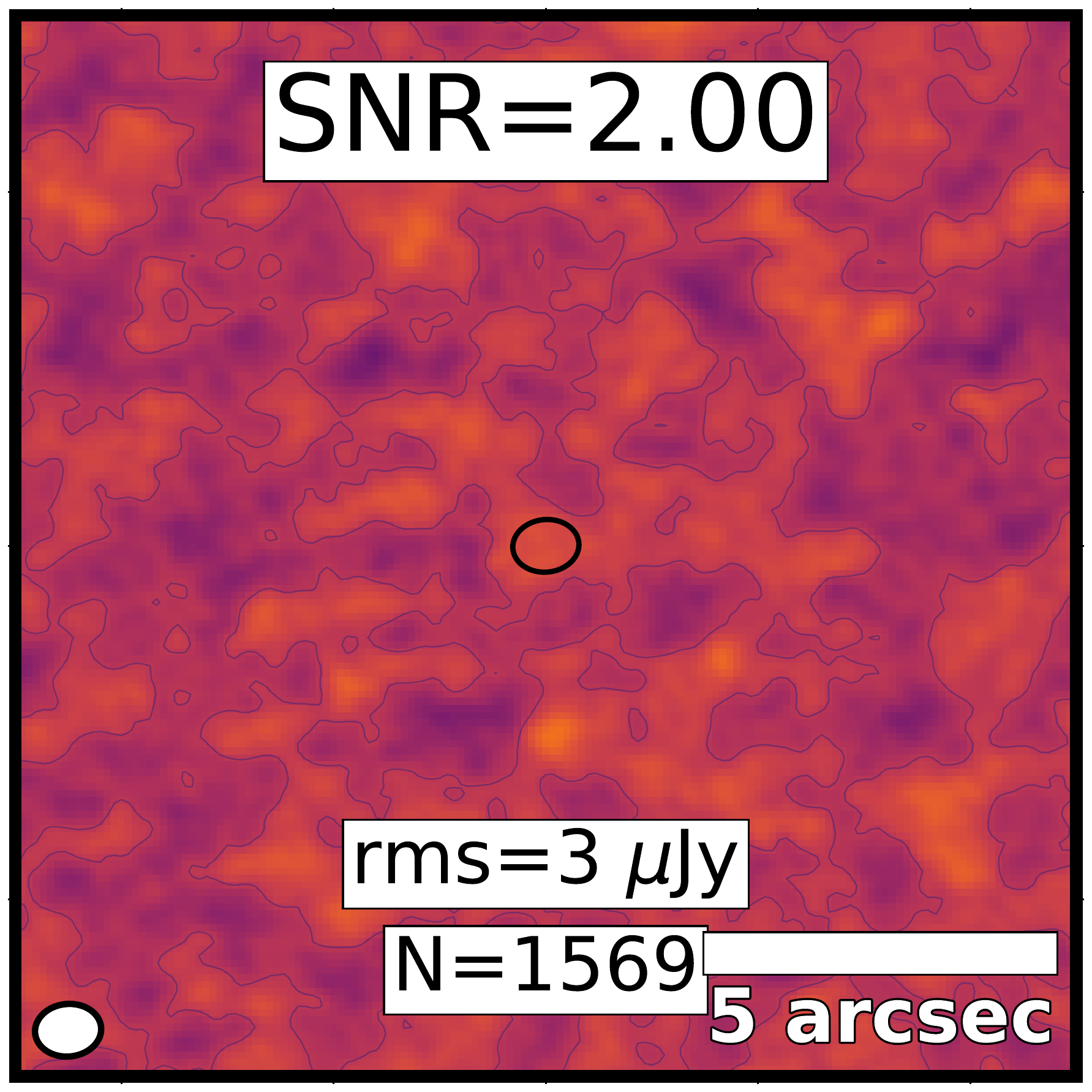}
		\end{subfigure}\\%
		\begin{subfigure}[t]{.245\linewidth}
			\centering
			\includegraphics[width=1.\textwidth]{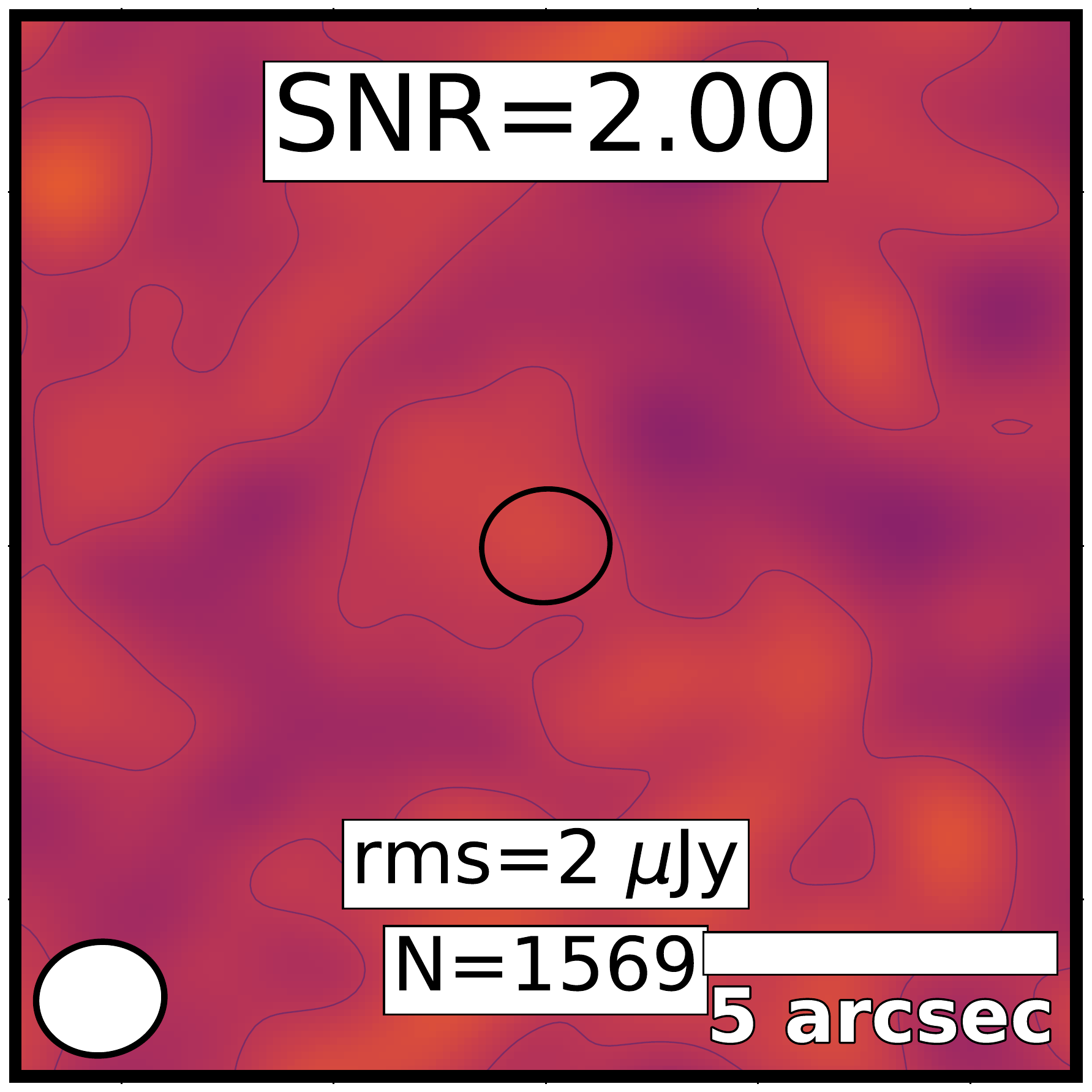}
			\caption{Weight: equal}\label{fig:all_uv_nw}
		\end{subfigure}%
		\begin{subfigure}[t]{.245\linewidth}
			\centering
			\includegraphics[width=1.\textwidth]{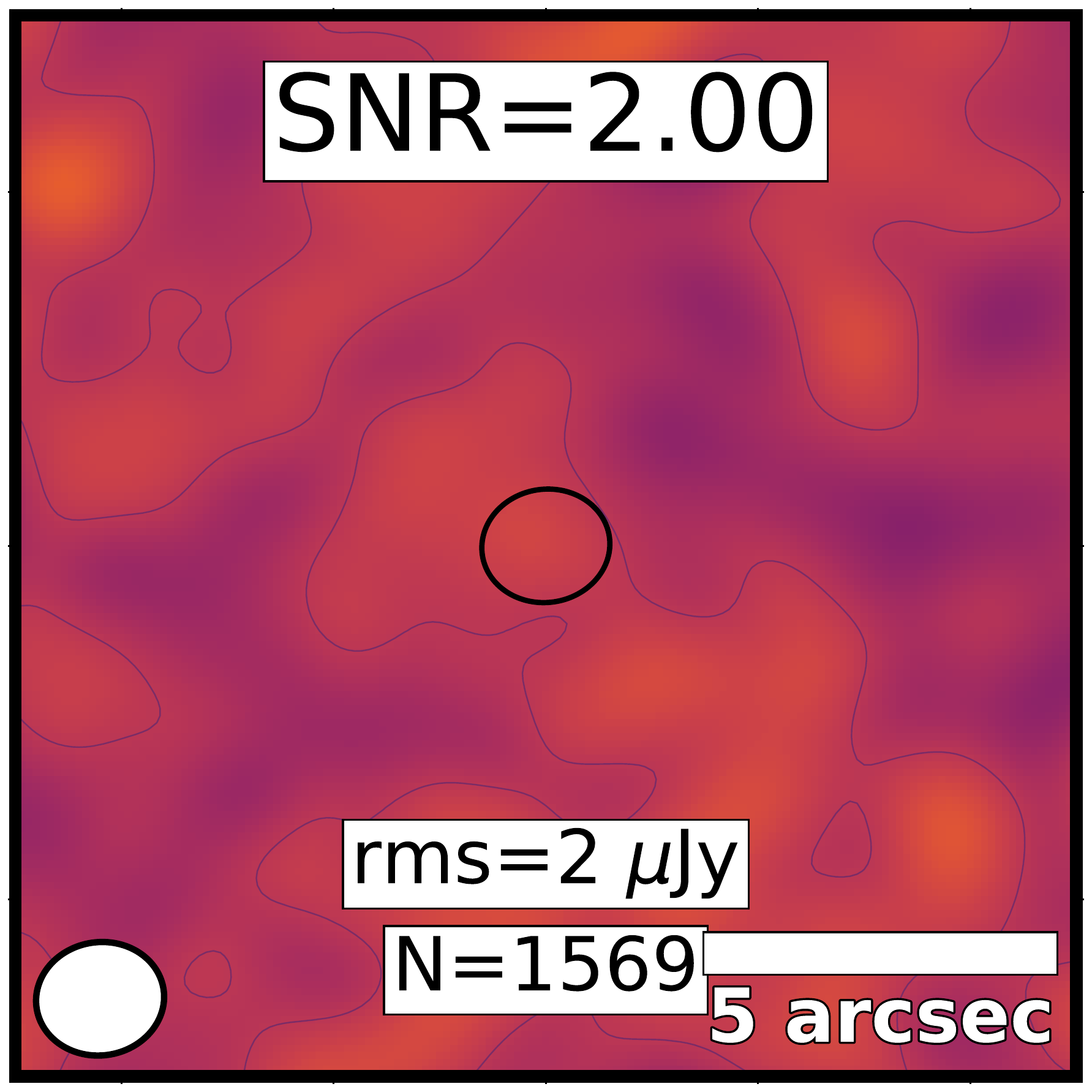}
			\caption{Weight: $pbcor$}\label{fig:all_uv_pb}
		\end{subfigure}%
		\begin{subfigure}[t]{.245\linewidth}
			\centering
			\includegraphics[width=1.\textwidth]{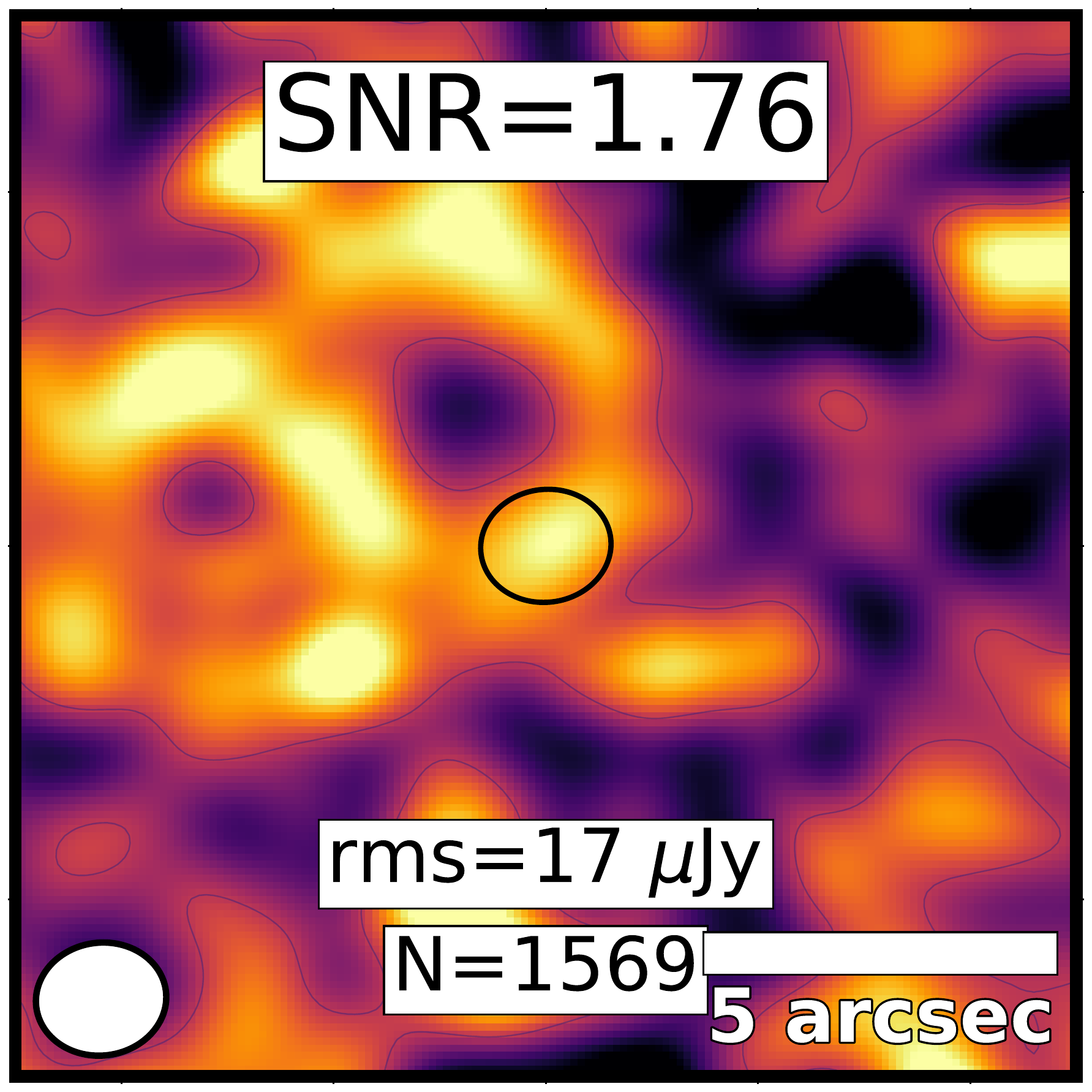}
			\caption{Weight: $F_{\rm UV}$}\label{fig:all_uv_fuv}
		\end{subfigure}%
		\begin{subfigure}[t]{.245\linewidth}
			\centering
			\includegraphics[width=1.\textwidth]{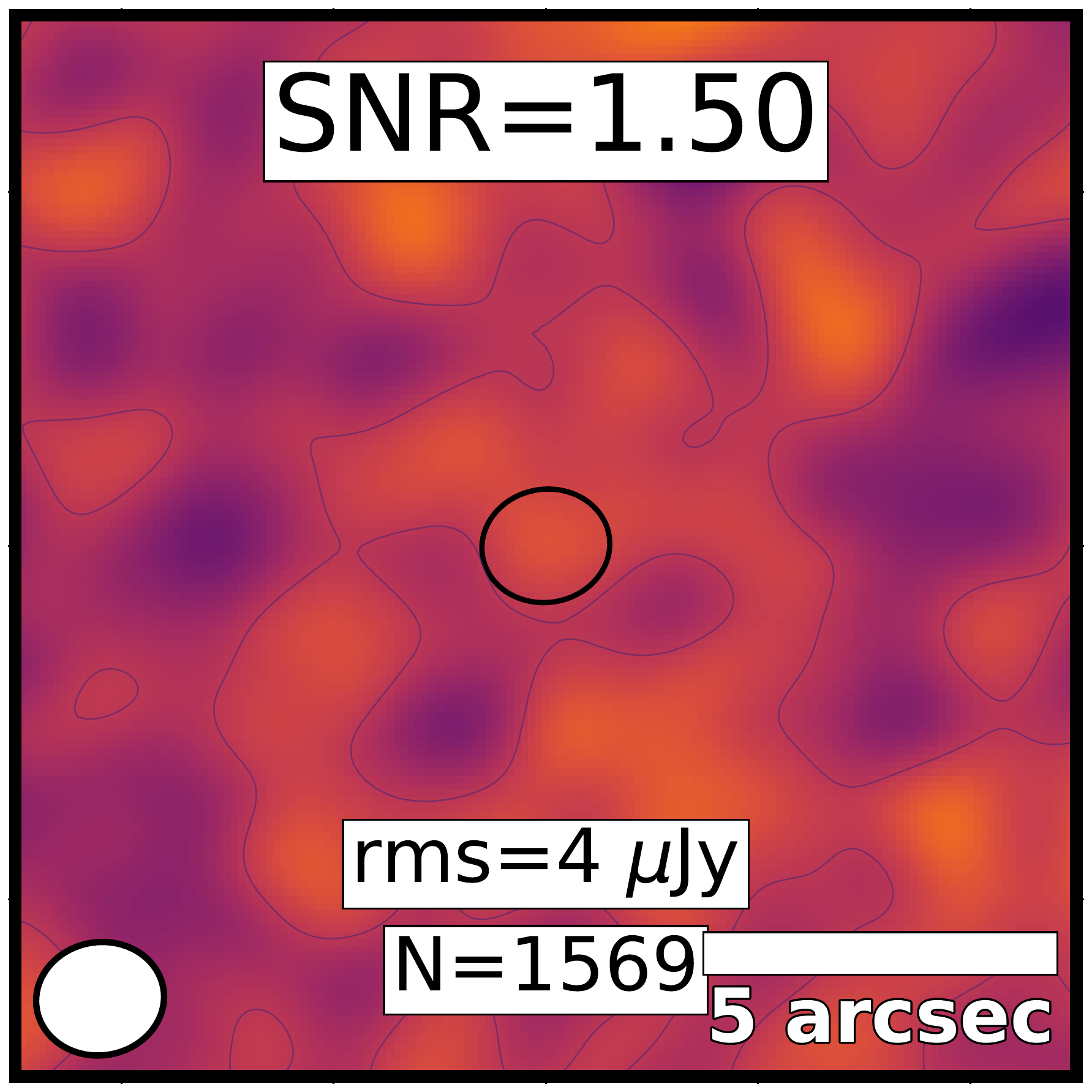}
			\caption{Weight: $\mu$}\label{fig:all_uv_mu}
		\end{subfigure}%
		\caption{$u$--$v$ stacked stamps for the $1569$ upper limits with stellar masses in excess of $10^{6}\,\mathrm{M}_{\bigstar}$. Details same as Fig.~\ref{fig:stack_FFI_uv}. Color scale spans ${-}30\,\mu$Jy to ${+}30\,\mu$Jy range.}\label{fig:stack_all_im}
	\end{figure*}

	As a second test, we stacked all $1569$ sources (upper limits with stellar masses higher than $10^{6}\mathrm{M}_{\bigstar}$) into a single image, adopting the four weighting schemes (equal, $pbcor$, $F_{\mathrm{UV}}$, and $\mu$), obtaining rms levels down to ${\sim} 2 \mu$Jy. S/N values are S/N$_{no}{=} 1.5$, S/N$_{pbcor}{=} 1.5$, S/N$_{F_{\mathrm{UV}}}{=} 3.1$, and S/N$_{\mu}{=} 2.0$ for equal, $pbcor$, $F_{\mathrm{UV}}$, and $\mu$-weighting, respectively. In this case, the $\mu$-weighting achieves only slightly higher $rms$ values than the equal and $pbcor$-weighting, despite weighting far fewer sources. On the other hand, the $F_{\mathrm{UV}}$-weighting achieves a much worse $rms$ yet a relatively high S/N since it is optimized for candidates with higher star formation rates. Stamps of the stacked images are shown in Fig.~\ref{fig:stack_all_im}.
	
\section{Stacking results}\label{app1}

	We include here the results of the stacking process for the LBG candidates as functions of UV-slope $\beta$ and stellar mass $M\mathrm{_{\bigstar}}$. For each configuration we list four results pairing the different stacking weights and CLEANing methods that were used.

\subsection{UV slope binning}\label{tab:beta_stack}

	LBG candidates were divided into five UV-slope ($\beta$) bins as described in $\S$\ref{sec:Methods}.	Targets with stellar masses below $10^{6}$\,M$_{\odot}$ were excluded to avoid poorly constrained sources with low-quality photometry and, hence, maintain consistency with stellar mass stacking. We list results in Table \ref{tab:stack_results_full_beta_z_bin} for both magnification ($\mu$) and UV flux ($F\mathrm{_{UV}}$) stacking weights, as well as natural and taper \textit{CLEAN}ing methods (see $\S$\ref{sec:Methods}).


\begin{table*}
\caption{$u$--$v$ stacking results for different $\beta$ and photometric redshift bins for our full sample.}\label{tab:stack_results_full_beta_z_bin}
\centering
\begin{tabular}{c c}
\begin{adjustbox}{width=0.45\linewidth}
\begin{tabular}{@{}c@{\hspace{1.25\tabcolsep}}c@{\hspace{0.5\tabcolsep}}c@{\hspace{0.75\tabcolsep}}c@{\hspace{0.75\tabcolsep}}c@{\hspace{0.75\tabcolsep}}c@{\hspace{0.75\tabcolsep}}S[separate-uncertainty=true,table-figures-decimal=0,table-text-alignment=center] S@{}}

\hline\hline
UV-slope	& $z_{ph}$	& Sources \#\tablefootmark{a}	& Weight\tablefootmark{b}   & $\log{(\mathrm{IRX})}$\tablefootmark{c}	& CLEAN\tablefootmark{d}	& {Flux\tablefootmark{e, f}}	& S/N\tablefootmark{g}\\
			 					& 		  					&  		   						& 	   						& 	   						& 	    					& {[$\mu$Jy]} 		&  \\
\hline
\multirow{24}{*}{$-4.0 \leq \beta < -3.0$}	&\multirow{8}{*}{$z < 4.0$}	& \multirow{8}{*}{22}	& \multirow{2}{*}{$\mathrm{equal}$}& \multirow{2}{*}{<2.86}	& Natural	& 9 \pm 15	& 0.60\\*
 	 	 	 	 	 	 	 	 	 	 	 	 	 	 	 	 	 	& 									& 						& 			& 		& Taper	& -22 \pm 22	& -1.00\\*
 	 	 	 	 	 	 	 	 	 	 	 	 	 	 	 	 	 	& 									& 						& \multirow{2}{*}{$pbcor$}	& \multirow{2}{*}{<2.84}	& Natural	& 14 \pm 16	& 0.88\\*
 	 	 	 	 	 	 	 	 	 	 	 	 	 	 	 	 	 	& 									& 						& 			& 		& Taper	& -19 \pm 23	& -0.83\\*
 	 	 	 	 	 	 	 	 	 	 	 	 	 	 	 	 	 	& 									& 						& \multirow{2}{*}{F$_{UV}$}	& \multirow{2}{*}{<3.10}	& Natural	& 13 \pm 20	& 0.65\\*
 	 	 	 	 	 	 	 	 	 	 	 	 	 	 	 	 	 	& 									& 						& 			& 		& Taper	& -23 \pm 28	& -0.82\\*
 	 	 	 	 	 	 	 	 	 	 	 	 	 	 	 	 	 	& 									& 						& \multirow{2}{*}{$\mu$}	& \multirow{2}{*}{<2.82}	& Natural	& 1 \pm 23	& 0.04\\*
 	 	 	 	 	 	 	 	 	 	 	 	 	 	 	 	 	 	& 									& 						& 			& 		& Taper	& -48 \pm 31	& -1.55\\*
\cline{2-8}
 	 	 	 	 	 	 	 	 	 	 	 	 	 	 	 	 	 	&\multirow{8}{*}{$4.0 \leq z < 7.0$}	& \multirow{8}{*}{14}	& \multirow{2}{*}{$\mathrm{equal}$}	& \multirow{2}{*}{<2.41}	& Natural	& 49 \pm 19	& 2.58\\*
 	 	 	 	 	 	 	 	 	 	 	 	 	 	 	 	 	 	& 									& 						& 			& 		& Taper	& 29 \pm 27	& 1.07\\*
 	 	 	 	 	 	 	 	 	 	 	 	 	 	 	 	 	 	& 									& 						& \multirow{2}{*}{$pbcor$}	& \multirow{2}{*}{<2.42}	& Natural	& 47 \pm 19	& 2.47\\*
 	 	 	 	 	 	 	 	 	 	 	 	 	 	 	 	 	 	& 									& 						& 			& 		& Taper	& 26 \pm 28	& 0.93\\*
 	 	 	 	 	 	 	 	 	 	 	 	 	 	 	 	 	 	& 									& 						& \multirow{2}{*}{F$_{UV}$}	& \multirow{2}{*}{<2.34}	& Natural	& 46 \pm 29	& 1.59\\*
 	 	 	 	 	 	 	 	 	 	 	 	 	 	 	 	 	 	& 									& 						& 			& 		& Taper	& 24 \pm 42	& 0.57\\*
 	 	 	 	 	 	 	 	 	 	 	 	 	 	 	 	 	 	& 									& 						& \multirow{2}{*}{$\mu$}	& \multirow{2}{*}{<2.57}	& Natural	& 63 \pm 25	& 2.52\\*
 	 	 	 	 	 	 	 	 	 	 	 	 	 	 	 	 	 	& 									& 						& 			& 		& Taper	& 39 \pm 35	& 1.11\\*
\cline{2-8}
 	 	 	 	 	 	 	 	 	 	 	 	 	 	 	 	 	 	&\multirow{8}{*}{$7.0 \leq z$}	& \multirow{8}{*}{2}	& \multirow{2}{*}{$\mathrm{equal}$}	& \multirow{2}{*}{<2.81}	& Natural	& 90 \pm 42	& 2.14\\*
 	 	 	 	 	 	 	 	 	 	 	 	 	 	 	 	 	 	& 									& 						& 			& 		& Taper	& 124 \pm 73	& 1.70\\*
 	 	 	 	 	 	 	 	 	 	 	 	 	 	 	 	 	 	& 									& 						& \multirow{2}{*}{$pbcor$}	& \multirow{2}{*}{<2.81}	& Natural	& 91 \pm 42	& 2.17\\*
 	 	 	 	 	 	 	 	 	 	 	 	 	 	 	 	 	 	& 									& 						& 			& 		& Taper	& 129 \pm 73	& 1.77\\*
 	 	 	 	 	 	 	 	 	 	 	 	 	 	 	 	 	 	& 									& 						& \multirow{2}{*}{F$_{UV}$}	& \multirow{2}{*}{<2.94}	& Natural	& 118 \pm 58	& 2.03\\*
 	 	 	 	 	 	 	 	 	 	 	 	 	 	 	 	 	 	& 									& 						& 			& 		& Taper	& 63 \pm 97	& 0.65\\*
 	 	 	 	 	 	 	 	 	 	 	 	 	 	 	 	 	 	& 									& 						& \multirow{2}{*}{$\mu$}	& \multirow{2}{*}{<2.90}	& Natural	& 120 \pm 53	& 2.26\\*
 	 	 	 	 	 	 	 	 	 	 	 	 	 	 	 	 	 	& 									& 						& 			& 		& Taper	& 202 \pm 91	& 2.22\\*
\hline
\multirow{24}{*}{$-3.0 \leq \beta < -2.0$}	&\multirow{8}{*}{$z < 4.0$}	& \multirow{8}{*}{570}	& \multirow{2}{*}{$\mathrm{equal}$}& \multirow{2}{*}{<2.42}	& Natural	& 2 \pm 3	& 0.67\\*
 	 	 	 	 	 	 	 	 	 	 	 	 	 	 	 	 	 	& 									& 						& 			& 		& Taper	& 1 \pm 4	& 0.25\\*
 	 	 	 	 	 	 	 	 	 	 	 	 	 	 	 	 	 	& 									& 						& \multirow{2}{*}{$pbcor$}	& \multirow{2}{*}{<2.42}	& Natural	& 3 \pm 3	& 1.00\\*
 	 	 	 	 	 	 	 	 	 	 	 	 	 	 	 	 	 	& 									& 						& 			& 		& Taper	& 2 \pm 4	& 0.50\\*
 	 	 	 	 	 	 	 	 	 	 	 	 	 	 	 	 	 	& 									& 						& \multirow{2}{*}{F$_{UV}$}	& \multirow{2}{*}{<1.67}	& Natural	& 79 \pm 28	& 2.82\\*
 	 	 	 	 	 	 	 	 	 	 	 	 	 	 	 	 	 	& 									& 						& 			& 		& Taper	& 53 \pm 41	& 1.29\\*
 	 	 	 	 	 	 	 	 	 	 	 	 	 	 	 	 	 	& 									& 						& \multirow{2}{*}{$\mu$}	& \multirow{2}{*}{<2.73}	& Natural	& 8 \pm 5	& 1.60\\*
 	 	 	 	 	 	 	 	 	 	 	 	 	 	 	 	 	 	& 									& 						& 			& 		& Taper	& 2 \pm 7	& 0.29\\*
\cline{2-8}
 	 	 	 	 	 	 	 	 	 	 	 	 	 	 	 	 	 	&\multirow{8}{*}{$4.0 \leq z < 7.0$}	& \multirow{8}{*}{193}	& \multirow{2}{*}{$\mathrm{equal}$}	& \multirow{2}{*}{<2.17}	& Natural	& 9 \pm 5	& 1.80\\*
 	 	 	 	 	 	 	 	 	 	 	 	 	 	 	 	 	 	& 									& 						& 			& 		& Taper	& 6 \pm 7	& 0.86\\*
 	 	 	 	 	 	 	 	 	 	 	 	 	 	 	 	 	 	& 									& 						& \multirow{2}{*}{$pbcor$}	& \multirow{2}{*}{<2.17}	& Natural	& 11 \pm 5	& 2.20\\*
 	 	 	 	 	 	 	 	 	 	 	 	 	 	 	 	 	 	& 									& 						& 			& 		& Taper	& 10 \pm 7	& 1.43\\*
 	 	 	 	 	 	 	 	 	 	 	 	 	 	 	 	 	 	& 									& 						& \multirow{2}{*}{F$_{UV}$}	& \multirow{2}{*}{<1.71}	& Natural	& -2 \pm 20	& -0.10\\*
 	 	 	 	 	 	 	 	 	 	 	 	 	 	 	 	 	 	& 									& 						& 			& 		& Taper	& -19 \pm 28	& -0.68\\*
 	 	 	 	 	 	 	 	 	 	 	 	 	 	 	 	 	 	& 									& 						& \multirow{2}{*}{$\mu$}	& \multirow{2}{*}{<2.16}	& Natural	& 9 \pm 8	& 1.12\\*
 	 	 	 	 	 	 	 	 	 	 	 	 	 	 	 	 	 	& 									& 						& 			& 		& Taper	& 10 \pm 11	& 0.91\\*
\cline{2-8}
 	 	 	 	 	 	 	 	 	 	 	 	 	 	 	 	 	 	&\multirow{8}{*}{$7.0 \leq z$}	& \multirow{8}{*}{14}	& \multirow{2}{*}{$\mathrm{equal}$}	& \multirow{2}{*}{<2.56}	& Natural	& 14 \pm 19	& 0.74\\*
 	 	 	 	 	 	 	 	 	 	 	 	 	 	 	 	 	 	& 									& 						& 			& 		& Taper	& 23 \pm 26	& 0.88\\*
 	 	 	 	 	 	 	 	 	 	 	 	 	 	 	 	 	 	& 									& 						& \multirow{2}{*}{$pbcor$}	& \multirow{2}{*}{<2.57}	& Natural	& 18 \pm 19	& 0.95\\*
 	 	 	 	 	 	 	 	 	 	 	 	 	 	 	 	 	 	& 									& 						& 			& 		& Taper	& 26 \pm 26	& 1.00\\*
 	 	 	 	 	 	 	 	 	 	 	 	 	 	 	 	 	 	& 									& 						& \multirow{2}{*}{F$_{UV}$}	& \multirow{2}{*}{<2.43}	& Natural	& 53 \pm 49	& 1.08\\*
 	 	 	 	 	 	 	 	 	 	 	 	 	 	 	 	 	 	& 									& 						& 			& 		& Taper	& 90 \pm 70	& 1.29\\*
 	 	 	 	 	 	 	 	 	 	 	 	 	 	 	 	 	 	& 									& 						& \multirow{2}{*}{$\mu$}	& \multirow{2}{*}{<2.64}	& Natural	& 32 \pm 28	& 1.14\\*
 	 	 	 	 	 	 	 	 	 	 	 	 	 	 	 	 	 	& 									& 						& 			& 		& Taper	& 12 \pm 38	& 0.32\\*
\hline
\multirow{24}{*}{$-2.0 \leq \beta < -1.0$}	&\multirow{8}{*}{$z < 4.0$}	& \multirow{8}{*}{448}	& \multirow{2}{*}{$\mathrm{equal}$}& \multirow{2}{*}{<2.39}	& Natural	& 6 \pm 3	& 2.00\\*
 	 	 	 	 	 	 	 	 	 	 	 	 	 	 	 	 	 	& 									& 						& 			& 		& Taper	& 12 \pm 5	& 2.40\\*
 	 	 	 	 	 	 	 	 	 	 	 	 	 	 	 	 	 	& 									& 						& \multirow{2}{*}{$pbcor$}	& \multirow{2}{*}{<2.38}	& Natural	& 6 \pm 3	& 2.00\\*
 	 	 	 	 	 	 	 	 	 	 	 	 	 	 	 	 	 	& 									& 						& 			& 		& Taper	& 10 \pm 5	& 2.00\\*
 	 	 	 	 	 	 	 	 	 	 	 	 	 	 	 	 	 	& 									& 						& \multirow{2}{*}{F$_{UV}$}	& \multirow{2}{*}{<1.78}	& Natural	& 19 \pm 11	& 1.73\\*
 	 	 	 	 	 	 	 	 	 	 	 	 	 	 	 	 	 	& 									& 						& 			& 		& Taper	& 21 \pm 17	& 1.24\\*
 	 	 	 	 	 	 	 	 	 	 	 	 	 	 	 	 	 	& 									& 						& \multirow{2}{*}{$\mu$}	& \multirow{2}{*}{<2.61}	& Natural	& 12 \pm 6	& 2.00\\*
 	 	 	 	 	 	 	 	 	 	 	 	 	 	 	 	 	 	& 									& 						& 			& 		& Taper	& 21 \pm 9	& 2.33\\*
\cline{2-8}
 	 	 	 	 	 	 	 	 	 	 	 	 	 	 	 	 	 	&\multirow{8}{*}{$4.0 \leq z < 7.0$}	& \multirow{8}{*}{227}	& \multirow{2}{*}{$\mathrm{equal}$}	& \multirow{2}{*}{<2.69}	& Natural	& 2 \pm 5	& 0.40\\*
 	 	 	 	 	 	 	 	 	 	 	 	 	 	 	 	 	 	& 									& 						& 			& 		& Taper	& -7 \pm 7	& -1.00\\*
 	 	 	 	 	 	 	 	 	 	 	 	 	 	 	 	 	 	& 									& 						& \multirow{2}{*}{$pbcor$}	& \multirow{2}{*}{<2.73}	& Natural	& 0 \pm 5	& 0.00\\*
 	 	 	 	 	 	 	 	 	 	 	 	 	 	 	 	 	 	& 									& 						& 			& 		& Taper	& -8 \pm 7	& -1.14\\*
 	 	 	 	 	 	 	 	 	 	 	 	 	 	 	 	 	 	& 									& 						& \multirow{2}{*}{F$_{UV}$}	& \multirow{2}{*}{<1.87}	& Natural	& 140 \pm 33	& 4.24\\*
 	 	 	 	 	 	 	 	 	 	 	 	 	 	 	 	 	 	& 									& 						& 			& 		& Taper	& 149 \pm 49	& 3.04\\*
 	 	 	 	 	 	 	 	 	 	 	 	 	 	 	 	 	 	& 									& 						& \multirow{2}{*}{$\mu$}	& \multirow{2}{*}{<3.05}	& Natural	& 3 \pm 7	& 0.43\\*
 	 	 	 	 	 	 	 	 	 	 	 	 	 	 	 	 	 	& 									& 						& 			& 		& Taper	& -5 \pm 10	& -0.50\\*
\cline{2-8}
 	 	 	 	 	 	 	 	 	 	 	 	 	 	 	 	 	 	&\multirow{8}{*}{$7.0 \leq z$}	& \multirow{8}{*}{12}	& \multirow{2}{*}{$\mathrm{equal}$}	& \multirow{2}{*}{<2.26}	& Natural	& -2 \pm 21	& -0.10\\*
 	 	 	 	 	 	 	 	 	 	 	 	 	 	 	 	 	 	& 									& 						& 			& 		& Taper	& -16 \pm 28	& -0.57\\*
 	 	 	 	 	 	 	 	 	 	 	 	 	 	 	 	 	 	& 									& 						& \multirow{2}{*}{$pbcor$}	& \multirow{2}{*}{<2.11}	& Natural	& -7 \pm 21	& -0.33\\*
 	 	 	 	 	 	 	 	 	 	 	 	 	 	 	 	 	 	& 									& 						& 			& 		& Taper	& -27 \pm 29	& -0.93\\*
 	 	 	 	 	 	 	 	 	 	 	 	 	 	 	 	 	 	& 									& 						& \multirow{2}{*}{F$_{UV}$}	& \multirow{2}{*}{<1.55}	& Natural	& 21 \pm 35	& 0.60\\*
 	 	 	 	 	 	 	 	 	 	 	 	 	 	 	 	 	 	& 									& 						& 			& 		& Taper	& -62 \pm 56	& -1.11\\*
 	 	 	 	 	 	 	 	 	 	 	 	 	 	 	 	 	 	& 									& 						& \multirow{2}{*}{$\mu$}	& \multirow{2}{*}{<2.32}	& Natural	& 15 \pm 35	& 0.43\\*
 	 	 	 	 	 	 	 	 	 	 	 	 	 	 	 	 	 	& 									& 						& 			& 		& Taper	& -16 \pm 46	& -0.35\\*
\hline
\end{tabular}
\end{adjustbox}%

&

\begin{adjustbox}{width=0.45\linewidth}
\begin{tabular}{@{}c@{\hspace{1.25\tabcolsep}}c@{\hspace{0.5\tabcolsep}}c@{\hspace{0.75\tabcolsep}}c@{\hspace{0.75\tabcolsep}}c@{\hspace{0.75\tabcolsep}}c@{\hspace{0.75\tabcolsep}}S[separate-uncertainty=true,table-figures-decimal=0,table-text-alignment=center] S@{}}

\hline\hline
UV-slope	& $z_{ph}$	& Sources \#\tablefootmark{a}	& Weight\tablefootmark{b}   & $\log{(\mathrm{IRX})}$\tablefootmark{c}	& CLEAN\tablefootmark{d}	& {Flux\tablefootmark{e, f}}	& S/N\tablefootmark{g}\\
			 					& 		  					&  		   						& 	   						& 	   						& 	    					& {[$\mu$Jy]} 		&  \\
\hline
\multirow{24}{*}{$-1.0 \leq \beta < 0.0$}	&\multirow{8}{*}{$z < 4.0$}	& \multirow{8}{*}{28}	& \multirow{2}{*}{$\mathrm{equal}$}& \multirow{2}{*}{<2.80}	& Natural	& 0 \pm 13	& -0.00\\*
 	 	 	 	 	 	 	 	 	 	 	 	 	 	 	 	 	 	& 									& 						& 			& 		& Taper	& -7 \pm 19	& -0.37\\*
 	 	 	 	 	 	 	 	 	 	 	 	 	 	 	 	 	 	& 									& 						& \multirow{2}{*}{$pbcor$}	& \multirow{2}{*}{<2.82}	& Natural	& 0 \pm 13	& 0.00\\*
 	 	 	 	 	 	 	 	 	 	 	 	 	 	 	 	 	 	& 									& 						& 			& 		& Taper	& -6 \pm 19	& -0.32\\*
 	 	 	 	 	 	 	 	 	 	 	 	 	 	 	 	 	 	& 									& 						& \multirow{2}{*}{F$_{UV}$}	& \multirow{2}{*}{<1.98}	& Natural	& 3 \pm 37	& 0.08\\*
 	 	 	 	 	 	 	 	 	 	 	 	 	 	 	 	 	 	& 									& 						& 			& 		& Taper	& -20 \pm 48	& -0.42\\*
 	 	 	 	 	 	 	 	 	 	 	 	 	 	 	 	 	 	& 									& 						& \multirow{2}{*}{$\mu$}	& \multirow{2}{*}{<2.65}	& Natural	& 9 \pm 25	& 0.36\\*
 	 	 	 	 	 	 	 	 	 	 	 	 	 	 	 	 	 	& 									& 						& 			& 		& Taper	& 42 \pm 38	& 1.11\\*
\cline{2-8}
 	 	 	 	 	 	 	 	 	 	 	 	 	 	 	 	 	 	&\multirow{8}{*}{$4.0 \leq z < 7.0$}	& \multirow{8}{*}{38}	& \multirow{2}{*}{$\mathrm{equal}$}	& \multirow{2}{*}{<2.64}	& Natural	& 28 \pm 12	& 2.33\\*
 	 	 	 	 	 	 	 	 	 	 	 	 	 	 	 	 	 	& 									& 						& 			& 		& Taper	& 23 \pm 15	& 1.53\\*
 	 	 	 	 	 	 	 	 	 	 	 	 	 	 	 	 	 	& 									& 						& \multirow{2}{*}{$pbcor$}	& \multirow{2}{*}{<2.65}	& Natural	& 31 \pm 12	& 2.58\\*
 	 	 	 	 	 	 	 	 	 	 	 	 	 	 	 	 	 	& 									& 						& 			& 		& Taper	& 24 \pm 15	& 1.60\\*
 	 	 	 	 	 	 	 	 	 	 	 	 	 	 	 	 	 	& 									& 						& \multirow{2}{*}{F$_{UV}$}	& \multirow{2}{*}{<2.46}	& Natural	& 34 \pm 36	& 0.94\\*
 	 	 	 	 	 	 	 	 	 	 	 	 	 	 	 	 	 	& 									& 						& 			& 		& Taper	& 16 \pm 48	& 0.33\\*
 	 	 	 	 	 	 	 	 	 	 	 	 	 	 	 	 	 	& 									& 						& \multirow{2}{*}{$\mu$}	& \multirow{2}{*}{<2.56}	& Natural	& 24 \pm 16	& 1.50\\*
 	 	 	 	 	 	 	 	 	 	 	 	 	 	 	 	 	 	& 									& 						& 			& 		& Taper	& 6 \pm 23	& 0.26\\*
\cline{2-8}
 	 	 	 	 	 	 	 	 	 	 	 	 	 	 	 	 	 	&\multirow{8}{*}{$7.0 \leq z$}	& \multirow{8}{*}{12}	& \multirow{2}{*}{$\mathrm{equal}$}	& \multirow{2}{*}{<2.52}	& Natural	& 24 \pm 20	& 1.20\\*
 	 	 	 	 	 	 	 	 	 	 	 	 	 	 	 	 	 	& 									& 						& 			& 		& Taper	& 52 \pm 29	& 1.79\\*
 	 	 	 	 	 	 	 	 	 	 	 	 	 	 	 	 	 	& 									& 						& \multirow{2}{*}{$pbcor$}	& \multirow{2}{*}{<2.59}	& Natural	& 33 \pm 21	& 1.57\\*
 	 	 	 	 	 	 	 	 	 	 	 	 	 	 	 	 	 	& 									& 						& 			& 		& Taper	& 62 \pm 30	& 2.07\\*
 	 	 	 	 	 	 	 	 	 	 	 	 	 	 	 	 	 	& 									& 						& \multirow{2}{*}{F$_{UV}$}	& \multirow{2}{*}{<1.70}	& Natural	& 14 \pm 57	& 0.25\\*
 	 	 	 	 	 	 	 	 	 	 	 	 	 	 	 	 	 	& 									& 						& 			& 		& Taper	& -5 \pm 86	& -0.06\\*
 	 	 	 	 	 	 	 	 	 	 	 	 	 	 	 	 	 	& 									& 						& \multirow{2}{*}{$\mu$}	& \multirow{2}{*}{<3.03}	& Natural	& 106 \pm 31	& 3.42\\*
 	 	 	 	 	 	 	 	 	 	 	 	 	 	 	 	 	 	& 									& 						& 			& 		& Taper	& 119 \pm 46	& 2.59\\*
\hline
\multirow{24}{*}{$0.0 \leq \beta < 1.5$}	&\multirow{8}{*}{$z < 4.0$}	& \multirow{8}{*}{0}	& \multirow{2}{*}{$\mathrm{equal}$}& \multirow{2}{*}{$\cdots$}	& Natural	& $\cdots$ 	& $\cdots$ \\*
 	 	 	 	 	 	 	 	 	 	 	 	 	 	 	 	 	 	& 									& 						& 			& 		& Taper	& $\cdots$ 	& $\cdots$ \\*
 	 	 	 	 	 	 	 	 	 	 	 	 	 	 	 	 	 	& 									& 						& \multirow{2}{*}{$pbcor$}	& \multirow{2}{*}{$\cdots$}	& Natural	& $\cdots$ 	& $\cdots$ \\*
 	 	 	 	 	 	 	 	 	 	 	 	 	 	 	 	 	 	& 									& 						& 			& 		& Taper	& $\cdots$ 	& $\cdots$ \\*
 	 	 	 	 	 	 	 	 	 	 	 	 	 	 	 	 	 	& 									& 						& \multirow{2}{*}{F$_{UV}$}	& \multirow{2}{*}{$\cdots$}	& Natural	& $\cdots$ 	& $\cdots$ \\*
 	 	 	 	 	 	 	 	 	 	 	 	 	 	 	 	 	 	& 									& 						& 			& 		& Taper	& $\cdots$ 	& $\cdots$ \\*
 	 	 	 	 	 	 	 	 	 	 	 	 	 	 	 	 	 	& 									& 						& \multirow{2}{*}{$\mu$}	& \multirow{2}{*}{$\cdots$}	& Natural	& $\cdots$ 	& $\cdots$ \\*
 	 	 	 	 	 	 	 	 	 	 	 	 	 	 	 	 	 	& 									& 						& 			& 		& Taper	& $\cdots$ 	& $\cdots$ \\*
\cline{2-8}
 	 	 	 	 	 	 	 	 	 	 	 	 	 	 	 	 	 	&\multirow{8}{*}{$4.0 \leq z < 7.0$}	& \multirow{8}{*}{0}	& \multirow{2}{*}{$\mathrm{equal}$}	& \multirow{2}{*}{$\cdots$}	& Natural	& $\cdots$ 	& $\cdots$ \\*
 	 	 	 	 	 	 	 	 	 	 	 	 	 	 	 	 	 	& 									& 						& 			& 		& Taper	& $\cdots$ 	& $\cdots$ \\*
 	 	 	 	 	 	 	 	 	 	 	 	 	 	 	 	 	 	& 									& 						& \multirow{2}{*}{$pbcor$}	& \multirow{2}{*}{$\cdots$}	& Natural	& $\cdots$ 	& $\cdots$ \\*
 	 	 	 	 	 	 	 	 	 	 	 	 	 	 	 	 	 	& 									& 						& 			& 		& Taper	& $\cdots$ 	& $\cdots$ \\*
 	 	 	 	 	 	 	 	 	 	 	 	 	 	 	 	 	 	& 									& 						& \multirow{2}{*}{F$_{UV}$}	& \multirow{2}{*}{$\cdots$}	& Natural	& $\cdots$ 	& $\cdots$ \\*
 	 	 	 	 	 	 	 	 	 	 	 	 	 	 	 	 	 	& 									& 						& 			& 		& Taper	& $\cdots$ 	& $\cdots$ \\*
 	 	 	 	 	 	 	 	 	 	 	 	 	 	 	 	 	 	& 									& 						& \multirow{2}{*}{$\mu$}	& \multirow{2}{*}{$\cdots$}	& Natural	& $\cdots$ 	& $\cdots$ \\*
 	 	 	 	 	 	 	 	 	 	 	 	 	 	 	 	 	 	& 									& 						& 			& 		& Taper	& $\cdots$ 	& $\cdots$ \\*
\cline{2-8}
 	 	 	 	 	 	 	 	 	 	 	 	 	 	 	 	 	 	&\multirow{8}{*}{$7.0 \leq z$}	& \multirow{8}{*}{0}	& \multirow{2}{*}{$\mathrm{equal}$}	& \multirow{2}{*}{$\cdots$}	& Natural	& $\cdots$ 	& $\cdots$ \\*
 	 	 	 	 	 	 	 	 	 	 	 	 	 	 	 	 	 	& 									& 						& 			& 		& Taper	& $\cdots$ 	& $\cdots$ \\*
 	 	 	 	 	 	 	 	 	 	 	 	 	 	 	 	 	 	& 									& 						& \multirow{2}{*}{$pbcor$}	& \multirow{2}{*}{$\cdots$}	& Natural	& $\cdots$ 	& $\cdots$ \\*
 	 	 	 	 	 	 	 	 	 	 	 	 	 	 	 	 	 	& 									& 						& 			& 		& Taper	& $\cdots$ 	& $\cdots$ \\*
 	 	 	 	 	 	 	 	 	 	 	 	 	 	 	 	 	 	& 									& 						& \multirow{2}{*}{F$_{UV}$}	& \multirow{2}{*}{$\cdots$}	& Natural	& $\cdots$ 	& $\cdots$ \\*
 	 	 	 	 	 	 	 	 	 	 	 	 	 	 	 	 	 	& 									& 						& 			& 		& Taper	& $\cdots$ 	& $\cdots$ \\*
 	 	 	 	 	 	 	 	 	 	 	 	 	 	 	 	 	 	& 									& 						& \multirow{2}{*}{$\mu$}	& \multirow{2}{*}{$\cdots$}	& Natural	& $\cdots$ 	& $\cdots$ \\*
 	 	 	 	 	 	 	 	 	 	 	 	 	 	 	 	 	 	& 									& 						& 			& 		& Taper	& $\cdots$ 	& $\cdots$ \\*
\hline
\end{tabular}
\end{adjustbox}
\end{tabular}

\tablefoot{
\tablefoottext{a}{When Source \# is listed as zero (0), no stacking was performed.}
\tablefoottext{b}{Weight scheme applied to each candidate, as explained in $\S$\ref{subsec:Stacking}.}
\tablefoottext{c}{$3{-}\sigma$ upper limits of weighted average for each bin (Eq.~\ref{eq:stack_irx_avg}).}
\tablefoottext{d}{\textit{CLEAN}ing method used in \texttt{CASA} to obtain final image.}
\tablefoottext{e}{Maximum value from a 0\farcs5$\times$0\farcs5 box in the stacked images.}
\tablefoottext{f}{$rms$ errors from Eq. \ref{eq:rms_detect}.}
\tablefoottext{g}{S/N $=$ Flux$_{\mathrm{peak}}$ / $rms$}
}

\end{table*}

\subsection{Stellar mass binning}\label{tab:mass_stack}

    The values obtained after stacking each different configuration 
    are shown in Tables \ref{tab:stack_results_mid_mass_z_bin} and \ref{tab:stack_results_full_mass_z_bin}.

\subsubsection{Candidates with \texorpdfstring{$\log{(M_{\bigstar} / M_{\odot})} < 6.0$}{log(Mstar/Msun)<6.0}}\label{subsubsec:low_mass_disc}

    Among the $1580$ non-detected LBG candidates in our initial sample, 
    $11$ have stellar masses below $10^{6} M_{\odot}$, which were not 
    considered for the results presented in $\S$\ref{subsec:Stack_results}. 
    Regarding photometric redshifts, eight ($8$) low-mass candidates 
    lie at $z_{ph} {<} 4.0$, three ($3$) between  $4 {\leq} z_{ph} {<} 7.0$ 
    and none at $z_{ph} {\geq} 7$. With so few candidates, and considering 
    their relatively low UV luminosities, we do not expect to detect anything then
    via stacking. Nonetheless, for completeness, we report results of their 
    \texttt{uv}-stacking in Table \ref{tab:ALMA_props_low_mass}. The 
    highest absolute S/N obtained is $1.18$ for low-redshift candidates with 
    taper \textit{CLEAN}ing, confirming our expectation.

\begin{table*}
\caption{UV Stacking results for candidates in stellar mass bins $6.0 \leq \log{(M_{\star} / M_{\odot})} < 8.0$ across all photometric redshift bins from our full sample.}\label{tab:stack_results_mid_mass_z_bin}
\centering
\begin{tabular}{c c}
\begin{adjustbox}{width=0.45\linewidth}
\begin{tabular}{@{}c@{\hspace{1.25\tabcolsep}}c@{\hspace{0.5\tabcolsep}}c@{\hspace{0.75\tabcolsep}}c@{\hspace{0.75\tabcolsep}}c@{\hspace{0.75\tabcolsep}}c@{\hspace{0.75\tabcolsep}}S[separate-uncertainty=true,table-figures-decimal=0,table-text-alignment=center] S@{}}
\hline\hline
Stellar Mass	& $z_{ph}$	& Sources \#	& Weight\tablefootmark{a}	& $\log{(\mathrm{IRX})}$\tablefootmark{b}	& CLEAN\tablefootmark{c}	& {Flux\tablefootmark{d, e}}	& S/N\tablefootmark{f}\\
			 					& 		  		& 		   						& 	   						& 	    					& 	    					& {[$\mu$Jy]} 		&  \\
\hline

\multirow{24}{*}{$6.0 \leq \log{(M_{\star} / M_{\odot})} < 6.5$}	&\multirow{8}{*}{$z < 4.0$}	& \multirow{8}{*}{41}	& \multirow{2}{*}{$\mathrm{equal}$}	& \multirow{2}{*}{<2.65}	& Natural	& 16 \pm 11	& 1.45\\*
 	 	 	 	 	 	 	 	 	 	 	 	 	 	 	 	 	 	& 									& 						& 			& 		& Taper	& 4 \pm 14	& 0.29\\*
 	 	 	 	 	 	 	 	 	 	 	 	 	 	 	 	 	 	& 									& 						& \multirow{2}{*}{$pbcor$}	& \multirow{2}{*}{<2.65}	& Natural	& 17 \pm 11	& 1.55\\*
 	 	 	 	 	 	 	 	 	 	 	 	 	 	 	 	 	 	& 									& 						& 			& 		& Taper	& 6 \pm 14	& 0.43\\*
 	 	 	 	 	 	 	 	 	 	 	 	 	 	 	 	 	 	& 									& 						& \multirow{2}{*}{F$_{UV}$}	& \multirow{2}{*}{<3.21}	& Natural	& 15 \pm 16	& 0.94\\*
 	 	 	 	 	 	 	 	 	 	 	 	 	 	 	 	 	 	& 									& 						& 			& 		& Taper	& 12 \pm 22	& 0.55\\*
 	 	 	 	 	 	 	 	 	 	 	 	 	 	 	 	 	 	& 									& 						& \multirow{2}{*}{$\mu$}	& \multirow{2}{*}{<2.64}	& Natural	& 28 \pm 15	& 1.87\\*
 	 	 	 	 	 	 	 	 	 	 	 	 	 	 	 	 	 	& 									& 						& 			& 		& Taper	& 4 \pm 20	& 0.20\\*
\cline{2-8}
 	 	 	 	 	 	 	 	 	 	 	 	 	 	 	 	 	 	&\multirow{8}{*}{$4.0 \leq z < 7.0$}	& \multirow{8}{*}{45}	& \multirow{2}{*}{$\mathrm{equal}$}	& \multirow{2}{*}{<2.54}	& Natural	& 16 \pm 10	& 1.60\\*
 	 	 	 	 	 	 	 	 	 	 	 	 	 	 	 	 	 	& 									& 						& 			& 		& Taper	& 21 \pm 14	& 1.50\\*
 	 	 	 	 	 	 	 	 	 	 	 	 	 	 	 	 	 	& 									& 						& \multirow{2}{*}{$pbcor$}	& \multirow{2}{*}{<2.54}	& Natural	& 16 \pm 10	& 1.60\\*
 	 	 	 	 	 	 	 	 	 	 	 	 	 	 	 	 	 	& 									& 						& 			& 		& Taper	& 22 \pm 14	& 1.57\\*
 	 	 	 	 	 	 	 	 	 	 	 	 	 	 	 	 	 	& 									& 						& \multirow{2}{*}{F$_{UV}$}	& \multirow{2}{*}{<2.67}	& Natural	& 30 \pm 16	& 1.88\\*
 	 	 	 	 	 	 	 	 	 	 	 	 	 	 	 	 	 	& 									& 						& 			& 		& Taper	& 33 \pm 20	& 1.65\\*
 	 	 	 	 	 	 	 	 	 	 	 	 	 	 	 	 	 	& 									& 						& \multirow{2}{*}{$\mu$}	& \multirow{2}{*}{<2.60}	& Natural	& 8 \pm 12	& 0.67\\*
 	 	 	 	 	 	 	 	 	 	 	 	 	 	 	 	 	 	& 									& 						& 			& 		& Taper	& 25 \pm 17	& 1.47\\*
\cline{2-8}
 	 	 	 	 	 	 	 	 	 	 	 	 	 	 	 	 	 	&\multirow{8}{*}{$7.0 \leq z$}	& \multirow{8}{*}{8}	& \multirow{2}{*}{$\mathrm{equal}$}	& \multirow{2}{*}{<2.41}	& Natural	& 10 \pm 25	& 0.40\\*
 	 	 	 	 	 	 	 	 	 	 	 	 	 	 	 	 	 	& 									& 						& 			& 		& Taper	& 16 \pm 37	& 0.43\\*
 	 	 	 	 	 	 	 	 	 	 	 	 	 	 	 	 	 	& 									& 						& \multirow{2}{*}{$pbcor$}	& \multirow{2}{*}{<2.42}	& Natural	& 10 \pm 25	& 0.40\\*
 	 	 	 	 	 	 	 	 	 	 	 	 	 	 	 	 	 	& 									& 						& 			& 		& Taper	& 13 \pm 37	& 0.35\\*
 	 	 	 	 	 	 	 	 	 	 	 	 	 	 	 	 	 	& 									& 						& \multirow{2}{*}{F$_{UV}$}	& \multirow{2}{*}{<2.63}	& Natural	& 7 \pm 26	& 0.27\\*
 	 	 	 	 	 	 	 	 	 	 	 	 	 	 	 	 	 	& 									& 						& 			& 		& Taper	& 0 \pm 39	& -0.00\\*
 	 	 	 	 	 	 	 	 	 	 	 	 	 	 	 	 	 	& 									& 						& \multirow{2}{*}{$\mu$}	& \multirow{2}{*}{<2.40}	& Natural	& 13 \pm 28	& 0.46\\*
 	 	 	 	 	 	 	 	 	 	 	 	 	 	 	 	 	 	& 									& 						& 			& 		& Taper	& 23 \pm 44	& 0.52\\*
\hline
\multirow{24}{*}{$6.5 \leq \log{(M_{\star} / M_{\odot})} < 7.0$}	&\multirow{8}{*}{$z < 4.0$}	& \multirow{8}{*}{107}	& \multirow{2}{*}{$\mathrm{equal}$}	& \multirow{2}{*}{<2.83}	& Natural	& 8 \pm 7	& 1.14\\*
 	 	 	 	 	 	 	 	 	 	 	 	 	 	 	 	 	 	& 									& 						& 			& 		& Taper	& 8 \pm 9	& 0.89\\*
 	 	 	 	 	 	 	 	 	 	 	 	 	 	 	 	 	 	& 									& 						& \multirow{2}{*}{$pbcor$}	& \multirow{2}{*}{<2.84}	& Natural	& 10 \pm 7	& 1.43\\*
 	 	 	 	 	 	 	 	 	 	 	 	 	 	 	 	 	 	& 									& 						& 			& 		& Taper	& 11 \pm 9	& 1.22\\*
 	 	 	 	 	 	 	 	 	 	 	 	 	 	 	 	 	 	& 									& 						& \multirow{2}{*}{F$_{UV}$}	& \multirow{2}{*}{<2.58}	& Natural	& 19 \pm 11	& 1.73\\*
 	 	 	 	 	 	 	 	 	 	 	 	 	 	 	 	 	 	& 									& 						& 			& 		& Taper	& 14 \pm 15	& 0.93\\*
 	 	 	 	 	 	 	 	 	 	 	 	 	 	 	 	 	 	& 									& 						& \multirow{2}{*}{$\mu$}	& \multirow{2}{*}{<3.15}	& Natural	& 5 \pm 9	& 0.56\\*
 	 	 	 	 	 	 	 	 	 	 	 	 	 	 	 	 	 	& 									& 						& 			& 		& Taper	& -3 \pm 13	& -0.23\\*
\cline{2-8}
 	 	 	 	 	 	 	 	 	 	 	 	 	 	 	 	 	 	&\multirow{8}{*}{$4.0 \leq z < 7.0$}	& \multirow{8}{*}{94}	& \multirow{2}{*}{$\mathrm{equal}$}	& \multirow{2}{*}{<2.41}	& Natural	& 19 \pm 7	& 2.71\\*
 	 	 	 	 	 	 	 	 	 	 	 	 	 	 	 	 	 	& 									& 						& 			& 		& Taper	& 20 \pm 10	& 2.00\\*
 	 	 	 	 	 	 	 	 	 	 	 	 	 	 	 	 	 	& 									& 						& \multirow{2}{*}{$pbcor$}	& \multirow{2}{*}{<2.43}	& Natural	& 23 \pm 8	& 2.88\\*
 	 	 	 	 	 	 	 	 	 	 	 	 	 	 	 	 	 	& 									& 						& 			& 		& Taper	& 23 \pm 10	& 2.30\\*
 	 	 	 	 	 	 	 	 	 	 	 	 	 	 	 	 	 	& 									& 						& \multirow{2}{*}{F$_{UV}$}	& \multirow{2}{*}{<2.50}	& Natural	& 42 \pm 14	& 3.00\\*
 	 	 	 	 	 	 	 	 	 	 	 	 	 	 	 	 	 	& 									& 						& 			& 		& Taper	& 39 \pm 19	& 2.05\\*
 	 	 	 	 	 	 	 	 	 	 	 	 	 	 	 	 	 	& 									& 						& \multirow{2}{*}{$\mu$}	& \multirow{2}{*}{<2.42}	& Natural	& 20 \pm 10	& 2.00\\*
 	 	 	 	 	 	 	 	 	 	 	 	 	 	 	 	 	 	& 									& 						& 			& 		& Taper	& 14 \pm 14	& 1.00\\*
\cline{2-8}
 	 	 	 	 	 	 	 	 	 	 	 	 	 	 	 	 	 	&\multirow{8}{*}{$7.0 \leq z$}	& \multirow{8}{*}{12}	& \multirow{2}{*}{$\mathrm{equal}$}	& \multirow{2}{*}{<2.65}	& Natural	& 36 \pm 22	& 1.64\\*
 	 	 	 	 	 	 	 	 	 	 	 	 	 	 	 	 	 	& 									& 						& 			& 		& Taper	& 39 \pm 32	& 1.22\\*
 	 	 	 	 	 	 	 	 	 	 	 	 	 	 	 	 	 	& 									& 						& \multirow{2}{*}{$pbcor$}	& \multirow{2}{*}{<2.66}	& Natural	& 31 \pm 22	& 1.41\\*
 	 	 	 	 	 	 	 	 	 	 	 	 	 	 	 	 	 	& 									& 						& 			& 		& Taper	& 33 \pm 31	& 1.06\\*
 	 	 	 	 	 	 	 	 	 	 	 	 	 	 	 	 	 	& 									& 						& \multirow{2}{*}{F$_{UV}$}	& \multirow{2}{*}{<2.73}	& Natural	& 55 \pm 33	& 1.67\\*
 	 	 	 	 	 	 	 	 	 	 	 	 	 	 	 	 	 	& 									& 						& 			& 		& Taper	& 54 \pm 47	& 1.15\\*
 	 	 	 	 	 	 	 	 	 	 	 	 	 	 	 	 	 	& 									& 						& \multirow{2}{*}{$\mu$}	& \multirow{2}{*}{<2.92}	& Natural	& 76 \pm 35	& 2.17\\*
 	 	 	 	 	 	 	 	 	 	 	 	 	 	 	 	 	 	& 									& 						& 			& 		& Taper	& 48 \pm 46	& 1.04\\*
\hline
\end{tabular}
\end{adjustbox}%

&

\begin{adjustbox}{width=0.45\linewidth}
\begin{tabular}{@{}c@{\hspace{1.25\tabcolsep}}c@{\hspace{0.5\tabcolsep}}c@{\hspace{0.75\tabcolsep}}c@{\hspace{0.75\tabcolsep}}c@{\hspace{0.75\tabcolsep}}c@{\hspace{0.75\tabcolsep}}S[separate-uncertainty=true,table-figures-decimal=0,table-text-alignment=center] S@{}}

\hline\hline
Stellar Mass	& $z_{ph}$	& Sources \#	& Weight\tablefootmark{a}	& $\log{(\mathrm{IRX})}$\tablefootmark{b}	& CLEAN\tablefootmark{c}	& {Flux\tablefootmark{d, e}}	& S/N\tablefootmark{f}\\
			 					& 		  		& 		   						& 	   						& 	    					& 	    					& {[$\mu$Jy]} 		&  \\
\hline

\multirow{24}{*}{$7.0 \leq \log{(M_{\star} / M_{\odot})} < 7.5$}	&\multirow{8}{*}{$z < 4.0$}	& \multirow{8}{*}{211}	& \multirow{2}{*}{$\mathrm{equal}$}	& \multirow{2}{*}{<2.52}	& Natural	& 6 \pm 5	& 1.20\\*
 	 	 	 	 	 	 	 	 	 	 	 	 	 	 	 	 	 	& 									& 						& 			& 		& Taper	& 11 \pm 7	& 1.57\\*
 	 	 	 	 	 	 	 	 	 	 	 	 	 	 	 	 	 	& 									& 						& \multirow{2}{*}{$pbcor$}	& \multirow{2}{*}{<2.50}	& Natural	& 6 \pm 5	& 1.20\\*
 	 	 	 	 	 	 	 	 	 	 	 	 	 	 	 	 	 	& 									& 						& 			& 		& Taper	& 11 \pm 7	& 1.57\\*
 	 	 	 	 	 	 	 	 	 	 	 	 	 	 	 	 	 	& 									& 						& \multirow{2}{*}{F$_{UV}$}	& \multirow{2}{*}{<2.50}	& Natural	& 15 \pm 8	& 1.88\\*
 	 	 	 	 	 	 	 	 	 	 	 	 	 	 	 	 	 	& 									& 						& 			& 		& Taper	& 11 \pm 11	& 1.00\\*
 	 	 	 	 	 	 	 	 	 	 	 	 	 	 	 	 	 	& 									& 						& \multirow{2}{*}{$\mu$}	& \multirow{2}{*}{<2.51}	& Natural	& 19 \pm 8	& 2.38\\*
 	 	 	 	 	 	 	 	 	 	 	 	 	 	 	 	 	 	& 									& 						& 			& 		& Taper	& 32 \pm 12	& 2.67\\*
\cline{2-8}
 	 	 	 	 	 	 	 	 	 	 	 	 	 	 	 	 	 	&\multirow{8}{*}{$4.0 \leq z < 7.0$}	& \multirow{8}{*}{102}	& \multirow{2}{*}{$\mathrm{equal}$}	& \multirow{2}{*}{<2.94}	& Natural	& 0 \pm 7	& 0.00\\*
 	 	 	 	 	 	 	 	 	 	 	 	 	 	 	 	 	 	& 									& 						& 			& 		& Taper	& -13 \pm 10	& -1.30\\*
 	 	 	 	 	 	 	 	 	 	 	 	 	 	 	 	 	 	& 									& 						& \multirow{2}{*}{$pbcor$}	& \multirow{2}{*}{<2.99}	& Natural	& 0 \pm 7	& 0.00\\*
 	 	 	 	 	 	 	 	 	 	 	 	 	 	 	 	 	 	& 									& 						& 			& 		& Taper	& -11 \pm 10	& -1.10\\*
 	 	 	 	 	 	 	 	 	 	 	 	 	 	 	 	 	 	& 									& 						& \multirow{2}{*}{F$_{UV}$}	& \multirow{2}{*}{<2.12}	& Natural	& 12 \pm 16	& 0.75\\*
 	 	 	 	 	 	 	 	 	 	 	 	 	 	 	 	 	 	& 									& 						& 			& 		& Taper	& -5 \pm 24	& -0.21\\*
 	 	 	 	 	 	 	 	 	 	 	 	 	 	 	 	 	 	& 									& 						& \multirow{2}{*}{$\mu$}	& \multirow{2}{*}{<3.39}	& Natural	& 3 \pm 10	& 0.30\\*
 	 	 	 	 	 	 	 	 	 	 	 	 	 	 	 	 	 	& 									& 						& 			& 		& Taper	& -5 \pm 14	& -0.36\\*
\cline{2-8}
 	 	 	 	 	 	 	 	 	 	 	 	 	 	 	 	 	 	&\multirow{8}{*}{$7.0 \leq z$}	& \multirow{8}{*}{4}	& \multirow{2}{*}{$\mathrm{equal}$}	& \multirow{2}{*}{<2.63}	& Natural	& 31 \pm 33	& 0.94\\*
 	 	 	 	 	 	 	 	 	 	 	 	 	 	 	 	 	 	& 									& 						& 			& 		& Taper	& 2 \pm 48	& 0.04\\*
 	 	 	 	 	 	 	 	 	 	 	 	 	 	 	 	 	 	& 									& 						& \multirow{2}{*}{$pbcor$}	& \multirow{2}{*}{<2.64}	& Natural	& 42 \pm 35	& 1.20\\*
 	 	 	 	 	 	 	 	 	 	 	 	 	 	 	 	 	 	& 									& 						& 			& 		& Taper	& 8 \pm 50	& 0.16\\*
 	 	 	 	 	 	 	 	 	 	 	 	 	 	 	 	 	 	& 									& 						& \multirow{2}{*}{F$_{UV}$}	& \multirow{2}{*}{<2.46}	& Natural	& 77 \pm 63	& 1.22\\*
 	 	 	 	 	 	 	 	 	 	 	 	 	 	 	 	 	 	& 									& 						& 			& 		& Taper	& 95 \pm 90	& 1.06\\*
 	 	 	 	 	 	 	 	 	 	 	 	 	 	 	 	 	 	& 									& 						& \multirow{2}{*}{$\mu$}	& \multirow{2}{*}{<2.80}	& Natural	& 53 \pm 48	& 1.10\\*
 	 	 	 	 	 	 	 	 	 	 	 	 	 	 	 	 	 	& 									& 						& 			& 		& Taper	& 20 \pm 71	& 0.28\\*
\hline
\multirow{24}{*}{$7.5 \leq \log{(M_{\star} / M_{\odot})} < 8.0$}	&\multirow{8}{*}{$z < 4.0$}	& \multirow{8}{*}{237}	& \multirow{2}{*}{$\mathrm{equal}$}	& \multirow{2}{*}{<2.48}	& Natural	& 3 \pm 4	& 0.75\\*
 	 	 	 	 	 	 	 	 	 	 	 	 	 	 	 	 	 	& 									& 						& 			& 		& Taper	& 5 \pm 6	& 0.83\\*
 	 	 	 	 	 	 	 	 	 	 	 	 	 	 	 	 	 	& 									& 						& \multirow{2}{*}{$pbcor$}	& \multirow{2}{*}{<2.47}	& Natural	& 2 \pm 5	& 0.40\\*
 	 	 	 	 	 	 	 	 	 	 	 	 	 	 	 	 	 	& 									& 						& 			& 		& Taper	& 3 \pm 6	& 0.50\\*
 	 	 	 	 	 	 	 	 	 	 	 	 	 	 	 	 	 	& 									& 						& \multirow{2}{*}{F$_{UV}$}	& \multirow{2}{*}{<2.23}	& Natural	& 8 \pm 10	& 0.80\\*
 	 	 	 	 	 	 	 	 	 	 	 	 	 	 	 	 	 	& 									& 						& 			& 		& Taper	& 0 \pm 13	& -0.00\\*
 	 	 	 	 	 	 	 	 	 	 	 	 	 	 	 	 	 	& 									& 						& \multirow{2}{*}{$\mu$}	& \multirow{2}{*}{<2.74}	& Natural	& 8 \pm 8	& 1.00\\*
 	 	 	 	 	 	 	 	 	 	 	 	 	 	 	 	 	 	& 									& 						& 			& 		& Taper	& 16 \pm 11	& 1.45\\*
\cline{2-8}
 	 	 	 	 	 	 	 	 	 	 	 	 	 	 	 	 	 	&\multirow{8}{*}{$4.0 \leq z < 7.0$}	& \multirow{8}{*}{91}	& \multirow{2}{*}{$\mathrm{equal}$}	& \multirow{2}{*}{<2.33}	& Natural	& 5 \pm 7	& 0.71\\*
 	 	 	 	 	 	 	 	 	 	 	 	 	 	 	 	 	 	& 									& 						& 			& 		& Taper	& -2 \pm 10	& -0.20\\*
 	 	 	 	 	 	 	 	 	 	 	 	 	 	 	 	 	 	& 									& 						& \multirow{2}{*}{$pbcor$}	& \multirow{2}{*}{<2.31}	& Natural	& 6 \pm 8	& 0.75\\*
 	 	 	 	 	 	 	 	 	 	 	 	 	 	 	 	 	 	& 									& 						& 			& 		& Taper	& -2 \pm 11	& -0.18\\*
 	 	 	 	 	 	 	 	 	 	 	 	 	 	 	 	 	 	& 									& 						& \multirow{2}{*}{F$_{UV}$}	& \multirow{2}{*}{<1.92}	& Natural	& 36 \pm 19	& 1.89\\*
 	 	 	 	 	 	 	 	 	 	 	 	 	 	 	 	 	 	& 									& 						& 			& 		& Taper	& 20 \pm 25	& 0.80\\*
 	 	 	 	 	 	 	 	 	 	 	 	 	 	 	 	 	 	& 									& 						& \multirow{2}{*}{$\mu$}	& \multirow{2}{*}{<2.47}	& Natural	& 11 \pm 11	& 1.00\\*
 	 	 	 	 	 	 	 	 	 	 	 	 	 	 	 	 	 	& 									& 						& 			& 		& Taper	& 2 \pm 15	& 0.13\\*
\cline{2-8}
 	 	 	 	 	 	 	 	 	 	 	 	 	 	 	 	 	 	&\multirow{8}{*}{$7.0 \leq z$}	& \multirow{8}{*}{1}	& \multirow{2}{*}{$\mathrm{equal}$}	& \multirow{2}{*}{<2.24}	& Natural	& 58 \pm 52	& 1.12\\*
 	 	 	 	 	 	 	 	 	 	 	 	 	 	 	 	 	 	& 									& 						& 			& 		& Taper	& 3 \pm 88	& 0.03\\*
 	 	 	 	 	 	 	 	 	 	 	 	 	 	 	 	 	 	& 									& 						& \multirow{2}{*}{$pbcor$}	& \multirow{2}{*}{<2.24}	& Natural	& 58 \pm 52	& 1.12\\*
 	 	 	 	 	 	 	 	 	 	 	 	 	 	 	 	 	 	& 									& 						& 			& 		& Taper	& 3 \pm 88	& 0.03\\*
 	 	 	 	 	 	 	 	 	 	 	 	 	 	 	 	 	 	& 									& 						& \multirow{2}{*}{F$_{UV}$}	& \multirow{2}{*}{<2.24}	& Natural	& 58 \pm 52	& 1.12\\*
 	 	 	 	 	 	 	 	 	 	 	 	 	 	 	 	 	 	& 									& 						& 			& 		& Taper	& 3 \pm 88	& 0.03\\*
 	 	 	 	 	 	 	 	 	 	 	 	 	 	 	 	 	 	& 									& 						& \multirow{2}{*}{$\mu$}	& \multirow{2}{*}{<2.24}	& Natural	& 58 \pm 52	& 1.12\\*
 	 	 	 	 	 	 	 	 	 	 	 	 	 	 	 	 	 	& 									& 						& 			& 		& Taper	& 3 \pm 88	& 0.03\\*
\hline
\end{tabular}
\end{adjustbox}
\end{tabular}
\tablefoot{
\tablefoottext{a}{Weight scheme associated to each candidate, as explained in $\S$\ref{subsec:Stacking}.}
\tablefoottext{b}{$3{-}\sigma$ upper limits of weighted average for each bin (Eq.~\ref{eq:stack_irx_avg}).}
\tablefoottext{c}{\textit{CLEAN}ing method used in \texttt{CASA} to obtain final image.}
\tablefoottext{d}{Maximum value from a 0\farcs5$\times$0\farcs5 box in the stacked images.}
\tablefoottext{e}{$rms$ errors from Eq. \ref{eq:rms_detect}.}
\tablefoottext{f}{S/N $=$ Flux$_{\mathrm{peak}}$ / $rms$}
}
\end{table*}


\begin{table*}
\caption{UV Stacking results for candidates in stellar mass bins 
$\log{(M_{\bigstar} / M_{\odot})} \geq 8.0$ across all photometric redshift bins from our full sample.}\label{tab:stack_results_full_mass_z_bin}
\centering
\begin{tabular}{c c}
\begin{adjustbox}{width=0.45\linewidth}
\begin{tabular}{@{}c@{\hspace{1.25\tabcolsep}}c@{\hspace{0.5\tabcolsep}}c@{\hspace{0.75\tabcolsep}}c@{\hspace{0.75\tabcolsep}}c@{\hspace{0.75\tabcolsep}}c@{\hspace{0.75\tabcolsep}}S[separate-uncertainty=true,table-figures-decimal=0,table-text-alignment=center] S@{}}
\hline\hline
Stellar Mass	& $z_{ph}$	& Sources \#	& Weight\tablefootmark{a}	& $\log{(\mathrm{IRX})}$\tablefootmark{b}	& CLEAN\tablefootmark{c}	& {Flux\tablefootmark{d, e}}	& S/N\tablefootmark{f}\\
			 					& 		  		& 		   						& 	   						& 	    					& 	    					& {[$\mu$Jy]} 		&  \\
\hline

\multirow{24}{*}{$8.0 \leq \log{(M_{\star} / M_{\odot})} < 8.5$}	&\multirow{8}{*}{$z < 4.0$}	& \multirow{8}{*}{246}	& \multirow{2}{*}{$\mathrm{equal}$}	& \multirow{2}{*}{<2.37}	& Natural	& 2 \pm 5	& 0.40\\*
 	 	 	 	 	 	 	 	 	 	 	 	 	 	 	 	 	 	& 									& 						& 			& 		& Taper	& -1 \pm 6	& -0.17\\*
 	 	 	 	 	 	 	 	 	 	 	 	 	 	 	 	 	 	& 									& 						& \multirow{2}{*}{$pbcor$}	& \multirow{2}{*}{<2.34}	& Natural	& 2 \pm 5	& 0.40\\*
 	 	 	 	 	 	 	 	 	 	 	 	 	 	 	 	 	 	& 									& 						& 			& 		& Taper	& -1 \pm 7	& -0.14\\*
 	 	 	 	 	 	 	 	 	 	 	 	 	 	 	 	 	 	& 									& 						& \multirow{2}{*}{F$_{UV}$}	& \multirow{2}{*}{<2.06}	& Natural	& 22 \pm 15	& 1.47\\*
 	 	 	 	 	 	 	 	 	 	 	 	 	 	 	 	 	 	& 									& 						& 			& 		& Taper	& 32 \pm 22	& 1.45\\*
 	 	 	 	 	 	 	 	 	 	 	 	 	 	 	 	 	 	& 									& 						& \multirow{2}{*}{$\mu$}	& \multirow{2}{*}{<2.55}	& Natural	& 11 \pm 8	& 1.38\\*
 	 	 	 	 	 	 	 	 	 	 	 	 	 	 	 	 	 	& 									& 						& 			& 		& Taper	& 8 \pm 12	& 0.67\\*
\cline{2-8}
 	 	 	 	 	 	 	 	 	 	 	 	 	 	 	 	 	 	&\multirow{8}{*}{$4.0 \leq z < 7.0$}	& \multirow{8}{*}{68}	& \multirow{2}{*}{$\mathrm{equal}$}	& \multirow{2}{*}{<2.34}	& Natural	& 6 \pm 9	& 0.67\\*
 	 	 	 	 	 	 	 	 	 	 	 	 	 	 	 	 	 	& 									& 						& 			& 		& Taper	& 2 \pm 12	& 0.17\\*
 	 	 	 	 	 	 	 	 	 	 	 	 	 	 	 	 	 	& 									& 						& \multirow{2}{*}{$pbcor$}	& \multirow{2}{*}{<2.32}	& Natural	& 8 \pm 9	& 0.89\\*
 	 	 	 	 	 	 	 	 	 	 	 	 	 	 	 	 	 	& 									& 						& 			& 		& Taper	& 4 \pm 13	& 0.31\\*
 	 	 	 	 	 	 	 	 	 	 	 	 	 	 	 	 	 	& 									& 						& \multirow{2}{*}{F$_{UV}$}	& \multirow{2}{*}{<1.71}	& Natural	& 1 \pm 24	& 0.04\\*
 	 	 	 	 	 	 	 	 	 	 	 	 	 	 	 	 	 	& 									& 						& 			& 		& Taper	& -13 \pm 36	& -0.36\\*
 	 	 	 	 	 	 	 	 	 	 	 	 	 	 	 	 	 	& 									& 						& \multirow{2}{*}{$\mu$}	& \multirow{2}{*}{<2.38}	& Natural	& 14 \pm 14	& 1.00\\*
 	 	 	 	 	 	 	 	 	 	 	 	 	 	 	 	 	 	& 									& 						& 			& 		& Taper	& 10 \pm 20	& 0.50\\*
\cline{2-8}
 	 	 	 	 	 	 	 	 	 	 	 	 	 	 	 	 	 	&\multirow{8}{*}{$7.0 \leq z$}	& \multirow{8}{*}{8}	& \multirow{2}{*}{$\mathrm{equal}$}	& \multirow{2}{*}{<2.59}	& Natural	& 21 \pm 22	& 0.95\\*
 	 	 	 	 	 	 	 	 	 	 	 	 	 	 	 	 	 	& 									& 						& 			& 		& Taper	& 7 \pm 33	& 0.21\\*
 	 	 	 	 	 	 	 	 	 	 	 	 	 	 	 	 	 	& 									& 						& \multirow{2}{*}{$pbcor$}	& \multirow{2}{*}{<2.64}	& Natural	& 23 \pm 22	& 1.05\\*
 	 	 	 	 	 	 	 	 	 	 	 	 	 	 	 	 	 	& 									& 						& 			& 		& Taper	& 10 \pm 34	& 0.29\\*
 	 	 	 	 	 	 	 	 	 	 	 	 	 	 	 	 	 	& 									& 						& \multirow{2}{*}{F$_{UV}$}	& \multirow{2}{*}{<1.61}	& Natural	& 37 \pm 58	& 0.64\\*
 	 	 	 	 	 	 	 	 	 	 	 	 	 	 	 	 	 	& 									& 						& 			& 		& Taper	& -65 \pm 77	& -0.84\\*
 	 	 	 	 	 	 	 	 	 	 	 	 	 	 	 	 	 	& 									& 						& \multirow{2}{*}{$\mu$}	& \multirow{2}{*}{<3.07}	& Natural	& 121 \pm 34	& 3.56\\*
 	 	 	 	 	 	 	 	 	 	 	 	 	 	 	 	 	 	& 									& 						& 			& 		& Taper	& 118 \pm 56	& 2.11\\*
\hline
\multirow{24}{*}{$8.5 \leq \log{(M_{\star} / M_{\odot})} < 9.0$}	&\multirow{8}{*}{$z < 4.0$}	& \multirow{8}{*}{145}	& \multirow{2}{*}{$\mathrm{equal}$}	& \multirow{2}{*}{<2.16}	& Natural	& -2 \pm 6	& -0.33\\*
 	 	 	 	 	 	 	 	 	 	 	 	 	 	 	 	 	 	& 									& 						& 			& 		& Taper	& 2 \pm 8	& 0.25\\*
 	 	 	 	 	 	 	 	 	 	 	 	 	 	 	 	 	 	& 									& 						& \multirow{2}{*}{$pbcor$}	& \multirow{2}{*}{<2.10}	& Natural	& -2 \pm 6	& -0.33\\*
 	 	 	 	 	 	 	 	 	 	 	 	 	 	 	 	 	 	& 									& 						& 			& 		& Taper	& 0 \pm 9	& 0.00\\*
 	 	 	 	 	 	 	 	 	 	 	 	 	 	 	 	 	 	& 									& 						& \multirow{2}{*}{F$_{UV}$}	& \multirow{2}{*}{<1.68}	& Natural	& 72 \pm 26	& 2.77\\*
 	 	 	 	 	 	 	 	 	 	 	 	 	 	 	 	 	 	& 									& 						& 			& 		& Taper	& 50 \pm 40	& 1.25\\*
 	 	 	 	 	 	 	 	 	 	 	 	 	 	 	 	 	 	& 									& 						& \multirow{2}{*}{$\mu$}	& \multirow{2}{*}{<1.91}	& Natural	& 10 \pm 10	& 1.00\\*
 	 	 	 	 	 	 	 	 	 	 	 	 	 	 	 	 	 	& 									& 						& 			& 		& Taper	& 11 \pm 16	& 0.69\\*
\cline{2-8}
 	 	 	 	 	 	 	 	 	 	 	 	 	 	 	 	 	 	&\multirow{8}{*}{$4.0 \leq z < 7.0$}	& \multirow{8}{*}{43}	& \multirow{2}{*}{$\mathrm{equal}$}	& \multirow{2}{*}{<2.34}	& Natural	& 6 \pm 11	& 0.55\\*
 	 	 	 	 	 	 	 	 	 	 	 	 	 	 	 	 	 	& 									& 						& 			& 		& Taper	& 0 \pm 16	& 0.00\\*
 	 	 	 	 	 	 	 	 	 	 	 	 	 	 	 	 	 	& 									& 						& \multirow{2}{*}{$pbcor$}	& \multirow{2}{*}{<2.35}	& Natural	& 9 \pm 11	& 0.82\\*
 	 	 	 	 	 	 	 	 	 	 	 	 	 	 	 	 	 	& 									& 						& 			& 		& Taper	& 1 \pm 16	& 0.06\\*
 	 	 	 	 	 	 	 	 	 	 	 	 	 	 	 	 	 	& 									& 						& \multirow{2}{*}{F$_{UV}$}	& \multirow{2}{*}{<1.76}	& Natural	& 27 \pm 25	& 1.08\\*
 	 	 	 	 	 	 	 	 	 	 	 	 	 	 	 	 	 	& 									& 						& 			& 		& Taper	& -2 \pm 36	& -0.06\\*
 	 	 	 	 	 	 	 	 	 	 	 	 	 	 	 	 	 	& 									& 						& \multirow{2}{*}{$\mu$}	& \multirow{2}{*}{<2.29}	& Natural	& 4 \pm 19	& 0.21\\*
 	 	 	 	 	 	 	 	 	 	 	 	 	 	 	 	 	 	& 									& 						& 			& 		& Taper	& -28 \pm 27	& -1.04\\*
\cline{2-8}
 	 	 	 	 	 	 	 	 	 	 	 	 	 	 	 	 	 	&\multirow{8}{*}{$7.0 \leq z$}	& \multirow{8}{*}{4}	& \multirow{2}{*}{$\mathrm{equal}$}	& \multirow{2}{*}{<2.36}	& Natural	& 38 \pm 32	& 1.19\\*
 	 	 	 	 	 	 	 	 	 	 	 	 	 	 	 	 	 	& 									& 						& 			& 		& Taper	& 49 \pm 44	& 1.11\\*
 	 	 	 	 	 	 	 	 	 	 	 	 	 	 	 	 	 	& 									& 						& \multirow{2}{*}{$pbcor$}	& \multirow{2}{*}{<2.33}	& Natural	& 38 \pm 33	& 1.15\\*
 	 	 	 	 	 	 	 	 	 	 	 	 	 	 	 	 	 	& 									& 						& 			& 		& Taper	& 57 \pm 46	& 1.24\\*
 	 	 	 	 	 	 	 	 	 	 	 	 	 	 	 	 	 	& 									& 						& \multirow{2}{*}{F$_{UV}$}	& \multirow{2}{*}{<1.68}	& Natural	& 48 \pm 54	& 0.89\\*
 	 	 	 	 	 	 	 	 	 	 	 	 	 	 	 	 	 	& 									& 						& 			& 		& Taper	& 58 \pm 89	& 0.65\\*
 	 	 	 	 	 	 	 	 	 	 	 	 	 	 	 	 	 	& 									& 						& \multirow{2}{*}{$\mu$}	& \multirow{2}{*}{<2.56}	& Natural	& 66 \pm 44	& 1.50\\*
 	 	 	 	 	 	 	 	 	 	 	 	 	 	 	 	 	 	& 									& 						& 			& 		& Taper	& 80 \pm 59	& 1.36\\*
\hline
\multirow{24}{*}{$9.0 \leq \log{(M_{\star} / M_{\odot})} < 9.5$}	&\multirow{8}{*}{$z < 4.0$}	& \multirow{8}{*}{61}	& \multirow{2}{*}{$\mathrm{equal}$}	& \multirow{2}{*}{<1.83}	& Natural	& 24 \pm 9	& 2.67\\*
 	 	 	 	 	 	 	 	 	 	 	 	 	 	 	 	 	 	& 									& 						& 			& 		& Taper	& 26 \pm 13	& 2.00\\*
 	 	 	 	 	 	 	 	 	 	 	 	 	 	 	 	 	 	& 									& 						& \multirow{2}{*}{$pbcor$}	& \multirow{2}{*}{<1.80}	& Natural	& 25 \pm 10	& 2.50\\*
 	 	 	 	 	 	 	 	 	 	 	 	 	 	 	 	 	 	& 									& 						& 			& 		& Taper	& 30 \pm 14	& 2.14\\*
 	 	 	 	 	 	 	 	 	 	 	 	 	 	 	 	 	 	& 									& 						& \multirow{2}{*}{F$_{UV}$}	& \multirow{2}{*}{<1.70}	& Natural	& 33 \pm 15	& 2.20\\*
 	 	 	 	 	 	 	 	 	 	 	 	 	 	 	 	 	 	& 									& 						& 			& 		& Taper	& 27 \pm 20	& 1.35\\*
 	 	 	 	 	 	 	 	 	 	 	 	 	 	 	 	 	 	& 									& 						& \multirow{2}{*}{$\mu$}	& \multirow{2}{*}{<1.76}	& Natural	& 37 \pm 15	& 2.47\\*
 	 	 	 	 	 	 	 	 	 	 	 	 	 	 	 	 	 	& 									& 						& 			& 		& Taper	& 51 \pm 21	& 2.43\\*
\cline{2-8}
 	 	 	 	 	 	 	 	 	 	 	 	 	 	 	 	 	 	&\multirow{8}{*}{$4.0 \leq z < 7.0$}	& \multirow{8}{*}{19}	& \multirow{2}{*}{$\mathrm{equal}$}	& \multirow{2}{*}{<2.03}	& Natural	& 22 \pm 19	& 1.16\\*
 	 	 	 	 	 	 	 	 	 	 	 	 	 	 	 	 	 	& 									& 						& 			& 		& Taper	& 11 \pm 25	& 0.44\\*
 	 	 	 	 	 	 	 	 	 	 	 	 	 	 	 	 	 	& 									& 						& \multirow{2}{*}{$pbcor$}	& \multirow{2}{*}{<2.04}	& Natural	& 25 \pm 20	& 1.25\\*
 	 	 	 	 	 	 	 	 	 	 	 	 	 	 	 	 	 	& 									& 						& 			& 		& Taper	& 10 \pm 26	& 0.38\\*
 	 	 	 	 	 	 	 	 	 	 	 	 	 	 	 	 	 	& 									& 						& \multirow{2}{*}{F$_{UV}$}	& \multirow{2}{*}{<1.94}	& Natural	& 208 \pm 53	& 3.92\\*
 	 	 	 	 	 	 	 	 	 	 	 	 	 	 	 	 	 	& 									& 						& 			& 		& Taper	& 204 \pm 67	& 3.04\\*
 	 	 	 	 	 	 	 	 	 	 	 	 	 	 	 	 	 	& 									& 						& \multirow{2}{*}{$\mu$}	& \multirow{2}{*}{<2.02}	& Natural	& 24 \pm 23	& 1.04\\*
 	 	 	 	 	 	 	 	 	 	 	 	 	 	 	 	 	 	& 									& 						& 			& 		& Taper	& 11 \pm 30	& 0.37\\*
\cline{2-8}
 	 	 	 	 	 	 	 	 	 	 	 	 	 	 	 	 	 	&\multirow{8}{*}{$7.0 \leq z$\tablefootmark{g}}	& \multirow{8}{*}{1}	& \multirow{2}{*}{$\mathrm{equal}$}	& \multirow{2}{*}{$\cdots$}	& Natural	& -37 \pm 94	& -0.39\\*
 	 	 	 	 	 	 	 	 	 	 	 	 	 	 	 	 	 	& 									& 						& 			& 		& Taper	& 10 \pm 126	& 0.08\\*
 	 	 	 	 	 	 	 	 	 	 	 	 	 	 	 	 	 	& 									& 						& \multirow{2}{*}{$pbcor$}	& \multirow{2}{*}{$\cdots$}	& Natural	& -37 \pm 94	& -0.39\\*
 	 	 	 	 	 	 	 	 	 	 	 	 	 	 	 	 	 	& 									& 						& 			& 		& Taper	& 10 \pm 126	& 0.08\\*
 	 	 	 	 	 	 	 	 	 	 	 	 	 	 	 	 	 	& 									& 						& \multirow{2}{*}{F$_{UV}$}	& \multirow{2}{*}{$\cdots$}	& Natural	& -37 \pm 94	& -0.39\\*
 	 	 	 	 	 	 	 	 	 	 	 	 	 	 	 	 	 	& 									& 						& 			& 		& Taper	& 10 \pm 126	& 0.08\\*
 	 	 	 	 	 	 	 	 	 	 	 	 	 	 	 	 	 	& 									& 						& \multirow{2}{*}{$\mu$}	& \multirow{2}{*}{$\cdots$}	& Natural	& -37 \pm 94	& -0.39\\*
 	 	 	 	 	 	 	 	 	 	 	 	 	 	 	 	 	 	& 									& 						& 			& 		& Taper	& 10 \pm 126	& 0.08\\*
\hline
\end{tabular}
\end{adjustbox}%

&

\begin{adjustbox}{width=0.45\linewidth}
\begin{tabular}{@{}c@{\hspace{1.25\tabcolsep}}c@{\hspace{0.5\tabcolsep}}c@{\hspace{0.75\tabcolsep}}c@{\hspace{0.75\tabcolsep}}c@{\hspace{0.75\tabcolsep}}c@{\hspace{0.75\tabcolsep}}S[separate-uncertainty=true,table-figures-decimal=0,table-text-alignment=center] S@{}}
\hline\hline
Stellar Mass	& $z_{ph}$	& Sources \#	& Weight\tablefootmark{a}	& $\log{(\mathrm{IRX})}$\tablefootmark{d}	& CLEAN\tablefootmark{c}	& {Flux\tablefootmark{d, e}}	& S/N\tablefootmark{f}\\
			 					& 		  		& 		   						& 	   						& 	    					& 	    					& {[$\mu$Jy]} 		&  \\
\hline

\multirow{24}{*}{$9.5 \leq \log{(M_{\star} / M_{\odot})} < 10.0$}	&\multirow{8}{*}{$z < 4.0$}	& \multirow{8}{*}{11}	& \multirow{2}{*}{$\mathrm{equal}$}	& \multirow{2}{*}{<2.21}	& Natural	& 62 \pm 23	& 2.70\\*
 	 	 	 	 	 	 	 	 	 	 	 	 	 	 	 	 	 	& 									& 						& 			& 		& Taper	& 77 \pm 31	& 2.48\\*
 	 	 	 	 	 	 	 	 	 	 	 	 	 	 	 	 	 	& 									& 						& \multirow{2}{*}{$pbcor$}	& \multirow{2}{*}{<2.18}	& Natural	& 60 \pm 23	& 2.61\\*
 	 	 	 	 	 	 	 	 	 	 	 	 	 	 	 	 	 	& 									& 						& 			& 		& Taper	& 68 \pm 32	& 2.12\\*
 	 	 	 	 	 	 	 	 	 	 	 	 	 	 	 	 	 	& 									& 						& \multirow{2}{*}{F$_{UV}$}	& \multirow{2}{*}{<1.74}	& Natural	& 33 \pm 38	& 0.87\\*
 	 	 	 	 	 	 	 	 	 	 	 	 	 	 	 	 	 	& 									& 						& 			& 		& Taper	& 6 \pm 50	& 0.12\\*
 	 	 	 	 	 	 	 	 	 	 	 	 	 	 	 	 	 	& 									& 						& \multirow{2}{*}{$\mu$}	& \multirow{2}{*}{<2.12}	& Natural	& 60 \pm 24	& 2.50\\*
 	 	 	 	 	 	 	 	 	 	 	 	 	 	 	 	 	 	& 									& 						& 			& 		& Taper	& 65 \pm 33	& 1.97\\*
\cline{2-8}
 	 	 	 	 	 	 	 	 	 	 	 	 	 	 	 	 	 	&\multirow{8}{*}{$4.0 \leq z < 7.0$}	& \multirow{8}{*}{6}	& \multirow{2}{*}{$\mathrm{equal}$}	& \multirow{2}{*}{<1.81}	& Natural	& 27 \pm 30	& 0.90\\*
 	 	 	 	 	 	 	 	 	 	 	 	 	 	 	 	 	 	& 									& 						& 			& 		& Taper	& -14 \pm 45	& -0.31\\*
 	 	 	 	 	 	 	 	 	 	 	 	 	 	 	 	 	 	& 									& 						& \multirow{2}{*}{$pbcor$}	& \multirow{2}{*}{<1.51}	& Natural	& -2 \pm 34	& -0.06\\*
 	 	 	 	 	 	 	 	 	 	 	 	 	 	 	 	 	 	& 									& 						& 			& 		& Taper	& -30 \pm 52	& -0.58\\*
 	 	 	 	 	 	 	 	 	 	 	 	 	 	 	 	 	 	& 									& 						& \multirow{2}{*}{F$_{UV}$}	& \multirow{2}{*}{<1.82}	& Natural	& 21 \pm 49	& 0.43\\*
 	 	 	 	 	 	 	 	 	 	 	 	 	 	 	 	 	 	& 									& 						& 			& 		& Taper	& -4 \pm 67	& -0.06\\*
 	 	 	 	 	 	 	 	 	 	 	 	 	 	 	 	 	 	& 									& 						& \multirow{2}{*}{$\mu$}	& \multirow{2}{*}{<1.37}	& Natural	& 26 \pm 49	& 0.53\\*
 	 	 	 	 	 	 	 	 	 	 	 	 	 	 	 	 	 	& 									& 						& 			& 		& Taper	& 14 \pm 80	& 0.18\\*
\cline{2-8}
 	 	 	 	 	 	 	 	 	 	 	 	 	 	 	 	 	 	&\multirow{8}{*}{$7.0 \leq z$}	& \multirow{8}{*}{1}	& \multirow{2}{*}{$\mathrm{equal}$}	& \multirow{2}{*}{<1.81}	& Natural	& 71 \pm 68	& 1.04\\*
 	 	 	 	 	 	 	 	 	 	 	 	 	 	 	 	 	 	& 									& 						& 			& 		& Taper	& -23 \pm 117	& -0.20\\*
 	 	 	 	 	 	 	 	 	 	 	 	 	 	 	 	 	 	& 									& 						& \multirow{2}{*}{$pbcor$}	& \multirow{2}{*}{<1.81}	& Natural	& 71 \pm 68	& 1.04\\*
 	 	 	 	 	 	 	 	 	 	 	 	 	 	 	 	 	 	& 									& 						& 			& 		& Taper	& -23 \pm 117	& -0.20\\*
 	 	 	 	 	 	 	 	 	 	 	 	 	 	 	 	 	 	& 									& 						& \multirow{2}{*}{F$_{UV}$}	& \multirow{2}{*}{<1.81}	& Natural	& 71 \pm 68	& 1.04\\*
 	 	 	 	 	 	 	 	 	 	 	 	 	 	 	 	 	 	& 									& 						& 			& 		& Taper	& -23 \pm 117	& -0.20\\*
 	 	 	 	 	 	 	 	 	 	 	 	 	 	 	 	 	 	& 									& 						& \multirow{2}{*}{$\mu$}	& \multirow{2}{*}{<1.81}	& Natural	& 71 \pm 68	& 1.04\\*
 	 	 	 	 	 	 	 	 	 	 	 	 	 	 	 	 	 	& 									& 						& 			& 		& Taper	& -23 \pm 117	& -0.20\\*
\hline
\multirow{24}{*}{$\log{(M_{\star} / M_{\odot})} \geq 10.0$}	&\multirow{8}{*}{$z < 4.0$\tablefootmark{g}}	& \multirow{8}{*}{1}	& \multirow{2}{*}{$\mathrm{equal}$}	& \multirow{2}{*}{$\cdots$}	& Natural	& 201 \pm 678	& 0.30\\*
 	 	 	 	 	 	 	 	 	 	 	 	 	 	 	 	 	 	& 									& 						& 			& 		& Taper	& -354 \pm 808	& -0.44\\*
 	 	 	 	 	 	 	 	 	 	 	 	 	 	 	 	 	 	& 									& 						& \multirow{2}{*}{$pbcor$}	& \multirow{2}{*}{$\cdots$}	& Natural	& 201 \pm 678	& 0.30\\*
 	 	 	 	 	 	 	 	 	 	 	 	 	 	 	 	 	 	& 									& 						& 			& 		& Taper	& -354 \pm 808	& -0.44\\*
 	 	 	 	 	 	 	 	 	 	 	 	 	 	 	 	 	 	& 									& 						& \multirow{2}{*}{F$_{UV}$}	& \multirow{2}{*}{$\cdots$}	& Natural	& 201 \pm 678	& 0.30\\*
 	 	 	 	 	 	 	 	 	 	 	 	 	 	 	 	 	 	& 									& 						& 			& 		& Taper	& -354 \pm 808	& -0.44\\*
 	 	 	 	 	 	 	 	 	 	 	 	 	 	 	 	 	 	& 									& 						& \multirow{2}{*}{$\mu$}	& \multirow{2}{*}{$\cdots$}	& Natural	& 201 \pm 678	& 0.30\\*
 	 	 	 	 	 	 	 	 	 	 	 	 	 	 	 	 	 	& 									& 						& 			& 		& Taper	& -354 \pm 808	& -0.44\\*
\cline{2-8}
 	 	 	 	 	 	 	 	 	 	 	 	 	 	 	 	 	 	&\multirow{8}{*}{$4.0 \leq z < 7.0$}	& \multirow{8}{*}{1}	& \multirow{2}{*}{$\mathrm{equal}$}	& \multirow{2}{*}{<1.66}	& Natural	& 172 \pm 119	& 1.45\\*
 	 	 	 	 	 	 	 	 	 	 	 	 	 	 	 	 	 	& 									& 						& 			& 		& Taper	& 277 \pm 147	& 1.88\\*
 	 	 	 	 	 	 	 	 	 	 	 	 	 	 	 	 	 	& 									& 						& \multirow{2}{*}{$pbcor$}	& \multirow{2}{*}{<1.66}	& Natural	& 172 \pm 119	& 1.45\\*
 	 	 	 	 	 	 	 	 	 	 	 	 	 	 	 	 	 	& 									& 						& 			& 		& Taper	& 277 \pm 147	& 1.88\\*
 	 	 	 	 	 	 	 	 	 	 	 	 	 	 	 	 	 	& 									& 						& \multirow{2}{*}{F$_{UV}$}	& \multirow{2}{*}{<1.66}	& Natural	& 172 \pm 119	& 1.45\\*
 	 	 	 	 	 	 	 	 	 	 	 	 	 	 	 	 	 	& 									& 						& 			& 		& Taper	& 277 \pm 147	& 1.88\\*
 	 	 	 	 	 	 	 	 	 	 	 	 	 	 	 	 	 	& 									& 						& \multirow{2}{*}{$\mu$}	& \multirow{2}{*}{<1.66}	& Natural	& 172 \pm 119	& 1.45\\*
 	 	 	 	 	 	 	 	 	 	 	 	 	 	 	 	 	 	& 									& 						& 			& 		& Taper	& 277 \pm 147	& 1.88\\*
\cline{2-8}
 	 	 	 	 	 	 	 	 	 	 	 	 	 	 	 	 	 	&\multirow{8}{*}{$7.0 \leq z$}	& \multirow{8}{*}{1}	& \multirow{2}{*}{$\mathrm{equal}$}	& \multirow{2}{*}{<1.71}	& Natural	& 17 \pm 68	& 0.25\\*
 	 	 	 	 	 	 	 	 	 	 	 	 	 	 	 	 	 	& 									& 						& 			& 		& Taper	& 14 \pm 118	& 0.12\\*
 	 	 	 	 	 	 	 	 	 	 	 	 	 	 	 	 	 	& 									& 						& \multirow{2}{*}{$pbcor$}	& \multirow{2}{*}{<1.71}	& Natural	& 17 \pm 68	& 0.25\\*
 	 	 	 	 	 	 	 	 	 	 	 	 	 	 	 	 	 	& 									& 						& 			& 		& Taper	& 14 \pm 118	& 0.12\\*
 	 	 	 	 	 	 	 	 	 	 	 	 	 	 	 	 	 	& 									& 						& \multirow{2}{*}{F$_{UV}$}	& \multirow{2}{*}{<1.71}	& Natural	& 17 \pm 68	& 0.25\\*
 	 	 	 	 	 	 	 	 	 	 	 	 	 	 	 	 	 	& 									& 						& 			& 		& Taper	& 14 \pm 118	& 0.12\\*
 	 	 	 	 	 	 	 	 	 	 	 	 	 	 	 	 	 	& 									& 						& \multirow{2}{*}{$\mu$}	& \multirow{2}{*}{<1.71}	& Natural	& 17 \pm 68	& 0.25\\*
 	 	 	 	 	 	 	 	 	 	 	 	 	 	 	 	 	 	& 									& 						& 			& 		& Taper	& 14 \pm 118	& 0.12\\*
\hline
\end{tabular}
\end{adjustbox}
\end{tabular}
\tablefoot{
\tablefoottext{a}{Weight scheme associated to each candidate, as explained in $\S$\ref{subsec:Stacking}.}
\tablefoottext{b}{$3{-}\sigma$ upper limits of weighted average for each bin (Eq.~\ref{eq:stack_irx_avg}).}
\tablefoottext{c}{\textit{CLEAN}ing method used in \texttt{CASA} to obtain final image.}
\tablefoottext{d}{Maximum value from a 0\farcs5$\times$0\farcs5 box in the stacked images.}
\tablefoottext{e}{$rms$ errors from Eq. \ref{eq:rms_detect}.}
\tablefoottext{f}{S/N $=$ Flux$_{\mathrm{peak}}$ / $rms$}
\tablefoottext{g}{Since IRX values are obtained from ALMA images and stacked fluxes from visibilities, we do not have coverage in the image but we are able to $u$--$v$ stack that position.}
}
\end{table*}


\begin{table*}
\caption{Properties of low stellar mass [$\log{(M_{\bigstar} / M_{\odot})} \leq 6.0$] stacked LBG candidates}\label{tab:ALMA_props_low_mass}
\centering
\begin{tabular}{c c}
\begin{adjustbox}{width=0.45\linewidth}
\begin{tabular}{@{}c@{\hspace{0.5\tabcolsep}}c@{\hspace{0.75\tabcolsep}}c@{\hspace{0.75\tabcolsep}}c@{\hspace{0.75\tabcolsep}}c@{\hspace{0.75\tabcolsep}}S[separate-uncertainty=true,table-figures-decimal=0,table-text-alignment=center] S@{}}     
\hline\hline
$z_{ph}$							& Sources \#\tablefootmark{a}	& Weight\tablefootmark{b}	& $\log{(\mathrm{IRX})}$\tablefootmark{c}   & CLEAN\tablefootmark{d}	& {Flux\tablefootmark{e, f}}	& S/N\tablefootmark{g}\\
	  								& 		   						& 	   						& 	    					                & 	    					& {[$\mu$Jy]} 				&  \\
\hline
\multirow{8}{*}{$z < 4.0$}	& \multirow{8}{*}{8}	& \multirow{2}{*}{$\mathrm{equal}$}	& \multirow{2}{*}{$<3.12$}	& Natural	& -16 \pm 27	& -0.59\\*
 					& 			& 						& 		& Taper	& -41 \pm 36	& -1.14\\*
 					& 			& \multirow{2}{*}{$pbcor$}	& \multirow{2}{*}{$<3.11$}	& Natural	& -16 \pm 27	& -0.59\\*
 					& 			& 						& 		& Taper	& -41 \pm 36	& -1.14\\*
 					& 			& \multirow{2}{*}{F$_{UV}$}	& \multirow{2}{*}{$<2.94$}	& Natural	& -25 \pm 34	& -0.74\\*
 					& 			& 						& 		& Taper	& -40 \pm 44	& -0.91\\*
 					& 			& \multirow{2}{*}{$\mu$}	& \multirow{2}{*}{$<3.19$}	& Natural	& -22 \pm 30	& -0.73\\*
 					& 			& 						& 		& Taper	& -45 \pm 38	& -1.18\\*
\hline
\multirow{8}{*}{$4.0 \leq z < 7.0$}	& \multirow{8}{*}{3}	& \multirow{2}{*}{$\mathrm{equal}$}	& \multirow{2}{*}{$<2.87$}	& Natural	& 5 \pm 43	& 0.12\\*
 					& 			& 						& 		& Taper	& -30 \pm 67	& -0.45\\*
 					& 			& \multirow{2}{*}{$pbcor$}	& \multirow{2}{*}{$<2.87$}	& Natural	& 5 \pm 43	& 0.12\\*
 					& 			& 						& 		& Taper	& -30 \pm 67	& -0.45\\*
 					& 			& \multirow{2}{*}{F$_{UV}$}	& \multirow{2}{*}{$<2.87$}	& Natural	& -3 \pm 44	& -0.07\\*
 					& 			& 						& 		& Taper	& -23 \pm 68	& -0.34\\*
 					& 			& \multirow{2}{*}{$\mu$}	& \multirow{2}{*}{$<2.87$}	& Natural	& 5 \pm 43	& 0.12\\*
 					& 			& 						& 		& Taper	& -30 \pm 67	& -0.45\\*
\hline
\end{tabular}
\end{adjustbox}%

&

\begin{adjustbox}{width=0.45\linewidth}
\begin{tabular}{@{}c@{\hspace{0.5\tabcolsep}}c@{\hspace{0.75\tabcolsep}}c@{\hspace{0.75\tabcolsep}}c@{\hspace{0.75\tabcolsep}}c@{\hspace{0.75\tabcolsep}}S[separate-uncertainty=true,table-figures-decimal=0,table-text-alignment=center] S@{}}     
\hline\hline
$z_{ph}$							& Sources \#\tablefootmark{a}	& Weight\tablefootmark{b}	& $\log{(\mathrm{IRX})}$\tablefootmark{c}   & CLEAN\tablefootmark{d}	& {Flux\tablefootmark{e, f}}	& S/N\tablefootmark{g}\\
	  								& 		   						& 	   						& 	    					                & 	    					& {[$\mu$Jy]} 				&  \\
\hline
\multirow{8}{*}{$7.0 \leq z$}	& \multirow{8}{*}{0}	& \multirow{2}{*}{$\mathrm{equal}$}	& \multirow{2}{*}{$\cdots$}	& Natural	& $\cdots$ 	& $\cdots$ \\*
 					& 			& 						& 		& Taper	& $\cdots$ 	& $\cdots$ \\*
 					& 			& \multirow{2}{*}{$pbcor$}	& \multirow{2}{*}{$\cdots$}	& Natural	& $\cdots$ 	& $\cdots$ \\*
 					& 			& 						& 		& Taper	& $\cdots$ 	& $\cdots$ \\*
 					& 			& \multirow{2}{*}{F$_{UV}$}	& \multirow{2}{*}{$\cdots$}	& Natural	& $\cdots$ 	& $\cdots$ \\*
 					& 			& 						& 		& Taper	& $\cdots$ 	& $\cdots$ \\*
 					& 			& \multirow{2}{*}{$\mu$}	& \multirow{2}{*}{$\cdots$}	& Natural	& $\cdots$ 	& $\cdots$ \\*
 					& 			& 						& 		& Taper	& $\cdots$ 	& $\cdots$ \\*
\hline
\end{tabular}
\end{adjustbox}
\end{tabular}
\tablefoot{
\tablefoottext{a}{When Source \# is zero (0), no stacking was performed.}
\tablefoottext{b}{Weight scheme associated to each candidate, as explained in $\S$\ref{subsec:Stacking}.}
\tablefoottext{c}{$3{-}\sigma$ upper limits of weighted average for each bin (Eq.~\ref{eq:stack_irx_avg}).}
\tablefoottext{d}{\textit{CLEAN}ing method used in \texttt{CASA} to obtain final image.}
\tablefoottext{e}{Maximum value from a 0\farcs5$\times$0\farcs5 box in the stacked images.}
\tablefoottext{f}{$rms$ errors from Eq. \ref{eq:rms_detect}.}
\tablefoottext{g}{S/N $=$ Flux$_{\mathrm{peak}}$ / $rms$.\\}
}
\end{table*}

\section{Individual properties}\label{tab:indiv_props}

	The individual properties of the first $10$ of the $1582$ selected LBG 
    candidates are presented in Tables~\ref{tab:individual_lbg_obs} ({\it HST} photometry), 
    \ref{tab:ALMA_props} (ALMA properties), \ref{tab:model_props} (derived properties), \ref{tab:fast_props} (\texttt{FAST++} derived properties) and 
    \ref{tab:magphys_props} (observed luminosities and flux densities).
    Properties for the remaining targets are available online.

\begin{table*}
\caption{Demagnified {\it HST} photometry. Only the first 10 selected LBG candidates are shown; the full table is available online.}\label{tab:individual_lbg_obs}
\centering
\resizebox{\textwidth}{!}{\begin{tabular}{@{}c@{\hspace{1\tabcolsep}}c@{\hspace{1\tabcolsep}}c@{\hspace{1\tabcolsep}}c@{\hspace{1.25\tabcolsep}}c@{\hspace{1.25\tabcolsep}}c@{\hspace{1.25\tabcolsep}}c@{\hspace{1.25\tabcolsep}}c@{\hspace{1.25\tabcolsep}}c@{\hspace{1.25\tabcolsep}}c@{\hspace{1.25\tabcolsep}}c@{\hspace{1.25\tabcolsep}}c@{\hspace{1.25\tabcolsep}}c@{}}
\hline
$\mathrm{ID}$ & $\mathrm{R.A}$ [J2000] & $\mathrm{Dec}$ [J2000] & $z_{\mathrm{ph}}$ & {$F\mathrm{_{F275W}}$} & {$F\mathrm{_{F336W}}$} & {$F\mathrm{_{F435W}}$} & {$F\mathrm{_{F606W}}$} & {$F\mathrm{_{F814W}}$} & {$F\mathrm{_{F105W}}$} & {$F\mathrm{_{F125W}}$} & {$F\mathrm{_{F140W}}$} & {$F\mathrm{_{F160W}}$}\\
 & [hh:mm:ss.ss] & [$\pm$ dd:mm:ss.ss] &   & {[uJy]} & {[uJy]} & {[uJy]} & {[uJy]} & {[uJy]} & {[uJy]} & {[uJy]} & {[uJy]} & {[uJy]}\\
\hline
\noalign{\smallskip}
0001 & 00:14:23.61 & -30:24:53.28 & $2.321^{+0.12}_{-0.07}$ & $  0.1^{+11.800}_{-11.800}$ & $ 6.4^{+7.201}_{-7.200}$   & $ 8.8^{+2.605}_{-2.602}$ & $13.9^{+3.011}_{-3.004}$ & $12.4^{+2.013}_{-2.005}$ & $ 12.5^{+3.408}_{-3.403}$ & $ 19.0^{+4.913}_{-4.905}$ & $ 21.7^{+4.817}_{-4.806}$ & $ 26.3^{+4.725}_{-4.710}$\\
\noalign{\smallskip}
0002 & 00:14:24.58 & -30:24:48.97 & $2.291^{+0.12}_{-0.09}$ & $  0.9^{+12.800}_{-12.800}$ & $10.3^{+7.802}_{-7.801}$   & $16.7^{+3.312}_{-3.307}$ & $17.1^{+3.611}_{-3.607}$ & $14.6^{+2.313}_{-2.308}$ & $ 12.0^{+3.107}_{-3.104}$ & $ 18.1^{+4.810}_{-4.806}$ & $ 17.1^{+4.709}_{-4.705}$ & $ 22.1^{+4.615}_{-4.609}$\\
\noalign{\smallskip}
0004 & 00:14:22.41 & -30:24:47.81 & $2.426^{+0.07}_{-0.10}$ & $  2.9^{+9.300}_{-9.300}$   & $ 7.5^{+5.501}_{-5.502}$   & $12.5^{+1.907}_{-1.918}$ & $12.4^{+2.306}_{-2.315}$ & $10.5^{+1.507}_{-1.516}$ & $  8.3^{+2.403}_{-2.406}$ & $  9.3^{+3.702}_{-3.705}$ & $  7.4^{+3.601}_{-3.603}$ & $ 10.7^{+3.703}_{-3.707}$\\
\noalign{\smallskip}
0005 & 00:14:23.99 & -30:24:34.50 & $1.924^{+0.09}_{-0.12}$ & $  2.9^{+13.201}_{-13.201}$ & $10.0^{+7.812}_{-7.821}$   & $20.3^{+3.226}_{-3.313}$ & $24.1^{+3.561}_{-3.672}$ & $25.0^{+2.549}_{-2.715}$ & $ 37.6^{+3.425}_{-3.701}$ & $ 48.9^{+4.988}_{-5.311}$ & $ 57.2^{+5.154}_{-5.577}$ & $ 61.1^{+5.155}_{-5.635}$\\
\noalign{\smallskip}
0007 & 00:14:23.04 & -30:24:24.85 & $2.194^{+0.10}_{-0.08}$ & $ -0.6^{+7.500}_{-7.500}$   & $ 4.0^{+4.301}_{-4.301}$   & $ 4.5^{+2.303}_{-2.303}$ & $ 5.8^{+2.604}_{-2.605}$ & $ 4.9^{+1.705}_{-1.705}$ & $  4.6^{+1.804}_{-1.804}$ & $  5.6^{+2.804}_{-2.804}$ & $  9.1^{+2.710}_{-2.711}$ & $  6.6^{+2.606}_{-2.606}$\\
\noalign{\smallskip}
0008 & 00:14:22.67 & -30:24:23.45 & $2.041^{+0.51}_{-0.14}$ & $ -1.0^{+11.300}_{-11.300}$ & $ 3.7^{+7.001}_{-7.005}$   & $10.4^{+3.314}_{-3.383}$ & $ 8.4^{+3.908}_{-3.947}$ & $ 4.7^{+2.704}_{-2.721}$ & $  5.9^{+2.406}_{-2.437}$ & $  7.4^{+3.507}_{-3.540}$ & $  4.9^{+3.503}_{-3.517}$ & $  4.2^{+3.402}_{-3.413}$\\
\noalign{\smallskip}
0010 & 00:14:18.53 & -30:24:54.25 & $2.184^{+0.19}_{-0.09}$ & $  0.9^{+10.900}_{-10.900}$ & $11.1^{+6.501}_{-6.500}$   & $20.3^{+1.908}_{-1.903}$ & $14.9^{+2.304}_{-2.301}$ & $13.3^{+1.604}_{-1.602}$ & $ 17.6^{+3.703}_{-3.701}$ & $ 18.2^{+5.602}_{-5.601}$ & $ 28.7^{+5.606}_{-5.602}$ & $ 31.6^{+5.407}_{-5.403}$\\
\noalign{\smallskip}
0011 & 00:14:19.29 & -30:24:53.86 & $2.367^{+0.10}_{-0.11}$ & $  1.7^{+12.800}_{-12.800}$ & $13.9^{+7.400}_{-7.401}$   & $28.3^{+2.205}_{-2.208}$ & $30.3^{+2.705}_{-2.708}$ & $29.1^{+1.906}_{-1.910}$ & $ 25.4^{+2.803}_{-2.805}$ & $ 27.7^{+4.103}_{-4.104}$ & $ 31.1^{+4.103}_{-4.105}$ & $ 32.1^{+4.004}_{-4.006}$\\
\noalign{\smallskip}
0012 & 00:14:19.11 & -30:24:51.09 & $1.565^{+0.06}_{-0.09}$ & $ 21.2^{+23.701}_{-23.701}$ & $58.9^{+14.307}_{-14.317}$ & $78.2^{+4.442}_{-4.499}$ & $82.0^{+5.339}_{-5.390}$ & $99.5^{+3.683}_{-3.792}$ & $192.5^{+4.840}_{-5.146}$ & $234.4^{+7.623}_{-7.915}$ & $243.0^{+6.868}_{-7.214}$ & $265.5^{+6.823}_{-7.236}$\\
\noalign{\smallskip}
0013 & 00:14:18.87 & -30:24:50.40 & $2.545^{+0.13}_{-0.13}$ & $-10.3^{+10.600}_{-10.600}$ & $ 6.3^{+6.200}_{-6.200}$   & $16.3^{+1.902}_{-1.903}$ & $15.4^{+2.202}_{-2.202}$ & $11.5^{+1.501}_{-1.502}$ & $ 10.1^{+2.701}_{-2.701}$ & $  7.6^{+4.500}_{-4.500}$ & $  9.5^{+4.200}_{-4.200}$ & $  7.0^{+4.200}_{-4.200}$\\
\noalign{\smallskip}
\hline
\end{tabular}}
\tablefoot{
\tablefoottext{*}{Error values for fluxes derived from errors in photometry and magnfication factors combined.}
}
\end{table*}

\begin{table*}
\caption{ALMA properties. Only the first 10 selected LBG candidates are shown; the full table is available online.}\label{tab:ALMA_props}
\centering          
\resizebox{0.8\textwidth}{!}{\begin{tabular}{@{}c@{\hspace{1.25\tabcolsep}}c@{\hspace{1.25\tabcolsep}}c@{\hspace{1.25\tabcolsep}}c@{\hspace{1.25\tabcolsep}}S[table-figures-decimal=0,separate-uncertainty=true,table-text-alignment=center]@{\hspace{1.25\tabcolsep}}S[table-figures-decimal=0]@{\hspace{1.25\tabcolsep}}S@{\hspace{1.25\tabcolsep}}S@{}}     
\hline
 ID & $\mathrm{R.A}$ [J2000]\tablefootmark{a} & $\mathrm{Dec}$ [J2000] & Cluster & $F\mathrm{_{ALMA,peak,pbcor}^{indiv,obs}}$\tablefootmark{b,c} &   $F_{\mathrm{ALMA,peak,pbcor}}^{\mathrm{indiv,obs,3{-}\sigma lim}}$\tablefootmark{d} & S/N\tablefootmark{e} & pbcor\tablefootmark{f}\\
 & [hh:mm:ss.ss] & [$\pm$ dd:mm:ss.ss] &  & [uJy] & [uJy] & & \\
 \hline
0001 & 00:14:23.60 & -30:24:53.10 & A2744 &       63 \pm 72 &      277 & 0.87 & 0.77\\
0002 & 00:14:24.56 & -30:24:49.20 & A2744 &       81 \pm 65 &      277 & 1.25 & 0.85\\
0004 & 00:14:22.41 & -30:24:48.00 & A2744 &      166 \pm 58 &      340 & 2.87 & 0.95\\
0005 & 00:14:23.98 & -30:24:34.30 & A2744 &       48 \pm 56 &      216 & 0.85 & 0.98\\
0007 & 00:14:23.04 & -30:24:25.00 & A2744 &       57 \pm 56 &      226 & 1.01 & 0.97\\
0008 & 00:14:22.66 & -30:24:23.50 & A2744 &       76 \pm 56 &      245 & 1.35 & 0.97\\
0010 & 00:14:18.53 & -30:24:54.30 & A2744 &      105 \pm 57 &      277 & 1.83 & 0.96\\
0011 & 00:14:19.28 & -30:24:53.80 & A2744 &      117 \pm 58 &      290 & 2.01 & 0.95\\
0012 & 00:14:19.10 & -30:24:51.30 & A2744 &        3 \pm 56 &      172 & 0.05 & 0.97\\
0013 & 00:14:18.87 & -30:24:50.60 & A2744 &       65 \pm 56 &      234 & 1.16 & 0.98\\
\hline                  
\end{tabular}}
\tablefoot{
\tablefoottext{a}{Position of the ALMA peak of each LBG candidate}
\tablefoottext{b}{Calculated from Eq. \ref{eq:peak_flux_detect}.}
\tablefoottext{c}{Error represented by $rms_{\mathrm{ALMA,pbcor}}^{\mathrm{indiv}}$ and obtained with Eq. \ref{eq:rms_detect}.}
\tablefoottext{d}{$3{-}\sigma$ upper limits for the ALMA observed fluxes (Eq. \ref{eq:ALMA_flux_nsigma}).}
\tablefoottext{e}{S/N = $F\mathrm{_{ALMA,peak,pbcor}^{indiv,obs}} / rms_{\mathrm{ALMA,pbcor}}^{\mathrm{indiv}}$}
\tablefoottext{f}{Primary beam correction for each position in \texttt{.flux} files.}
}
\end{table*}

\begin{table*}
\caption{Derived properties. Only the first 10 selected LBG candidates are shown; the full table is available online.}\label{tab:model_props}
\centering          
\resizebox{0.39\textwidth}{!}{\begin{tabular}{@{}c c c c@{}}     
\hline
ID & {$\beta$\tablefootmark{a}} & $\mu$\tablefootmark{b,c,d} & D$_{l}$\tablefootmark{e}\\
   &          &        & [Mpc]  \\
\hline
\noalign{\smallskip}
0001 & $-1.74^{+0.23}_{-0.22}$ & $3.49^{+0.06}_{-0.04} \pm 0.14$ & $18630^{+ 1230}_{-  700}$\\
\noalign{\smallskip}
0002 & $-2.34^{+0.19}_{-0.19}$ & $3.28^{+0.06}_{-0.04} \pm 0.24$ & $18340^{+ 1180}_{-  850}$\\
\noalign{\smallskip}
0004 & $-2.38^{+0.12}_{-0.12}$ & $4.92^{+0.07}_{-0.10} \pm 0.19$ & $19660^{+  650}_{-  970}$\\
\noalign{\smallskip}
0005 & $-1.29^{+0.06}_{-0.06}$ & $6.44^{+0.28}_{-0.37} \pm 0.26$ & $14820^{+  880}_{- 1130}$\\
\noalign{\smallskip}
0007 & $-2.07^{+0.27}_{-0.26}$ & $4.75^{+0.12}_{-0.12} \pm 0.25$ & $17400^{+  920}_{-  790}$\\
\noalign{\smallskip}
0008 & $-2.73^{+0.33}_{-0.34}$ & $2.93^{+0.09}_{-0.21} \pm 0.49$ & $15930^{+ 4970}_{- 1300}$\\
\noalign{\smallskip}
0010 & $-2.40^{+0.21}_{-0.21}$ & $1.82^{+0.02}_{-0.01} \pm 0.18$ & $17300^{+ 1800}_{-  900}$\\
\noalign{\smallskip}
0011 & $-2.07^{+0.13}_{-0.13}$ & $2.08^{+0.01}_{-0.01} \pm 0.16$ & $19080^{+ 1000}_{- 1050}$\\
\noalign{\smallskip}
0012 & $-0.83^{+0.07}_{-0.07}$ & $1.92^{+0.02}_{-0.02} \pm 0.19$ & $11490^{+  600}_{-  810}$\\
\noalign{\smallskip}
0013 & $-2.56^{+0.23}_{-0.23}$ & $2.00^{+0.01}_{-0.01} \pm 0.18$ & $20840^{+ 1270}_{- 1290}$\\
\noalign{\smallskip}
\hline                  
\end{tabular}}
\tablefoot{
\tablefoottext{a}{UV-slope calculated as stated in $\S$\ref{subsec:UV-slope}.}
\tablefoottext{b}{Magnification factors following \citet{2015ApJ...800...84C} without capping.}
\tablefoottext{c}{Magnification error values obtained from evaluating the limits of the $1{-}\sigma$ confidence intervals of the $z_{\mathrm{ph}}$ values.}
\tablefoottext{d}{Second error value corresponds to systematic uncertainties from different magnification models ($\S$\ref{subsec:mufactor}).}
\tablefoottext{e}{Luminosity distances after \citet{2006PASP..118.1711W}.}
}
\end{table*}

\begin{table*}
\caption{Properties obtained from \texttt{FAST++}. Only the first 10 selected LBG candidates are shown; the full table is available online.}\label{tab:fast_props}
\centering          
\resizebox{0.41\textwidth}{!}{\begin{tabular}{@{}c@{\hspace{1.25\tabcolsep}}c@{\hspace{1.25\tabcolsep}}c@{\hspace{1.25\tabcolsep}}c@{}}     
\hline
ID & $\log_{10}{(M_{\bigstar} / M_{\odot})}$ & $\log{(\mathrm{SFR} / \mathrm{M_{\odot} yr^{-1}})}$ & $\log{(\mathrm{sSFR} / \mathrm{ yr^{-1}})}$\\
   &                                            &                                                         &   \\
\hline
\noalign{\smallskip}
0001 & $ 7.93^{+0.48}_{-0.45}$ & $-0.47^{+0.49}_{-0.49}$ & $-8.36^{+0.19}_{-0.36}$\\
\noalign{\smallskip}
0002 & $ 7.56^{+0.46}_{-0.45}$ & $-0.80^{+0.44}_{-0.44}$ & $-8.36^{+0.28}_{-0.18}$\\
\noalign{\smallskip}
0004 & $ 6.79^{+0.49}_{-0.45}$ & $-0.68^{+0.47}_{-0.47}$ & $-7.44^{+0.28}_{-0.46}$\\
\noalign{\smallskip}
0005 & $ 8.15^{+0.46}_{-0.42}$ & $-0.02^{+0.44}_{-0.44}$ & $-8.17^{+0.18}_{-0.19}$\\
\noalign{\smallskip}
0007 & $ 7.12^{+0.47}_{-0.46}$ & $-1.11^{+0.48}_{-0.48}$ & $-8.26^{+0.36}_{-0.37}$\\
\noalign{\smallskip}
0008 & $ 6.25^{+0.56}_{-0.42}$ & $-0.82^{+0.46}_{-0.46}$ & $-6.98^{+0.00}_{-0.74}$\\
\noalign{\smallskip}
0010 & $ 8.07^{+0.48}_{-0.49}$ & $-0.91^{+0.44}_{-0.44}$ & $-9.00^{+0.18}_{-0.18}$\\
\noalign{\smallskip}
0011 & $ 7.44^{+0.49}_{-0.45}$ & $ 0.05^{+0.54}_{-0.54}$ & $-7.35^{+0.28}_{-0.55}$\\
\noalign{\smallskip}
0012 & $ 8.74^{+0.44}_{-0.44}$ & $ 0.17^{+0.46}_{-0.46}$ & $-8.54^{+0.28}_{-0.28}$\\
\noalign{\smallskip}
0013 & $ 6.56^{+0.48}_{-0.44}$ & $-0.54^{+0.45}_{-0.45}$ & $-6.98^{+0.00}_{-0.37}$\\
\noalign{\smallskip}
\hline                  
\end{tabular}}
\tablefoot{
\tablefoottext{*}{Magnification-corrected values from \texttt{FAST++}.\\}
}
\end{table*}

\begin{table*}
\caption{Luminosities from \textit{HST} photometry and modified blackbody (graybody) spectrum for first 10 selected LBG candidates. Full table is available online}\label{tab:magphys_props}
\centering
\resizebox{0.65\textwidth}{!}{\begin{tabular}{@{}c@{\hspace{1.25\tabcolsep}}c@{\hspace{1.25\tabcolsep}}c@{\hspace{1.25\tabcolsep}}c@{\hspace{1.25\tabcolsep}}c@{\hspace{1.25\tabcolsep}}c@{}} 
\hline
ID & {$\log{(\mathrm{L_{UV}} / \mathrm{L_{\odot}})}$} & $\log{(\mathrm{L_{IR}}^{3-\sigma} / \mathrm{L_{\odot}})}$ & $\log{(\mathrm{L_{IR}}^{3-\sigma} / \mathrm{L_{UV}})}$ & {$\log{(\mathrm{F_{UV}} / \mu \mathrm{Jy})}$} & $\log{(\mathrm{F_{IR}}^{3-\sigma} / \mu \mathrm{Jy})}$\\
 & & & & & \\
\hline
\noalign{\smallskip}
0001 & $ 8.96^{+0.32}_{-0.29}$ &  11.81 &  2.91 & $-2.40^{+0.10}_{-0.10}$ &  2.00\\
\noalign{\smallskip}
0002 & $ 9.04^{+0.32}_{-0.29}$ &  11.83 &  2.89 & $-2.28^{+0.09}_{-0.09}$ &  2.08\\
\noalign{\smallskip}
0004 & $ 8.94^{+0.32}_{-0.30}$ &  11.74 &  3.02 & $-2.60^{+0.09}_{-0.09}$ &  2.15\\
\noalign{\smallskip}
0005 & $ 9.05^{+0.30}_{-0.27}$ &  11.45 &  2.45 & $-2.43^{+0.17}_{-0.14}$ &  1.63\\
\noalign{\smallskip}
0007 & $ 8.54^{+0.32}_{-0.29}$ &  11.59 &  3.12 & $-2.91^{+0.20}_{-0.20}$ &  1.80\\
\noalign{\smallskip}
0008 & $ 8.65^{+0.37}_{-0.28}$ &  11.85 &  3.34 & $-2.54^{+0.22}_{-0.20}$ &  2.10\\
\noalign{\smallskip}
0010 & $ 8.94^{+0.33}_{-0.29}$ &  12.09 &  3.31 & $-2.09^{+0.07}_{-0.07}$ &  2.41\\
\noalign{\smallskip}
0011 & $ 9.31^{+0.32}_{-0.29}$ &  12.05 &  2.90 & $-1.84^{+0.04}_{-0.04}$ &  2.39\\
\noalign{\smallskip}
0012 & $ 9.40^{+0.28}_{-0.25}$ &  11.90 &  2.50 & $-1.39^{+0.03}_{-0.03}$ &  1.95\\
\noalign{\smallskip}
0013 & $ 9.08^{+0.33}_{-0.30}$ &  11.96 &  2.98 & $-2.11^{+0.06}_{-0.06}$ &  2.21\\
\noalign{\smallskip}
\hline
\end{tabular}}
\end{table*}

\end{appendix}


\end{document}